\documentclass[12pt]{article}
\usepackage{heck}
\usepackage{graphicx}
\usepackage[cmtip,arrow]{xy}
\usepackage{pb-diagram,pb-xy}
\input xy
\xyoption{all}
\input xy
\xyoption{all}
\usepackage{pifont}
\usepackage{float}
\usepackage{subfig}
\usepackage{multicol}
\usepackage{amsfonts}
\usepackage{amssymb}
\usepackage{mathrsfs}
\usepackage{amsmath}
\usepackage{amsfonts}
\usepackage{color}
\usepackage{hyperref}
\hypersetup{colorlinks=true}
\hypersetup{linkcolor=black}
\hypersetup{citecolor=black}
\hypersetup{urlcolor=black}
\usepackage{setspace}
\usepackage{wrapfig}
\usepackage{verbatim}
\numberwithin{equation}{section}
\providecommand{\U}[1]{\protect\rule{.1in}{.1in}}
\newcommand{\bea}{\begin{eqnarray}}
\newcommand{\eea}{\end{eqnarray}}
\newcommand{\be}{\begin{equation}}
\newcommand{\ee}{\end{equation}}

\newcommand{\bem}{\begin{pmatrix}}
\newcommand{\eem}{\end{pmatrix}}
\def\U{\Upsilon}

\def\ch{{\cal H}}

\def\cn{{\cal N}}

\def\cq{{\cal Q}}

\def \N {{\mathcal N}}
\def \Z {{\mathbb Z}}

\xyoption{arc}

\begin{document}    
\date{October, 2011}

\institution{SISSA}{\centerline{${}^{1}$Scuola Internazionale Superiore di Studi Avanzati, Via Bonomea 265 34100 Trieste, ITALY}}

\institution{HarvardU}{\centerline{${}^{2}$Jefferson Physical Laboratory, Harvard University, Cambridge, MA 02138, USA}}

\title{Braids, Walls, and Mirrors}

\authors{ Sergio Cecotti,\worksat{\SISSA}\footnote{e-mail: {\tt cecotti@sissa.it}} Clay C\'{o}rdova,\worksat{\HarvardU}\footnote{e-mail: {\tt cordova@physics.harvard.edu}}  and Cumrun Vafa\worksat{\HarvardU
}\footnote{e-mail: {\tt vafa@physics.harvard.edu}}}

\abstract{We construct 3d, ${\cal N}=2$ supersymmetric gauge theories by considering a one-parameter `R-flow' of 4d, ${\cal N}=2$ theories, where the central charges vary while preserving their phase order. Each BPS state in 4d leads to a BPS particle in 3d, and thus each chamber of the 4d theory leads to a distinct 3d theory.  Pairs of 4d chambers related by wall-crossing, R-flow to mirror pairs of 3d theories. In particular, the 2-3 wall-crossing for the $A_2$ Argyres-Douglas theory leads to 3d mirror symmetry for $N_f=1$ SQED and the XYZ model. Although our formalism applies to arbitrary ${\cal N}=2$ models, we focus on the case where the parent 4d theory consists of pairs of M5-branes wrapping a Riemann surface, and develop a general framework for describing 3d ${\cal N}=2$ theories engineered by wrapping pairs of M5-branes on three-manifolds. Each 4d chamber, which corresponds to a dual 3d description, maps to a particular tetrahedral decomposition of the UV 3d geometry.  In the IR the physics is captured by a single recombined M5-brane which is a branched double cover of the original UV three-manifold. The braiding of branch loci and the geometry of branch sheets play a key role in encoding the physics. }

\maketitle

\enlargethispage{\baselineskip}

\setcounter{tocdepth}{3}
\begin{spacing}{.9}
\tableofcontents
\end{spacing}
\section{Introduction}
The study of supersymmetric theories in various dimensions has revealed the importance of BPS states in probing the theory.  This includes the characterization of such states by their charges, degeneracies, and interactions.  In some cases the BPS states, which are often viewed as composites of the more elementary fields, can become light and constitute the fundamental constituents of the theory. A well known example of this occurs for ${\cal N} =4$ supersymmetric Yang-Mills in four dimensions:  for sufficiently strong coupling, the magnetic BPS states become light and lead to a dual magnetic description of the theory. ${\cal N}=2$ supersymmetric theories in four dimensions also enjoy a rich BPS structure, and in fact it appears that in many ways the BPS data completely characterizes the theory \cite{CV11}.   The picture that emerges is that for each set of BPS charges and degeneracies, there is at most one consistent $\mathcal{N}=2$ theory. Thus, the BPS structure, which can roughly be viewed as an IR data, appears powerful enough to reconstruct the full UV description of the theory.  Moreover, powerful techniques are now available for finding the BPS spectra in a wide class of 4d  $\mathcal{N}=2$ models \cite{GMN1, GMN2, GMN3, ACCERV1, ACCERV2}.  However, despite these developments, there is no simple explicit map from the four-dimensional BPS data to the UV description of the field theory.  In particular, as the BPS states carry both electric and magnetic charges, there is in general no local Lagrangian description of their interactions.

The story may be simpler for ${\cal N}=2$ theories in three dimensions.  These theories are close cousins of ${\cal N}=2$ theories in four dimensions, but they have half as much supersymmetry. They enjoy a real central charge in the BPS algebra.  Moreover they have the advantage, compared to their 4d cousin, that even abelian theories are UV-complete.  Thus, one can in principle hope that given an effective Lagrangian description of all the BPS states, the same Lagrangian may describe the theory in the UV.

In this paper, we provide a link between the study of 4d BPS states and 3d BPS states by constructing 3d $\mathcal{N}=2$ theories from parent 4d $\mathcal{N}=2$ theories.  One natural way to carry out such a reduction, is to consider a one-parameter family of 4d models, parameterized by an extra circle where as one goes around the circle one identifies the two sides up to some symmetry transformation. In \cite{Gaiottowall, yamawall}, the corresponding symmetry was an element of the S-duality group of the ${\cal N}=2$ theory.  Another choice, studied in \cite{Cecotti:2009uf,CNV} was to use the R-symmetry of conformal ${\cal N}=2$ theories to reduce the theory.  Our construction of 3d theories from 4d theories is close in spirit to \cite{Cecotti:2009uf,CNV}, except that the circle is replaced by real line with suitable boundary conditions at infinity.

In our description, the 3d theory will appear as a 3d domain wall inside a 4d theory.  This wall is characterized by a one-parameter flow of the 4d BPS central charges $Z_{i}$.  As we traverse the thickness of the 3d wall from one side to the other, the $Z_{i}$ vary along parallel lines while preserving their phase order, and the boundary conditions of the flow are such that asymptotically all central charges become infinitely large.  As a consequence of these boundary conditions, all degrees of freedom of the 4d bulk theory, except the massless $U(1)$ gauge multiples,  become infinitely heavy and decouple from the 3d wall theory.  However, the BPS particles of the 4d theory have finite mass on the wall and are trapped there.  Thus, the result of this construction is a 3d theory with $\mathcal{N}=2$ supersymmetry whose BPS states are inherited from the parent 4d and are potentially gauged under the $U(1)$ symmetries of the bulk.   We call this flow of the 4d theory the `R-flow'  due to the fact that at the two boundaries the central charges have flipped sign and hence have undergone an R-symmetry rotation,  $Z_i\rightarrow e^{i\theta}Z_i$,  by $\theta=\pi$.   The most important feature of this wall is that, because the R-flow respects the phase order of the 4d central charges, each BPS state of a given 4d chamber will give rise to a trapped particle on the 3d domain wall.

Our reduction of a given 4d theory to a 3d theory is not unique, as one could in principle start the R-flow from different chambers in 4d related by BPS wall-crossing.  This results in a different set of trapped modes on the 3d wall and hence determines a correspondence between 3d theories constructed by R-flow from a parent 4d model, and the BPS chambers of the parent.  We thus have the analog of induced wall-crossing in three dimensions, and as we will see, this 3d wall-crossing phenomenon can be interpreted as mirror symmetry.  As a result, we find a set of 3d dual theories which are labeled by chambers of the parent 4d theory.  And further, the 4d Kontsevich-Soibelman wall-crossing formula enforces partition function equality of these 3d dual theories.  For example, the simplest non-trivial ${\cal N}=2$ superconformal theory is the $A_2$ Argyres-Douglas theory \cite{AD}.  In this case we have two or three 4d BPS states depending on the choice of chamber, and under R-flow, these lead to two dual theories in 3d, known as $N_f=1$ SQED and the XYZ model \cite{XYZ}.  

In the process of reduction of 4d theories to 3d, supersymmetry demands that we vary all the central charges along parallel lines. However, this is not generally possible for arbitrary 4d theories, as the space of allowed central charges is a subspace of all allowed complex numbers.  An exception is the case of `complete' ${\cal N}=2$ theories, which are characterized by the property that all their central charges can be varied arbitrarily \cite{CV11}.  Thus, the reduction of complete ${\cal N}=2$ theories from 4d to 3d will constitute the main example of this paper.  For other theories we can vary all the central charges and hence carry out the R-flow only if we give up the notion of their UV completion, and view the corresponding 4d theory as an effective theory.\footnote{\ For example, a pure $SU(N)$ theory in 4d has $2(N-1)$ central charges, but only $N-1$ Coulomb parameters and 1 coupling constant.  On the other hand we can add $N-2$ additional coupling terms to the action of the form
\begin{equation*}
\int d^4\theta\, \lambda_i\, \mathrm{Tr}(\Phi^i)
\end{equation*}
for $i=3,...N$, where $\Phi$ denotes the vector superfield.  Such couplings $\lambda_i$ can be generated in principle by integrating out other degrees of freedom from a more fundamental UV complete theory.}  At any rate, our main examples in this paper will be limited to complete theories.

As shown in \cite{CV11}, with the exception of eleven cases, all complete theories (which have BPS quivers) come from two M5-branes wrapping a punctured Riemann surface as studied for example in \cite{GMN1, GMN2, GMN3, WittenM, Gaiottodual}. Thus, their reduction to 3d will correspond to two M5-branes wrapping a one-parameter family of Riemann surfaces. In other words, it is a 3d theory determined to two M5-branes wrapping a 3d geometry $M$. For this class of theories we make contact with the recent work \cite{DGG}.   The BPS data of the 4d parent theory is governed by a triangulation of the associated Riemann surface.  During the R-flow, this triangulation evolves by a sequence of flips each of which corresponds to a 4d BPS state.  Remarkably, exactly the sequence of flips prescribed by R-flow determines a decomposition of the 3d geometry $M$ into tetrahedra, where each tetrahedron is in direct correspondence with a 3d BPS particle.  This picture leads to an explanation and extension of the rules proposed in \cite{DGG}.\footnote{ For instance we find that some, but not all superpotential terms arise from tetrahedra sharing an edge.}  Each 4d chamber, which corresponds to a dual 3d description, maps to a particular tetrahedral decomposition $M$, and 4d wall-crossings, reinterpreted as 3d mirror symmetries, manifest themselves as changes in the number of tetrahedra. 

The 3d geometry described by $M$, together with its decomposition into tetrahedra, encodes the physics of two M5-branes and hence can be viewed as a non-abelian UV data of the theory.  However, in the IR, this non-abelian structure is higgsed to an abelian one.  As a result, the physics is captured by the geometry of a \emph{single} recombined M5-brane $\widetilde{M}$ which is a double cover of $M$ branched along a knot.   The IR geometry $\widetilde{M}$ is the direct 3d analog of the Seiberg-Witten curve for 4d $\mathcal{N}=2$ theories and we develop its properties in detail.  We find that the R-flow of the parent 4d theory naturally determines a braid diagram presentation of the branching knot.  The geometry of this branching braid is the key to deciphering the 3d theory.  Each intersection of the branching braids describes a massless 3d particle.  Giving the particle mass resolves the intersection and, in simple cases, determines a correspondence between general 3d particles and braid moves.  Further, the geometry of the braid also encodes the existence of certain superpotentials.  These superpotentials are generated by M2-brane instantons ending on the M5-brane, and are seen as primitive polygons in the braid diagram.

We illustrate these ideas in the context of the $ADE$ Argyres-Douglas theories.  For example, for $A_n$ theories, there are various chambers ranging from $n$ particles to $n(n+1)/2$ particles.  This in turn translates to a UV 3d geometry with a minumum number of $n$ tetrahedra and a maximum of $n(n+1)/2$ tetrahedra.  In the IR this same theory is described by a branching braid on $n+1$ strands, with particles described by braid moves, and generically cubic and quartic superpotentials.  For the R-flow of the $E$-case, as we will demonstrate, we can still obtain the resulting 3d theory.  However, since these do not correspond to multiple M5-branes, the corresponding 3d theories are not captured by a 3d geometry.

Perhaps the most exciting new 3d theories correspond to the case where the bulk 4d theory has infinitely many BPS particles. This for example, happens for the weak coupling phase of pure $SU(2)$. In this case there are infinitely many dyonic BPS states.  However, unlike the 4d case where the dyons have unbounded masses, their reduction to 3d can lead to nearly equal and finite mass for the trapped dyons. Moreover the 4d vector W-bosons can also be trapped on the 3d wall.  In this way it appears that all of the infinitely many BPS states of the 4d theory assemble themselves into a representation of $SL(2,\mathbb{R})$, and it is natural to conjecture that the trapped W-bosons mean that the $SL(2,\mathbb{R})$ symmetry is gauged.  What is remarkable, is that this theory also has a strongly coupled phase with only two particles which should describe its 3d dual.  In terms of the 3d geometry, this phenomenon corresponds to situations where, as the hyperbolic structure is varied, the manifold $M$ goes from having a finite ideal tetrahedralization, to a decomposition into infinitely many accumulating tetrahedra.   It appears that similar phenomena have been studied in the math literature \cite{hodgson, hodgson2}.

The organization of this paper is as follows:  In section 2 we study the geometry of M5-branes wrapping special Lagrangian cycles of Calabi-Yau threefolds, leading to ${\cal N}=2$ theories in their three uncompactified directions.  We explore the emergence of the 3d recombined M5-brane geometry which encodes the 3d gauge theory, as a direct 3d analog of Seiberg-Witten geometry for ${\cal N}=2$ theories in 4d.   We focus on the main example of the paper which involves two M5-branes.  In this case we explain how the geometry of the branching knot encodes the 3d physics. In particular we show how the Seifert surface of the knot encodes the description of the $U(1)^k$ Chern-Simons gauge theories, with the Seifert matrix giving the matrix of the Chern-Simons levels.  In section 3 we introduce the main notion of R-flow and explain our reduction of $\mathcal{N}=2$ theories in 4d to $\mathcal{N}=2$ theories in 3d.  In section 4 we recall some basic facts about 4d BPS states, and how quivers and their mutations encode them. In section 5 we provide some concrete examples, and study the R-flow of 4d $A_n$ Argyres-Douglas theories and their resulting braids.  In particular, we explain how the $A_1$ theory (a 4d free hypermultiplet) maps to two M5-branes wrapping the tetrahedron geometry and show how the double cover of the tetrahedron is a special Lagrangian lens space in $\mathbb{C}^3$.  We also show how the chambers of the $A_2$ theory map to the $XYZ$ model and $N_{f}=1$ SQED, and explain how the 4d wall-crossing leads to 3d mirror symmetry.  We also discuss some aspects of other $A_n$ theories and show that they correspond to UV 3d geometries comprised of $n$-piramids (in the minimal chamber). In section 6 we discuss the case where we have infinitely many particles corresponding to weak coupling limit of $SU(2)$.  In section 7 we discuss the partition function of the resulting theories compactified on 3d Melvin cigar and relate it to the partition function on the squashed $S^3$.  In section 8 we compute the partition function for the R-flow of $ADE$ Argyres-Douglas theories and confirm the geometric predictions for the resulting 3d theory for the $A_n$ case.

\section{Five-Branes on Three-Manifolds}
\label{sec:M5}
One purpose of this paper is to describe a class of three-dimensional $\mathcal{N}=2$ quantum field theories which can be engineered by wrapping M5-branes on three-manifolds which we generically denote by $M$.  In later sections of the paper our primary applications will be to the case involving two five-branes though the geometry described in this section applies more generally.

\subsection{Three-Dimensional $\mathcal{N}=2$ Gauge Theories}
\label{sec:3dfield}
Let us begin by recalling the basic parameters and properties of the field theories in question \cite{XYZ}.  We will be focused on describing the degrees of freedom in the infrared on the Coulomb branch where all non-abelian gauge symmetries have been higgsed to a product of $U(1)$ factors.  The data of such a field theory is then:
\begin{itemize}
\item A gauge group $U(1)^{N}$. 
\item A flavor group $U(1)^{F}$ with an associated real mass parameter $m_{i}$ for each $U(1)$ factor.
\item A symmetric matrix of $k_{ij}$ of Chern-Simons terms.  
\item A spectrum of charged chiral matter multiplets $X_{i}$.
\item A superpotential, $\mathcal{W}$, a holomorphic function of chiral fields.
\end{itemize}
An important fact is that in three dimensions, abelian gauge fields with field strength $F$ are dual to scalars $\gamma$ via the relation
\begin{equation}
* F =d\gamma. \label{dual}
\end{equation}
Charge quantization means that $\gamma$ is periodic, and in simple cases the resulting theory after duality enjoys a flavor $U(1)$ which acts on the dual photon $\gamma$ as a shift.  Under this duality, the real mass parameter $m$ of the dual flavor symmetry can be interpreted as the real FI parameter $\zeta$ of the original gauge group.  However, in general it is not always true that shifts of the dual photon appear as flavor symmetries of the theory.  If $\sigma$ denotes expectation value of the real scalar in the $U(1)$ gauge multiplet, then after duality the monopole operator
\begin{equation}
\mathcal{M}=\exp(\sigma + i \gamma)
\end{equation}
is a chiral field which carries charge under the candidate $U(1)$ flavor symmetry.  In particular, if say $\mathcal{M}$ appears in the superpotential, then the flavor symmetry will be broken and correspondingly there is no real mass parameter, or equivalently no FI-term for the original gauge theory.

Next we consider the central charge of 3d $\mathcal{N}=2$ theories.  Just as in four-dimensional $\mathcal{N}=2$ theories, the superalgebra admits the appearance of a central term $Z$ which sets the BPS bound for the masses of particles carrying $U(1)$ charges.  However, unlike the situation in four dimensions where $Z$ is complex, in three dimensions the central charge is real.  If $q_{j}$ and $f_{i}$ denote gauge and flavor charges respectively, then the total central charge of a particle is 
\begin{equation}  
Z(q,f)=\sum_{j}q_{j}\sigma_{j}+\sum_{i}f_{i}m_{i}. \label{zdef}
\end{equation}
Where in \eqref{zdef} we have implicitly included FI terms as real masses to dual flavor groups.  Then, as stated above, charged particles satisfy a bound on their mass
\begin{equation}
m\geq|Z|.
\end{equation}
Charged BPS states saturate the above and, in the simplest case of minimal spin, form chiral multiplets.

Finally, we take a moment to discuss Chern-Simons terms.  In general, we study theories involving fermions and thus $\mathbb{R}^{1,2}$ (or any other manifold on which we study a three-dimensional field theory) is equipped with a choice of spin structure.  In this situation, the correct quantization condition for the level matrix is half-integral units, $k_{ij}\in \frac{1}{2}\mathbb{Z}$ \cite{DWl}.  For convenience, we therefore introduce the notation $\hat{k}_{ij}\equiv 2k_{ij}$.  Then, $\hat{k}_{ij}$ is integrally quantized.  Concretely, given a collection of $U(1)$ gauge fields $A_{i}$ with canonically normalized kinetic terms, $\hat{k}_{ij}$ appears in the action as
\begin{equation}
\sum_{ij}\frac{\hat{k}_{ij}}{4\pi}\int A_{i} \wedge dA_{j}.
\end{equation}
From now on we will always work with the quantity $\hat{k}_{ij},$ and refer to it as the level.

We also note that in three dimensions, CS levels receive anomalous contributions from integrating massive fermions at one loop.  Specifically, if $(q_{F})_{i}$ denotes the vector of $U(1)$ gauge charges of a chiral fermion $F$ with mass $m_{F}$ then the effective levels are related to the bare ones as
\begin{equation}
(\hat{k}_{ij})_{\mathrm{eff}}=(\hat{k}_{ij})_{\mathrm{bare}}+\sum_{F}(q_{F})_{i}(q_{F})_{j}\mathrm{sign}(m_{F}). \label{csanomaly}
\end{equation}
For answering questions about the physics in the extreme IR it is the effective levels which are the relevant ones.  Indeed, assuming that all matter fields are massive, they may be integrated out leaving a pure Yang-Mills-CS theory with level matrix $(\hat{k}_{ij})_{\mathrm{eff}}$.  However, from the right-hand-side of $\eqref{csanomaly}$ we can see that the effective levels depend on the masses of fields which in turn depend on the parameters and moduli $(m_{i},\zeta_{j}, \sigma_{k})$.  By contrast the bare CS terms are a globally well-defined property of a theory.  Thus, in the following, when we compute CS terms we will always have in mind the bare contribution.  The effective levels can then be determined from a knowledge of the spectrum and an application of \eqref{csanomaly}.
\subsection{One Five-Brane}
\label{sec:OM5}
Now we study a class of three-dimensional $\mathcal{N}=2$ field theories that can be constructed from M-theory.  We let $Q$ denote a Calabi-Yau threefold and consider M-theory on the spacetime 
\begin{equation}
\mathbb{R}^{1,4}\times Q.
\end{equation}  
We pick a linear subspace $\mathbb{R}^{1,2}\subset \mathbb{R}^{1,4}$ and a consider a three-manifold $M$ embedded inside $Q$ as a special Lagrangian.  We then consider the effective three-dimensional field theory determined by a single M5 brane on
\begin{equation}
\mathbb{R}^{1,2}\times M.
\end{equation}
In the field theory limit, which is all that is relevant for this paper, we are interested only in the local dynamics near $M$ inside the Calabi-Yau $Q$. Then, we may consider a scaling limit where $Q$ is taken to be non-compact and gravity is decoupled from the degrees of freedom determined by the five-brane.  By construction, the resulting field theory admits four supercharges and hence has $\mathcal{N}=2$ supersymmetry in the three-dimensional sense.  We will see that the structure of this field theory is intimately related to the geometry of $M$.

\subsubsection{Geometry of the Coulomb Branch}

A basic observation is that there are scalar degrees of freedom describing the small fluctuations of the special Lagrangian $M$ inside the local Calabi-Yau threefold $Q$.  To characterize these, we first note that near $M,$ $Q$ can be modeled by the contangent bundle $T^{*}M$, and hence to describe deformations it suffices to think of $M$ embedded inside its own cotangent bundle as the special Lagrangian zero section.  To be explicit, we may introduce a system of local coordinates $x_{i}$ on $M$.  Then, any one-form can be expressed locally as $y_{i}dx_{i},$  and hence the $y_{i}$ provide a natural set of coordinates on the cotangent directions to $M$.  In terms of these our starting point for studying deformations is therefore the special Lagrangian
\begin{equation}
M=\left\{(x,0)\in T^{*}M\right\}.
\end{equation}

Consider a deformation of $M$.  Since $M$ deforms in its cotangent bundle its local motion is described by activating a certain one-form $\lambda$.  In other words, $M$ has deformed to the locus of points $M'$ of the form
\begin{equation}
M'=\left\{(x,\lambda(x))\right\}\subset T^{*}M.
\end{equation}
To minimize the energy, the deformation $M'$ must also be special Lagrangian.  In the linear approximation, such deformations are canonically identified with the space of harmonic one-forms on $M$.  To see this we note that in terms of the local coordinates $(x,y)$ on $T^{*}M$ the symplectic form $\omega$ has the canonical expression
\begin{equation}
\omega=dy_{1}\wedge dx_{1}+dy_{2}\wedge dx_{2}+dy_{3}\wedge dx_{3}=d(y_{i}dx_{i}).
\end{equation}
Therefore on the deformed locus $M'$, the symplectic form restricts to
\begin{equation}
\omega|_{M'}=d(\lambda). \label{dlambda}
\end{equation}  
Since we wish $M'$ to be Lagrangian, the restriction of $\omega$ to $M'$ must vanish and hence $\lambda$ must be closed. 

We can perform a similar calculation with the local holomorphic three-form $\Omega$ on $T^{*}M$.  Restricted to the deformation locus $M'$ the imaginary part of $\Omega$ appears to first order in $\lambda$ as
\begin{equation}
\Im\left(\Omega|_{M'}\right)=d(*\lambda).
\end{equation}
To ensure that the deformation is special, the imaginary part of $\Omega$ must vanish when restricted to $M'$.  This implies that the one-form $\lambda$ is co-closed and hence harmonic on the original three-manifold $M$.

Thus, in the linear approximation, the classical moduli space of special Lagrangian deformations of the three-manifold $M$ can be identified with the vector space of harmonic one-forms which can in turn be identified with the cohomology group $H^{1}(M,\mathbb{R})$ via Hodge theory.  To generalize beyond the linear approximation we now invoke a theorem of Mclean \cite{Mclean} which ensures that every first order supersymmetric fluctuation of $M$ can in fact be integrated to a supersymmetric deformation of finite size.  Hence, the full non-linear classical moduli space of deformations of the special Lagrangian $M$ inside $Q$ can be identified with a manifold whose tangent space at $M$ is canonically the space of harmonic one-forms.

Now, supersymmetry dictates that all fields must appear in representations of the $\mathcal{N}=2$ superalegebra. In particular, this means that the real scalars we have found must in fact be paired with other bosons.  To find the remaining half of the bosonic fields, we recall that the five-brane theory supports a two-form field $B$ propagating on its worldvolume.  This field can be activated for zero cost in energy provided that the field strength is vanishing $dB=0$.  On the other hand, $B$ itself is only defined up to gauge transformations which shift its value by an exact two-form.  Hence, flat $B$ fields on $M$ yield a space of deformations of dimension $h^{2}(M,\mathbb{R})$.  To be completely precise we should also note that as a gauge field, $B$ is naturally a periodic variable and hence the correct cohomology measuring $B$ is valued in $\mathbb{R}/\mathbb{Z}$.  If we combine these scalars with those arising from fluctuations of $M$ we find that locally, the classical five-brane moduli space can be parameterized by
\begin{equation}
H^{1}(M,\mathbb{R}) \times H^{2}(M,\mathbb{R}/\mathbb{Z}).
\end{equation}
Three-dimensional Poincar\'{e} duality ensures that the two vector spaces introduced above are of equal dimension and implies that these scalars fill out $\mathcal{N}=2$ chiral multiplets.
 
To a low-energy three-dimensional observer in $\mathbb{R}^{1,2}$, the scalar degrees of freedom that we have identified have a natural interpretation in terms of the classical coordinates on the Coulomb branch of an effective $U(1)^{b_{1}(M)}$ gauge theory.  Indeed the one-form $\lambda$ is characterized by its periods and describes the expectation values of the real adjoint scalars $\sigma$ appearing in $\mathcal{N}=2$ vector multiplets.  Meanwhile the circle valued variables described by the periods of $B$ are the expectation values of dual photons $\gamma$.  If we introduce a basis of one-cycles $\alpha_{i}$ and a Poincar\'{e} dual basis of two-cycles $\beta_{j}$ then an explicit set of local coordinates along the moduli space is given by
\begin{equation}
\int_{\alpha_{i}}\lambda=\sigma_{i}, \hspace{.5in} \int_{\beta_{j}}B=\gamma_{j}.
\end{equation}
The fact that the moduli space can be coordinatized in terms of periods is the starting point for a kind of real special geometry which governs the classical effective action.\footnote{\ Indeed, in generalizing beyond the linear approximation, one finds a real prepotential $F(\sigma)$ characterized by the condition that $\frac{\partial{F}}{\partial \sigma_{i}}=\int_{\beta_{i}} *\lambda.$  In terms of $F$ the full non-linear metric on the classical moduli space is then \cite{Hitchin}
\begin{equation}
ds^{2}=\frac{\partial^{2}F}{\partial \sigma_{i}\partial\sigma_{j}}d\sigma_{i} \otimes d\sigma_{j}+\left(\frac{\partial^{2}F}{\partial \sigma_{i}\partial \sigma_{j}} \right)^{-1}d\gamma_{i}\otimes d\gamma_{j}. \nonumber
\end{equation}
}
This real special geometry is the three-dimensional counterpart to the holomorphic special geometry of four-dimensional $\mathcal{N}=2$ systems.  However unlike the situation there where non-renormalization theorems protect the form of the metric from quantum corrections, a three-dimensional $\mathcal{N}=2$ system has only four supercharges and hence the metric is subject to quantum corrections.  Nevertheless,  the observation that the central charges of particles can be characterized in terms of the periods of a one-form $\lambda$ will play a crucial role in the remainder of this paper.  These central charges are protected from quantum corrections involving chiral multiplets \cite{XYZ, SeibP}, and hence the periods of $\lambda$ will remain meaningful when we study the quantum behavior of the theory.

There is an important subtlety in the above description of the Coulomb branch which arises due to the fact that three-dimensional Yang-Mills theories admit Chern-Simons terms.  In the presence of a non-vanishing level $\hat{k}$ the equation of motion for a three-dimensional $U(1)$ gauge field with field strength $F$ and Yang-Mills coupling $e$ is modified to
\begin{equation}
\Delta F\sim(\hat{k}e^{2})^{2}F.
\end{equation}
This equation means that the propagating photon has been given a non-zero mass $\hat{k}e^{2}$.  In particular this implies that the expectation value of the dual photon $\gamma$ is frozen to zero.  By supersymmetry the same is in fact true for the adjoint scalar, $\sigma$.  In our geometric context this has the following significance.  The quantities $\gamma$ and $\sigma$ are measured by periods of the two form $B$ and the one-form $\lambda$ over a certain two-cycle $\beta$ and dual one-cycle $\alpha$.  If these periods are frozen to zero then at the level of cohomology valued in $\mathbb{R}$ the associated cycles cannot be detected, and hence the Betti numbers $b_{1}(M)$ and $b_{2}(M)$ have each been decreased by one unit.  

However, the cycle can still be detected by the more refined data of the integer valued homology.  In the presence of a non-zero level $\hat{k}$ for the $U(1),$ there are observables given by the holonomy of the gauge field along cycles $C$ in $\mathbb{R}^{1,2}$
\begin{equation}
\exp\left(i q \oint_{C} A\right). \label{wilsonlinedef}
\end{equation}
For such an operator, the charge $q$ is naturally valued in $\mathbb{Z}_{\hat{k}}$.  Indeed, given two such observables, the correlation function is \cite{Polyakov}
\begin{equation}
\left \langle  \exp\left(i q_{1} \oint_{C_{1}} A\right) \exp\left(i q_{2} \oint_{C_{2}} A\right)\right \rangle \sim \exp\left(\frac{2\pi i q_{1}q_{2}}{\hat{k}}[C_{1}, C_{2}]\right).
\end{equation}
Here the quantity $[C_{1}, C_{2}]$ denotes the integer valued linking number between the curves $C_{i}$.  From the form of this correlation function, we see that if $q$ vanishes mod $\hat{k}$ then the Wilson line \eqref{wilsonlinedef} has trivial correlation functions thus illustrating that $q$ is valued in $\mathbb{Z}_{\hat{k}}$.  

In our context, gauge charges for the theory are captured by $H_{1}(M,\mathbb{Z})$.  Then, if the CS level is $\hat k$ we see from the above discussion that we only expect mod $\hat{k}$ charges.  In other words the CS level is $\hat{k}$ if and only if $H_{1}(M,\mathbb{Z})=\mathbb{Z}_{\hat{k}}$.  This is our desired result: CS levels for the $U(1)$ gauge theory are encoded in geometry by torsion classes in $H_{1}(M,\mathbb{Z})$.  We can extend this observation to the case where we have many $U(1)$'s. In full generality, the relationship between the homology of $M$ and the gauge theory in $\mathbb{R}^{1,2}$ is as follows.  Let the gauge theory be that of $n$ $U(1)$ gauge fields with a level matrix $\hat{k}_{ij}$.  Then $H_{1}(M,\mathbb{Z})$ is generated by $n$ elements $\Gamma_{i}$ modulo relations defined by the image of $\hat{k}_{ij}$
\begin{equation}
H_{1}(M,\mathbb{Z})\cong \bigoplus_{i=1}^{n}\mathbb{Z}[\Gamma_{i}]\left / \left(\hat{k}_{ij}\Gamma_{j}=0\right)\right. . \label{levelgen}
\end{equation}
This equation has the key feature that $b_{1}(M)$ counts the number of zero-eigenvalues of $\hat{k}_{ij}$ and hence captures the number of propagating massless gauge fields.  The remaining non-degenerate part of $\hat{k}_{ij}$ encodes the torsion structure of the homology.  The fact that the charges of $U(1)^n$ Chern-Simons theories are captured by \eqref{levelgen} is well known (see \textit{e.g.}\! \cite{BSM, MooreCS}), and we return to concrete applications of this formula in our study of examples in section \ref{sec:TM5}.
  
Finally, to complete our geometric description of the massless sector of the Coulomb branch, we will now describe how to include FI terms and real masses into the description.  Both of these deformations are naturally associated with activating bulk moduli of the ambient Calabi-Yau $Q$.  In fact both arise from a variation in the K\"{a}hler class of $Q$.  To see this let us suppose that the symplectic form $\omega$ is varied to a new class
\begin{equation}
\omega\rightarrow \omega +\delta \omega.
\end{equation}
Then, since $\omega$ enters in determining the Lagrangian condition on submanifolds, the deformation above enters our description as a modification in the behavior of the one-form $\lambda$ as in \eqref{dlambda}
\begin{equation}
d\lambda =\delta \omega|_{M}\neq 0.
\end{equation}
The interpretation of the above modification depends on the behavior of $\delta \omega $ restricted to $M$.  Specifically, since $\delta \omega$ is closed its restriction to $M$ can be in general a sum of terms which are cohomologicaly trivial or non-trivial. We examine the effects of each of these:
\begin{itemize} 
\item  $\delta \omega$ restricts to $M$ to an exact form $d\eta$.  Then, the one-form $\lambda$ is modified to include a contribution from $\eta$
\begin{equation}
\lambda \rightarrow \lambda +\eta.
\end{equation}
Such a modification permits $\lambda$ to develop periods over contractible one-cycles in $M$ and it is these periods which are interpreted as real mass parameters.  They are well defined as a consequence of the fact that the symplectic form is closed.

To see the connection of the modulus $\delta \omega $ to a flavor symmetry, we note that this modulus is in the same $\mathcal{N}=2$ multiplet as the bulk $U(1)$ gauge field $A$ which descends from the reduction of the M-theory three-form $C$ as
\begin{equation}
C=\delta\omega \wedge A.
\end{equation}
From the point of view of the five-brane, the field $A$ is non-dynamical, and therefore fields charged under $A$ carry a flavor charge.  The expectation value of the scalar modulus $\delta \omega$ then determines the associated mass.   

To be precise, one should view the non-vanishing contribution to $d\lambda$ as being supported at infinity in the Calabi-Yau $Q$, and the real mass as a kind of residue.  This is analogous to how mass parameters appear in 4d $\mathcal{N}=2$ theories described by wrapping an M5-brane on a Riemann surface.  There is a one-form $\phi$ on the Riemann surface, the Seiberg-Witten differential, which characterizes the normal motion of the brane.  The embedding of the Riemann surface in $Q$ is non-compact and has ends which appear asymptotically as $\mathbb{R}\times S^{1}$.  The periods of $\phi$ over these asymptotic circles are then the mass parameters of the theory \cite{GMN1, GMN2, GMN3, WittenM, Gaiottodual, DGG}.  We can equivalently describe this feature by compactifying the Riemann surface, and allowing $\phi$ to have residues.  This means that $\phi$ is no longer closed as $d\phi \sim \delta(x)$.

Similarly in our three-dimensional context, the embedding of $M$ in $Q$ can have ends which appear asymptotically as $\mathbb{R}\times \mathcal{C}$ for some Riemann surface $\mathcal{C}$.  Then, the one-form $\lambda$ can have periods over cycles in $\mathcal{C}$ which encode the real masses.  Compactifying $M$, simply means that $\lambda$ is no longer closed as above.

\item $\delta \omega$ restricts to $M$ to a non-exact form.  In that case we make use of a basis $\beta_{i}$ of cohomologically non-trivial two-forms and expand $\delta \omega$.  Then, the modification of the equation defining $\lambda$ is
\begin{equation}
d\lambda =\sum_{i}\zeta_{i}\beta_{i}. \label{FIdefearly}
\end{equation}
Here, the real coefficients $\zeta_{i}$ appearing in the above expansion are naturally interpreted as FI parameters.  Observe that, there is one such constant for each two-cycle, dual to $\beta_{i}$ which are non-trivial not only inside $M$ but also in $Q$.  Later, when we describe M2-branes we will see that it is exactly these cycles which give rise to dual flavor symmetries.

The fact that these parameters are indeed the FI terms can be understood by noting that in the presence of non-vanishing $\zeta_{i}$ there is no solution to the above equation.  As in our description of Chern-Simons levels this is interpreted as a destruction of the two-cycle dual to $\beta_{i}$.  As a consequence of this we see that the parameter $\zeta_{i}$ has the correct physical effect of higgsing the associated $U(1)$ gauge group.  Again, as in the case of real masses, one can make $\lambda$ closed at the expense of deleting certain loci.
\end{itemize}

\subsubsection{BPS M2-Branes and Instantons}

The massless $U(1)$ gauge multiplets we have identified constitute an important subset of the information defining the Coulomb branch of the three-dimensional field theory determined by a five-brane on a three-manifold $M$.  To complete the description, we now incorporate charged chiral multiplets and superpotentials.  As we will see, all such objects arise from the possibility of M2-branes ending on the M5-brane and altering the physics.

\begin{figure}[here!]
  \centering
\includegraphics[width=0.49\textwidth, height=0.3\textwidth]{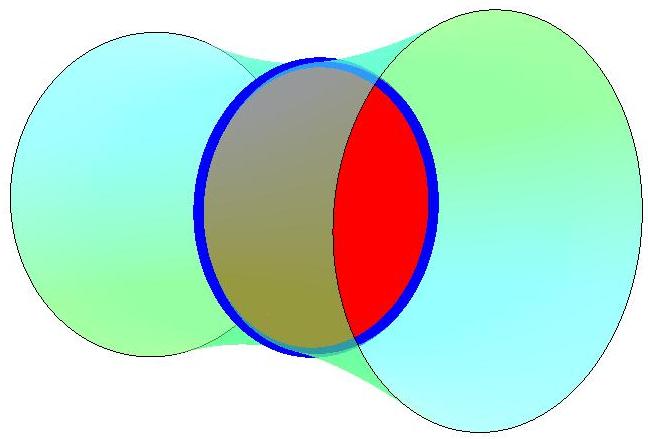}   
  \caption{A BPS M2 brane.  A three-manifold $M$, shown in green, sits inside the ambient Calabi-Yau $Q$, shown in white.  The three-manifold supports a non-trivial one-cycle $\Gamma$ shown in blue.  A minimal M2 disc, shown in red, can end on this cycle and describes a BPS particle in $\mathbb{R}^{1,2}$.}
  \label{fig:mspart}
\end{figure}

First, let us discuss the inclusion of chiral multiplets in the field theory.  We recall that because an M5 supports the two-form field $B$ an M2 may end on an M5 in two spacetime dimensions while remaining consistent with charge conservation.  Thus, to make a particle in three dimensions we may consider a two-brane whose worldvolume meets the five-brane along a timelike direction in $\mathbb{R}^{1,2}$ and a one-cycle $\Gamma$ in $M$, as illustrated in Figure \ref{fig:mspart}.  
Let $D$ denote the two-cycle in the Calabi-Yau $Q$ defined by the spatial directions of the M2.  Then the mass $m$ of the associated particle in $\mathbb{R}^{1,2}$ is determined by the volume of $D$.   However, since we are interested in chiral multiplets we are interested in short representations of the supersymmetry algebra and hence in BPS M2 branes.  Thus the cycle $D$ must be minimal in its homology class and is therefore holomorphic.  As a result the volume of $D$ is fixed by the K\"{a}hler form
\begin{equation}
m=\int_{D}\omega.
\end{equation}
However, locally near $M$ we may use \eqref{dlambda} to write $\omega=d\lambda$.  Then since $\partial D=\Gamma$ we use Stokes' theorem in the above to obtain
\begin{equation} 
m=\int_{\Gamma}\lambda =Z
\end{equation}
Where $Z$ is the central charge of the particle as measured by the periods of $\lambda$.  This fact clarifies why it is the periods of the one-form $\lambda$ which measures the central charges of charged particles.  In the far infrared, all matter particles in $\mathbb{R}^{1,2}$ can be described by two-branes, and hence the gauge charge lattice of the theory is naturally identified with the set of one-cycles $H_{1}(M,\mathbb{Z})$.  The one-form $\lambda$ pairs with these charges and hence its periods can encode the central charges of the field-theory.  

Geometrically, a non-trivial chiral multiplet of charge $\Gamma$ is described by a two-brane with topology of a disc.  The existence of this disc means that while the one-cycle $\Gamma$ may be non-trivial in $M$, when considered as a cycle in the ambient Calabi-Yau $Q$ it is homologically trivial.  It is exactly these cycles which become contractible in the ambient space that give rise to charged matter.  Those one-cycles in $M$ which remain non-contractible in $Q$ describe gauge groups which have no associated charged chiral particles.

The above discussion of chiral multiplets sets the stage for other ways in which two-branes can influence the three-dimensional physics.  Indeed, because an M2 can end on an M5 in two spacetime dimensions, its interpretation to a low-energy observer depends upon how many of the macroscopic dimensions the two-brane occupies.  If a two-brane ends along a compact two-cycle in $M$ then it occupies zero macroscopic dimensions and hence exists at a point in $\mathbb{R}^{1,2}$.  Such an object is naturally interpreted as an instanton.  One way to understand this is to examine the contributions to the action of this instanton.  Since the two-brane carries $B$ field charge, if it ends on the cycle $\beta_{j}$ in $M$ then its action will receive a contribution of the form
\begin{equation}
\exp \left(i\int_{\beta_{j}}B\right)=\exp \left(\hspace{-.2in}\phantom{\int_{\beta_{j}}}i\gamma_{j}\right).
\end{equation}
Thus, the instanton action is weighted by a phase determined by the expectation value of the dual photon.  This is familiar from the general structure three-dimensional field theories.  It also serves to illustrate why it is the periods of $B$ which measure the expectation values of the dual photons.  The charges of possible instantons are naturally labeled by two-cycles, and it is with these objects that $B$ can naturally pair.

Similar to the case of M2-brane particles, the presence of the instanton ending on the cycle $\beta$ implies that while $\beta$ is a non-trivial cycle in $M$ it is homologically trivial in $Q$.  This in turn implies that the associated monopole operator $\mathcal{M}$ is present in the Lagrangian of the theory, and hence the dual flavor symmetry is broken.  Note that this further clarifies why the parameters appearing in \eqref{FIdefearly} are indeed the FI parameters.  Dual flavor symmetries appear only for those non-trivial two-cycles in $M$ which remain non-trivial in $Q$.  For those two-cycles in $M$ which are contractible in $Q$ there are M2-brane instantons, and the dual flavor symmetry is broken.

\begin{figure}[here!]
  \centering
  \subfloat[Superpotential Geometry for Massive Fields]{\label{fig:m2w1}\includegraphics[width=0.45\textwidth, height=0.35\textwidth]{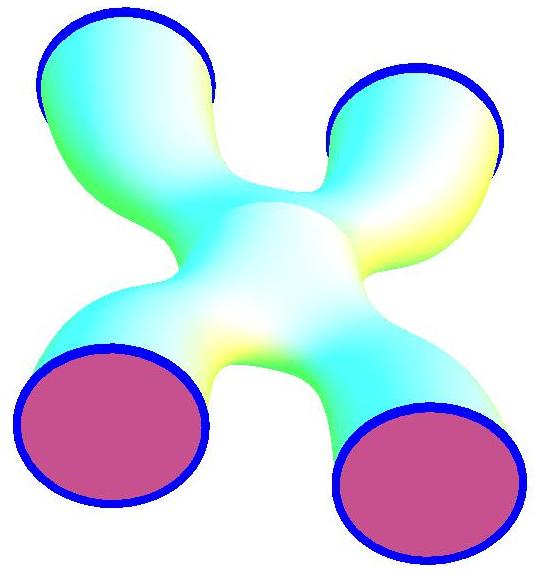}}     
  \hspace{.4in}    
  \subfloat[Superpotential Geometry for Massless Fields]{\label{fig:m2w2}\includegraphics[width=0.46\textwidth, height=0.35\textwidth]{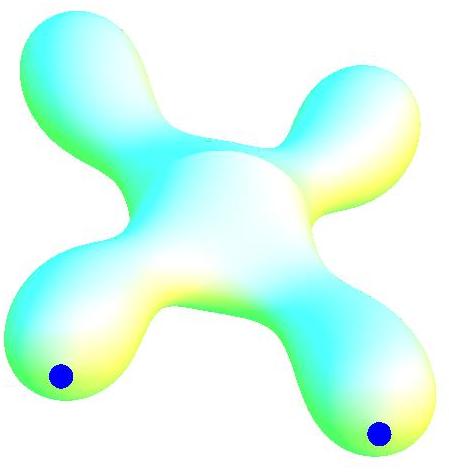}}
  \caption{M2 brane contributions to the superpotential.  In (a), we have four massive BPS states described by the pink M2 discs ending on blue one-cycles in the three-manifold.  A three-dimensional closed M2 has boundary on these discs and along a two-dimensional locus in $M$ and mediates a quartic interaction between the BPS particles. In (b), the BPS states become massless and the membrane geometry degenerates to a solid ball with four marked points, whose boundary lies entirely in $M$.}
  \label{fig:m2w}
\end{figure}

Finally, to construct a superpotential for the chiral fields we may consider an two-brane geometry which is a  hybrid of the two elementary geometries described above.  We fix background chiral particles $X_{i}$ described geometrically by M2-brane discs ending on a collection of one-cycles $\Gamma_{i}\subset M$.  Then, we find an M2 world-membrane that mediates an interaction between these objects.  Topologically the worldvolume of this membrane is a three-manifold with boundary.  This three-manifold lies entirely in the internal geometry $Q$ and has boundary along the M2 discs describing the particles and along a two dimensional surface in $M$ whose boundary is the union of the $\Gamma_{i}$.   An example of this geometry is illustrated in Figure \ref{fig:m2w1}. When the world-volume of this membrane is minimal, it describes a supersymmetric interaction and hence can give a contribution to the superpotential for the chiral fields. 
In practice, the most relevant case of this phenomenon occurs in the limit where the masses of the chiral particles become small and the superpotential is important.  In that limit, the one-cycles labeling the charge of the chiral fields collapse to points and the two-brane we are describing is a handlebody whose boundary Riemann surface lies on a two-cycle in $M$ and has a number of marked points corresponding to the insertion of massless chiral fields as shown in Figure \ref{fig:m2w2}.  That such instanton-like brane geometries make contributions to the superpotential is familiar from a variety of similar situations.  

\subsection{Many Five-Branes}
\label{sec:MM5}
When multiple five-branes wrap a three-manifold $M$, the resulting non-abelian dynamics gives rise to a strongly interacting field theory in three dimensions.  Nevertheless in the IR on the Coulomb branch, we can still make use of the geometry described in the previous section to encode the physics.  The key observation is again to recognize the effective scalar degrees of freedom.  Just as for the case of a single five-brane, the tranverse motion of the branes can be viewed as taking place in the cotangent bundle $T^{*}M$.  If there are a total of $n$ five-branes wrapping $M$ then there are naively $n$ independent one-forms $\lambda_{i}$ on $M$ which describe the motion of each individual five-brane.  The reason that this assertion is naive is that it fails to account for the possibility that, after activating fields, the $n$ distinct branes will recombine into a single connected object.  In fact, a generic point on the Coulomb branch of the field theory is described by a geometry of this sort, and thus this possibility must be taken into account.  

Fortunately, there is an elementary way to allow for brane recombination. We simply permit the possibility that the $n$ objects $\lambda_{i}$ are not individually globally well-defined but instead permute amongst themselves as we traverse the manifold $M$ \cite{TBranes}.  Said differently, the one-forms $\lambda_{i}$ are permitted to have one-dimensional branch loci and, under circling the branch locus, they are acted on by $S_{n}$, the permutation group on $n$ letters.  Such a structure naturally encodes brane recombination and gives rise to a three-manifold $\widetilde{M}$ which is a $n$-sheeted cover of $M$.  By definition, $\widetilde{M}$ is exactly the three-manifold where the $n$ branched one-forms $\lambda_{i}$ glue together to yield a single, globally well-defined, harmonic one-form $\lambda$.  We can encode this condition in equations by noting that $\lambda$ defines completely the locus of the three-manifold cover $\widetilde{M}$ inside the cotangent bundle of the base $T^{*}M$.  Thus, knowledge of $\lambda$ is equivalent to knowledge of the induced metric on the special Lagrangian $\widetilde{M}$ and hence defines a hodge star operation $*_{\lambda}$.  Then, the supersymmetric equations defining the IR geometry are
\begin{equation}
d\lambda =d*_{\lambda}\lambda=0.
\end{equation}
These are a set of non-linear relations on $\lambda$ or equivalently the $\lambda_{i}$.  They state that $\lambda$ is harmonic in the induced metric which it determines.

Conceptually, the advantage of passing to the cover $\widetilde{M}$ is that in the infrared all of the physics that is described by $n$ five-branes wrapping $M$ is completely encoded by the recombined brane $\widetilde{M}$.  The virtue of this description is that the effective description is that of a \emph{single} five-brane on $\widetilde{M}$.   It is therefore naturally abelian and described by the geometry of the previous section.  For example, it is the periods of the harmonic one-form $\lambda$ on $\widetilde{M}$ which determine the real central charges of the three dimensional field theory.  Thus all of the non-abelian dynamics of multiple five-branes is encoded in the geometry of the covering manifold.

It is natural to interpret the existence of the cover $\widetilde{M}$, and its central role in the field theory, as a parallel to a similar structure which occurs in four-dimensional $\mathcal{N}=2$ gauge theories which arise from placing $n$ five-branes on a Riemann surface $\Sigma$.  Just as above, the infrared dynamics of that theory are determined by brane recombination.  The transverse motion of a single five brane is again identified with fluctuations in the contangent directions to the compactification manifold, namely $\Sigma$.  Thus, for each brane we expect a holomorphic one-form $\phi_{i}$ which parametrizes the position of the $i$-th brane.  Brane recombination implies that the $\phi_{i}$ are not well-defined and instead have branch cuts where they mix.  On passing to an $n$-fold cover $\widetilde{\Sigma}$ these one-forms glue together to a single globally well-defined object $\phi$.  This cover $\widetilde{\Sigma}$ is the Seiberg-Witten curve and $\phi$ is the Seiberg-Witten differential \cite{WittenM}.  Their geometry and periods completely encode the low-energy action \cite{SW1, SW2}.  The manifold $\widetilde{M}$, whose abstract existence we have eluded to in this section plays a similar role in the three-dimensional physics, and in later sections where we study explicit examples, our primary task will be to determine $\widetilde{M}$.

\subsection{Two Five-Branes}
\label{sec:TM5}
For most of our explicit examples in later sections, we will be interested in the specialization to the case where the number of five-branes, before recombination, is two.  Then, the IR five-brane geometry is that of a branched double cover $\widetilde{M}\rightarrow M$.  In this section we discuss in more detail the resulting three-dimensional topology and its relation to the physics.\footnote{\ Much of this geometry is classical.  For an introduction see \cite{Lickorish}.}  In practice our primary examples will be to the case where $M$ is a three-sphere and for the remainder of this section we make that restriction.\footnote{\ In terms of the topology of the cover $\widetilde{M}$ this is no restriction.  Indeed, every orientable compact three-manifold can be presented as a double branched cover of $S^{3}$ \cite{Lickorish}.}  Our specific goal will be to determine the homology group $H_{1}(\widetilde{M},\mathbb{Z})$.  As we have argued in equation \eqref{levelgen}, complete knowledge of this homology is equivalent to a description of the gauge boson sector of the field theory on $\mathbb{R}^{1,2}$, with propagating fields captured by the Betti number $b_{1}(\widetilde{M})$ and non-trivial levels $\hat{k}_{ij}$ encoded in the torsion classes of $H_{1}(\widetilde{M},\mathbb{Z})$.  Further, in section \ref{BPSM2proj}, we also illustrate how the M2 brane geometries discussed in previous sections can be visualized more directly in the case of a double cover.

\subsubsection{Seifert Surfaces}
\begin{figure}[here!]
  \centering
   \subfloat[Pretzel Knot]{\label{fig:seifert1}\includegraphics[width=0.4\textwidth]{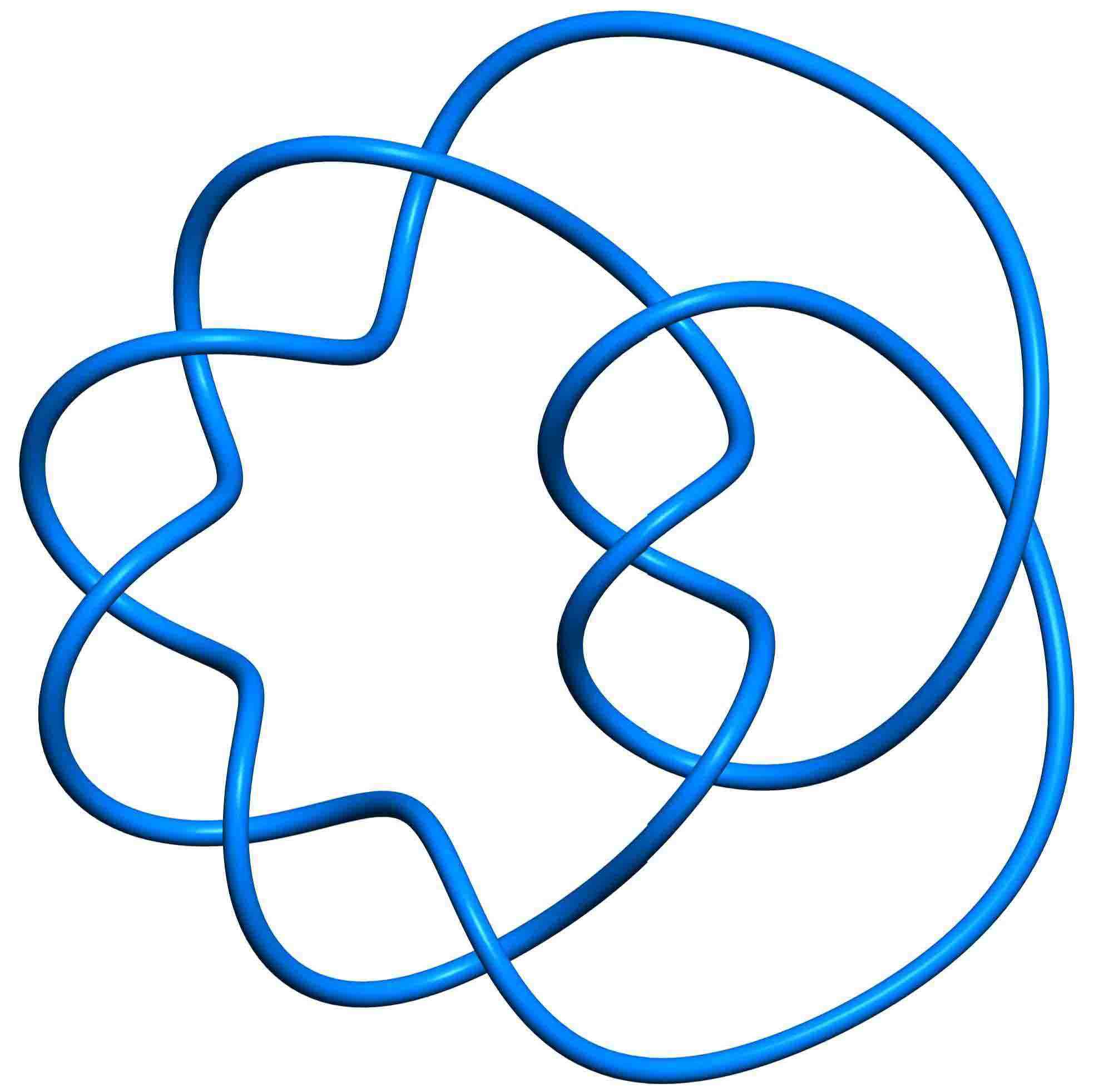}}
   \hspace{.5in}          
  \subfloat[Seifert Surface]{\label{fig:seifert2}\includegraphics[width=0.4\textwidth]{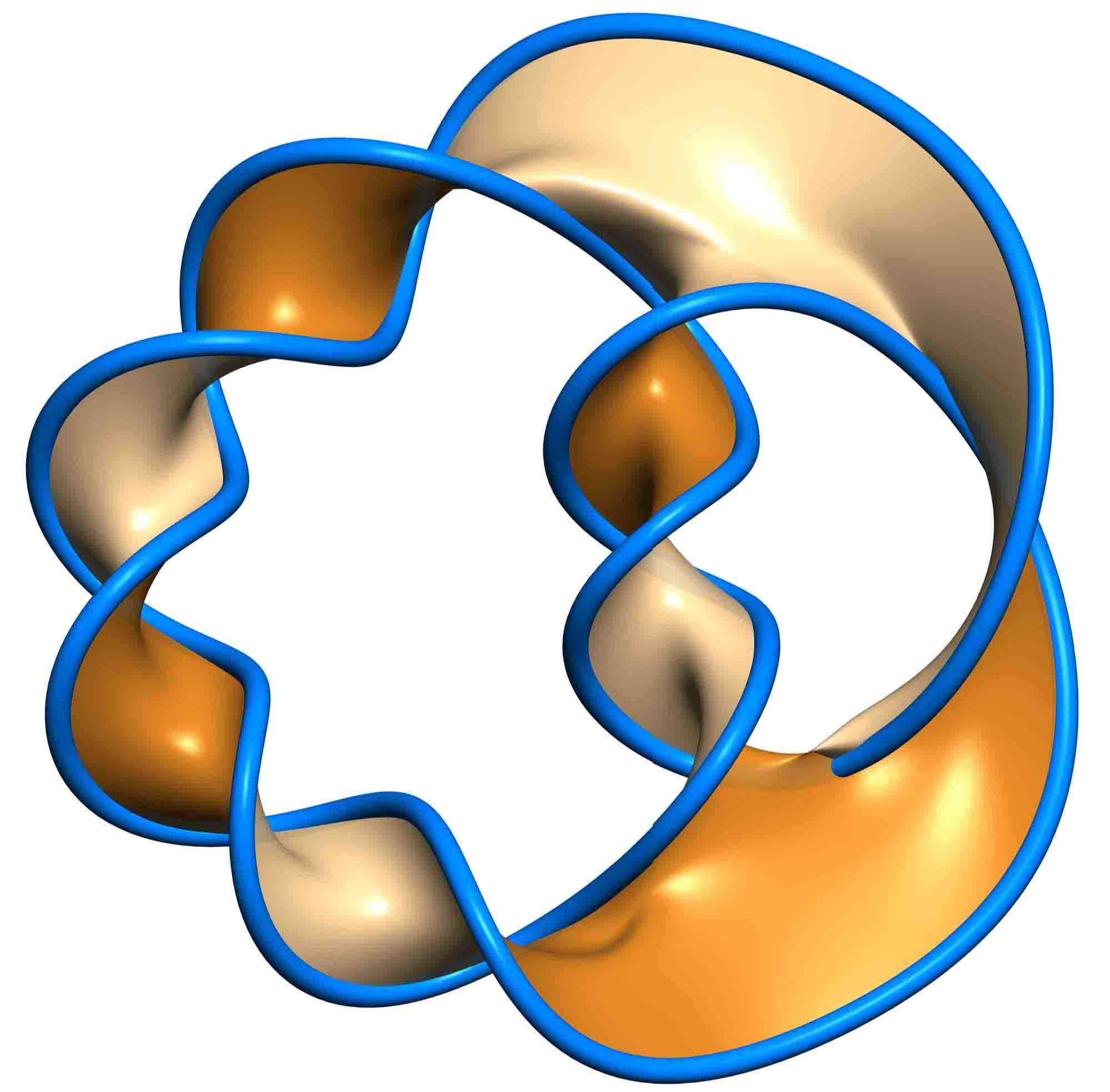}}\\
  \hspace{.5in}
  \subfloat[Borromean Rings]{\label{fig:seifert3}\includegraphics[width=0.4\textwidth]{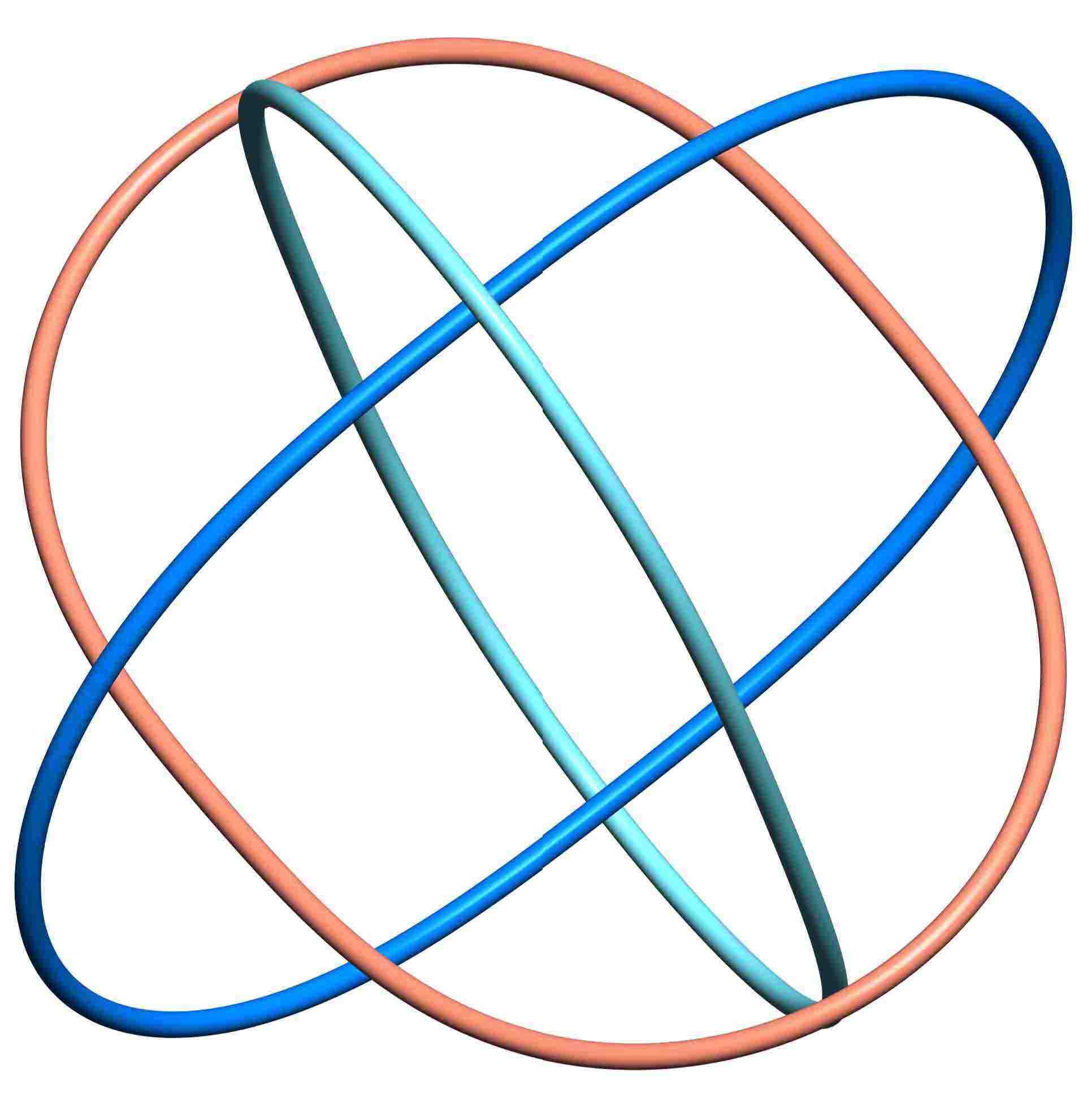}}
  \hspace{.5in}          
  \subfloat[Seifert Surface]{\label{fig:seifert4}\includegraphics[width=0.4\textwidth]{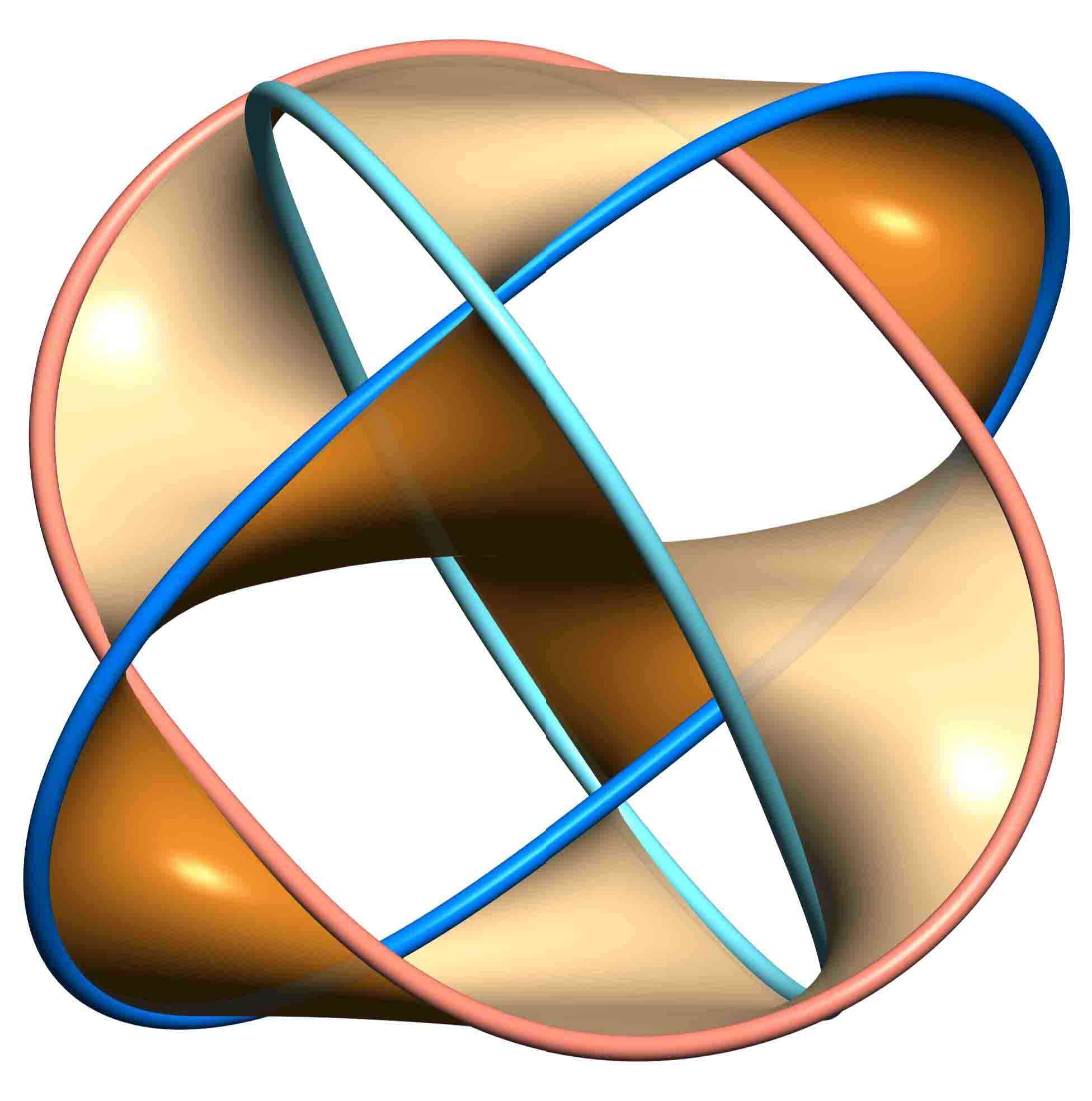}}
  \caption{Knots and associated Siefert surfaces.  In (a) and (c) we see non-trivial knots.  In (b) and (d), we see Seifert surfaces whose boundaries are the given knots.  The knots are an example of possible branch loci for a double cover of $S^{3}$.  The Seifert surfaces are then the branch sheets defining the cover geometry.}
  \label{fig:seifert}
\end{figure}

We begin with the elementary observation that $S^{3}$ has trivial topology.  From this it follows that all the resulting topology of the cover is encoded in its branch structure over $S^{3}$. Since $\widetilde{M}$ is smooth, the branch locus is required to be a smooth embedded closed submanifold of $M$ dimension one.  Topologically, the branch locus is therefore a union of circles.  However, the circles may be embedded in $S^{3}$ in a complicated fashion and hence form a non-trivial knot $\mathcal{K}$.\footnote{\ In this paper by the term \emph{knots} we will refer to both knots and links, and whenever we really mean ``knot" we shall emphasize it. }  The topology of the cover $\widetilde{M}$ is completely fixed by $\mathcal{K}$.  To construct the cover we first proceed by drawing a branch sheet.  This is a smooth two-dimensional surface $F$ whose boundary is the given knot $\mathcal{K}$.  A classical theorem of Seifert, Frankl, and Pontrajgin ensures that such a surface always exists, and that further one may assume $F$ to be orientable.  When this is so, $F$ is referred to as a  \emph{Seifert surface} for the knot $\mathcal{K}$.  Some examples are illustrated in Figure \ref{fig:seifert}.  

Once a Seifert surface has been specified, the double cover $\widetilde{M}$ can be constructed explicitly.  We take two copies of $S^{3}$ and cut them both along $F$.  In each three-sphere this creates a boundary which is topologically two copies of the Seifert surface, $F_{+}$ and $F_{-}$, intersecting along $\mathcal{K}$.   Then we glue $F_{+}$ in one $S^{3}$ to $F_{-}$ in the other $S^{3}$ and vice versa to create $\widetilde{M}$ which is a two-to-one cover of $S^{3}$ except over the knot $\mathcal{K}$.  Of course, as is the case with branched covers of Riemann surfaces, there are in general many choices of branch sheets, and so given a knot $\mathcal{K}$ its Seifert surface is not unique.  However, the topology of the cover $\widetilde{M}$ is independent of this choice and thus any convenient Seifert surface can be used for the branch sheet.  In practice this construction is useful since many properties of the cover can be deduced directly from any given Seifert surface.

One useful quantity that we may extract from the Seifert surface $F$ is the homology $H_{1}(\widetilde{M})$.  Indeed, since $S^{3}$ has no non-trivial one-cycles, all cycles in $\widetilde{M}$ can be localized to a neighborhood of $\mathcal{K}$ and must involve the knot if they are to be non-trivial.  This is quite similar to the case of the homology of a branched cover of the two-sphere.  There, the branch locus is a number of points, and the branch lines are segments $I$ connecting pairs of these points.  The homology of the double cover is then generated by the classes in the complement of the branch lines, $H_{1}(S^{2}-\cup I)$.  Our situation is exactly parallel, save for the fact that it takes place in one higher dimension.  The homology of $\widetilde{M}$, a branched double cover of $S^{3}$, is generated by the cycles in the complement of the branch sheet $H_{1}(S^{3}-F)$.

We can make a further simplification by observing that cycles in $S^{3}-F$ and cycles in $F$ are naturally dual.  Indeed, given $\alpha \in H_{1}(S^{3}-F)$ and $\beta \in H_{1}(F)$ we represent $\alpha$ and $\beta$ by simple, oriented, closed curves and compute the linking number $[\alpha, \beta] \in \mathbb{Z}$.  This determines a pairing
\begin{equation}
[\cdot, \cdot]: H_{1}(S^{3}-F)\times H_{1}(F)\longrightarrow \mathbb{Z}.
\end{equation}
And in fact the pairing is an isomorphism.  This implies that a natural spanning set of one-cycles generating $H_{1}(\widetilde{M})$ is  given by the homology classes on the Seifert surface itself, $H_{1}(F)$.  

Now, although the homology $H_{1}(F)$ generates the homology of the cover, typically when considered in $\widetilde{M}$, many of these cycles are in fact homologically trivial.  Thus, $H_{1}(F)$ is generally an overcomplete set of cycles, and our task is to determine which classes on the Seifert surface become trivial in $\widetilde{M}$.  To do so, we introduce the concept of a Seifert matrix.  This is a $b_{1}(F)\times b_{1}(F)$ integral matrix defined as follows.
\begin{itemize}
\item Definition:  Let $\Gamma_{i}$ be a one-cycle in $F$.  Since $F$ is oriented we can define $\Gamma^{+}_{i}$ as a small pushoff of $\Gamma_{i}$ out of $F$ in the positive direction.  Then the Seifert matrix $A$ is the matrix of linking numbers $A_{ij}$ between the one-cycle $\Gamma_{i}$ and the pushoff $\Gamma_{j}^{+}$.
\end{itemize}

Armed with this matrix we can then say that the first homology of the cover $\widetilde{M}$ is generated by $H_{1}(F)$ modulo the relations defined by the symmetrized Seifert matrix
\begin{equation}
H_{1}(\widetilde{M})\cong H_{1}(F)/\mathrm{Image}(A+A^{T}). \label{seiferthom}
\end{equation}
Thus for example, the rank of the kernel of the map $A+A^{T}$ computes the first Betti number of $\widetilde{M}$.  Meanwhile if the homology of $\widetilde{M}$ is a finite abelian group, then the order of this group is computed by $|\det(A+A^{T})|$.\footnote{\ Incidentally, the Seifert matrix can also be used to define the Alexander polynomial of the knot by the definition ${\cal A}_{\cal K}(t)=\det(A-tA^t)$.  Then the determinant above, which computes the order of $H_{1}(\widetilde{M})$ when finite, is the Alexander polynomial evaluated at $t=-1$.}

The fact \eqref{seiferthom} should be directly interpreted in terms of our general discussion of $U(1)$ gauge fields and CS terms in \eqref{levelgen}.  In general, the gauge multiplet sector of the theory is described by a collection of abelian gauge fields and a matrix  $\hat{k}_{ij}$ of levels.  This data translates into a description of the homology $H_{1}(\widetilde{M})$.  The $U(1)$ gauge fields are a generating set of classes in the homology, and the matrix $\hat{k}_{ij}$ describes the relations among these generators.  In \eqref{seiferthom} we see exactly this description and hence we propose that:
\begin{itemize}    
\item $U(1)$ gauge fields are given by generators $\Gamma_{i}$ of $H_{1}(F)$. 
\item CS levels $\hat{k}_{ij}$ are given by the symmetrized Seifert matrix $A+A^{T}$.
\end{itemize}
In this description, the fact that the Seifert surface is non-unique translates to a statement about equivalence of various CS theories.  Any Seifert surface may be used to describe the physics, and distinct surfaces give distinct sets of $U(1)$'s and level matrices $\hat{k}_{ij}$ all of which determine the same theory.

To derive the above proposal, we first phrase the computation of linking numbers in a more familiar language of differential forms as follows.  Each homology class $\Gamma_{i}$ in $H_{1}(F)$ can be represented by a cycle that is topologically an unknot embedded in $S^{3}$.  Thus, the pushoff $\Gamma_{i}^{+}$ bounds an embedded disc $D_{i} \subset S^{3}$.  The symmetrized Seifert matrix of linking numbers is then given by computing the intersection number of $\Gamma_{i}$ with $D_{j}$
\begin{equation}
(A+A^{T})_{ij}=\Gamma_{i}\cap D_{j}+\Gamma_{j}\cap D_{i}.
\end{equation}
However, the intersection number on the right of the above can alternatively be computed in terms of the Poincar\'{e} dual form to the disc $D_{i}$ \cite{griffiths}.  Specifically, since the disc $D_{i}$ are bounded by the cycles $\Gamma_{i}^{+}$ the above is
\begin{equation}
(A+A^{T})_{ij}=\int _{M}\alpha_{i}\wedge d\alpha_{j}. \label{cslevelclas}
\end{equation}

Equation \eqref{cslevelclas} gives a direct way of seeing that the symmetrized Seifert matrix computes the levels $\hat{k}_{ij}$.  For each of the one-forms $\alpha_{i}$ introduced above we may consider an associated $U(1)$ gauge field $A_{i}$ by decomposing the two-form field $B$ propagating on the fivebrane $\widetilde{M}$.  This gauge field may be massless, or massive depending on the resulting equation of motion.  To examine this issue, we consider the self-dual 3-form field strength $T=dB$ written as
\begin{equation}
T=\alpha_{i}\wedge F_{i}+* \alpha_{i}\wedge * F_{i}.
\end{equation}
Then, the equation of motion $dT=0$ implies in particular
\begin{equation}
d\alpha_{i}\wedge F_{i}+*\alpha_{i}\wedge d *F_{i}=0.
\end{equation}
Wedge the above equation with $\alpha_{j}$ and integate over $\widetilde{M}$ to find
\begin{equation}
\left(\int_{\widetilde{M}}\alpha_{j}\wedge *\alpha_{i}\right)d *F_{i}+(A+A^{T})_{ji}F_{i}=0. \label{kkredux}
\end{equation}
The normalization matrix $\int_{\widetilde{M}}\alpha_{j}\wedge *\alpha_{i}$ determines the metric on the space of $U(1)$ gauge fields, and with this identification, \eqref{kkredux} is exactly the equation for a collection of $U(1)$ vectors with a level matrix $\hat{k}_{ij}$ given by the symmetrized Seifert matrix $A+A^{T}$.

\begin{figure}[here!]
  \centering
 \includegraphics[width=0.4\textwidth, height=0.4\textwidth]{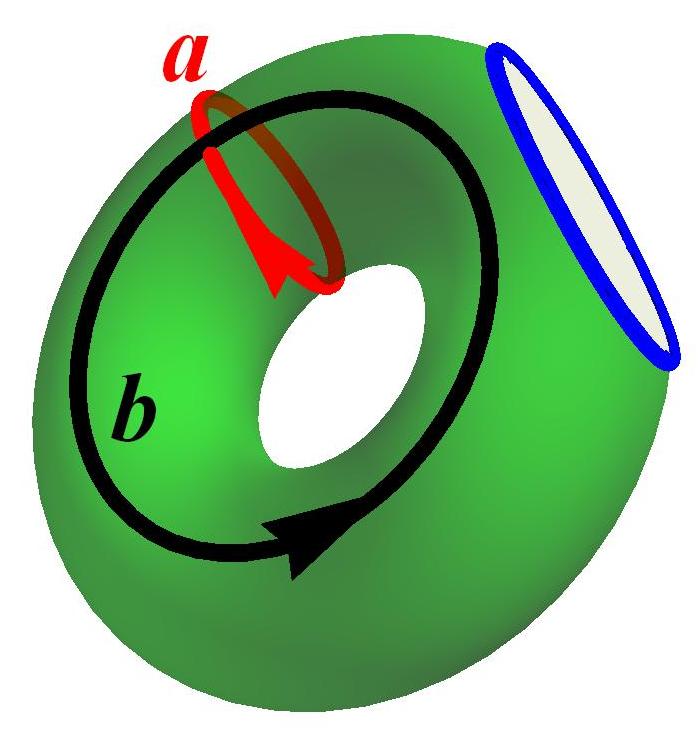}     
  \caption{A Seifert surface $F$ for the unknot.  The green torus minus a disc bounds the unknot shown in blue.  The red and black cycles, $a$ and $b$, are a basis for the homology of $F$.  The pushoff $a^{+}$ is linked with $b$.}
  \label{fig:torus}
\end{figure}

As a sample application of these ideas, we consider the case of a cover branched over the unknot.  Of course, because the unknot bounds a disc we may choose this as the Seifert surface.  Then glueing together two $S^{3}$'s with a  branching sheet given by a disc is simply taking the connected sum of the two $S^{3}$'s.  This means that the cover $\widetilde{M}$ is again an $S^{3}$ and hence has trivial homology.  However for a more interesting choice we can take as the Seifert surface $F$ a torus minus a disc as shown in Figure \ref{fig:torus}.
A basis of $H_{1}(F)$ is then the two one-cycles $a$ and $b$ shown in the Figure.  Taking the outward direction of the torus to be the positive orientation we then conclude that there is a non-vanishing linking number $+1$  between $b$ and the pushoff $a^{+}$.  Hence in this case we find that the symmetrized Seifert matrix is
\begin{equation}
A+A^{T}=\left(\begin{array}{cc} 0 & 1 \\ 1 & 0 \end{array}\right). \label{torusres}
\end{equation}
From this we deduce that both $a$ and $b$ are in the image $A+A^{T}$ and hence trivial in the homology of $\widetilde{M}$.  Thus we recover the correct result that $H_{1}(\widetilde{M})$ is vanishing.

Physically, the example given above describes a known duality \cite{WittenSL2}.  A $U(1)\times U(1)$ gauge theory together with level matrix
\begin{equation} 
A+A^{T}=\hat{k}_{ij}=\left(\begin{array}{cc} 0 & 1 \\ 1 & 0 \end{array}\right)
\end{equation}
is equivalent to a trivial theory of no gauge group whatsoever.  In fact, this example suffices to understand the general result that the gauge multiplet theory is independent of the choice of $F$.  Indeed, given a fixed knot $\mathcal{K}$ any two topologically distinct Seifert surfaces $F$ and $F'$ with the genus of $F'$ larger than $F$, differ by excising some number $n$ of discs from $F$ and attaching $n$ handles like those shown in Figure \ref{fig:torus} to obtain $F'$.  At the level of the Seifert matrices this has the effect
\begin{equation}
(A+A^{T})|_{F'}=(A+A^{T})|_{F} \bigoplus_{i=1}^{n} \left(\begin{array}{cc} 0 & 1 \\ 1 & 0 \end{array}\right).
\end{equation}
In other words, adding an irrelevant handle simply adds a trivial theory in the form of a $U(1)\times U(1)$ with off-diagonal level matrix \eqref{torusres} and hence does not modify the physics.

\subsubsection{Checkerboards and Tait Graphs}

\begin{figure}[here!]
  \centering
  \subfloat[Knot Crossing]{\label{fig:cross}\includegraphics[width=0.25\textwidth, height=.2\textwidth]{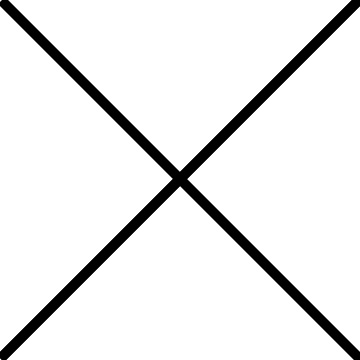}}     
  \hspace{.75in}         
  \subfloat[Checkerboard Shading ]{\label{fig:crossfill}\includegraphics[width=0.25\textwidth, height=.2\textwidth]{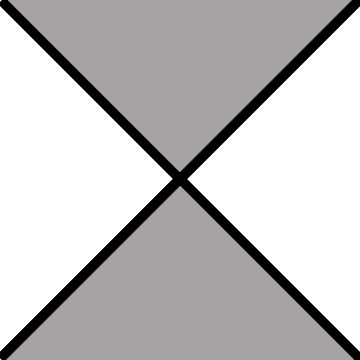}}
  \caption{Local definition of checkerboard coloring.  In (a) a planar projection of a crossing in a knot diagram.  In (b) a checkerboard coloring at the crossing.}
  \label{fig:checkerdef}
\end{figure}

In the forthcoming applications of this paper, it will be important for us to have a more explicit method for determining a Seifert surface for a given knot and computing Chern-Simons levels.  One way to produce such a surface is to use a so-called \emph{checkerboard coloring} of the knot.  This is an assignment of black versus white to each region enclosed by the planar projection of the knot.  It has the property that regions which share an edge have opposite colors.  Given any knot, there is no obstruction to constructing a checkerboard coloring.  Indeed, we simply consider the local structure of the knot near a given crossing.  If we forget the data of which component passes over versus under, the crossing appears as the intersection of two lines, and separates the plane into four regions as shown in Figure \ref{fig:cross}.  Then, we choose a pair of non-adjacent regions and shade them as shown in Figure \ref{fig:crossfill}.  We continue this shading procedure consistently to all the remaining crossing in the knot.  At the conclusion we have constructed a checkerboard coloring.  

The relation of checkerboard colorings to Seifert surfaces is simply that the shaded regions of the coloring define a two-dimensional surface $F$ whose boundary is the knot $\mathcal{K}$.  The interior of each shaded region is a disc and at the crossings, these discs are glued together by twisted bands. Thus the shaded regions of a checkerboard diagram determine a surface $F$ whose boundary is the knot $\mathcal{K}$.\footnote{\ In many situations this surface is orientable and hence meets the requirements to be called a Seifert surface for $\mathcal{K}$.  Sometimes, however the surface is non-orientable and is technically therefore not a Seifert surface.  This causes no problem from the point of view of using such a surface as a branch sheet to construct a cover.  Further as we describe below, all of the pertinent results of the previous section, go through unmodified.}

\begin{figure}[here!]
  \centering
  \subfloat[Trefoil]{\label{fig:trefoil}\includegraphics[width=0.3\textwidth]{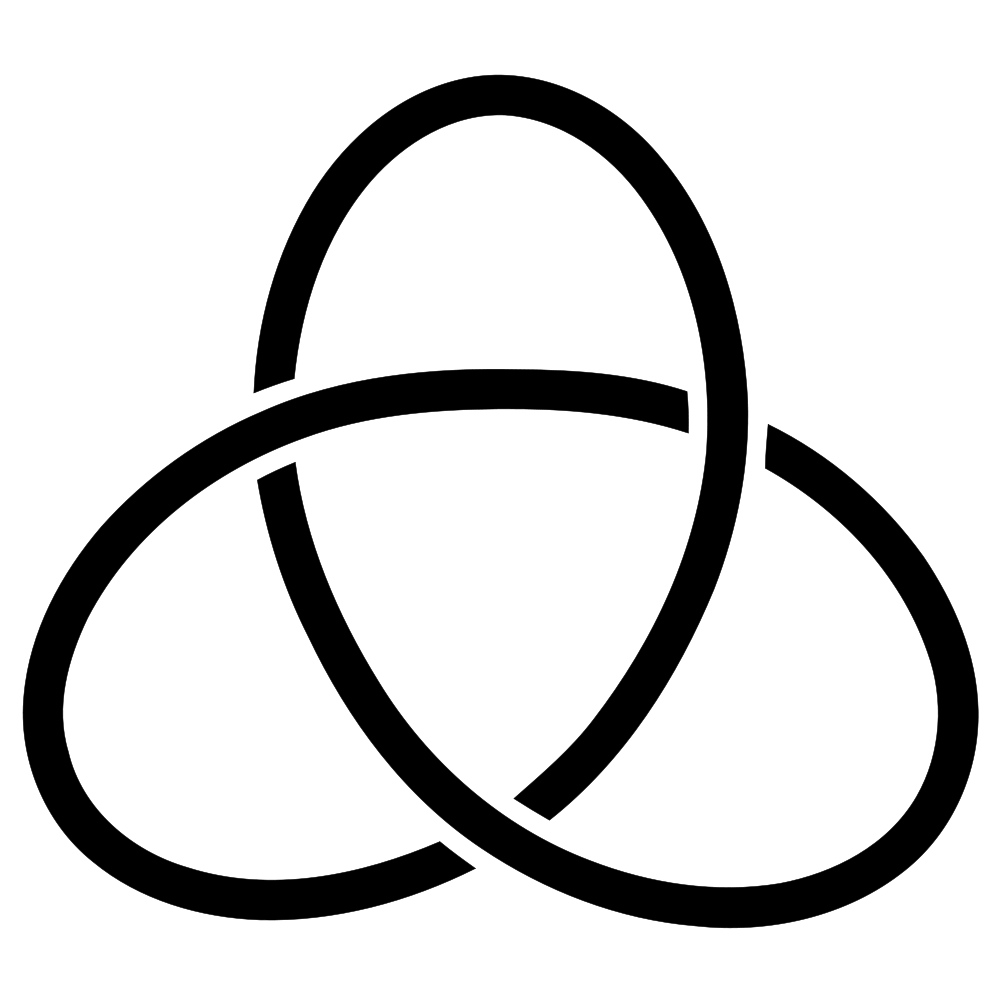}}     
  \hspace{.25in}        
  \subfloat[Checkerboard]{\label{fig:wtrefoil}\includegraphics[width=0.3\textwidth]{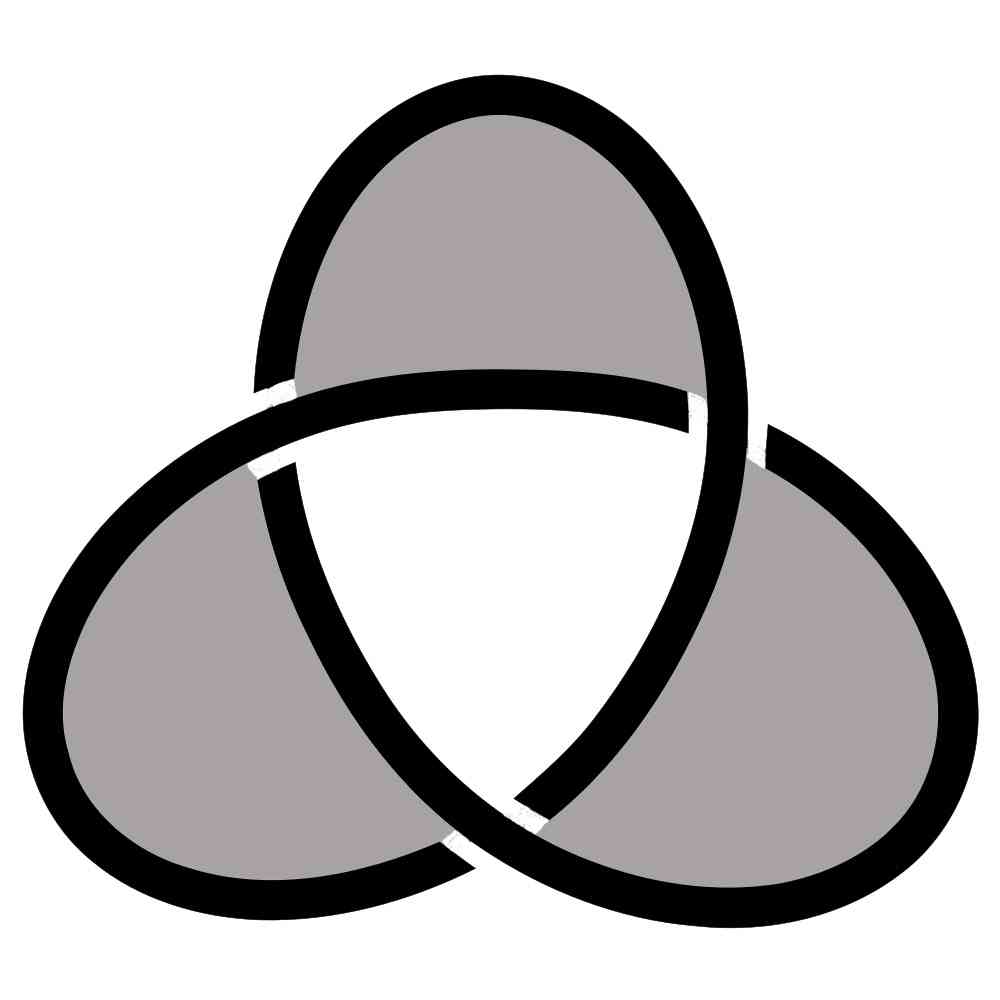}}
    \hspace{.25in}        
  \subfloat[Dual Checkerboard]{\label{fig:btrefoil}\includegraphics[width=0.3\textwidth]{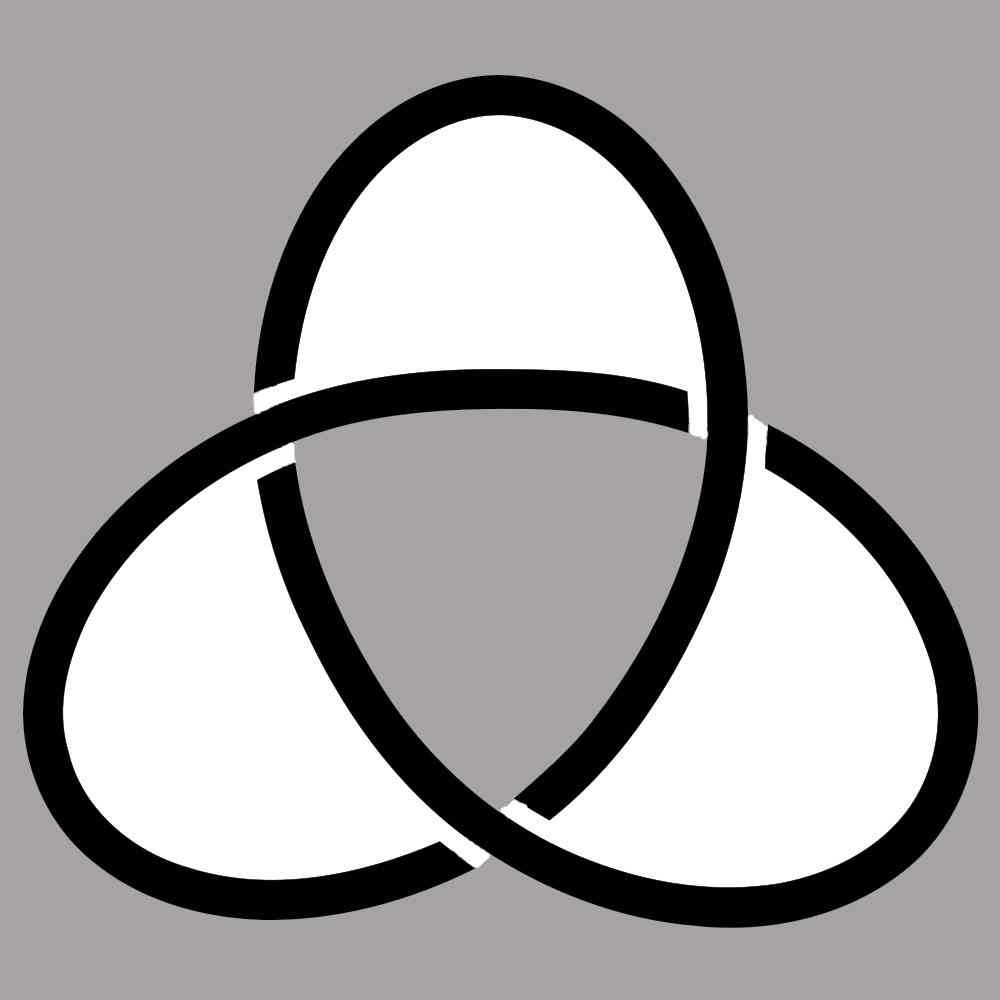}}
  \caption{Checkerboard colorings for the trefoil knot.  In (a) the trefoil knot.  In (b) and (c) its two dual colorings.  The shaded regions can be interpreted as surfaces with boundary the trefoil.}
  \label{fig:tchecker}
\end{figure}

Notice that as a consequence of the construction, there is a natural notion of duality among checkerboard colorings of the knot.  Given such a coloring, we may exchange the black and white regions to produce a new coloring.  A specific example of this is given for the trefoil knot in Figure \ref{fig:tchecker}.
In later sections we will see that this basic black-white duality of checkerboard colorings has an interesting physical interpretation in terms of duality of 3d field theories.

However, for our present purposes our main interest in checkerboard colorings is simply that they provide a convenient way of determining the homology of the cover $\widetilde{M}$, and therefore a method for determining a set of $U(1)$ gauge fields and a level matrix $\hat{k}_{ij}$.

\begin{figure}[here!]
  \centering
  \subfloat[Checkerboard]{\label{fig:8black}\includegraphics[width=0.45\textwidth,]{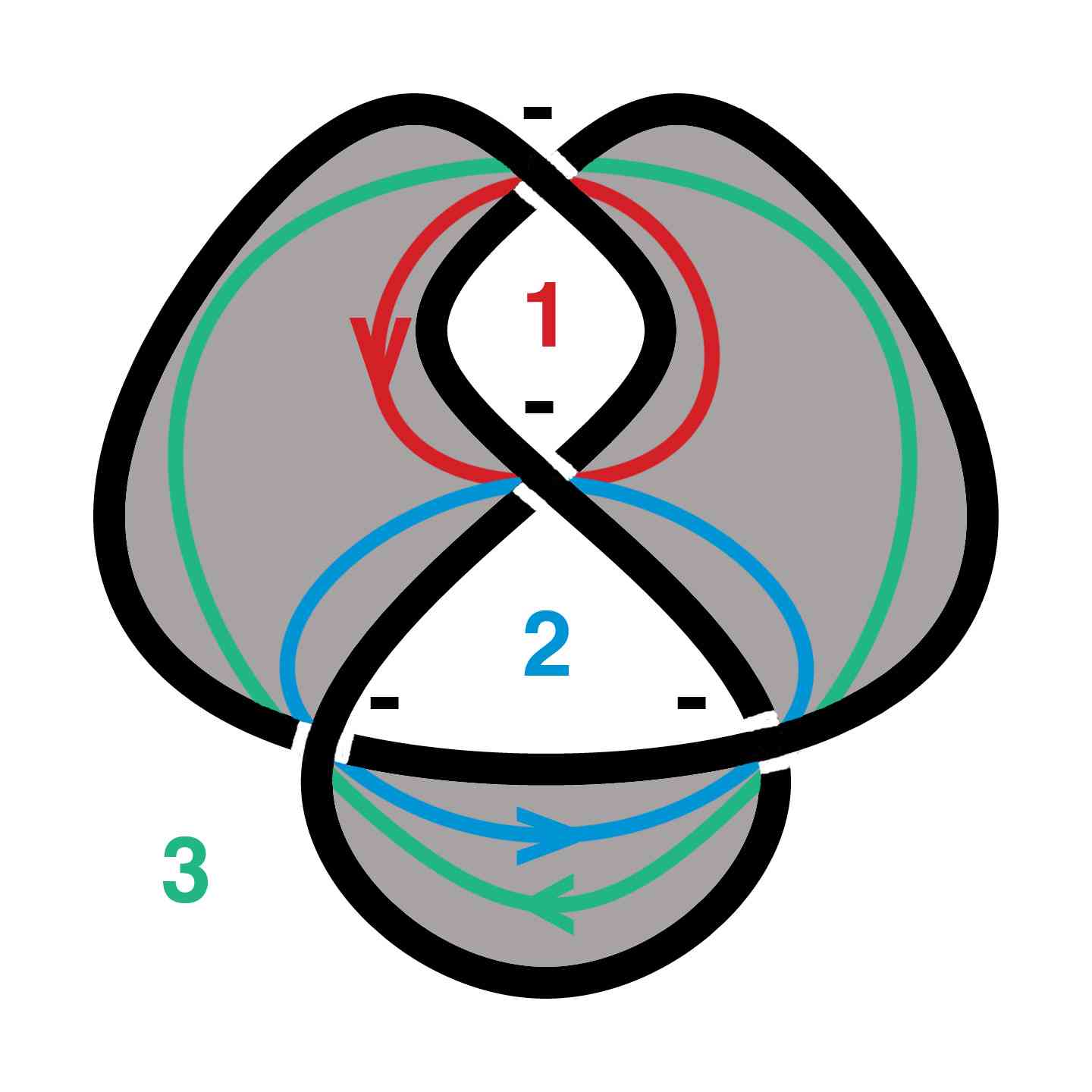}}     
  \hspace{.25in}  
  \subfloat[Dual Checkerboard]{\label{fig:8white}\includegraphics[width=0.45\textwidth]{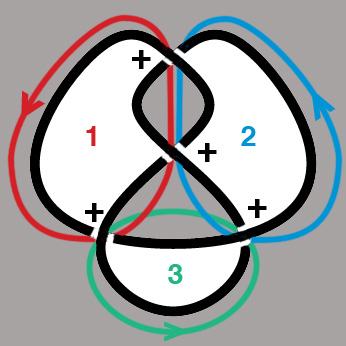}}
  \caption{Dual checkerboard colorings for the figure eight knot.  The gray regions denote the Seifert surface of the black knot.  The  colored cycles denote the spanning set of the homology of the surface given by the enclosed white regions of the checkerboard.  The signs at each crossing indicate the contribution to the Chern-Simons levels.}
  \label{fig:8checker}
\end{figure}

Let us first fix the basis of cycles.  These are manifest in the checkerboard coloring.  Each white region of the diagram is, by construction, a hole in the surface defined by the shaded regions of the coloring.  Therefore there is a non-trivial cycle defined by simply moving the boundary of the given white region slightly into the shaded region.  Thus, if $R_{1} ,\cdots R_{n+1}$ denote the white regions of the checkerboard there are associated cycles $\Gamma_{i}$ encircling $R_{i}$.  Examples are illustrated in Figure \ref{fig:8checker}.  One of these cycles, say $\Gamma_{n+1}$ can be generated in homology of the surface $F$ by the remaining $n$.  The remaining cycles $\Gamma_{1}, \cdots \Gamma_{n}$ are an explicit basis for the homology of the surface.  In physical language these are the defining generators for a $U(1)$ gauge theory.

\begin{figure}[here!]
  \centering
  \subfloat[$\zeta(c)=-1$]{\label{fig:zetap}\includegraphics[width=0.25\textwidth, height=.2\textwidth]{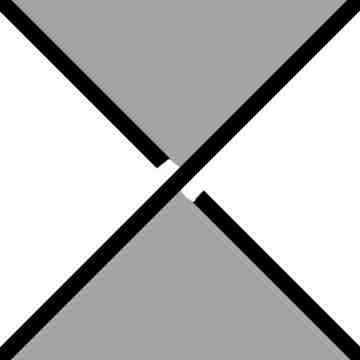}}     
  \hspace{.75in}         
  \subfloat[$\zeta(c)=+1$ ]{\label{fig:zetam}\includegraphics[width=0.25\textwidth, height=.2\textwidth]{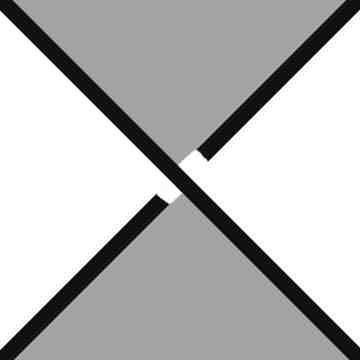}}
  \caption{Positive and negative crossings in a knot.}
  \label{fig:zetadef}
\end{figure}

Next we extract the CS levels.  As in our general discussion, these levels are determined by a linking number computation.  However in the case of the checkerboard coloring there is a simple more algorithmic way of determining the levels.  First we associate to each crossing $c$ in the diagram a sign function $\zeta(c)=\pm1$ determined by whether the overstrand or understrand is to the left or the right of the surface $F$ as illustrated in Figure \ref{fig:zetadef}.  
The CS matrix is then most easily determined by keeping track of all the $n+1$ cycles associated to all the white regions, as opposed to just the $n$ in a spanning set.  Specifically, we construct an $(n+1)\times (n+1)$ matrix whose rows and columns index the white regions.  Then:
\begin{itemize}
\item The off-diagonal elements $\hat{k}_{ij}$ for $i\neq j$ are determined by summing over all crossings where regions $i$ and $j$ meet taken with sign.
\begin{equation}
\hat{k}_{ij}=\sum_{i,j  \  \mathrm{crossings}}   \zeta(c) \label{offdiagchecker}
\end{equation}
\item The diagonal elements $\hat{k}_{ii}$ are determined by the condition that the sum of all entries in any row vanishes
\begin{equation}
\hat{k}_{ii}=-\sum_{j\neq i}\hat{k}_{ji}.
\end{equation}
\item At the conclusion of the computation, eliminate the $(n+1)$-st row and column to obtain an $n\times n$ matrix of levels $\hat{k}_{ij}$ involving the $U(1)$'s related to the white regions $R_{1}, \cdots, R_{n}.$
\end{itemize}
In this context of checkerboard colorings, the matrix $\hat{k}_{ij}$ is known as the \emph{Goerizt form}, and the equations above provide us with an algorithmic recipe for determining CS levels.

For concreteness, let us now apply this construction to the case of the knot shown in Figure \ref{fig:8checker}.  As illustrated, there are two dual checkerboards each of which has three white regions.
Then the the $2\times 2$ level matrix for the cycles defined by regions $1$ and $2$ are given respectively by
\begin{equation}
\hat{k}_{ij}=\left(\begin{array}{cc}2 & -1 \\ -1 & 3\end{array}\right), \hspace{.5in}\mathrm{and}  \hspace{.5in} \hat{k}_{ij}=\left(\begin{array}{cc}-3 & 2 \\ 2 & -3\end{array}\right).
\end{equation}
As a consistency check on this computation, note that the two level matrices determined by the checkerboard and its dual have identical determinants.  Indeed as we have previously described, when finite, $|\det(\hat{k}_{ij})|$ computes the order of the first homology group of the cover $\widetilde{M}$, and hence is invariant to the choice of surface.

\begin{figure}[here!]
  \centering
  \includegraphics[width=.48\textwidth]{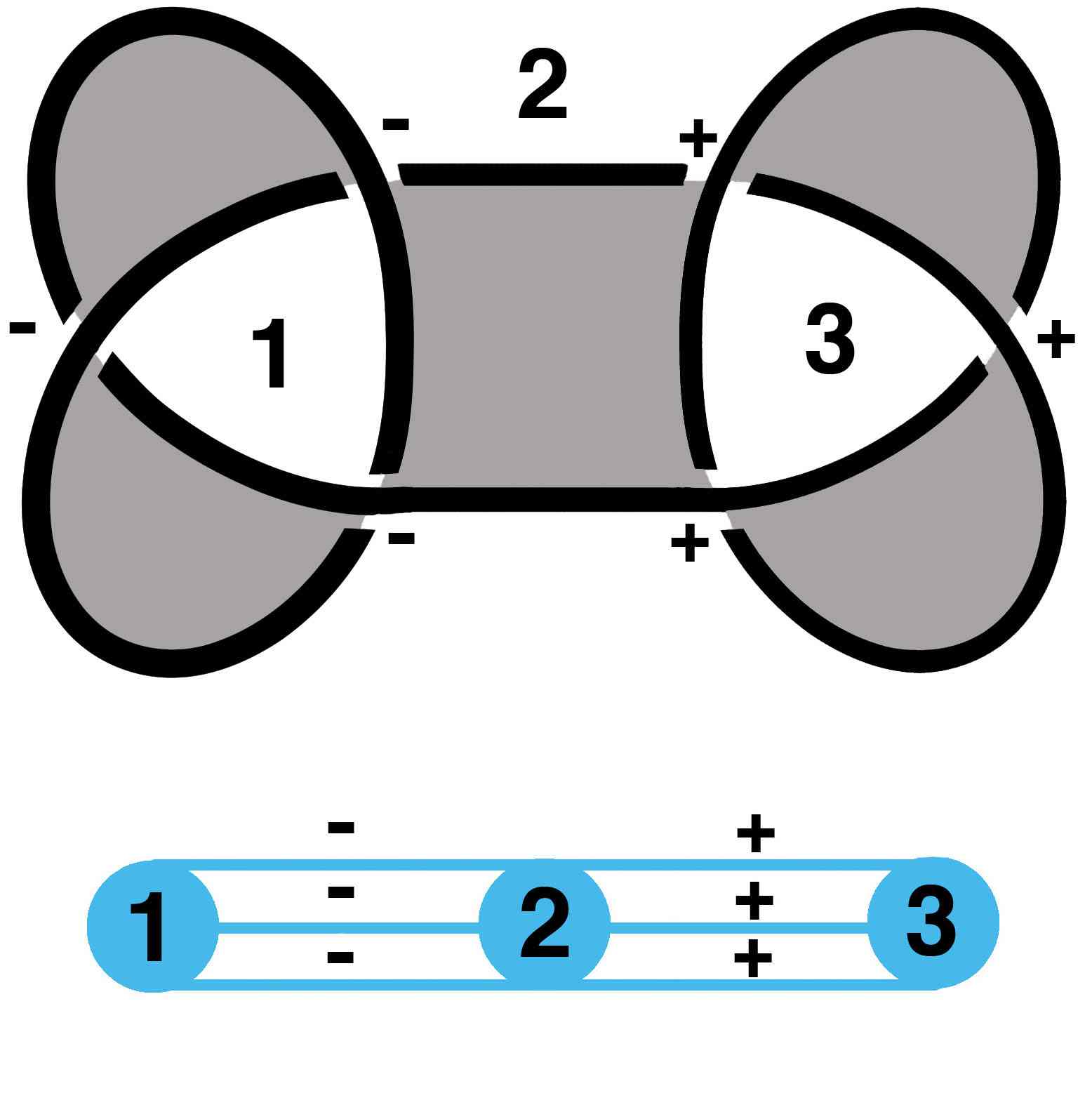}     
  \hspace{.01in}      
\includegraphics[width=0.48\textwidth]{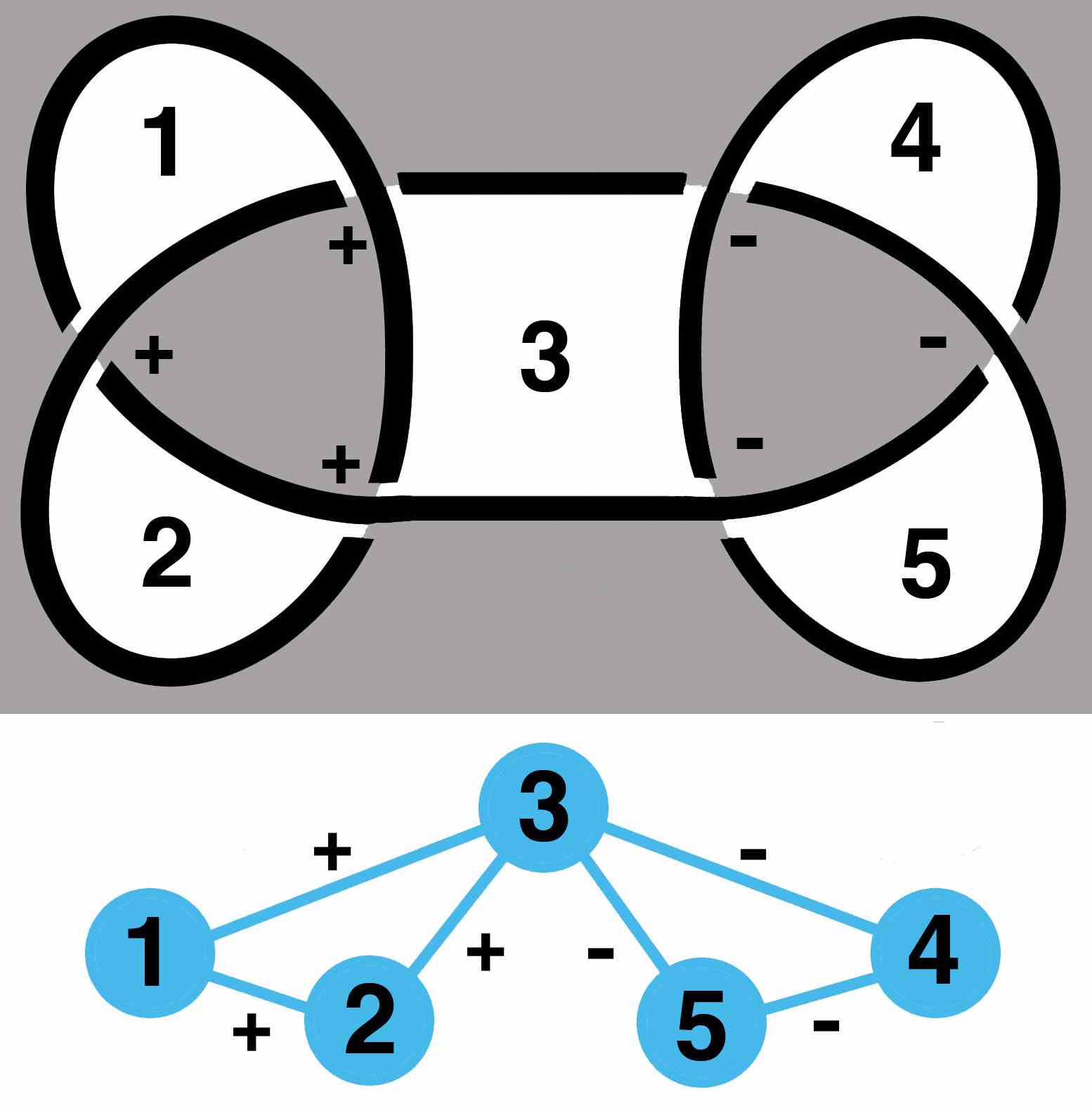}
  \caption{Checkerboards and associated Tait graphs for the square knot.  The two checkerboards are related by a black white exchange.   And correspondingly, the two Tait graphs are dual. }
  \label{fig:taitsquare}
\end{figure}

The combinatorics of checkerboard colorings can also be conveniently encoded in a so-called \emph{Tait graph}.  Given a coloring we extract the graph as follows:
\begin{itemize}
\item For each white region $R_{i}$ draw a node of the diagram.
\item For each crossing connecting white regions $i$ and $j$ connect the corresponding nodes with a link.
\item Attach a sign $\pm1$ to each link by evaluating $\zeta$ of the corresponding crossing.
\end{itemize}
As a sample application of this notion, we consider the two checkerboards of the knot illustrated in Figure \ref{fig:taitsquare}.  

Notice from this example that the black-white duality between checkerboard colorings of the knot maps to duality of the corresponding Tait graphs.  Specifically, given a graph $G$, to construct the dual $\hat{G}$ we simply:
\begin{itemize}
\item Replace each cell of $G$ with a dual vertex of $\hat{G}$.
\item Replace each link in $G$ with a dual link in $\hat{G}$.
\item Change the sign of each link relative to its dual.
\end{itemize}

The Tait graph encodes completely the gauge content of the theory.  Each node describes a white region, and hence corresponds to a cycle in the surface $F$.  Up to removing one such cycle or equivalently one node in the graph, these are exactly the $U(1)$ gauge fields.  Similarly, the CS level matrix $\hat{k}_{ij}$ is determined by summing over the links connecting nodes $i$ and $j$ weighted as in \eqref{offdiagchecker} by the sign of the link.  This structure is completely universal and depends only on the topology of the double cover $\widetilde{M}$.  In subsequent sections however, we will see examples where the Tait diagram encodes more than merely the gauge group and levels.  Indeed, in Section \ref{sec:T+B}, after determining the matter content of the theory, we will see that the Tait diagram plays the role of the quiver diagram for the gauge theory in $\mathbb{R}^{1,2}$.

\subsubsection{BPS M2-Branes, Instantons and Double Covers}
\label{BPSM2proj}
\begin{figure}[here!]
  \centering
  \subfloat[Projection of BPS Particle]{\label{fig:bpsproj1}\includegraphics[width=0.4\textwidth, height=0.3\textwidth]{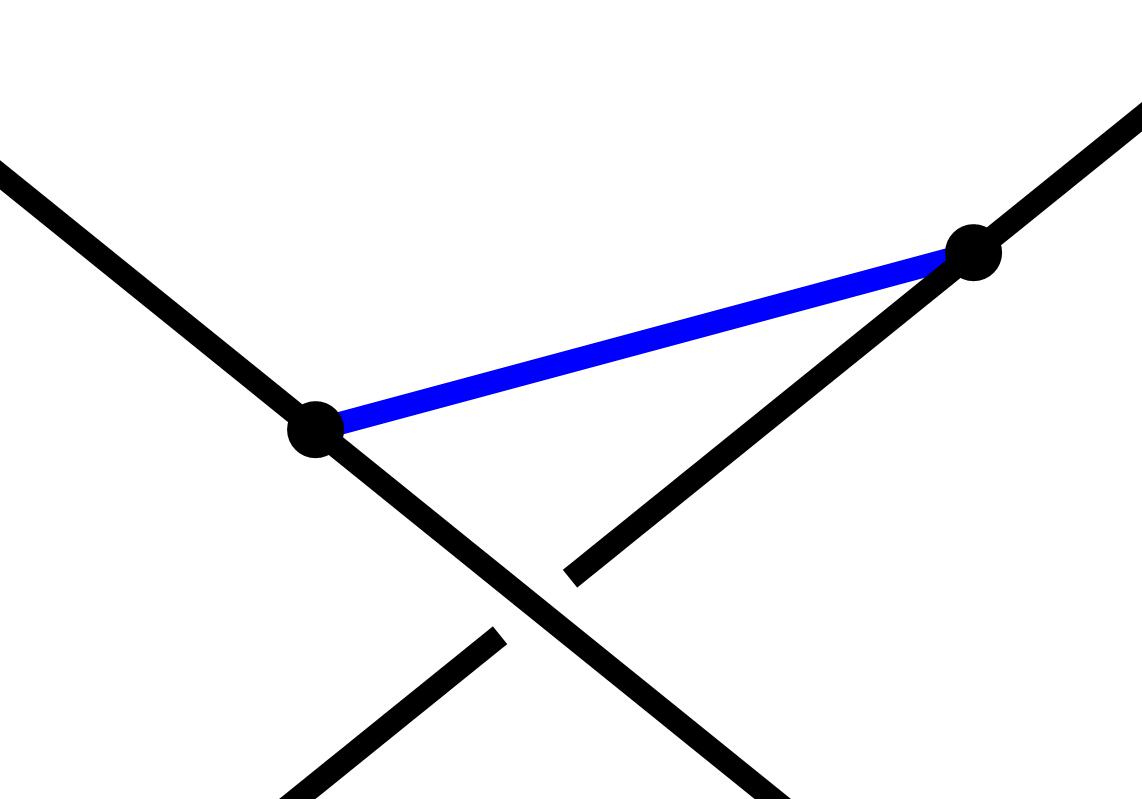}}     
  \hspace{.25in}       
  \subfloat[Lift to $Q$]{\label{fig:bpsproj2}\includegraphics[width=0.4\textwidth, height=0.3\textwidth]{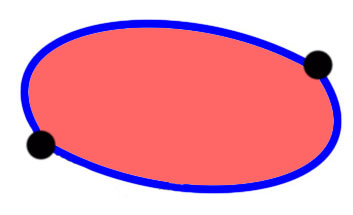}}
  \caption{Projections of BPS M2-brane particles to the base.  A portion of the branching knot $\mathcal{K}$ is shown in black.  In (a), the boundary of an M2 disc appears as a blue segment connecting two pieces of the knot.  In (b), the blue segment is doubled in $\widetilde{M}$ to make a closed cycle $S^{1}$.  The red disc, which lies in $Q-\widetilde{M}$, shows how $\Gamma$ is filled in to make the full M2-brane geometry. }
  \label{fig:m2proj}
\end{figure}

In addition to the homology of the cover, which encodes the gauge sector of the theory, there are other physical quantities of interest that can be read directly from the knot.  Of particular importance to us in later sections will be the possible BPS M2 brane geometries.  Let us begin with the case of an M2 brane disc describing a particle in three dimensions.  The boundary of this disc is a circle $\Gamma$ inside the double cover $\widetilde{M}$.  Now project $\Gamma$ to the base $S^{3}$.  If the projection is a circle then, since $S^{3}$ has vanishing homology, the cycle is contractible and the particle carries no gauge charges.  We will therefore ignore this case.  The remaining possibility is that the projection is an interval connecting two pieces of the branching knot $\mathcal{K}$ as shown in Figure \ref{fig:bpsproj1}.  Such a particle can in principle carry gauge charges depending on whether or not the cycle $\Gamma$ is non-trivial in $\widetilde{M}$.  

Geometrically, $\Gamma$ is partitioned into two pieces, one on each sheet of the cover $\widetilde{M}_{\pm}$, both of which project to the given segment in $S^{3}$.   The two segments are then glued together at their intersection with $\mathcal{K}$ which occupies both sheets.  This description also makes elementary why the mass of such a particle is determined by $\lambda$ as
\begin{equation}
m=\int_{\Gamma} |\lambda|.
\end{equation}
In this case, $|\lambda|$ is the physical height separating the two sheets $\widetilde{M}_{\pm}$ of the cover, and the above integral, by definition, computes the area of the disc illustrated in Figure \ref{fig:bpsproj2}.

\begin{figure}[here!]
  \centering
  \subfloat[Projection of BPS Instanton]{\label{fig:bpsinstproj1}\includegraphics[ width=0.4\textwidth]{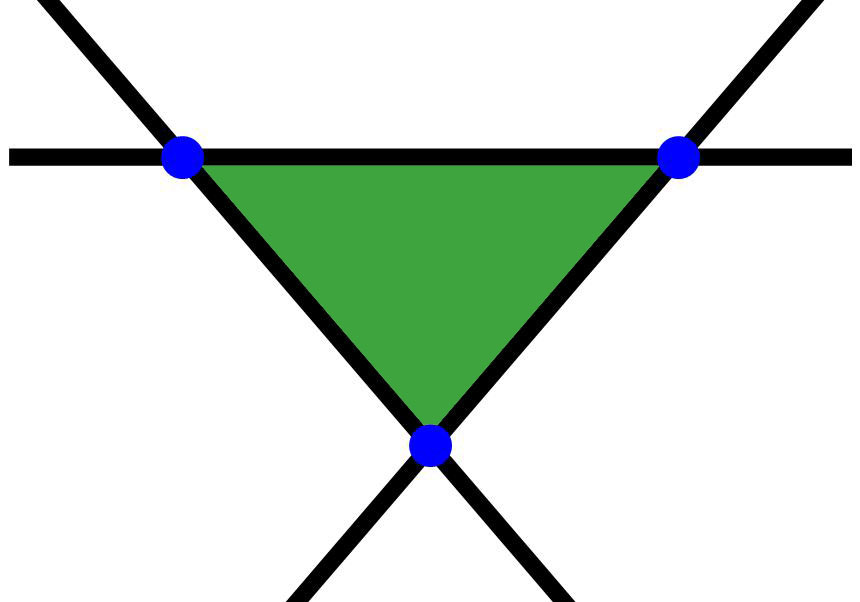}}     
  \hspace{.25in}       
  \subfloat[Lift to $Q$]{\label{fig:bpsinstproj2}\includegraphics[ width=0.4\textwidth]{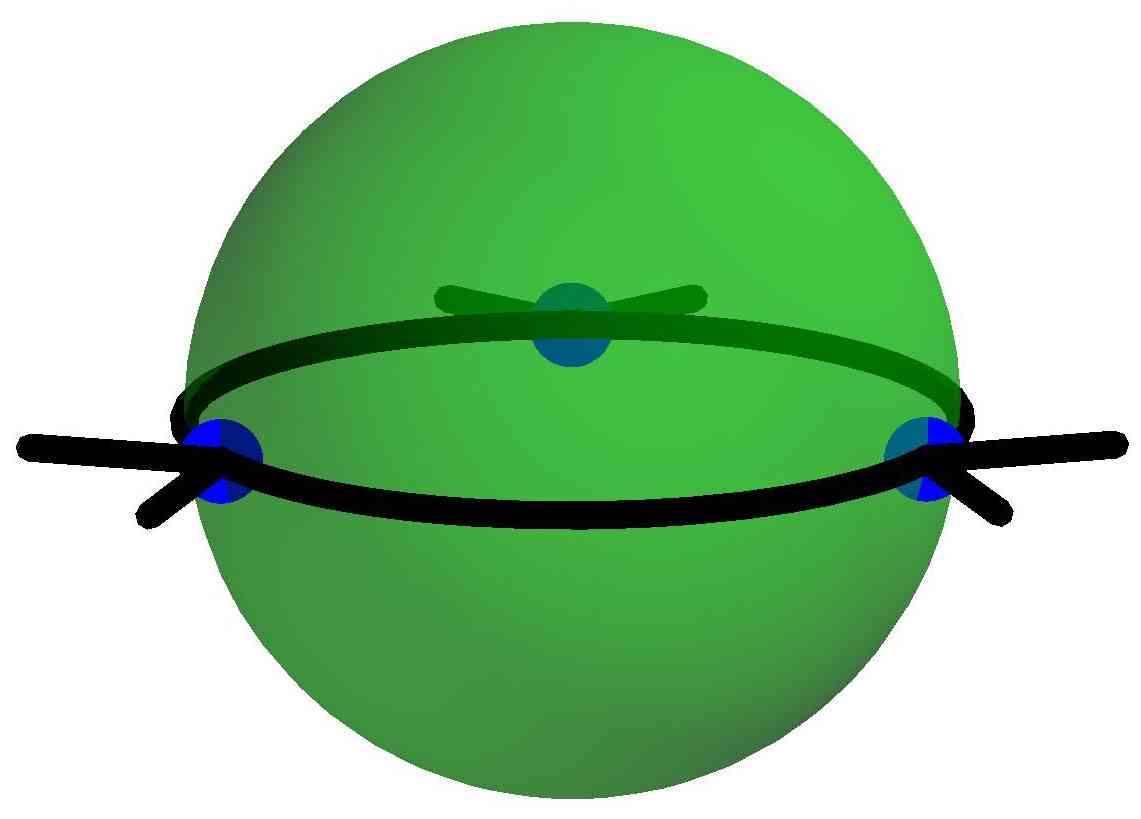}}
  \caption{Projections of BPS M2-brane instanton to the base.  A portion of the branching knot $\mathcal{K}$ is shown in black.  The knot has self-intersections supporting massless particles shown in blue.  In (a), the interior of the polygon, shown in green, is the projection to $\widetilde{M}$, of the boundary of an M2 instanton.  In (b), we see the lift of the M2 instanton to the Calabi-Yau.  Its boundary is doubled to an $S^{2}$ presented as two hemispheres glued along the knot.  In the interior, this $S^{2}$ is filled in to make a three-ball.}
  \label{fig:m2instproj}
\end{figure}

M2 brane contributions to the superpotential can also be seen from the knot diagram.   Suppose first that we consider the case of interactions among \emph{massless} particles.  According to the geometry described in the previous paragraph, this means that the segment projections shown in blue in Figure \ref{fig:bpsproj1} have all collapsed to points, and as a result the knot $\mathcal{K}$ has developed self-intersections. Let us further assume that these self-crossings form the vertices of a closed polygon whose boundary lies entirely in $\mathcal{K}$.  Then, there is an M2 brane instanton in $Q$ whose boundary projects to the polygon and which can give rise to interactions among the massless fields at the vertices as illustrated in Figure \ref{fig:m2instproj}.

To be explicit, in the Calabi-Yau $Q$, the polygon lifts to a three-ball $B^{3}$ whose boundary $S^{2}$ is $P_{\pm}$, the interior of the polygons on each of the sheets $\widetilde{M}_{\pm}$ of the cover.   These two polygons have been glued together along their common boundary in the branching knot $\mathcal{K}$
whose boundary is
\begin{equation}
\partial B^3=S^2=P_{+} \bigcup_{\mathcal{K}} P_{-}\subset {\widetilde M}.
\end{equation}
And a  wrapped M2 brane over $B^3$ leads to a superpotential term involving the product of the massless chiral fields, one for each of the vertices of the polygon.  Deforming the geometry and making the chiral fields massive, resolves the self-crossings of the knot diagram.  This can be described by an M2 brane instanton, whose boundary will also include a disc `plug' for each massive chiral field as in the general description of Figure \ref{fig:m2w}.

\section{R-flow, Domain Walls and a 4d-3d Link}
\label{sec:4d3dL}
The geometry described in the previous section gives an abstract prescription for extracting the effective three-dimensional $\mathcal{N}=2$ system that arises when multiple five-branes wrap a three-manifold.  We first determine the IR covering geometry $\widetilde{M}$, then we compute the spectrum of BPS M2-brane particles which give rise to chiral multiplets, and finally determine their interactions from the various M2-brane contributions to the superpotential.  However, in practice it may be difficult to carry out this procedure.  The first difficulty is that we have no general method for determining $\widetilde{M}$, or equivalently the branching knot $\mathcal{K}\subset M$.  Moreover, even if the topology of $\widetilde{M}$ were deduced, this only suffices to describe the gauge groups and flavor symmetries but not the BPS states.  In the infrared, it is the BPS states which describe the charged chiral multiplets of the theory, and thus extracting the spectrum of these particles is a crucial step in determining the behavior of the quantum field theory.  

In principle, the BPS states are completely encoded by the one-form $\lambda$ on the cover.  Indeed, given that $\lambda$ defines the local central charge density, it follows that the boundary of a BPS M2 brane is an integral curve of the flow defined by $\lambda$.  In equations, if $\gamma$ denotes this boundary one-cycle and $s$ is a local coordinate on $\gamma$, then the BPS condition requires
\begin{equation}
\lambda|_{\gamma}=ds. \label{bpsstring}
\end{equation}
For example, a chiral field arises as a solution to the above equation whose endpoints are on the branch knot as described in the previous section.  This is the three-dimensional analog of the flow equation defining BPS states in 4d $\mathcal{N}=2$ theories \cite{SelfDual}.  However, without explicit knowledge of $\lambda$, there is no method for determining the BPS particles in the theory and hence no way of directly fixing the IR behavior of the model. 

\begin{figure}[here!]
  \centering
  \subfloat[$t>>0$]{\label{fig:zflow1}\includegraphics[width=0.23\textwidth]{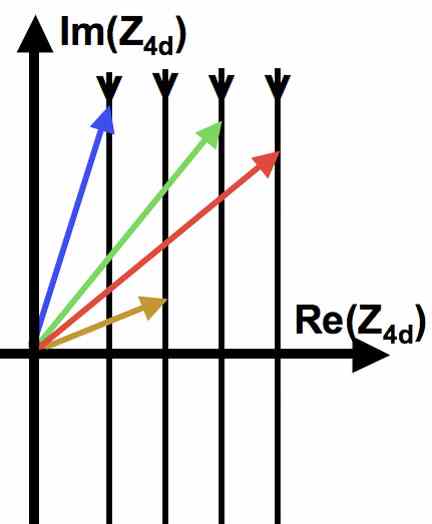}}     
  \hspace{.1in}       
  \subfloat[$t>0$]{\label{fig:zflow2}\includegraphics[width=0.23\textwidth]{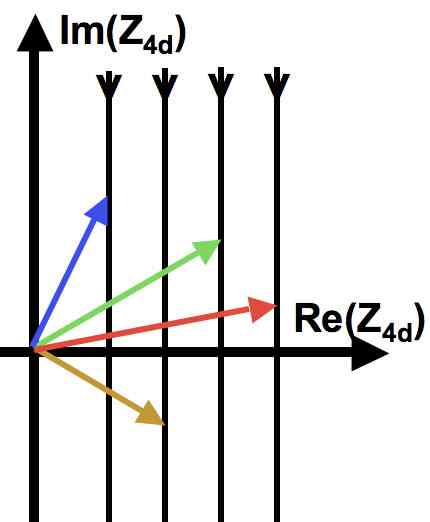}}     
  \hspace{.1in}       
  \subfloat[$t<0$]{\label{fig:zflow2}\includegraphics[width=0.23\textwidth]{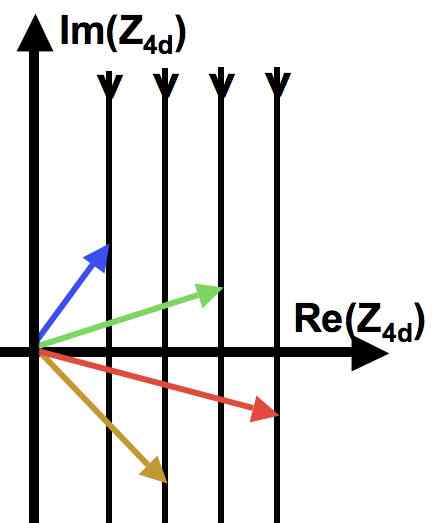}}    
  \hspace{.1in}       
  \subfloat[$t<<0$]{\label{fig:zflow2}\includegraphics[width=0.23\textwidth]{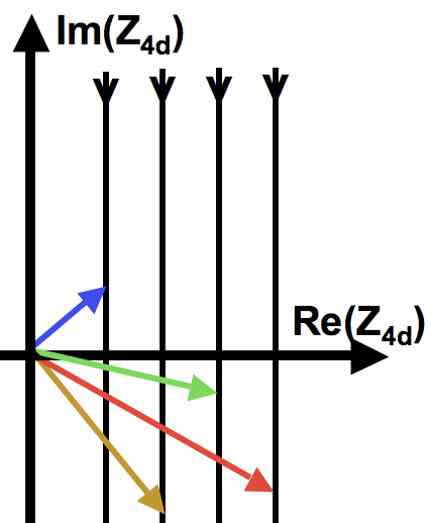}}    
  \caption{Evolution of the cetral charges along an R-flow.  The colored rays denote the central charges of the 4d theory.  As we move along the flow, these rays flow along parallel lines and maintain their phase order.}
  \label{fig:zflow}
\end{figure}

In this section, we introduce our main technique for circumventing these difficulties.  We consider the special case where the M5-branes are related to a flow of a parent 4d theory, and use the knowledge of the spectrum of the 4d BPS states and Seiberg-Witten geometry to find the answer for the resulting 3d theory.  Let us recall that in the 4d case, instead of \eqref{bpsstring}, the BPS spectrum is determined by integral curves of the Seiberg-Witten differential
\begin{equation}
\phi|_{\gamma}  =e^{i\theta}ds, \label{4dm2flowline}
\end{equation}
where there is a solution only if $\theta $ is chosen to equal the phase of the central charge $Z$ of the BPS state
\begin{equation}
Z=|Z| e^{i\theta}.
\end{equation}
Now let us consider the 3d case.  Our ansatz, of viewing the 3d theory as a one-parameter variation of the 4d theory, implies that we are studying a domain wall, where roughly
\begin{equation}
\lambda= e^{i\theta(t)}\, \phi+c.c. +\dots \label{RT}
\end{equation}
In this way, we will have solutions to the 3d BPS  equation \eqref{bpsstring} at specific `times' $t_i$ during the one-parameter variation, such that $\theta(t_i)=\theta(\mathrm{BPS})$ for some BPS particle. This is exactly the characteristic feature of the time evolution defined by R-twisting \cite{Cecotti:2009uf, CNV}.  However, there is one important difference in our context:  unlike the case in R-twisting where the central charges rotate in phase as we evolve in time, in order to preserve a standard 3d supersymmetry, we need to make the central charges {\it flow along parallel lines}.  This can be achieved by the suitable choice of the $\dots$ terms in equation \eqref{RT}, as will be discussed later in this section.  The result, which we shall call the `R-flow,' is a one-parameter variation of the parent 4d theory, characterized in terms of the variation of the 4d central charges $Z_{i}$ by two simple features:
\begin{itemize}
\item The real part of the $Z_{i}$ is constant along the flow.
\item Along the flow, the $Z_{i}$ retain their phase order.
\end{itemize}
These features are shown in Figure \ref{fig:zflow}.  
As we will see, these two properties mean that the time evolution in the R-flow respects the BPS spectrum of the 4d theory, and ultimately implies that the 3d BPS spectrum of chiral fields living on the domain wall is in one-to-one correspondence with the BPS spectrum of the ambient 4d theory.

Finally, a crucial aspect of the domain wall construction is the observation that, in general, such a domain wall field theory couples non-trivially to the bulk four-dimensional physics.  However, our interest is in constructing honest three-dimensional theories which have an independent existence.  Thus, in addition to the construction of the domain walls, we must also take a decoupling limit in which the four-dimensional theory decouples and the three-dimensional theory on the wall remains.  This decoupling limit amounts to a specification of boundary conditions for the R-flow, where as $|t|\rightarrow \infty$ we also have $|Z| \rightarrow \infty$.  Then, the full trajectories of the 4d central charges during an R-flow are infinite parallel lines.
  
\subsection{Domain Wall Geometry}
\label{sec:WallGeo}

\begin{figure}[here!]
  \centering
 \includegraphics[width=0.6\textwidth]{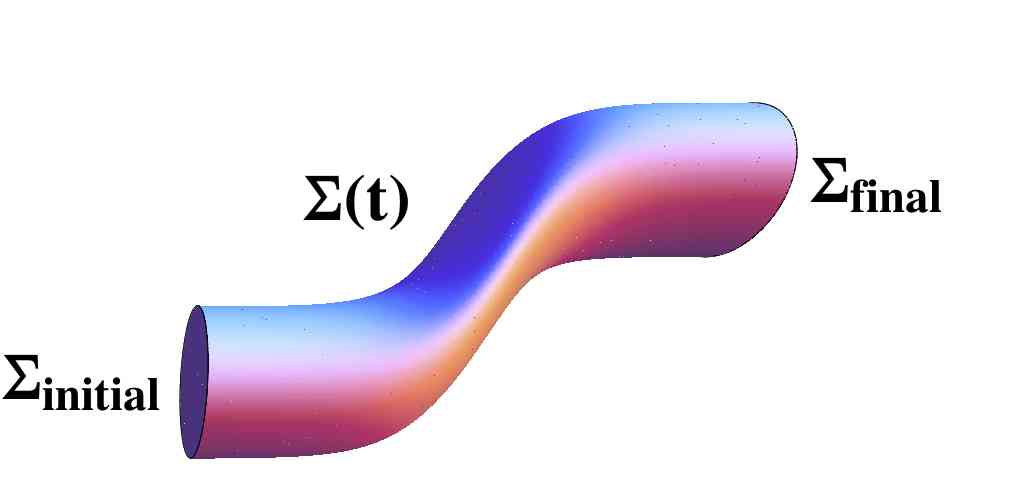}     
  \caption{The domain wall geometry.  Asymptotically, the five-branes wrap the Riemann surfaces $\Sigma_{\mathrm{initial}}$ and $\Sigma_{\mathrm{final}}$ an give rise to theories with four macroscopic dimensions.  In the interior, the parameters of $\Sigma$ vary and describe a domain wall.}
  \label{fig:wall}
\end{figure}

Consider any number five-branes which wrap a Riemann surface $\Sigma$ and give rise to a four-dimensional $\mathcal{N}=2$ gauge theory.  The local geometry is then
\begin{equation}
\mathbb{R}^{1,4}\times T^{*}\Sigma \times \mathbb{C}.
\end{equation}
To engineer a macroscopically four-dimensional system we choose linear subspaces $\mathbb{R}^{1,2}\subset \mathbb{R}^{1,4}$ and $\mathbb{R}\subset \mathbb{C}$ and place the five-branes on 
\begin{equation}
\mathbb{R}^{1,2}\times \Sigma \times \mathbb{R}.
\end{equation}
Such a geometry supports a natural class of defects which describe domain walls.  The linear subspace $\mathbb{R}\subset \mathbb{C}$ is replaced by a non-trivial path $\gamma(t)$.  Asymptotically for $t\longrightarrow \pm \infty$  this path approaches horizontal asymptotes, and combines with the fixed $\mathbb{R}^{1,2}$ dimensions to make a macroscopically four dimensional theory described by five-branes on $\Sigma(\pm \infty)$.  However in the interior of the path there is a non-trivial kink along which we allow the parameters describing the Riemann surface, and hence the parameters of the four-dimensional field theory, to vary, $\Sigma \rightarrow \Sigma(t)$.  Since this defect is codimension one in space it describes a domain wall in the four-dimensional $\mathcal{N}=2$ system \cite{Gaiottowall}.  The total geometry is illustrated in Figure \ref{fig:wall}.

In terms of the geometry of the previous section, we can phrase the domain wall construction as follows.  The three-manifold, which supports the five-branes is $\Sigma \times \mathbb{R}$.  We coordinatize $\mathbb{R}$ by $t$ and loosely refer to as ``time," and we allow the parameters describing the Riemann surface $\Sigma$ to vary with $t$.   The asymptotic boundaries of the three-manifold $M$, namely $\Sigma \times \{-\infty\}$ and $\Sigma \times \{+\infty\}$ encode the fact that this domain wall theory does not have an independent existence but couples to the bulk four-dimensional theory.

Now, if the variation of parameters of the Riemann surface $\Sigma$ is done in an arbitrary way, then the domain wall will break all the supersymmetry of the problem.  If we wish to preserve 3d $\mathcal{N}=2$ supersymmetry, then the domain wall should be half-BPS, and the supersymmetries preserved in 3d are embedded inside the 4d $\mathcal{N}=2$ superalgebra as a subalgebra.  Such 3d $\mathcal{N}=2$ subalgebras are labeled by a choice of angle.  Note that this also matches the central charge structure.   A 3d theory with ${\cal N}=2$ supersymmetry has a real central charge.  To get a reduction from ${\cal N}=2$ theory in 4d, which has a complex central charge, to the one in 3d, with a real central charge, we must choose a `real' subspace in the 4d complex central charge plane.   Let us choose this direction to correspond to the real axis in the complex plane of the 4d central charges.  Then, the four-dimensional and three-dimensional central charges obey by the relation
\begin{equation}
Z^i_{3d}={\rm Re}(Z^i_{4d}). \label{zrel}
\end{equation}

In terms of the IR Coulomb branch geometry, the domain wall construction means that there is a relationship between the SW curve $\widetilde{\Sigma}$ of the parent 4d $\mathcal{N}=2$ model and the IR three-manifold $\widetilde{M}$.  Specifically, $\widetilde{M}$ is a one parameter thickening of the SW curve
\begin{equation}
\widetilde{M}=\widetilde{\Sigma}(t)\times \mathbb{R}_{t}. \label{newthick}
\end{equation}
This means that every non-trivial one-cycle $\Gamma$ in $\widetilde{M}$ is inherited from $\widetilde{\Sigma}$.  As a result, \eqref{zrel} yields a simple relationship between the periods of the Seiberg-Witten defferential $\phi$ on the Seiberg-Witten curve $\widetilde{\Sigma}$,  and the periods of the harmonic one-form $\lambda$ on $\widetilde{M}$

\begin{equation}
\int_{\Gamma} \lambda =\Re\!\left(\int_{\Gamma}\phi\right).
\end{equation}

In the primary case of interest in this paper, the parent 4d theory can be described by two M5-branes on $\Sigma$.  Hence, the SW curve $\widetilde{\Sigma}$ is a branched double cover of $\Sigma$, where the branch points of the cover are exactly the zeros of the SW differential.  In this case, the presentation \eqref{newthick} of the IR three-manifold implies that $\widetilde{M}$ is a branched double cover of $\Sigma \times \mathbb{R},$ where the branch locus is exactly the one-dimensional \emph{strands} swept out by the zeros of the SW differential during the time evolution.  This fact will be of crucial importance to us throughout the remainder of this work. 

\subsubsection{An Elementary Example}

Let us now turn to the most basic example of this construction.  We consider an ${\cal N}=2$ theory in 4d which is the theory of a free massive hypermultiplet.  We can model this geometrically as above by taking $\Sigma$ to be simply the complex plane with coordinate $x$ and placing a pair of five-branes there with suitable boundary conditions at infinity.  Then the Seiberg-Witten geometry is given by the following curve and differential
\begin{equation}
y^2=x^2-m, \hspace{.5in}\phi=y\,dx.
\end{equation}
In the above, the function $y(x)=\pm \sqrt{ x^2-m}$ describes the separation between the two branes.  The sign ambiguity in $y(x)$ is consistent with the fact that the two five-branes are indistinguishable.  At $y(x)=0$ 
the two branches of the function $y(x)$ exchange and hence the two M5 branes connect up into a single smooth object.  This is consistent with the general geometry described in section 2: the IR Coulomb branch physics is governed by the geometry of a single smooth five-brane related to the UV description by brane recombination.  In this case, the recombined brane is an infinite cylinder which is a branched double cover of the complex plane.

The BPS hypermultiplet of the theory can also be seen from the general analysis in section 2.  The non-trivial one-cycle $\Gamma$ in the cylinder, describes a charge in the 4d IR physics.  However in the ambient Calabi-Yau geometry, this cycle is contractible.  Physically this means that there is an M2 brane disc which ends on the cycle $\Gamma$.  The boundary of the disc is a circle made up of two halves, each half corresponds to an interval on each of the two M5 branes stretched between the two branch points.  It gives rise to a BPS particle with central charge $m$ in four-dimensions.

\begin{figure}[here!]
  \centering  
\includegraphics[width=0.3\textwidth]{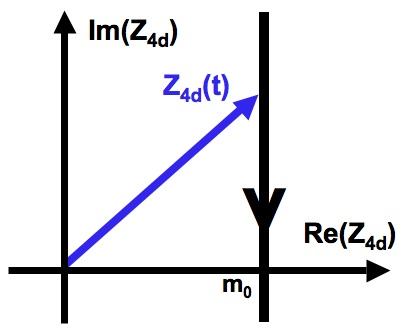}
  \caption{The flow of the 4d central charge for the $A_{1}$ theory.  The blue line indicates the trajectory of $Z_{4\mathrm{d}}(t)$ shown in blue.  The 3d central charge $m_{0}$, is the real part of $Z_{4\mathrm{d}}$.}
  \label{fig:a1zflow}
\end{figure}

Now we would like to construct a domain wall in this theory by considering a one-parameter family of these SW geometries.  Thus, we let $m$ vary as a function of a parameter $t$ as
\begin{equation}
m(t)=m_0 +it
\end{equation}
Further, we take $m_{0}$ to be real, and this will be the resulting three-dimensional central charge.  The flow of $Z_{4\mathrm{d}}$ is illustrated in Figure \ref{fig:a1zflow}.

As a result of this one-parameter variation, the UV description of the theory is two five-branes which wrap the three dimensional space $(x,t)$.  In the IR, the theory is described by a single five-brane described as a branched double cover over $(x,t)$ and given by the equation 
\begin{equation}
y^2=x^2-(m_0+it).
\end{equation}
Further, the one-form $\lambda$ and the SW differential are related as
\begin{equation}
\lambda =y(t)dx+\overline{y}(t)d\overline{x}+f dt.
\end{equation}
Where $f$ denotes separation of the M5 branes in the cotangent direction to $t$ and is chosen so that $d\lambda =0$.  Notice that this satisfies the key requirement \eqref{zrel} for preserving three-dimensional $\mathcal{N}=2$ supersymmetry, namely the periods of $\lambda$ over one-cycles at constant time $t$ are simply the real parts of the periods of $\phi,$ and hence the three-dimensional central charge is simply $m_{0}$.

\begin{figure}[here!]
  \centering
\subfloat[$m_{0}>>0$]{\includegraphics[width=0.15\textwidth, height=0.15\textwidth]{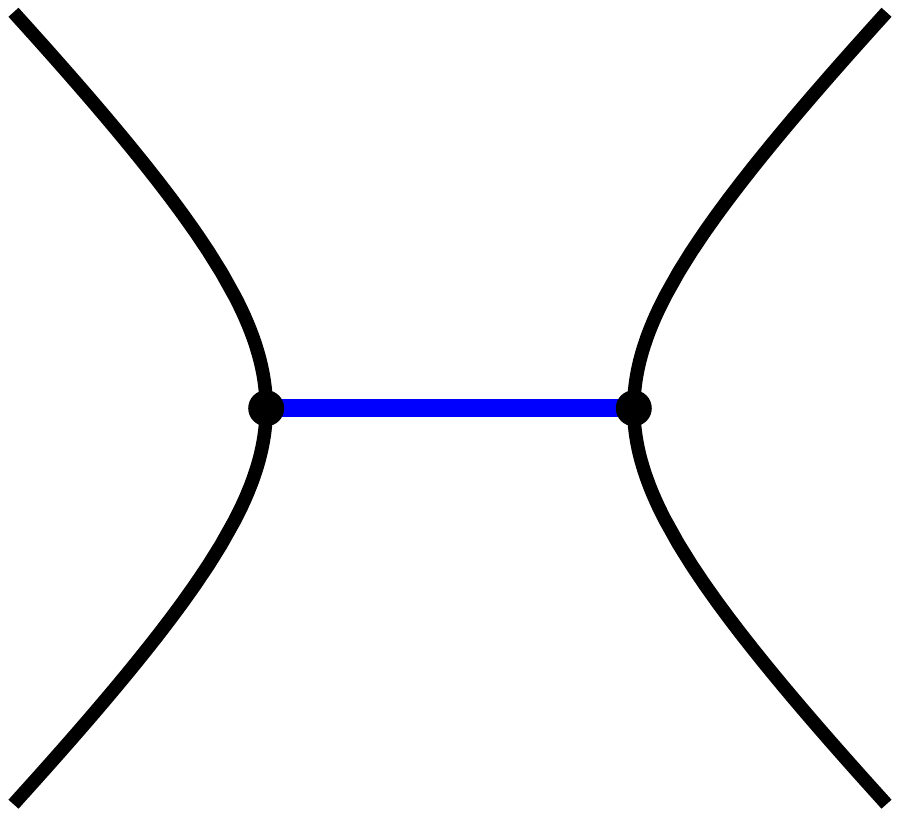}}
   \hspace{.15in}          
\subfloat[$m_{0}>0$]{\includegraphics[width=0.15\textwidth, height=0.15\textwidth]{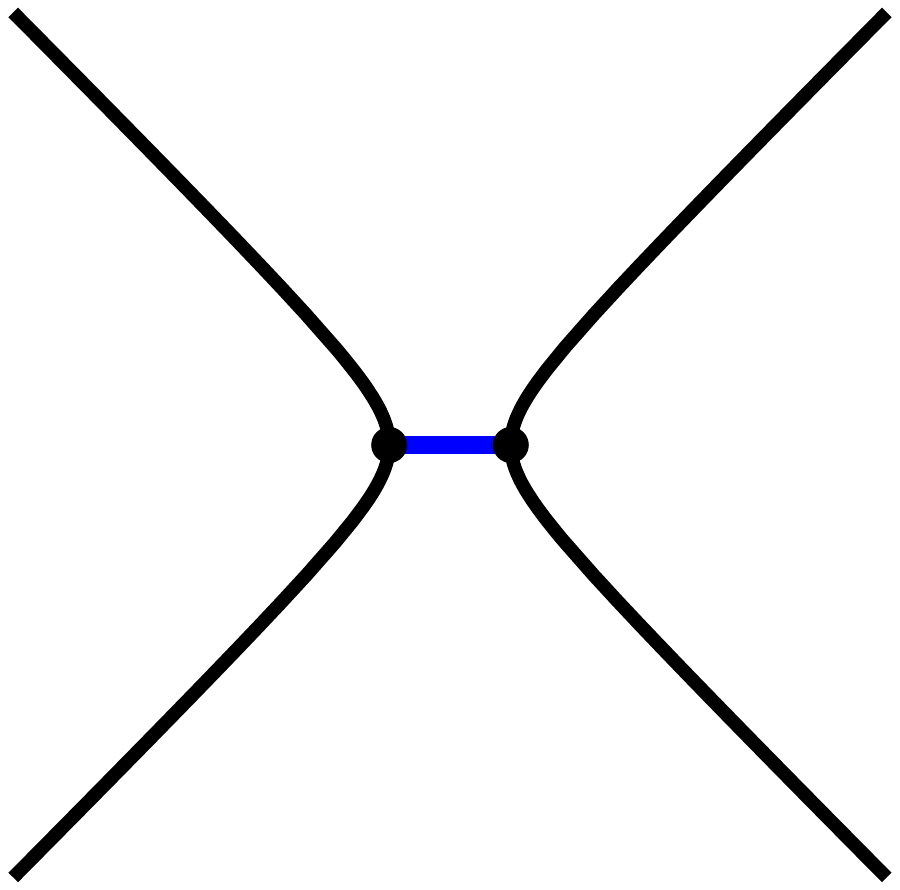}}
   \hspace{.15in}     
\subfloat[$m_{0}=0$]{\includegraphics[width=0.15\textwidth, height=0.15\textwidth]{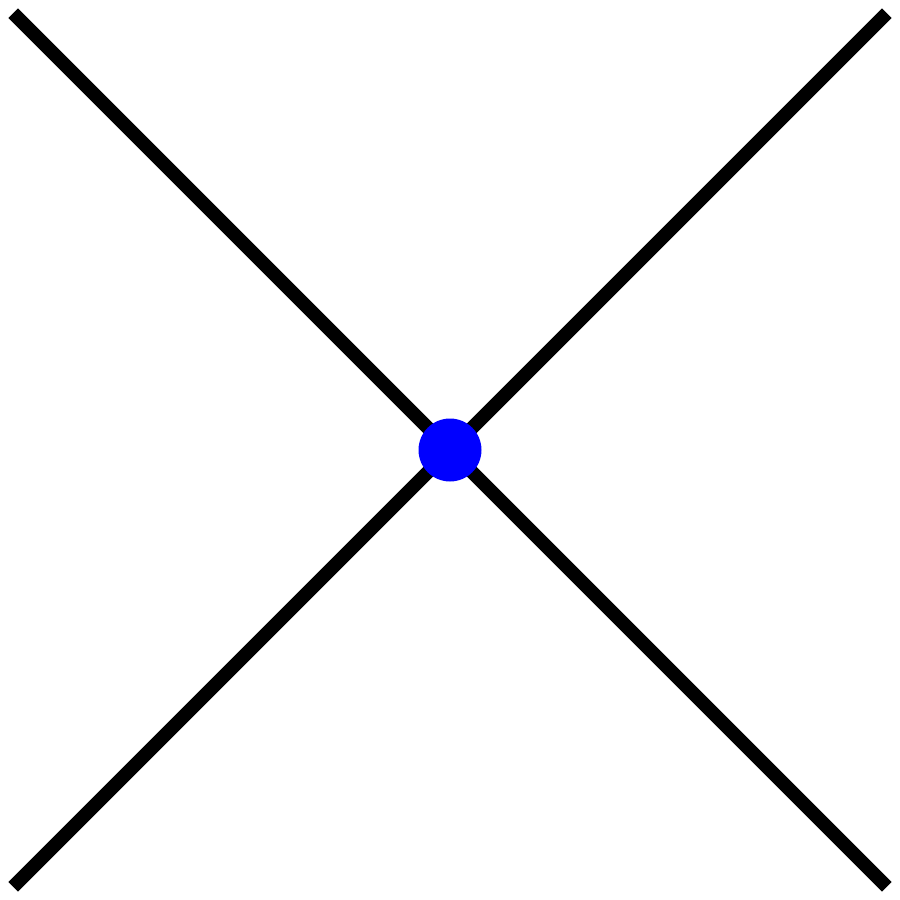}}
   \hspace{.15in}             
\subfloat[$m_{0}<0$]{\includegraphics[width=0.15\textwidth, height=0.15\textwidth]{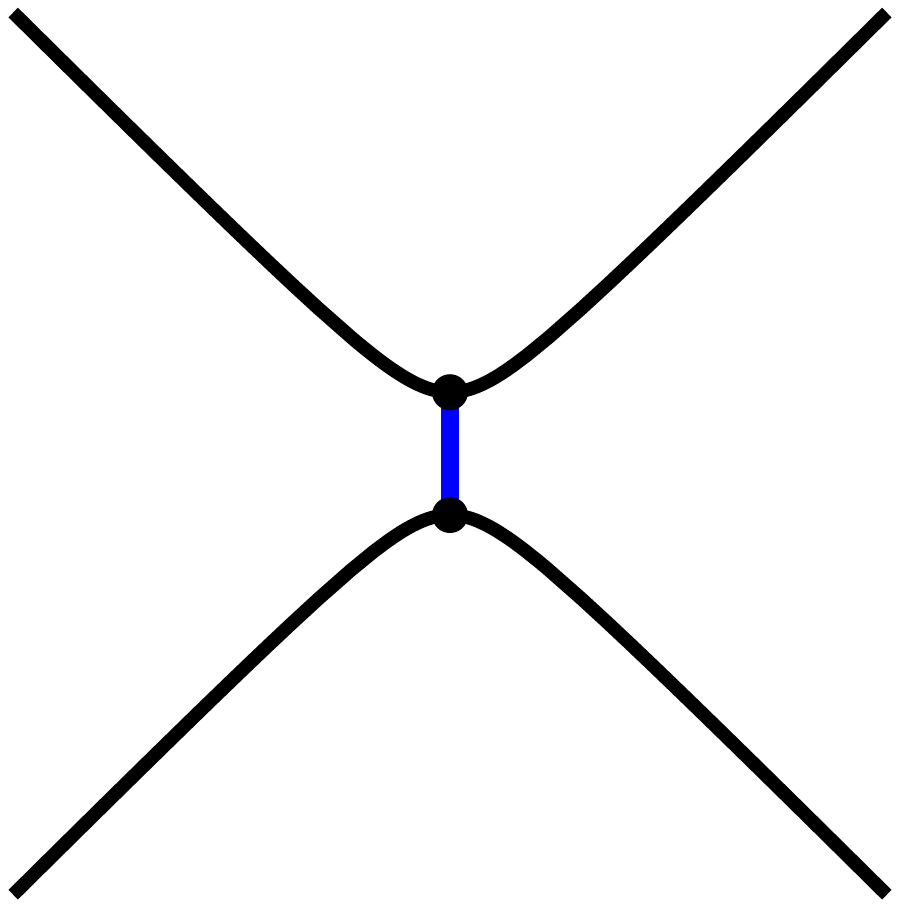}}
   \hspace{.15in}     
   \subfloat[$m_{0}<<0$]{\includegraphics[width=0.15\textwidth, height=0.15\textwidth]{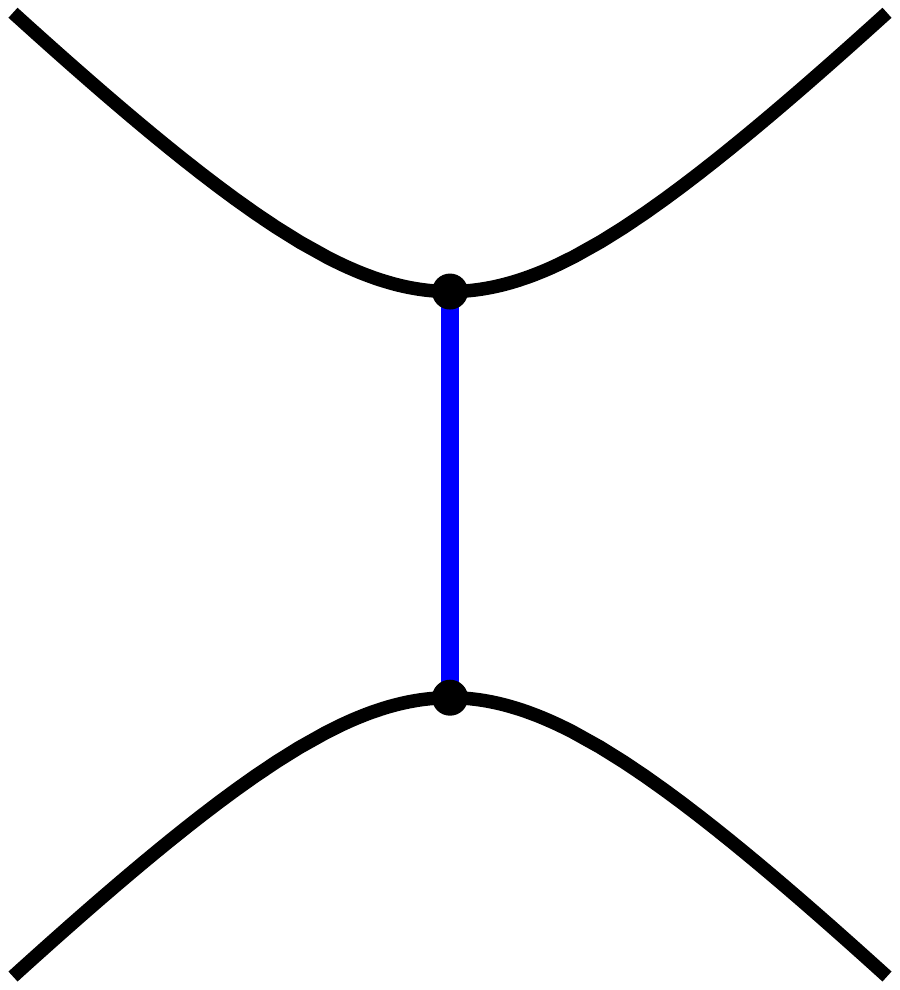}}
  \caption{The reconnection process.  The strands are illustrated in black and the blue line indicates the projection of the boundary of the BPS M2 brane.   In (c), when the mass $m_{0}$ of the particle vanishes, the two strands touch and their individual identity is ambiguous.  As $m_{0}$ becomes negative the strands reconnect.}
  \label{fig:recon}
\end{figure}

An important fact is that already in this simple example we can see topology changing transitions occurring in the IR geometry as 3d parameters are varying.  Specifically, consider the branch loci of the cover.  These are given by the two curves
\begin{equation}
x_{\pm}=\pm \sqrt{m_0+it}
\end{equation}
Note that when $m_0=0$ the two branch lines meet at $(x,t)=(0,0)$.  Also note that the branch lines have reconnected as $m_0$ goes from positive values to negative.  This reconnection is illustrated in Figure \ref{fig:recon}.
  In terms of the IR geometry the topology of the cover is jumping as $m_{0}$ passes through zero.  In later sections we will interpret these topology changes in terms of mirror symmetries. 

\subsubsection{General Flows}

Let us return to the general discussion of domain walls.   We start with an $\mathcal{N}=2$ theory in four dimensions given by a SW geometry.  This can be an arbitrary ${\cal N}=2$ theory, and not necessarily one arising from M5 branes.  However, for simplicity of exposition here we describe it for the case of M5-branes.  We start with an IR M5-brane geometry in 4d given by the Seiberg-Witten curve $\widetilde{\Sigma}$. We then consider a one real parameter family of these theories to yield the IR three-manifold $\widetilde{M}$.  The real one-form $\lambda$ and the SW differential are related as
\begin{equation}
\lambda =\phi (t)+ {\overline \phi(t)} + g dt 
\end{equation}
We require that $\lambda$ is closed, which in particular requires
\begin{equation}
d\lambda =0 \rightarrow \Big({d\phi\over dt}+c.c.\Big)dt+ d(g\, dt) =0.
\end{equation}
This means that $d( \phi+{\overline \phi})/dt$ is cohomologically trivial, and hence has no periods on the SW curve, which in turn implies that the periods of ${\rm Re}(\phi)$ do not change with $t$.  Thus to preserve supersymmetry, all the central charges will have to move along straight vertical lines as we move through the flow parameter $t$.  Can this be arranged?  

In general the answer is simply ``no.''   The various central charges of the 4d theory will be related in an intricate way determined by special geometry.  In particular they are not independent parameters and hence there is no reason to believe that they can be varied in any particular prescribed way.  However, there does exist a class of $\mathcal{N}=2$ theories, the so-called \emph{complete theories} which have exactly enough moduli and coupling constants to treat the central charges as independent parameters.  Further, except for a finite number of  exceptional cases, all complete $\mathcal{N}=2$ theories can be described by two M5-branes wrapping a punctured Riemann surface with suitable boundary conditions at the punctures \cite{CV11}.  Our primary examples will be in the case involving \emph{two} five-branes where we can preserve 3d $\mathcal{N}=2$ supersymmetry via a flow of 4d central charges in straight vertical lines.  However, in section 8 we also discuss exceptional complete theories which are not M5-brane theories.

For non-complete theories, we cannot vary the central charges arbitrarily if we wish to maintain having a UV-complete theory.  However, even for non-complete theories the central charges can vary arbitrarily if one does not insist on a UV completion and views them as effective theories which are UV incomplete, but can in principle be embedded in a UV complete theory.  An example of this is pure $SU(N)$ gauge theory, where the non-renormalizable terms $\mathrm{tr}\,\Phi^k$ with $k>2$ can in principle be generated if $SU(N)$ is embedded in a bigger UV complete theory.

In any case, our primary examples in this paper will all be associated to domain walls in theories described by pairs of five-branes wrapping punctured Riemann surfaces.  In fact our main focus discussed in section 5 will all be generalizations of the free hypermultiplet theory, where the UV five-brane Riemann surface is again the complex plane $\mathbb{C}$.  In that case, the total internal Calabi-Yau threefold is simply $\mathbb{C}^{3}$ with its standard symplectic, and holomorphic structure.  The abstract flow of Riemann surfaces studied in this section is then a specific instance of a Joyce-Harvey flow construction \cite{Joyceflow1, Joyceflow2,harvey} of special Lagrangians.\footnote{\ The general structure of this flow equation is as follows.  We consider a Riemann surface $\Sigma$ and a one parameter family of real analytic embeddings $\psi_{t}:\Sigma \longrightarrow \mathbb{C}^{3}$. Then given any positive bivector $X$ on $\Sigma$ one studies the following flow equation
\begin{equation*}
\frac{\partial \psi^{i}_{t}}{\partial{t}}=g^{ij}(\psi_{t *}X)^{kl}\Re\left(\Omega\right)_{jkl}.
\end{equation*}  
It is then a fact that if the symplectic form vainishes on the initial surface $\psi_{0}(\Sigma)$ then the three-manifold swept out by $\Sigma$ as one varies through time is special Lagrangian.} These equations turn out to be difficult to solve.  Luckily, many features of what we need are independent of the detailed solution.

For another perspective on the domain wall geometry we can ask for the dual description for these theories in type IIB.  The dual to an M5 brane is a local ALE fibration of the form
\begin{equation}
uv=P(x,y,t). \label{g2}
\end{equation} 
Where in the above the equation $P(x,y,t)=0$ defines the locus of five-branes in the original geometry and as $t$ varies describes a one-parameter family of SW geometries.  Abstractly, the equation \eqref{g2} defines a  a one-parameter family of Calabi-Yau threefolds, and supersymmetry demands that the resulting seven-dimensional total space have $G_{2}$ holonomy.  Then, the one-form $\lambda$ we have discussed is promoted to the three-form $\rho$ which determines the $G_{2}$ structure.  This three-form fixes the metric completely and hence satisfies the equation
\begin{equation}
d \rho =d *_{\rho}\rho=0.
\end{equation}  
These are the analogs of the harmonicity of the one-form $\lambda$.  If we fix the boundary conditions for the flow, then the $G_{2}$ version of Yau's theorem implies that the metric is characterized completely by the three-dimensional real central charges.

\subsection{Decoupling Limits and R-Twisting}
\label{sec:Decoupling}
In the previous section we have described a class of domain walls which exist in four-dimensional $\mathcal{N}=2$ systems described by five-branes on Riemann surfaces.  Such domain walls are characterized by the fact that the flow of the 4d central charges is on vertical straight lines.  In general such walls will have complicated interactions with the ambient 4d field theories.  In this section, we take the key step of decoupling the bulk physics leaving only the remaining 3d $\mathcal{N}=2$ system.  In the process, we see how the domain wall geometries described in this section can be interpreted in terms of R-twisting.

The most important observation is simply the BPS bound in the bulk 4d theory.  This states that all charged particles have a mass $m$ which satisfies
\begin{equation}
m\geq |Z_{4\mathrm{d}}|. \label{4dbps}
\end{equation}
Consider this bound applied to the bulk 4d theories living at the endpoints of the flow defining the domain wall.  To decouple the 4d charged particles from the low-energy physics, it suffices to make them infinitely massive.  On the other hand according to the BPS bound \eqref{4dbps} this will be achieved, provided that the initial and final central charges of the flow have infinite length.  Thus, decoupling demands that for both the initial and final condition
\begin{equation}
|Z^{i}_{4\mathrm{d}}|\longrightarrow \infty. \label{decoup1}
\end{equation}
Since the flow demands that the 4d central charges evolve along straight lines, the above equation implies that the trajectories in the complex $Z_{4\mathrm{d}}$ plane swept out by the central charges during the flow are infinite vertical lines which cross the real axis at the values dictated by the three dimensional real central charges.

If these boundary conditions for the flow are satisfied then all 4d charged particles have infinite mass in the bulk and decouple from the domain wall.  However the massless 4d neutral gauge multiplets remain unaffected by this limit.  For these fields, which have independent $U(1)$ coupling constants, we are free to chose their three-dimensional physics. We can take these coupling constants to be finite in which case we are left with dynamical gauge field in three dimensions, or we may dial these constants to zero in which case the resulting $U(1)$ appears as a flavor symmetry in three dimensions.  Thus, what the decoupling limit \eqref{decoup1} naturally produces is in fact a class of, in general distinct, 3d theories labeled by a choice of whether the $U(1)'s$ are gauged or ungauged.  We will examine this freedom in detail in section 5.  It turns out, that there are some additional $U(1)$'s coupled to the chiral fields which are massed up by Chern-Simons terms, but are nevertheless necessary for describing the full content of the theory.

Finally, we come to a crucial ansatz of our theory of domain walls.  We have succeeded in producing a decoupled 3d $\mathcal{N}=2$ theory, but so far there is no simple relationship between the spectrum of BPS chiral multiplets on the wall and the spectrum of the bulk four-dimensional theory.  As the central charges of the 4d theory flow in general they flow at different speeds and cross each other at various times.  Such crossings lead to the wall-crossing phenomenon.  If they occur they imply that the effective spectrum of the 4d theory is changing, and hence during the flow the 4d theory is crossing walls of marginal stability in its moduli space.  As a key simplifying assumption, we will now assume that such crossings do not happen at any time during the evolution.  Thus our assumption for the rates of flow of central charge is:
\begin{itemize}
\item During the flow the central charges retain whatever phase order they started with.
\end{itemize}
This is a natural assumption for solutions to the Joyce flow equations.  For example in the context of Janus domain walls, such BPS walls do indeed exist \cite{Janus1, Janus2}.  One can see that a simple ansatz satisfying the above assumption is given by taking the 4d central charges to flow linearly in some coordinate with a speed controlled by their real part
\begin{equation}
Z_i(t)=Z_i^0-i \Re(Z_i^0) t.
\end{equation}
Given such an anastz for the flow, one can see that the central charges retain their phase order and have constant real part. 

Let us now take stock of the resulting properties of the domain walls we have described. 
\begin{itemize}
\item They are characterized by a phase ordered flow of the central charges along vertical lines.
\item In the decoupling limit, the central charges begin at $i \infty$ and terminate at $-i\infty$.
\end{itemize}
Notice that if we ignore the length of the central charges, which varies during the flow in time, the first property is identical to the evolution of the central charges generated by an R-symmetry rotation $Z \rightarrow e^{i\theta}Z.$   Further, the decoupling limit boundary conditions can also naturally be interpreted as saying that as time evolves, the central charges rotate through angle of $\pi$.  

Thus, we have in a sense succeeded in making the R-twisting compactification physical.  To preserve the standard supersymmetry, the central charges flow along straight vertical lines and hence their lengths during the evolution are not constant.  In this sense, the time evolution we have constructed is not merely a phase rotation on $Z$.  However, when our ansatz for the rates of flow is satisfied, the central charges of 4d flow in way which respects their phase order and in this sense the time coordinate we have constructed can be interpreted as, essentially, the phase of $Z$.  Further, our decoupling limit boundary conditions mean that under the complete time evolution the phase rotates by $\pi$. As in the general story of R-twisting this leads to a simple relationship between the 4d BPS spectrum and the 3d BPS spectrum, and for this reason we refer to the flow as the `R-flow.'

\subsubsection{3d BPS Spectrum from Trapped 4d BPS States}

Now we come to the central consequence of the decoupling limit developed in the previous section:
\begin{itemize}
\item The 3d BPS chiral spectrum is in one-to-one correspondence with the 4d BPS spectrum.
\end{itemize}
To see this fundamental fact, we observe that each 4d central charge, $Z_{4\mathrm{d}},$ is the central charge of a certain BPS particle in 4d.  However for a typical point along the flow such a particle is not BPS in the three-dimensional sense.  Indeed, to be BPS in three dimensions $Z_{4\mathrm{d}}(t)$ must align with real direction defining the 3d central charges. This means that at the time $t=t_i$ when $Z^{i}_{4\mathrm{d}}(t_i)=\mathrm{real}$ the corresponding BPS state will be a 3d BPS state with the same central charge.  In other words, the 4d BPS state with central charge $Z_i$ is trapped in the wall at $t=t_i$. Note that this is physically sensible, in the sense that the 4d mass $m(t)=|Z_i(t)|$ is minimized at $t_i$, where the length of $Z_i(t)$ is minimized.  Thus, the particle is trapped at $t=t_i$ simply by energy considerations.  If the boundary condition of the flow were such that the asymptotic central charges had finite length, then the difference in length between the 3d central charge and the 4d central charge at either side of the wall would be finite and the 3d BPS chiral particle, while trapped on the wall, could escape out to the bulk for a finite cost in energy.  However, in the decoupling limit where the 4d central charges become parametrically large as $|t|\rightarrow \infty,$ the potential energy well trapping the 3d particle becomes infinitely deep, and the 4d bulk physics decouples.

Finally if we now invoke our ansatz where the order of the phases of the central charges do not change, it follows from our discussion above that {\it for each chamber of the 4d theory, we get a chamber of a 3d theory, where the corresponding 4d BPS states are trapped and become BPS states of the 3d theory.}  Given that by changes of parameters in the 4d theory we can go from one chamber to another (passing through walls of marginal stability), it suggests that the same should be true for the 3d theory, at least as far as the IR behaviour is concerned.  In particular the initial conditions for the R-flow which can vary continuously, should not affect the IR dynamics. In other words we should get  {\it a family of dual 3d theories labeled by chambers of the parent 4d theory}.  In the remainder of this paper we provide evidence for this claim through a study of explicit examples.  We aim to illustrate that, via this correspondence, the three-dimensional version of wall-crossing is mirror symmetry.

\section{4d BPS States Reviewed}
\label{sec:4dBPS}
At the conclusion of the previous section, we have arrived at a class of domain wall theories whose 3d BPS particles are in one-to-one correspondence with the ambient 4d BPS particles.  In order to apply this useful fact to the study of 3d field theories, we will need to make use of various methods for counting 4d BPS states.  In this section we present a brief review of two such techniques: ideal triangulations, and BPS quivers.  Our goal is simply to develop the necessary statements to apply these technologies in our problem, and we refer the reader to the original papers \cite{GMN1, GMN2, GMN3, ACCERV1, ACCERV2, SelfDual, SV} for a complete treatment.
\subsection{Ideal Triangulations}
\label{sec:Triangle}
The first method we describe is that of ideal triangulations developed in detail in \cite{GMN1}.  We consider a pair of M5 branes wrapping a Riemann surface $\Sigma$.  The Seiberg-Witten geometry is described by a quadratic differential $\phi^{2}$ on $\Sigma$.  As in previous sections, $\phi^{2}$ defines a double cover $\widetilde{\Sigma}$ of $\Sigma$.  This double cover is the Seiberg-Witten curve and on $\widetilde{\Sigma}$, the one-form $\phi$ is the Seiberg-Witten differential.\footnote{\ We are being a bit loose with notation here.  On the base $\Sigma$ the quantity $\phi^{2}$ is not the square of anything.  Only on passing to $\widetilde{\Sigma}$ does it have a globally well defined square root.}  The key idea in this method is to recognize that a BPS M2-brane, describing a BPS particle in four-dimensions, must have minimal area.  In particular the boundary of this two-brane defines a one-cycle in the Seiberg-Witten curve and its length must be minimal.  

It is straightforward to translate this idea into concrete equations formulated on the original curve $\Sigma$ itself.  The boundary one-cycle of the M2-brane projects to $\Sigma$ and defines a curve $\gamma$ parametrized by a variable $s$.   Since the central charge is measured by $\phi$ it is this quantity which characterizes the notion of minimal length and hence $\gamma(s)$ solves the differential equation \eqref{4dm2flowline} \cite{SelfDual}.
\begin{equation}
\phi|_{\gamma}=e^{i\theta}ds. \label{4dflow}
\end{equation}
As discussed in the previous section, the angle $\theta$ entering the equation above is the angle of the central charge of the particle defined by the two-brane.  The geometry of the one-cycle $\gamma$ determines the geometry of the two-brane and hence the kind of BPS particle observed in four dimensions as:
\begin{itemize}
\item A two-brane disc describes a hypermultiplet.  The boundary one-cycle is a circle half of which is on each of the two sheets of the double cover $\widetilde{\Sigma}\rightarrow \Sigma$.  The cycle $\gamma \subset \Sigma$ is then an interval stretching between the branch points of the cover which are the zeros of $\phi$. This is identical to the BPS states described in detail in section 2.
\item A two-brane cylinder describes a vector multiplet.  Its boundary one-cycle is a pair of closed circles, one on each sheet.  The cycle $\gamma \subset \Sigma$ is a single closed circle.  
\end{itemize}
For most of the remainder of this paper, we will be focused on a simple class of examples involving theories which have a finite number of BPS hypermultiplets and no BPS vector multiplets.  These examples are the so-called Argyres-Douglas $ADE$ theories \cite{AD}.  The $A_n$ case, which will
be our main focus in this section, is characterized by a particularly simple geometry.  The curve $\Sigma$ which supports a pair of five-branes is just the complex plane $\mathbb{C}$.  Giving this plane the complex coordinate $x$, the Seiberg-Witten differntial $\phi$ defining the central charge density is given by a polynomial in $x$ of degree $n+1$
\begin{equation}
\phi=\sqrt{P_{n+1}(x)}\,dx.
\end{equation}
In these cases the BPS counting problem is particularly simple as we will see below.  

The main observation is that for the vast majority of angles $\theta$, there is no BPS state whose central charge has that given angle, and hence no finite length solution to the flow equation \eqref{4dflow}.  In this case, we can draw a simple combinatorial picture which characterizes the global asymptotic properties of the flow.  Then, if we perturb the angle $\theta$ by a small amount, we we will not encounter any BPS states and hence the combinatorial diagram will be stable.  On the other hand, if we tune $\theta$ a large amount past a critical angle which supports a BPS hypermultiplet, the global flow diagram will jump in a definite way.  As $\theta$ varies from $0$ to $\pi$ we encounter all BPS particles in the theory and thus the BPS spectrum is realized geometrically as a sequence of moves in the flow diagram.

\begin{figure}[here!]
  \centering
  \subfloat[An Ideal Trangulation]{\label{fig:flip1}\includegraphics[width=0.25\textwidth]{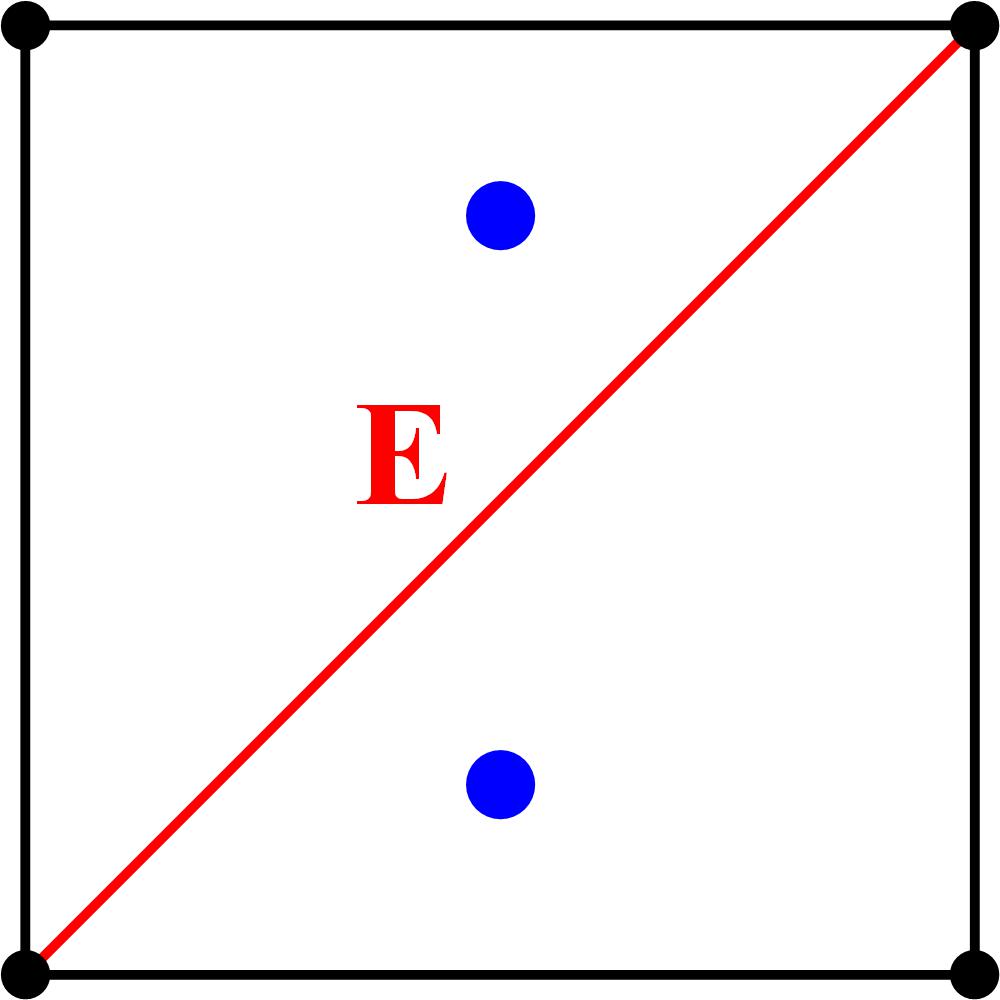}}     
  \hspace{.5in}         
  \subfloat[A Flip]{\label{fig:flip2}\includegraphics[width=0.25\textwidth]{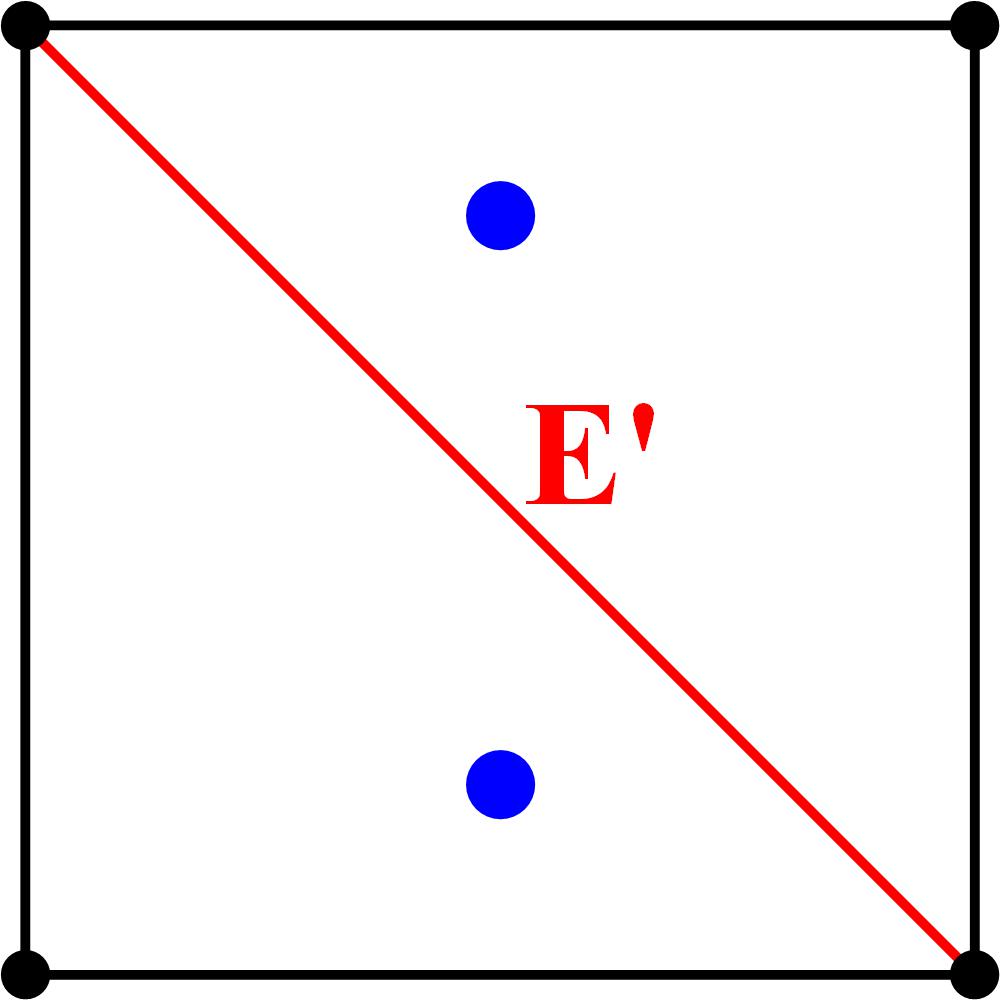}}
  \caption{A sample ideal triangulation.  In (a), we have an ideal triangulation of the square used for counting BPS states in the $A_{1}$ model.  The red line denotes an interior edge.  The blue points the zeros of the differential $\phi^{2}$.  In (b), the edge $E$ has flipped to $E'$. }
  \label{fig:flips}
\end{figure}

In the context of our simple $A_{n}$ theories, let us now be more specific and introduce the asymptotic flow diagram, an \emph{ideal triangulation}, and the operation on it a \emph{flip}, determined by a BPS state.  We first draw a large circle the $x$ plane which defines the asymptotic boundary of $\mathbb{C}$.  Then, on this circle we mark the $(n+3)$-rd roots of unity, which makes the boundary circle into an $(n+3)$-sided polygon.  The complex plane is represented by the interior of this polygon.  We then triangulate this space using only lines that end at the vertices of the polygon.  So defined, we have constructed an ideal trangulation of the $(n+3)$-gon.  This triangulation has the important property that each traingle contains exactly one zero of the differential $\phi$.  An example for the case of $A_{1}$ is shown in Figure \ref{fig:flip1}. 

Now that we have introduced ideal triangulations, it remains to explain how BPS states are visualized in this setup.  As we have described above, the BPS states are sudden changes in the flow as we rotate $\theta$.  This means that they are described by operations, known as flips, which change the triangulation.  The flip operation can be performed on any internal (\textit{i.e.}\! non-boundary) edge $E$ in the triangulation.  We first delete $E$ making a quadrilateral, and then replace $E$ with $E'$, the unique other edge in the quadrilateral which forms a triangulation as shown in Figure \ref{fig:flip2}.  The name of the operation, a flip, is justified by the fact that the new triangulation is related to the old one by rotating the edge $E$.  

In terms of the flow equation \eqref{4dflow}, the significance of a flip is easy to explain. Each triangle in the triangulation contains exactly one zero of $\phi$ and the trajectories of the flow equation $\eqref{4dflow}$ emanating from $\phi$ asymptotically terminate on the vertices of the triangle containing the zero.  Meanwhile, an internal edge $E$ is an object at the interface of two triangles, and hence equivalently two zeros.  As the BPS angle is rotated towards a critical value, a pair of trajectories, one from each zero, become near to each other.  Exactly at the critical angle the trajectories connect, leading to a BPS hypermultiplet described by a segment which crosses the edge $E$.  Just after the critical angle the trajectories again separate and the edge $E$ is replaced with $E'$.  

Given that an individual BPS state appears as a flip, the complete BPS spectrum is then characterized by a sequence of flips.  To describe the sequence, we note that as the BPS angle rotates from 0 to $\pi$, all BPS particles will be seen by the flow and hence all flips will occur.  On the other hand, as $\theta$ rotates through $\pi$ the quadratic differential returns to itself, except that the asymptotic vertices rotate counterclockwise by an $(n+3)$-rd root of unity.  In other words, the vertices of the polygon rotate one unit to the left.  These facts determine how a BPS spectrum is encoded in a sequence of flips:
\begin{itemize}
\item A BPS spectrum of $A_{n}$ is a sequence of flips on the internal edges of a triangulation such that, after all flips have occurred, the ideal triangulation has returned to itself up to a rotation by an $(n+3)$-rd root of unity.
\end{itemize}  
The allowed sequences of flips also satisfy several minimality properties.  Namely, if the edge $E$ flips to $E'$ then the edge $E'$ is not the next edge to flip, and if the sequence at any point reaches the initial triangulation rotated by an $(n+3)$-rd root of unity, then it must terminate.

\begin{figure}[here!]
  \centering
\includegraphics[width=0.9\textwidth]{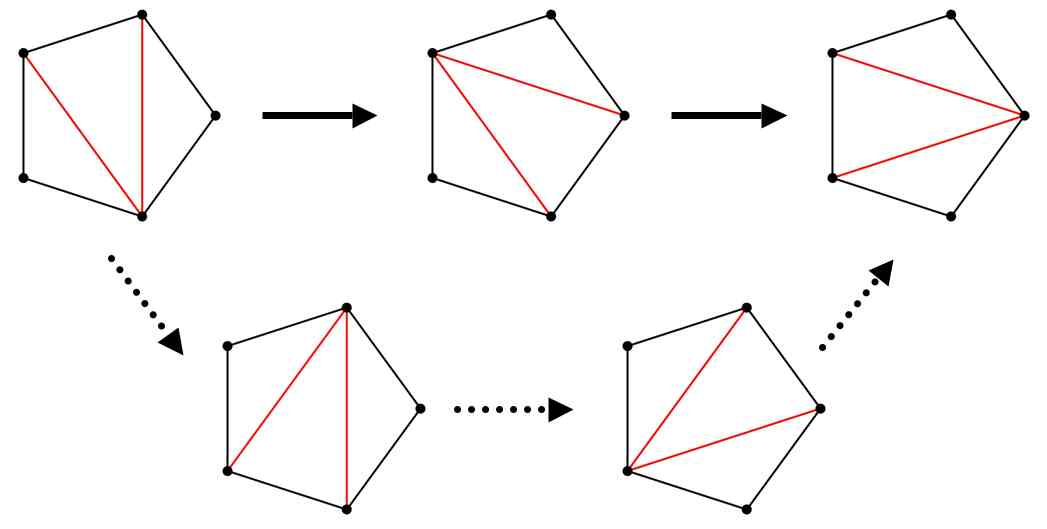}    
  \caption{The two BPS chambers of the $A_{2}$ model realized as a sequence of flips.  The upper-left pentagon, and upper-right pentagons, are the initial and final triangulations respectively.  Each arrow denotes a flip on one of the internal edges shown in red.  Following the solid arrows we find a chamber with two BPS states.  Following the dashed arrows we find a chamber with three BPS states. }
  \label{fig:pentagonid}
\end{figure}

As developed in detail in \cite{GMN1}, the most fascinating and useful aspect of this description of BPS spectra is the ease with which one can describe wall-crossing.  In this context, the fact that there exists more than one chamber of BPS states is simply reflected in the fact that there exists more than one sequence of flips satisfying the above criteria.  Indeed, in the simplest example, the $A_{1}$ model, there is exactly one possible sequence of flips shown in Figure \ref{fig:flips}, and hence the BPS spectrum consists of exactly one BPS hypermultiplet as described in section 3.  However  the $A_{2}$ model, corresponding to the pentagon, already exhibits two such sequences and hence two chambers of BPS spectra, as illustrated in Figure \ref{fig:pentagonid}.  This geometric fact will be significant for us in our study of 3d field theories in section 5 and beyond.
\subsection{Quivers and Mutation}
\label{sec:Quiver}
A second, equivalent method for studying BPS states of the $A_{n}$ models is to make use of BPS quivers and mutations \cite{ACCERV1, ACCERV2}.  In this method, BPS states are described by studying a quiver quantum mechanics on the worldvolume of a BPS particle.  In the $A_{n}$ examples the quiver is given by an oriented version of the $A_{n}$ Dynkin diagram.
\begin{equation}
\xy 
(-45,0)*+{}*\cir<8pt>{}="a" ; (-15,0)*+{}*\cir<8pt>{}="b" ; (15,0)*+{\cdots}="c" ;(45,0)*+{}*\cir<8pt>{}="d" ;
(-45,-8)*+{\gamma_{1}}="e" ; (-15,-8)*+{\gamma_{2}}="f" ; (15,-8)*+{\cdots}="g" ;(45,-8)*+{\gamma_{n}}="h" ;
\ar @{->} "a"; "b" 
\ar @{->} "b"; "c" 
\ar @{->} "c"; "d" 
 \endxy
\label{anquiver}
\end{equation}
In \eqref{anquiver} $\gamma_{i}$ denote the charges of an elementary basis of hypermultiplet which are always stable states.  Since charges are defined by one-cycles on the SW-curve each $\gamma_{i}$ is associated to an element of $H_{1}(\widetilde{\Sigma})$.  The number of arrows between the nodes of the quiver is then fixed by computing the intersection product of cycles, $\gamma_{i}\circ \gamma_{j}$, or equivalently the four-dimensional electric-magnetic inner product of the associated 4d particles.  All remaining BPS particles in the spectrum can be viewed as supersymmetric  bounds states of these, which exist in the quiver quantum mechanics theory defined by \eqref{anquiver} \cite{D1, D2, D3}. 

In comparing the method of ideal triangulations to that of BPS quivers, the quiver diagram plays the role of the ideal triangulation of a polygon.  It is a basic combinatorial diagram which encodes information about the spectrum.  The analogous operation to a flip is then a so-called quiver \emph{mutation} which acts on the quiver to produce a new quiver.  This operation can be defined on any node $i$ of the quiver, and acts on the charges as follows:
\begin{eqnarray}\label{eqmutation}
\gamma_{i} & \longrightarrow & -\gamma_{i} \nonumber\\
\gamma_{j}  & \longrightarrow & 
  \begin{cases}
   \gamma_{j}+( \gamma_{i} \circ   \gamma_{j})\gamma_{1} & \text{if }  \gamma_{i} \circ   \gamma_{j} >0 \\
   \gamma_{j}      & \text{if }   \gamma_{i} \circ \gamma_{j} \leq0 \label{quivermut}
  \end{cases}
\end{eqnarray}
Thus, after mutation we can form a new quiver by computing the intersection of the charges on the right-hand-side of \eqref{quivermut}.

Now, in the method of ideal triangulations, each BPS state is associated to a flip.  Similarly in the method of BPS quivers each BPS state is associated to a mutation.  It then follows that the complete BPS spectrum is captured by certain sequences of mutations.  These sequences are defined by the following properties \cite{ACCERV2}:
\begin{itemize}
\item The initial quiver appears as in \eqref{anquiver} with node charges $\gamma_{i}$.
\item The final quiver has charges $-\gamma_{i}$.
\item At each step one may mutate on any node whose charge $\gamma$ can be expressed as
\begin{equation}
\gamma=\sum_{i}n_{i}\gamma_{i},
\end{equation}
where in the above the $n_{i}$ are non-negative integers.
\end{itemize}

Let us see how the two examples considered in the previous section, the $A_{1}$ and $A_{2}$ theories, are described using this method.  In the case of $A_{1},$ the quiver consists of one node and there is trivially one possible sequence of mutations. 
\begin{equation}
\xy 
 (-15,0)*+{}*\cir<8pt>{}="b" ; 
 (-15,-8)*+{\gamma_{1}}="c" ; 
 \endxy
\hspace{.2in}
\longrightarrow 
\hspace{.2in}
\xy 
 (15,0)*+{}*\cir<8pt>{}="b" ; 
 (15,-8)*+{-\gamma_{1}}="c" ; 
 \endxy
\end{equation}
This agrees with our identification of this theory as a single free hypermultiplet.  There are no interactions and hence no wall-crossing.  Meanwhile, in the case of the $A_{2}$ theory things are more interesting.  The two spectra described in Figure \ref{fig:pentagonid} map to two possible sequences of mutations.  The first sequence, with two BPS particles, is:
\begin{equation}
\xy 
 (-12,0)*+{}*\cir<8pt>{}="a" ; 
 (-12,-8)*+{\gamma_{1}}="c" ; 
(12,0)*+{}*\cir<8pt>{}="b" ; 
 (12,-8)*+{\gamma_{2}}="d" ; 
\ar @{->} "a"; "b" 
 \endxy
\hspace{.2in}
\longrightarrow 
\hspace{.2in}
\xy 
 (-12,0)*+{}*\cir<8pt>{}="a" ; 
 (-12,-8)*+{\gamma_{1}}="c" ; 
(12,0)*+{}*\cir<8pt>{}="b" ; 
 (12,-8)*+{-\gamma_{2}}="d" ; 
\ar @{->} "b"; "a" 
 \endxy
\hspace{.2in}
\longrightarrow 
\hspace{.2in}
\xy 
 (-12,0)*+{}*\cir<8pt>{}="a" ; 
 (-12,-8)*+{-\gamma_{1}}="c" ; 
(12,0)*+{}*\cir<8pt>{}="b" ; 
 (12,-8)*+{-\gamma_{2}}="d" ; 
\ar @{->} "a"; "b" 
 \endxy
\label{a21}
\end{equation}
While the second sequence describing the second chamber with three BPS particles is:
\begin{equation}
\xy 
 (-10,0)*+{}*\cir<8pt>{}="a" ; 
 (-10,-8)*+{\gamma_{1}}="c" ; 
(10,0)*+{}*\cir<8pt>{}="b" ; 
 (10,-8)*+{\gamma_{2}}="d" ; 
\ar @{->} "a"; "b" 
 \endxy
\hspace{.12in}
\longrightarrow 
\hspace{.12in}
\xy 
 (-10,0)*+{}*\cir<8pt>{}="a" ; 
 (-10,-8)*+{-\gamma_{1}}="c" ; 
(10,0)*+{}*\cir<8pt>{}="b" ; 
 (10,-8)*+{\gamma_{1}+\gamma_{2}}="d" ; 
\ar @{->} "b"; "a" 
 \endxy
\hspace{.12in}
\longrightarrow 
\hspace{.12in}
\xy 
 (-10,0)*+{}*\cir<8pt>{}="a" ; 
 (-10,-8)*+{\gamma_{2}}="c" ; 
(10,0)*+{}*\cir<8pt>{}="b" ; 
 (10,-8)*+{-\gamma_{1}-\gamma_{2}}="d" ; 
\ar @{->} "a"; "b" 
 \endxy
\hspace{.12in}
\longrightarrow 
\hspace{.12in}
\xy 
 (-10,0)*+{}*\cir<8pt>{}="a" ; 
 (-10,-8)*+{-\gamma_{2}}="c" ; 
(10,0)*+{}*\cir<8pt>{}="b" ; 
 (10,-8)*+{-\gamma_{1}}="d" ; 
\ar @{->} "a"; "b" 
 \endxy
\label{a22}
\end{equation}
One can easily generalize from these examples to determine the spectrum in the various chambers of $A_{n}$ theories for larger $n$.  In our applications of this method to 3d $\mathcal{N}=2$ theories in later sections, one detail of these calculations will be important to us:
\begin{itemize}
\item At the conclusion of a sequence of mutations the original quiver charges $\{\gamma_{i}\}$, as a set, have been changed to $\{-\gamma_{i}\}$.  However, they may have also undergone a non-trivial permutation by an element $\chi\in S_{n}$.  Indeed, in the case of the first chamber of $A_{2}$ described  by \eqref{a21} $\chi$ is the identity element, while in the case of the second sequence, descirbed by \eqref{a22} $\chi$ is the non-trivial element in $S_{2}$.
This permutation proves important for our considerations later in this paper.
\end{itemize}
\section{Tetrahedra and Braids}
\label{sec:T+B}
Armed with the technology of the previous section, we now return to our general discussion of 3d $\mathcal{N}=2$ theories constructed as domain walls in 4d $\mathcal{N}=2$ theories.  Our aim will be to apply the techniques of ideal triangulations and quiver mutations to develop a detailed geometrical toolkit for extracting the physics of the domain wall.

Throughout all of the examples discussed in this section, the 4d theory will be one of the $A_{n}$ models whose BPS spectra we have now described in some detail.  In the UV these theories are determined by a pair of five-branes wrapping the complex plane $\mathbb{C}$ and this leads to a particularly simple geometry of the associated three-manifold $M$ defining the domain wall theory.  To be specific, $M$ is simply a thickening of the complex plane to $\mathbb{C}\times \mathbb{R}$, where $\mathbb{R}$ describes the time parameter of the R-flow in section 3.  Along this flow all the central charges move in vertical straight lines, and central charges cross the real axis in phase order.  As we have previously noted this means that each 4d BPS state will appear as a 3d BPS chiral particle trapped along the wall.  Further, if we ignore the length of the 4d central charges along the flow and concentrate only on their angles, then we may interpret the fact that the particles cross in phase order as an identification of the time coordinate with the BPS angle $\theta$ of the 4d central charges.  In this section our aim will be to make use of this fact to determine a concrete Lagrangian description of the field theory on the domain wall.

\begin{figure}[here!]
  \centering
  \includegraphics[width=0.5\textwidth]{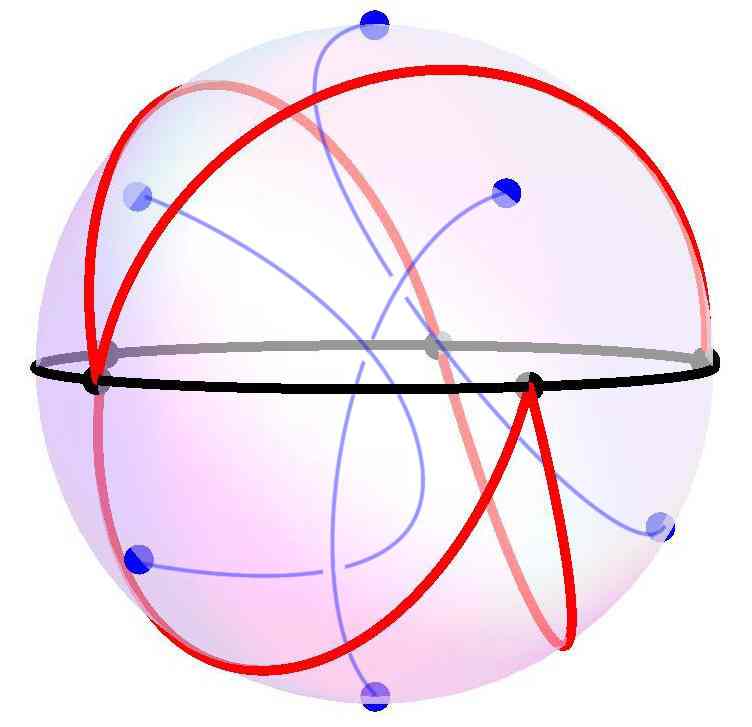}     
    \caption{The manifold $M$ and its boundary triangulation for the case of $A_{2}$.  $M$ is an infinite solid ball.  Its boundary two sphere has two faces given by the northern and southern hemispheres.  Each face is a triangulated pentagon.  The vertices of the pentagon are shown in black while the arcs in the triangulation are shown in red.  The blue dots are the zeros of the SW differential.  One such zero occurs in every triangle.  As we flow through time, the zeros on the front face interpolate to those on the back face.  }
  \label{fig:a2fb}
\end{figure}

First, we study the structure of the three-manifold $M$.  As we have described above, $M=\mathbb{C}\times \mathbb{R}$, however the boundary conditions on the circle at infinity in the complex plane are fixed for all time.  Thus, we will in fact work in a quotient three-manifold defined by identifying these asymptotic regions for all time.  It then follows that our three-manifold $M$ can be viewed as an infinite solid ball with an asymptotic $S^{2}$ boundary.  This boundary two sphere is naturally partitioned into two components, the northern hemisphere corresponding to the initial boundary condition, and the southern hemisphere corresponding to the final boundary condition.  We will refer to these hemispheres as the ``front'' and ``back'' face of our three-manifold respectively.  The equatorial circle of the boundary $S^{2}$ is where the front and back faces are glued together and is the boundary circle inside $\mathbb{C}^{2}$ that is identified for all time.  Further, both the front and back face of our three-manifold describe an $A_{n}$ theory, and as such these faces are naturally equipped with ideal triangulations of $(n+3)$-gons governing their BPS spectra.  Since the complete flow through time corresponds to a rotation of the BPS angle by $\pi$, the final triangulation differs from the initial triangulation by a rotation by $\frac{2\pi}{n+3}$.  An example of the geometry for the case of $A_{2}$ is shown in Figure \ref{fig:a2fb}.

As we flow through time, the initial triangulation will evolve by a sequence of flips as described in section 4.  We will see that this sequence of flips will naturally endow the three-manifold $M$ with a decomposition into tetrahedra.  Since the 4d BPS states correspond to both tetrahedra and trapped 3d BPS particles, we then learn that each tetrahedron in the manifold $M$ will encode the existence of a 3d BPS particle.  In this way we will make contact with the work of \cite{DGG}: the tetrahedron is a kind of basic BPS building block of these 3d theories.  Further as we will see, the fact that the ambient 4d theories can undergo wall crossing, and hence have different numbers of flips, becomes the statement that a given three-manifold admits many distinct decompositions into tetrahedra.  In our context, these distinct tetrahedral decompositions of $M$ will encode different dual descriptions of the same IR field theory.

\begin{figure}[here!]
  \centering
  \includegraphics[width=0.8\textwidth]{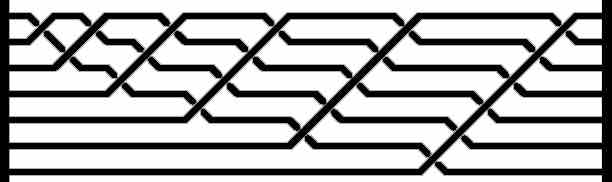}     
    \caption{A braid with seven strands describing a chamber in the $A_{6}$ theory.  Each strand follows the evolution in time of a zero of $\phi$.  3d BPS particles are described by braid moves. The endpoints are the zeros of the initial and final SW differential.}
  \label{fig:braidcart}
\end{figure}

Next in our analysis, we describe the IR geometry which is given by a branched double cover $\widetilde{M}\rightarrow M$.  Since $M$ is an infinite solid ball its topology is trivial.  Thus, up to data at infinity, the situation is exactly the same as that of double covers of $S^{3}$ described in section 2.  In particular, $\widetilde{M}$ is completely fixed by the associated branch locus knot in $M$.  In our context, this knot is exactly the set of zeros of the one-form $\lambda$, or equivalently the zeros of the evolving Seiberg-Witten differential $\phi$.  On the front face of $M$ the differential $\phi$ for the $A_{n}$ model has exactly $n+1$ zeros and each zero resides in a triangle in the ideal triangulation.  As we flow through time the zeros evolve continuously and sweep out a \emph{braid} composed of $n+1$ strands.  As we will argue, the structure of this braid completely determines the 3d physics with BPS particles in direct correspondence with the crossings in the braid diagram.  An example is shown in Figure \ref{fig:braidcart}.

To complete the description of the 3d theory from its braid diagram, there is one final step:  we must turn the braid into a knot; that is we choose a way of identifying the enpoints of the strands of the braid in pairs to turn all components of the braid into closed loops.  This step is physically natural from a number of perspectives.  First, our three-manifold $M$ is non-compact, and hence we must impose boundary conditions.  These boundary conditions involve specifying a choice of which cycles in $\widetilde{M}$ are contractible at infinity and which remain non-trivial.  Since all cycles in the cover $\widetilde{M}$ can be localized to a neighborhood of the branching link, this choice is equivalent to a specification of how the braid is capped off to form a closed knot.  Alternatively, from the perspective of the domain wall theory we can see the need for boundary conditions as follows.  At the conclusion of the decoupling limit described in section 3 all the massive BPS states of the ambient 4d theory have decoupled.  However there remains the coupling to the $U(1)$ gauge and flavor symmetries.  To completely specify the theory on the wall we must specify how we couple our 3d field theory to these vectors.  Since the coupling constants of these $U(1)$'s are arbitrary parameters, we can choose whether in three dimensions a given $U(1)$ appears as a gauge or global symmetry.  In fact such coupling choices for the $A_{n}$ are in direct correspondence elements of $Sp(2n,\mathbb{Z}),$ where the various $S$ transformations act by changing the set of gauged versus global $U(1)$'s and the $T$ transformations appear as changes in the Chern-Simons levels.  We will see how these facts are made geometrically manifest  in the course of our analysis.
\subsection{The Tetrahedron theory}
\label{1T}
\begin{figure}[here!]
  \centering
\includegraphics[width=0.6\textwidth, ]{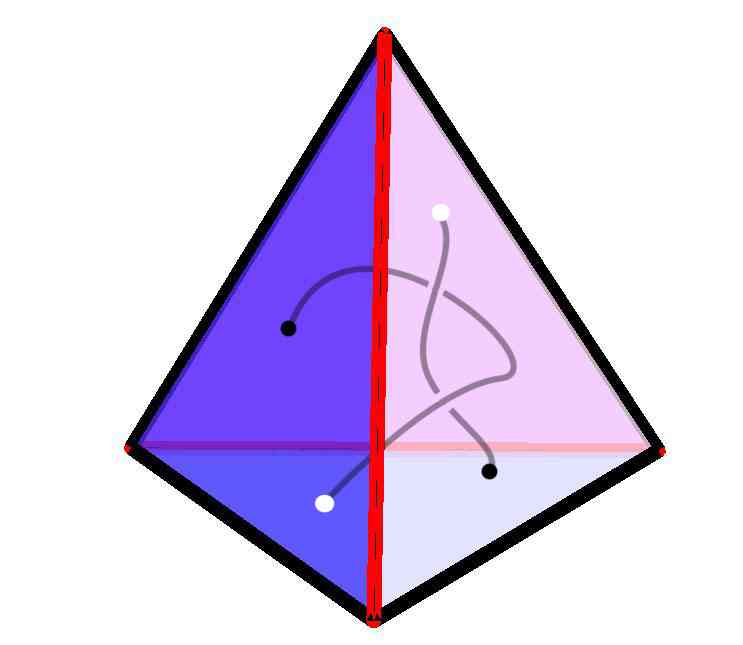}
  \caption{The tetrahedron associated to the $A_{1}$ domain wall.  The tetrahedron is viewed with its front face pointing out of the page.  The square of the $A_{1}$ theory is given by the black edges.  The red diagonal flips as one flows from the front to the back face.  The black dots denote the two zeros of the SW differential on the front face.  As we flow through time, these zeros evolve to the two zeros on the back face shown in white.  In the process they sweep out two strands.}
  \label{fig:tetra}
\end{figure}

We begin with the simplest example of domain walls in the $A_{1}$ theory.  In 4d, this is the theory of a free hypermultiplet, and the R-flow of central charges for this example was studied in section 3.  In this case, the boundary triangulations of the front and back face are squares, and as we flow through time the triangulation evolves by a single flip to produce a single tetrahedron shown in Figure \ref{fig:tetra}.  
We know that this flip is naturally associated to a 3d BPS chiral particle which has become trapped on the wall, and thus this theory of two five-branes on a tetrahedron supports exactly one BPS chiral particle.  The mass of this particle, $m_{0}$ is the real part of the 4d mass of the parent 4d hypermultiplet.

\begin{figure}[here!]
  \centering
  \subfloat[$A_{1}$ Braid]{\label{fig:a1braid}\includegraphics[width=.4\textwidth]{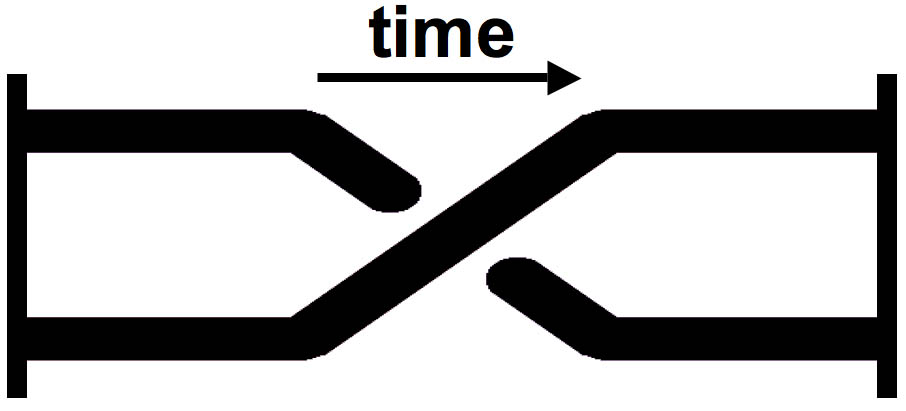}}     
  \hspace{.5in}          
  \subfloat[The $U(1)$ in the $A_{1}$ theory.]{\label{fig:a1braidcyc}\includegraphics[width=0.4\textwidth]{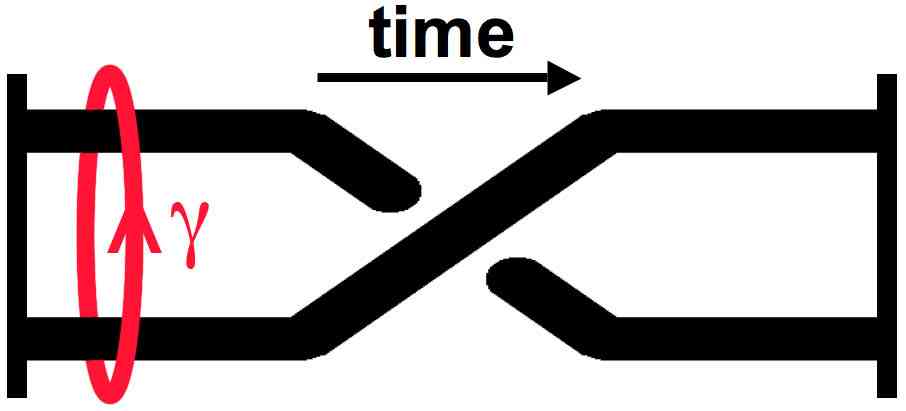}}
  \caption{In (a), the braid for the tetrahedron theory.  The single particle is encoded by the single braid move.  In (b), the red cycle $\gamma$ circles the two strands of the braid.  The particle at the crossing is charged under this cycle.}
  \label{fig:a1braids}
\end{figure}

To study the geometry in more detail, we track the evolving zeros of the SW differential as we move through the geometry of the tetrahedron.  In each triangle in both the front and back face there is one zero, and as time flows they determine a braid composed of two strands.  At exactly one critical time the strands of the braid become closest to each other and the BPS chiral particle in 3d appears.  We encode this fact in the braid diagram by drawing exactly one braid move as shown in Figure \ref{fig:a1braid}.

In terms of the geometry of section 2, the single BPS particle appears as a segment connecting the two strands of the braid.  Since the BPS particle is also associated to the one braid move we can view this BPS segment as being localized at the crossing in the braid diagram.  As explained in section 2 such a particle carries a $U(1)$ charge under the cycle $\gamma$ which wraps around the two strands of the braid illustrated in Figure \ref{fig:a1braids}.  Depending on boundary conditions to be specified, the cycle $\gamma$ may be non-contractible, in which case it is gauged, or it may be contractible at infinity, in which case the $U(1)$ will survive as a flavor symmetry of the theory.  Thus in either case, the BPS particle carries a unit charge under this $U(1)$.  Note that in the limit where the particle is massless, the two strands of the braid intersect.  Thus, we can view the separation between the strands as proportional to the mass of the particle and the effect of going from overcross to undercross corresponds to changing the sign of the mass for the chiral field.  Finally, we will always make the convention that time flows from left to right in the braid diagram.  So defined the configuration of Figure \ref{fig:a1braid} encodes a charged BPS particle with charge $+1$ under the cycle $\gamma$.

\begin{figure}[here!]
  \centering  
\includegraphics[width=0.8\textwidth]{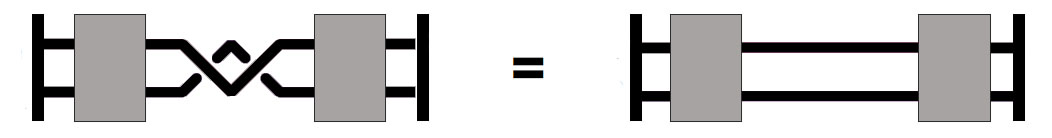}
  \caption{The braid group relation $b^{-1}b=1$.  In the gray region, the strands braid with each other in an arbitrary manner.}
  \label{fig:braidg1}
\end{figure}

Thus far, the braid we have introduced is simply a diagrammatic notation for the rather trivial particle content of the tetrahedron theory.  However, the reason that the braid is useful is that operations on the braid diagram have a natural physical interpretation.  We will illustrate this feature throughout the course of our analysis.  To begin, the first and most basic point we address is the proof that the field theory we have defined is canonically associated to the braid.  What this means is the following.  The braid group on two strands is an infinite cyclic group which is generated by a single element $b$ which acts on the two strands, as in Figure \ref{fig:a1braid}, by braiding the lower strand over the upper strand in time order.  Then the tautological relationship $b^{-1}b =1$ translates to the clear geometrical fact that an insertion of an overcross followed by and undercross at any point in the braid is trivial as illustrated in Figure \ref{fig:braidg1}.

\begin{figure}[here!]
  \centering  
\includegraphics[width=0.4\textwidth]{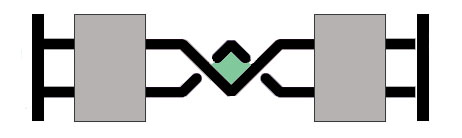}
  \caption{The superpotential coupling the fields corresponding to the insertion of $b b^{-1}$.  The green region lifts to the boundary of an M2-brane instanton which gives rise to a quadratic interaction between the particles.}
  \label{fig:a1poten}
\end{figure}

Now, in our physical context we may ask whether the relation in Figure \ref{fig:braidg1} is satisfied.  To address this we follow the tentative dictionary set in the previous paragraphs.  For each crossing in the diagram we add a single chiral particle to the theory.  Thus, in the left of Figure \ref{fig:braidg1}, the relevant region where $b^{-1}b$ has been inserted corresponds to two particles $X$ and $Y$.  However, as we will argue later in this section, these particles have opposite $U(1)$ charges.  This means that there is an invariant superpotential term
\begin{equation}
W=\mu XY. \label{cpxmass}
\end{equation}
Furthermore, we know from our general discussion of M2-brane contributions to the superpontential in section 2 that exactly in this situation we expect to find such a quadratic contribution to $W$.  Indeed, the region of the overcross followed by undercross bounds a disc which is precisely the projection of an M2 describing a quadratic interaction between the inserted particles as illustrated in Figure \ref{fig:a1poten}.

Now, in equation  \eqref{cpxmass} the parameter $\mu$ is a complex (as opposed to real) mass for the fields $X$ and $Y$.   Such a mass term means that the fields $X$ and $Y$ are irrelevant in the infrared and may be safely removed from the spectrum.  This should be contrasted with the case of particles with non-zero real masses.  In the latter case, even though such fields are massive, their real mass is detected by the partition function of the theory as we will discuss in sections 7 and 8.  By contrast, the partition function is independent of complex masses such as $\mu$ and thus we may freely take these to be parametrically large.  Doing so, we find that the insertion of $b^{-1}b$ in the braid diagram is physically equivalent to inserting the identity, i.e. no particles whatsoever.  In this way, we have verified the braid group relation described by Figure \ref{fig:braidg1}.
\subsubsection{Boundary Conditions and $SL(2,\mathbb{Z})$}
Next in our analysis, we turn to the discussion of boundary conditions for the theory of two M5-branes on the tetrahedron.  As we have previously discussed, what the domain wall and decoupling limit constructs for us is a 3d theory, together with an arbitrary choice of coupling to the background $U(1)$ multiplet.  On such field theories, there is a natural action of $SL(2,\mathbb{Z})$ \cite{KapusStrass, WittenSL2} defined by the action of its $S$ and $T$ generators as:
\begin{itemize}
\item $T$ acts to increase the Chern-Simons level of the background $U(1)$ by $\hat{k}\rightarrow \hat{k}+1$.
\item $S$ acts to gauge the $U(1)$ in three dimensions, and introduces a new background $U(1)$ which is the dual flavor group.
\end{itemize}
Thus, $SL(2,\mathbb{Z})$ does not act as a duality group, but simply acts on such a theory to produce a new one.  As we will see, in our context, this $SL(2,\mathbb{Z})$ is realized as acting on our choice of boundary conditions.

The simplest way to study the boundary conditions is to consider the IR geometry $\widetilde{M}\rightarrow M$.  This is a double cover of $M$ branched over the braid described in the previous section.  In particular, the boundary of $M$ as an $S^{2}$ which contains the four endpoints of the braid, two from the front face and another two from the back face.   It follows that the boundary of $\widetilde{M}$ is a double cover of $S^{2}$ branched over four points, and therefore $\partial \widetilde{M}$ is a torus.  The three-manifold $\widetilde{M}$ fills in this boundary smoothly, and is thus a solid torus.  

\begin{figure}[here!]
  \centering
  \subfloat[SW Curve]{\label{fig:cyl}\includegraphics[width=0.4\textwidth, height=0.2\textwidth]{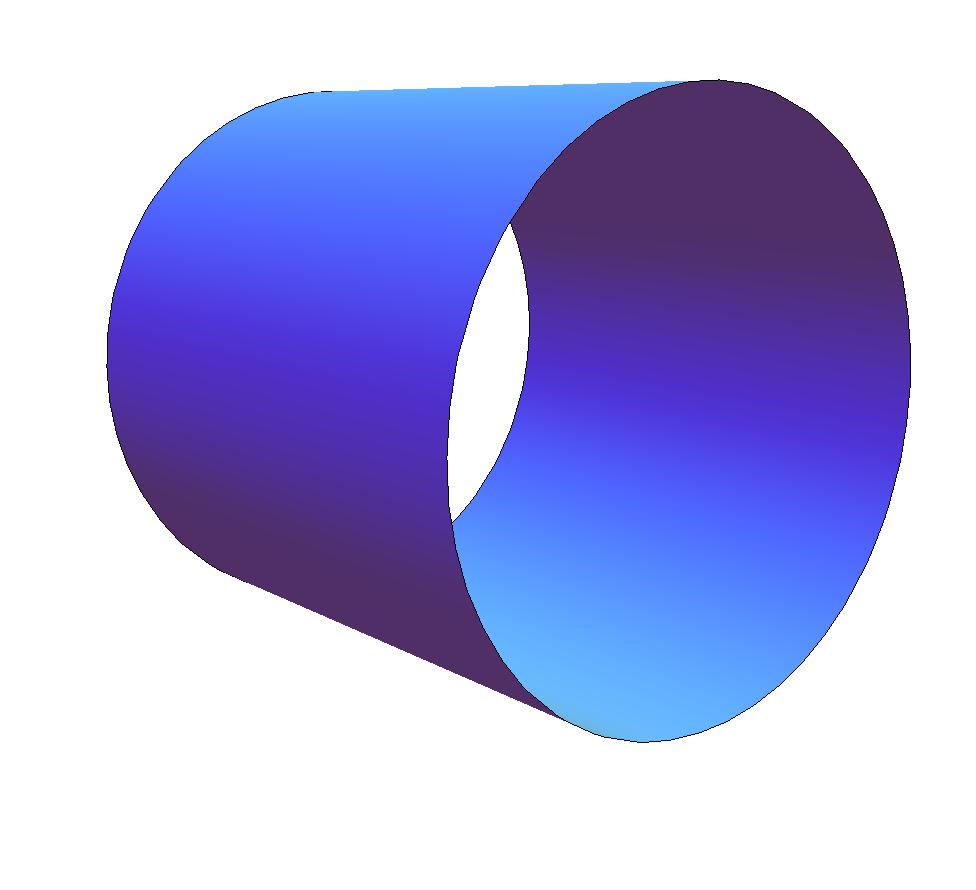}}     
  \hspace{.5in}           
  \subfloat[$\widetilde{M}$]{\label{fig:solidtorus}\includegraphics[width=0.4\textwidth, height=0.2\textwidth]{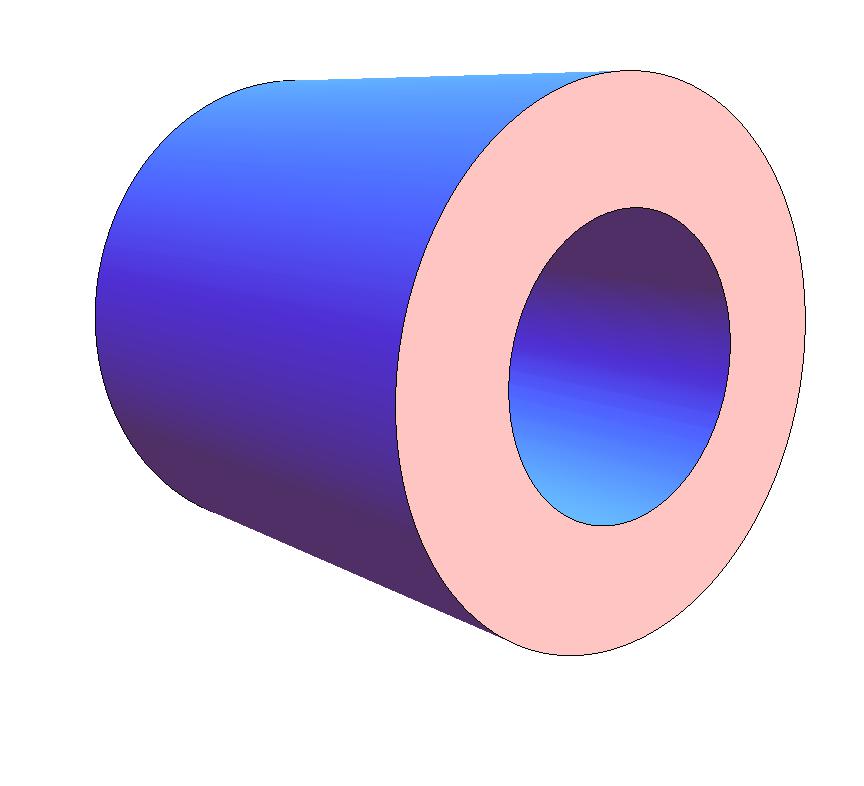}}
  \caption{The IR geometry for the tetrahedron theory.  In (a), we see the SW curve, in this case a cylinder, for the 4d $A_{1}$ theory.  In (b), the three-manifold $\widetilde{M}$ obtained as a thickening of the SW curve.  Topologically this thickened cylinder has an asymptotic boundary of a torus.}
  \label{fig:solidt}
\end{figure}

Alternatively, one can also see the fact that $\widetilde{M}$ is a solid torus by recalling that the tetrahedron theory is determined by a one-parameter thickening of the $A_{1}$ theory in 4d.  The Seiberg-Witten curve for the latter is a cylinder.  Then, $\widetilde{M}$ is a thickening of this cylinder.  It has as boundary the SW cylinders associated to the front and back face $A_{1}$ theories which are connected at their respective ends to make the surface $\partial M$ into a torus as illustrated in Figure \ref{fig:solidt}.

Now we are equipped to specify boundary conditions.  We will modify the manifold $\widetilde{M}$ by adding data at infinity which turns it into a closed manifold without boundary $\widetilde{M}_{c}$.  Then, all fields are required to be well-behaved on $\widetilde{M}_{c}$.  Since $\widetilde{M}$ has boundary given by a torus, to close $\widetilde{M}$ means to glue it to another three-manifold whose boundary is a torus, in other words we simply glue $\widetilde{M}$ to another solid torus.  From this description, we see that our choices of boundary conditions are labeled by the gluing map $g:T^{2}\rightarrow T^{2}$ that specifies how the boundary tori are glued.  Up to isotopy, such gluings $g$ are specified by their $SL(2,\mathbb{Z})$ action on the homology of the boundary of the torus.  The manifolds $\widetilde{M}_{c}$ that we obtain from such gluing are exactly the \emph{lens spaces}.  For example, gluing two solid tori with the identity map makes, $S^{2}\times S^{1}$, while gluing with the $S$ transformation produces $S^{3}$.  More generally, given $p$ and $q$ relatively prime, we consider the following element of $SL(2,\mathbb{Z}):$
\begin{equation}
g=\left(\begin{array}{cc}m & n \\
p & q\end{array}\right).
\end{equation}
Where in the above $m,n$ are chosen such that $g$ has determinant one.  Then, the three-manifold obtained by gluing two solid tori with the map $g$ is the Lens space $L(p,q)$.\footnote{\ Recall that, for any choice of signs $L(\pm p, \pm q)$ are all identical.  Thus we can be somewhat lax about signs in the following.}

One can see from this description that the $S$ and $T$ generators have the desired physical effect of gauging, and shifting the level $\hat{k}$ respectively.  Indeed, for example consider as a starting point the theory on $S^{3}=L(1,0)$.  This manifold has no homology and hence no gauge group.  Acting with $S$ changes the gluing to produce $S^{2}\times S^{1}$.  Since this has first Betti number one, the $U(1)$ has been gauged, which is indeed the appropriate action for the generator $S$.  Similarly, we can act on the $S^{3}$ theory with the transformation $T^{p}$.  This means that we are gluing two solid tori with the map
\begin{equation}
g=T^pS=\left(\begin{array}{cc}p & -1 \\
1 & 0\end{array}\right).
\end{equation}
This again produces the $S^{3}=L(1,0)$.  However, the integer $p$ in the above is physical as it encodes the CS level of the coupling of the theory to the background $U(1)$ flavor symmetry now given by $\hat{k}=p$.  Indeed, to make this manifest we can now further act by $S$.  This gauges the $U(1)$ which is now at level $p$.  It is specified by the gluing map
\begin{equation}
g=ST^pS=\left(\begin{array}{cc}1 & 0 \\
p & 1\end{array}\right),
\end{equation}
and hence results in the Lens space $L(p,1)$.  This space has first homology group that is pure torsion $H_{1}(L(p,1))\cong \mathbb{Z}_{p},$ and thus, as explained in section 2 describes a gauged $U(1)$ CS theory at level $\hat{k}=p$ as desired.

\begin{figure}[here!]
  \centering
  \subfloat[A Braid]{\label{fig:basicclose}\includegraphics[width=.4\textwidth]{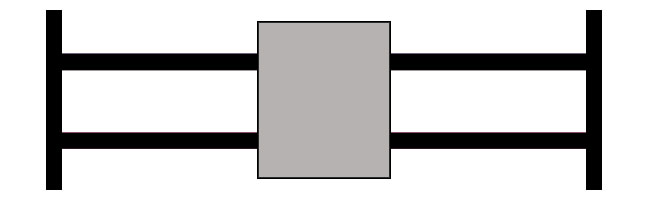}}   
  \hspace{.5in}          
  \subfloat[Closure of the Braid]{\label{fig:closedlink}\includegraphics[width=0.4\textwidth]{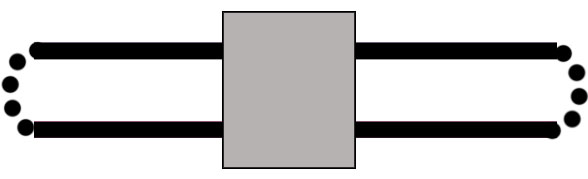}}
  \caption{Imposing boundary conditions.  In (a) we see a braid.  In the gray region the strands braid with each other in an arbitrary manner. In (b), the strands are connected by the dashed string to make a closed knot. }
  \label{fig:boundcondef}
\end{figure}

We can further illuminate this $SL(2,\mathbb{Z})$ structure by alternatively studying it from the point of view of the branching braid which encodes the structure of the cover $\widetilde{M}\rightarrow M$.  The $SL(2,\mathbb{Z})$ action on the homology of the boundary $T^{2}$ of $\widetilde{M}$ is obtained by motions involving the four branch points in the cover $T^{2}\rightarrow S^{2}$.  Since these four endpoints are precisely the endpoints of the braid, this means that the $SL(2,\mathbb{Z})$ action can be seen as acting on the braid.    To describe this action, we must first state how we specify boundary conditions at the level of a braid.   Our infinite tetrahedron can be compactified to $S^{3}$ by adding a point at infinity.  As discussed in section 2, a double branched cover of $S^{3}$ is completely specified by its branching knot $\mathcal{K}$.  Thus, to specify the boundary conditions we must close our braid into a knot.  We do this by identifying the endpoints of the braid in pairs.  Specifically, we glue the initial points at $t= -\infty$ together, and the final points at $t=+\infty$ together.  In this way make a closed knot as illustrated in Figure \ref{fig:boundcondef}.

In general, for the $A_{1}$ theory, we will always specify boundary conditions by gluing initial and final points as above.  We illustrate this diagrammatically, with the dashed string shown in Figure \ref{fig:closedlink} to emphasize that this gluing is boundary data at infinity.  With this prescription, we can now specify completely the geometry of the compactified double cover $\widetilde{M}_{c}$.  For example, in the case of the braid of Figure \ref{fig:a1braid} describing the basic tetrahedron theory, this procedure produces an unknot.   Then $\widetilde{M}_{c},$ is double cover of $S^{3}$ branched along the unknot and hence is also an $S^{3}$.

\begin{figure}[here!]
  \centering
  \subfloat[A Braid]{\label{fig:basict}\includegraphics[width=.4\textwidth]{Figures/bbraid.jpg}}     
  \hspace{.5in}      
  \subfloat[Action by $T$]{\label{fig:ttbraid}\includegraphics[width=0.4\textwidth]{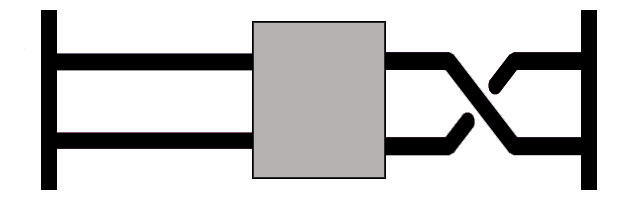}}
  \caption{The definition of the operator $T$.  In (a) we see a generic braid.  In (b) the action of $T$ on this braid.}
  \label{fig:tdef}
\end{figure}

Now we are equipped to discuss the action of $SL(2,\mathbb{Z})$ on closed braids.  Let us first consider the $T$ generator.  This is to act by increasing the CS level for the background $U(1)$ by $\hat{k}\rightarrow \hat{k}+1$.  We can interpret this action by making use of the quantum parity anomaly.   This states that upon integrating out a particle of mass $m>0$ with charge $\pm 1$ the CS level shifts as $\hat{k}\rightarrow \hat{k}+1$.  In terms of its action of CS levels, the operation of adding a massive particle is therefore identical to the desired $T$ operation.  In the above, we have associated the charged particles to the crossings in the braid diagram, that is to the action of the braid group generator $b^{\pm1}$.  Sticking to this principle, means that we simply identify the action of the $SL(2,\mathbb{Z})$ element $T$ with the insertion of $b^{-1}$ at the conclusion of the braid, as in Figure \ref{fig:tdef}.  
In order to only modify the CS level, the particles inserted by the $T$ transformation should be interpreted as having parametrically large mass.  This is natural if we view $T$ as acting on boundary conditions of the theory.  Then, the closure of the braid assocaited to $T^{p}$ acting on the basic tetrahedron braid in Figure \ref{fig:a1braid} is again an unknot.   However, the integer $p$ is physical and keeps track of the background CS level.  Thus, although all such examples produce covers $\widetilde{M}_{c}$ which are topologically $S^{3}$'s there is a physical integer ambiguity, namely the CS level, which is resolved by the braid diagram.

\begin{figure}[here!]
  \centering
  \subfloat[A Braid]{\label{fig:basic}\includegraphics[width=.48\textwidth]{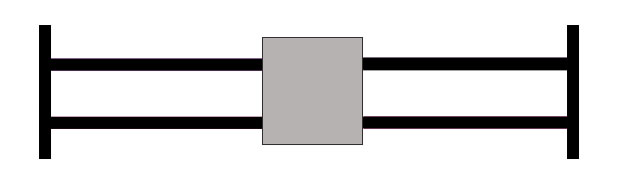}}     
  \hspace{.2in}         
  \subfloat[Action by $S$]{\label{fig:ssbraid}\includegraphics[width=0.48\textwidth]{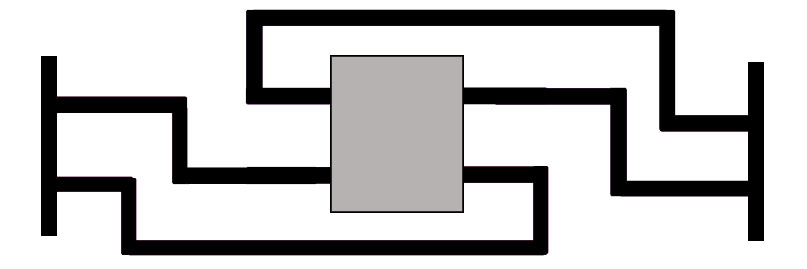}}
  \caption{The definition of the operator $S$.  In (a) we see a generic braid.  In the gray region the strands braid over each other in an arbitrary fashion.  In (b) the action of $S$ on this braid.}
  \label{fig:sdef}
\end{figure}

Having defined the generator $T$ let us now turn to the generator $S$.  In our braid diagrams time flows from left to right vertical slices define the notion  of space.  The operator $T$ respects this partition into space and time directions since it preserves the pairs of endpoints that appear as initial and final points of the braid.  By contrast, the operator $S$ will not respect this partition into space and time and mixes what were originally the initial and final endpoints of the braid.  Specifically, our definition of $S$ is to permute the endpoints of the braid as shown in Figure \ref{fig:sdef}.

\begin{figure}[here!]
  \centering
  \subfloat[Closure of the Trivial Braid]{\label{fig:basicclose}\includegraphics[width=.48\textwidth]{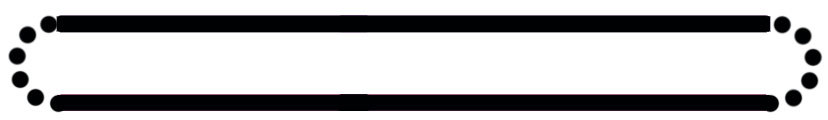}}     
  \hspace{.2in}         
  \subfloat[Closure of $S$]{\label{fig:sclose}\includegraphics[width=0.48\textwidth]{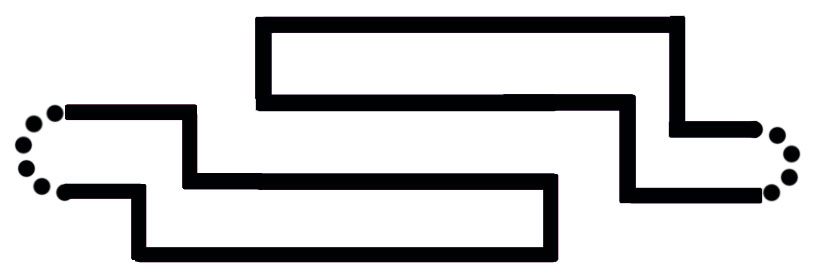}}
  \caption{The operator $S$ changes the gauging prescription.  In (a), the closure of a trivial braid leads to an unknot.  In (b) the closure of $S$ leads to two unlinked circles.  This changes the topology of $\widetilde{M}$ by increasing $b_{1}$. }
  \label{fig:closecompare}
\end{figure}

Given that $S$ creates no new crossings in the diagram, we will not associate the creation of new chrial particles with its action.  However, the operator $S$ does have the desired effect of gauging the background $U(1)$.  To illustrate this fact consider the comparison of the closure of the trivial braid with the closure of the braid defined by $S$ as shown in Figure \ref{fig:closecompare}.  
In the case of the trivial braid, the closure forms a connected unknot.  However, in the case of inserting $S$, the closure defines two unlinked circles.  In the first case, the cycle $\gamma$ encircling the two components of the braid, has become contractible at infinity and the associated $U(1)$ is not gauged.  Meanwhile in the case of the $S$ braid, $\gamma$ remains as a homologically non-trivial one-cycle and hence in this theory the $U(1)$ is gauged. Topologically, the compactified double cover geometry has changed to $\widetilde{M}_{c}\cong S^{1}\times S^{2}$.   

\begin{figure}[here!]
  \centering
\includegraphics[width=.6\textwidth]{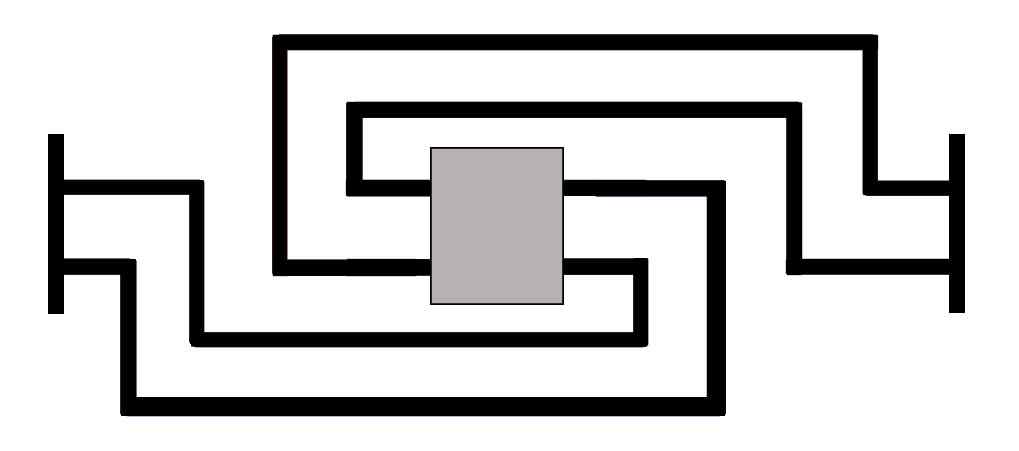}     
  \caption{The action of $S^{2}$.  This acts as time reversal on the braid.}
  \label{fig:s2}
\end{figure}

From these two definitions of $S$ and $T$, we may now see that they satisfy the required relations to generate an action of $SL(2,\mathbb{Z})$.   This means that $S^{2}$ must be a central element whose square is the identity (sometimes written as $S^{2}=-1$), and further that $(ST)^{3}=1$.  To begin consider the action of $S^{2}$ shown in Figure \ref{fig:s2}.  
As compared to the original braid, the action of $S^{2}$ has been to reverse the direction of time flow by changing the initial versus final conditions.  Thus, $S^{2}$ is simply time reversal along the R-flow and hence acts centrally.  Since reversing time twice is the identity operation, we conclude that $S^{4}=1$. 

\begin{figure}[here!]
  \centering
\includegraphics[width=.9\textwidth]{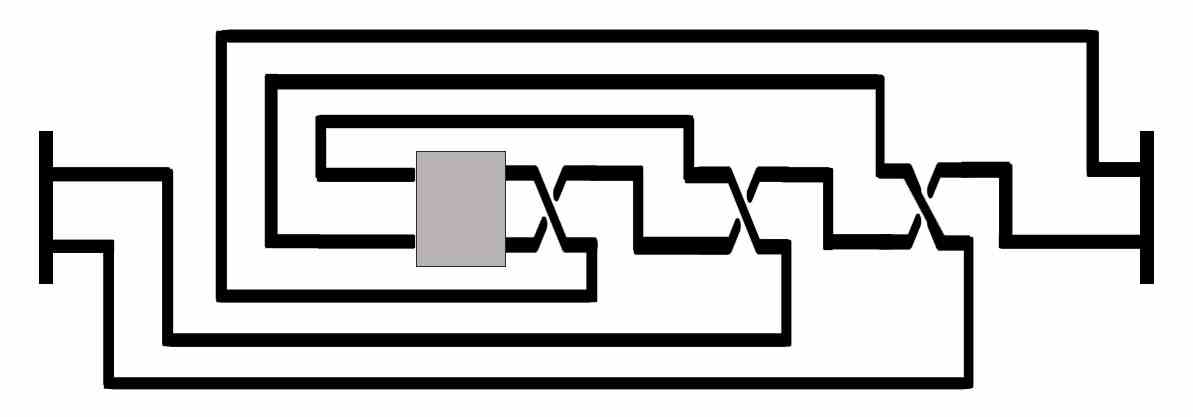}     
  \caption{The action of $(ST)^{3}$.  On braids this is the identity operator.}
  \label{fig:st3}
\end{figure}

Similarly, we may consider the action of $(ST)^{3}$ illustrated in Figure \ref{fig:st3}.  
One can see, manifestly from the above, that the operator $(ST)^{3}$ acts as the identity on the braid.  This completes the verification of the $SL(2,\mathbb{Z})$ group structure.  

Given that we have completely specified our choices of boundary data, we may now ask quite generally:  what are the possible IR geometries $\widetilde{M}_{c}$ which we obtain by these methods?  Since the geometry is determined by the resulting closed knot obtained from capping off the braid diagram, we may alternatively ask: what is the set of knots that we can obtain from the trivial braid  by repeated action of $S$ and $T$?  The answer to this question is exactly the set of \emph{rational knots}.  They are completely classified by their so-called Conway fraction, $z,$ which is valued in $\mathbb{Q}\cup \{\infty\}$.  To define this fraction, we first normalize $z$ by setting its value for the link defined by the closure of $S$ shown in Figure \ref{fig:sclose} to be 0.  Then, given any rational knot $\mathcal{K}_{1},$ constructed by action of $\rho\in SL(2,\mathbb{Z})$ from the rational knot $\mathcal{K}_{2},$ we set
\begin{equation} 
z(\mathcal{K}_{1})=\rho \left(z(\mathcal{K}_{2})\right).
\end{equation}
Where in the above the action of $\rho$ on $z$ is the usual action of $SL(2,\mathbb{Z})$ as fractional linear transformations.  

The result of this construction is thus an invariant fraction $z=p/q$ associated to each rational link.  We demand that the integers $p$ and $q$ are coprime.  Consider two such rational knots with Conway frations $z_{1}=p_{1}/q_{1}$ and $z_{2}=p_{2}/q_{2}$.  Then, a theorem due to Schubert asserts that the resulting knots are isotopic (that is equal as knots) if and only if 
\begin{equation}
p_{1}=p_{2}, \hspace{.5in}q_{1}\equiv q_{2}^{\pm1}  \mod p_{i}.
\end{equation}
This is exactly the same arithmetic conditions that occur in the classification of lens spaces $L(p,q)$.  This is not a coincidence.  The double branched cover of the $S^{3}$, branched over the rational knot with Conway fraction $p/q,$ is $L(p,q)$.  Thus we recover our original answer.  The IR geometries $\widetilde{M}_{c}$ for the tetrahedron theory are exactly the lens spaces.

\begin{figure}[here!]
  \centering
  \subfloat[Closure of the Tetrahedron Braid]{\label{fig:tetraclosed}\includegraphics[width=.35\textwidth]{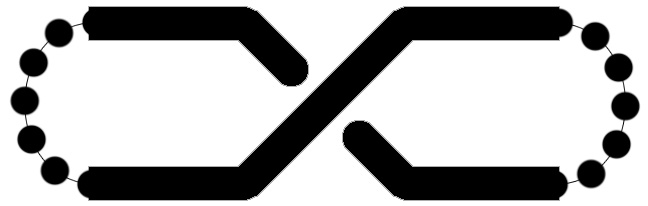}}     
  \hspace{.2in}         
  \subfloat[Action by $ST$]{\label{fig:sttetra}\includegraphics[width=0.6\textwidth]{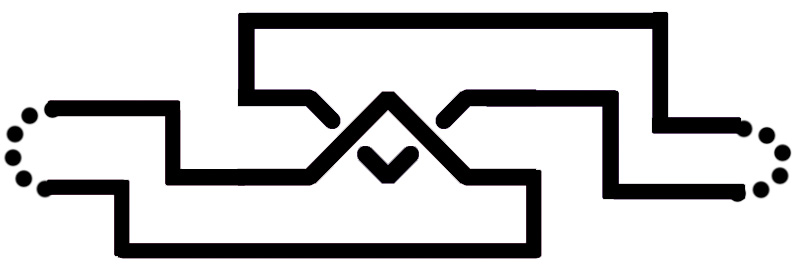}}
  \caption{Computing the CS level for the background $U(1)$ in the tetrahedron theory.  In (a) we see the closure of the basic tetrahedron braid.  In (b), the action of $ST$ on this braid changes the topology of the cover.}
  \label{fig:cscomp}
\end{figure}

Finally, let us note that these methods allow us to fully specify the basic theory associated to closing the tetrahedron braid shown in Figure \ref{fig:tetraclosed}.  Indeed, the closed knot shown there is the unknot and thus there is no gauged $U(1)$.  However, the chiral particle is still charged under a flavor $U(1)$.  To full specify the resulting theory, it remains to determine the background CS level $\hat{k}$ for this flavor $U(1)$.  If $\hat{k}$ is non-vanishing then, upon gauging the background $U(1)$, that is acting with the operator $S$, we obtain a three-manifold cover $\widetilde{M}_{c}$ which has $H_{1}(\widetilde{M}_{c},\mathbb{Z})\cong \mathbb{Z}_{\hat{k}}$.  Meanwhile, if $\hat{k}=0$ then acting with $S$ produces a geometry with non-vanishing first Betti number.

Now, we know that $T$ acts to change the CS level by one unit, and hence the operator $ST^{-\hat{k}}$ must act on Figure \ref{fig:tetraclosed} to produce a cover geometry with $b_{1}(\widetilde{M}_{c})=1$.  However, as is clear from Figure \ref{fig:sttetra}, the action of $ST$ on the tetrahedron braid produces topologically two unlinked circles.  The double cover of $S^{3}$ branched over two unlinked circles is precisely $S^{2}\times S^{1}$ which indeed has $b_{1}=1$.  Therefore we conclude that the basic tetrahedron theory defined by the closed knot in Figure \ref{fig:tetraclosed} has CS level $\hat{k}=-1$.  This is identical to the definition of the theory given in \cite{DGG}.

\subsubsection{Doubled Tetrahedron as a Special Lagrangian in $\mathbb{C}^3$}
We have now described a class of IR geometries relevant for the study of the $A_{1}$ domain walls.  These are special Lagrangians presented as double covers of the tetrahedron, and are given by the Joyce-Harvey flow of the SW geometry $y^2=x^2-m$ in $\mathbb{C}^3$.  It is therefore natural to try and identify these special Lagrangian subspaces of $\mathbb{C}^3$ more explicitly.  As already mentioned, before imposing boundary conditions, the IR special Lagrangian geometry $\widetilde{M}$ is a non-compact solid torus.  Furthermore, the geometry supports a unique BPS state described by an M2 brane ending on an $S^{1}$ inside the solid torus.  In other words, we should be looking for a non-compact special Lagrangian in $\mathbb{C}^3$ which has the topology of $T^2\times \mathbb{R}_+$ where at the origin of $\mathbb{R}_+$ one of the two circles of $T^2$ shrinks to a point.   The M2-brane boundary is then also located at the origin of $\mathbb{R}_{+}$, and is supported on the non-contracted $S^1\subset T^2$.  Precisely such special Lagrangian submanifolds have been constructed by Joyce \cite{JoyceCount}, and figure prominently in the study of open string mirror symmetry \cite{AVTOP}.  Here we will recall some facts about this class of special Lagrangians.

\begin{figure}[here!]
  \centering
\includegraphics[width=.48\textwidth]{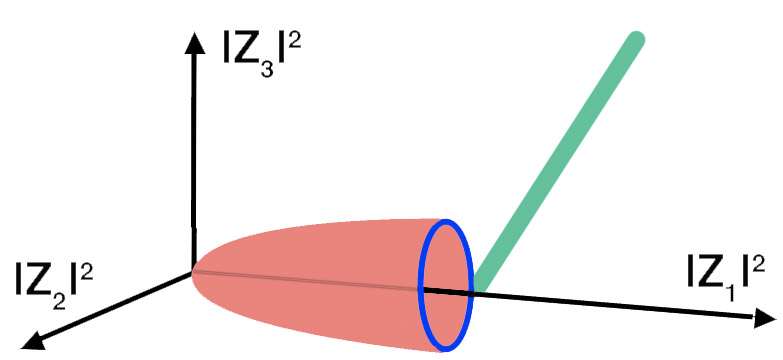}     
  \caption{Toric special Lagrangian in $\mathbb{C}^{3}$.  The green ray denotes the toric projection of the $T^{2}\times\mathbb{R}_{+}$ special Lagrangian.  The red disc is the worldvolume of a BPS M2-brane in $\mathbb{C}^{3}$ which ends on the special Lagrangian on the blue circle, and gives rise to a BPS particle in $\mathbb{R}^{1,2}$. }
  \label{fig:c3slag}
\end{figure}

Let $z_i$ for $i=1,2,3$ denote the three complex coordinates of $\mathbb{C}^{3}$.  Then, the special Lagrangians of interest can be depicted as follows:
\begin{equation}\label{Mtetra}
|z_3|^2=|z_2|^2=|z_1|^2-\frac{m_{0}}{2\pi}, \hspace{.5in} \theta_1+\theta_2+\theta_3=0
\end{equation}
where $m_{0}>0$.  Another way to characterize this subspace is as the locus where
\begin{equation}
{\overline z_3}={z_1z_2\over |z_1|},\hspace{.5in} |z_1|^2=|z_2|^2+\frac{m_{0}}{2\pi}. \label{toriclag}
\end{equation}
From the second description, we see that this subspace has the topology of $\mathbb{C}\times S^1$, parameterized by $\{ z_2, \theta_1\}$.  We can view this as a $T^2$ fibration over $\mathbb{R}_{+}$, where the torus is made of the angles $\theta_1,\theta_2$, $\mathbb{R}_+$ is parameterized by $|z_2|$, and at the origin of $\mathbb{R}_+$, the $\theta_2$ circle shrinks.  The projection of this special Lagrangian on the base of the toric representation of $\mathbb{C}^3$, given by $(|z_1|^2,|z_2|^2,|z_3|^2)$ is shown in Figure \ref{fig:c3slag}.  
Note that this special Lagrangian supports a unique M2-brane \cite{AVTOP}, which ends on the $\theta_1$ circle at the origin of $z_1$ space. In fact, one can show that if $M\subset\mathbb{C}^3$ is any special Lagrangian submanifold, then the harmonic form $\lambda$ is
\begin{equation}
 \lambda = \sum_i |z_i|^2\, d\theta_i\big|_M,
\end{equation}
 and from equation \eqref{Mtetra} we get
\begin{equation}
 \lambda = \frac{m_{0}}{2\pi}\, d\theta_1,
\end{equation}
and the mass of the corresponding BPS state is $m_{0}$. 
 
This geometry thus has all the characteristics we expect for the special Lagrangian corresponding to the double cover of the tetrahedron, and we conjecture that they are equal.   In fact, given our explicit description we can see how the double cover works: it is simply given by the complex conjugation action on $\mathbb{C}^3$
\begin{equation}
z_i\rightarrow {\overline z_i}.
\end{equation}
This is clearly a symmetry of the space defined by \eqref{toriclag}.  Furthermore, the fixed locus of this geometry are two strands given by 
\begin{equation}
(z_2\in {\mathbb{R}},\theta_1=0) \hspace{.5in} \mathrm{and} \hspace{.5in} (z_2\in {\mathbb{R}},\theta_1=\pi).
\end{equation}
A further check for the identification of the this subspace with the double cover $\widetilde{M}$, is that if we compactify the theory on $S^1$ then the moduli space of the special Lagrangian submanifold is given by the mirror geometry defined by a pair of complex variables $(u,v)$ subject to the relation
\begin{equation}
e^u+e^v=1.
\end{equation}
This is exactly the moduli space of $SL(2,\mathbb{C})$ Chern-Simons theory on the tetrahedron.   As we will explain in section 7, this is to be expected and demystifies some of the observations in \cite{tudor}, and explains why the partition function of $SL(2,\mathbb{C})$ Chern-Simons on a tetrahedron should be the same as that of the special Lagrangian brane on $\mathbb{C}^3$.  Furthermore, this shows why the partition function of the $SL(2,\mathbb{C})$ Chern-Simons on a tetrahedron should be that of the open topological string for this special Lagrangian A-brane.  

\begin{figure}[here!]
  \centering
\subfloat[Compactification to $S^{2}\times S^{1}$]{\label{fig:slags2s1}\includegraphics[width=.35\textwidth]{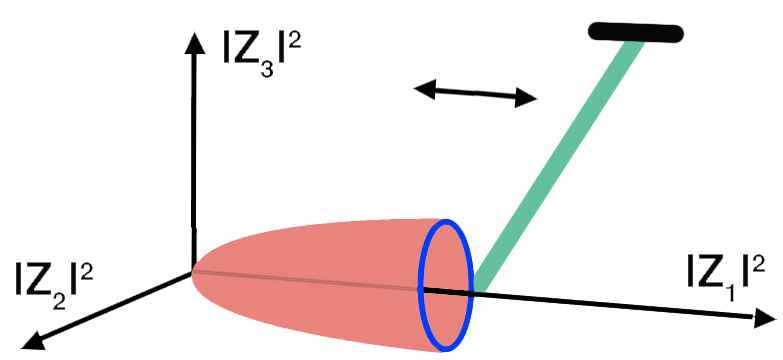}}  
\hspace{.5in}
\subfloat[Compactification to $S^{3}$]{\label{fig:slags3}\includegraphics[width=.35\textwidth]{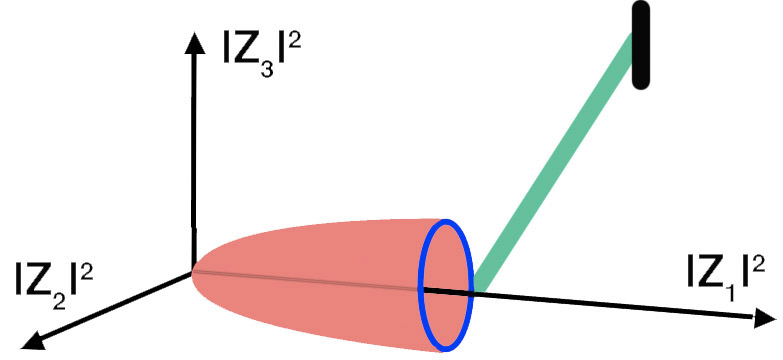}}     
  \caption{Toric compactifications of the special Lagrangian. In (a) the compactified geometry is $S^{2}\times S^{1}$ and has a modulus which is described by sliding it along the $|z_{1}|$ axis.  In (b) the compactified geometry is $S^{3}$ and the special Lagrangian is rigid. }
  \label{fig:c3slagcompact}
\end{figure}

Finally, let us note that the identification of the IR geometry $\widetilde{M}$ as an explicit special Lagrangian in $\mathbb{C}^{3}$ yields yet one more way to see the $SL(2,\mathbb{Z})$ action on boundary conditions, and to recover the fact that the compactified geometries $\widetilde{M}_{c}$ are lens spaces.  Specifically, we can consider toric compactifications of the subspace \eqref{toriclag}.  For example, we can complete the $\mathbb{C}\times S^1$ geometry to $S^2\times S^1$, which corresponds to having a locus where $\theta_2$ shrinks, depicted torically in Figure \ref{fig:slags2s1}.  Note that here the special Lagrangian has a modulus corresponding to `sliding' it along the $|z_1|$ axis.  Thus in this phase the $U(1)$ is gauged, and coupled to a charged chiral field described by BPS M2-brane.  Suppose instead we want to have the geometry of $S^3$.  This corresponds to shrinking the $\theta_1$ circle, which is depicted in Figure \ref{fig:slags3}.  In this case we have no gauged $U(1)$ but we still have a chiral field living on the M5-brane, again described by the M2-brane ending on the special Lagrangian. Similarly we can obtain lens space geometries.  For example,  $L(p,1)$ is obtained by having the circle $p[S^1]_1+[S^1]_2$ shrink at infinity.

\subsubsection{Black-White Duality, Mirror Symmetry and Geometric Transitions}
To summarize the results of the previous sections, we have obtained a class of 3d theories which are described in the IR by a single M5-brane on a lens space $L(p,q)$ together with a single BPS M2-brane charged under a gauged or global $U(1)$ symmetry of the theory.  To conclude our discussion of these theories, in this section we discuss simple examples of mirror symmetries.  

\begin{figure}[here!]
  \centering
  \subfloat[Checkerboard for the Basic Tetrahedron]{\label{fig:a1black}\includegraphics[width=.48\textwidth]{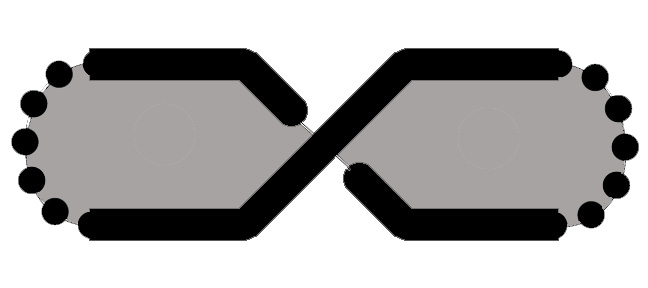}}     
  \hspace{.2in}         
  \subfloat[Dual Checkerboard for the Basic Tetrahedron]{\label{fig:a1white}\includegraphics[width=0.48\textwidth]{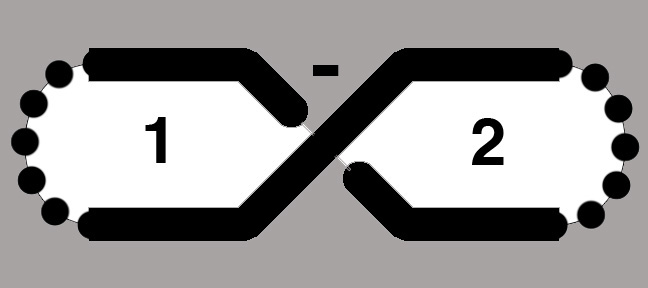}}
  \caption{Black-White duality for the tetrahedron theory.  In (a) we see a description of the theory with no gauge group.  In (b) there is a $U(1)$ with $\hat{k}=1$.  The two theories are dual. }
  \label{fig:bwda1}
\end{figure}

Let us revisit the basic tetrahedron theory.  We equip the resulting knot with a checkerboard coloring as shown in Figure \ref{fig:a1black}.
As explained in section 2, the checkerboard provides an algorithmic way to read off the data of the gauge multiplets on $\mathbb{R}^{1,2}$.  In Figure \ref{fig:a1black}, we see one white region, and hence no gauge field.    However, we may alternatively consider the dual checkerboard for the same knot shown in Figure \ref{fig:a1white}.  Now, there are two white regions and hence the gauge group is $U(1)$.  Further there is now a crossing connecting the white regions labeled 1 and 2 and correspondingly, the CS level for the $U(1)$ is $\hat{k}=1$.  

Thus, without changing any data about the knot, and hence without changing the field theory, we have found two distinct descriptions of the basic tetrahedron theory:
\begin{itemize}
\item A free chiral multiplet coupled to a background flavor $U(1)$ with level $\hat{k}=-1$.
\item A chiral multiplet coupled to a gauged $U(1)$ with level $\hat{k}=1$.
\end{itemize} 
Consistency of our formalism demands that these two descriptions are equivalent, and this is indeed a known mirror symmetry \cite{KapusStrass}.

\begin{figure}[here!]
  \centering
\includegraphics[width=.8\textwidth]{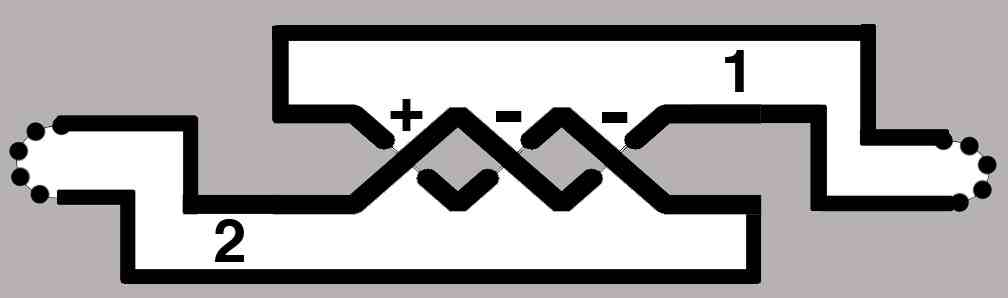}     
  \caption{The action of $ST^{2}$ on the basic tetrahedron theory.  This recovers the black-white duality of the theory as invariance under the operator $ST^{2}$.}
  \label{fig:st2}
\end{figure}

We can further investigate this basic duality by noting that the second description of the theory involving a gauged $U(1)$ is in fact identical to the action of $ST^{2}$ on the first description of the theory.  Thus, we can alternatively study this mirror symmetry by acting with the operator $ST^{2}$ on the knot in Figure \ref{fig:a1black}.  This produces the checkerboard shown in Figure \ref{fig:st2}.
Of course, geometrically one can clearly see that the knot defined by Figure \ref{fig:st2} is equivalent to that of Figure \ref{fig:a1white} as two of the crossing in the diagram are redundant and can be eliminated.  Nevertheless, it is still instructive to see that the algorithmic procedure of extracting the IR field content from the checkerboard produces the correct duality.  This is easily verified.  The two white regions yield one gauge field.  Summing over the crossings connecting the regions with the indicated sign then gives $\hat{k}=1$ and hence reproduces the black-white mirror symmetry above.

\begin{figure}[here!]
  \centering
\subfloat[$m_{0}>>0$]{\label{fig:recon1}\includegraphics[width=0.15\textwidth, height=0.15\textwidth]{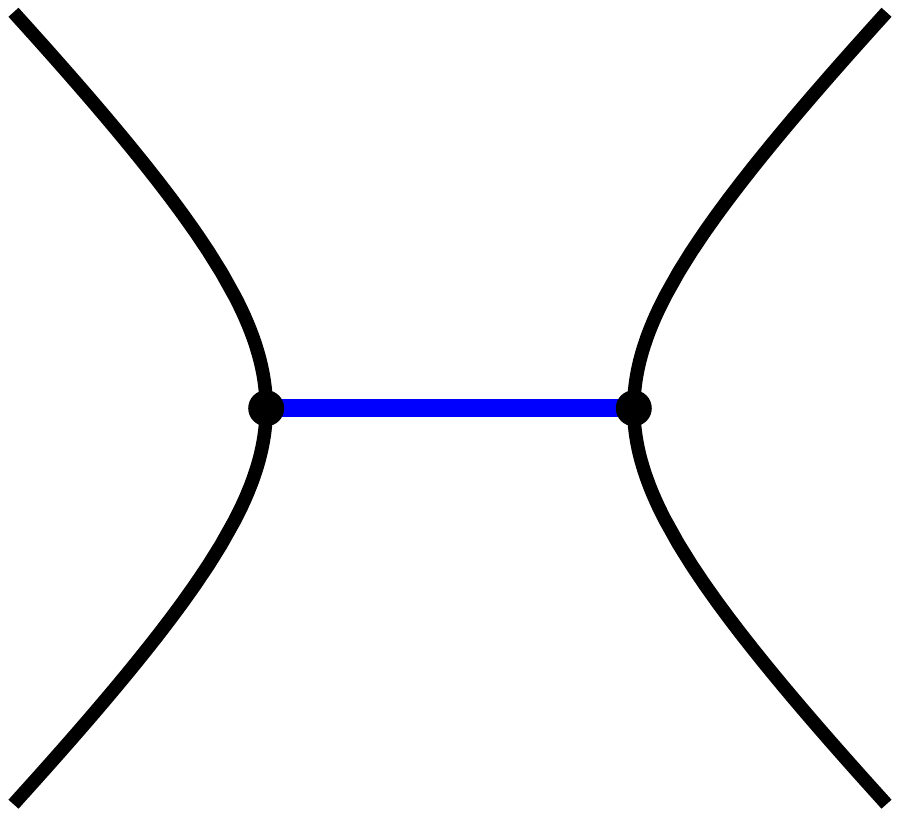}}
   \hspace{.15in}          
\subfloat[$m_{0}>0$]{\includegraphics[width=0.15\textwidth, height=0.15\textwidth]{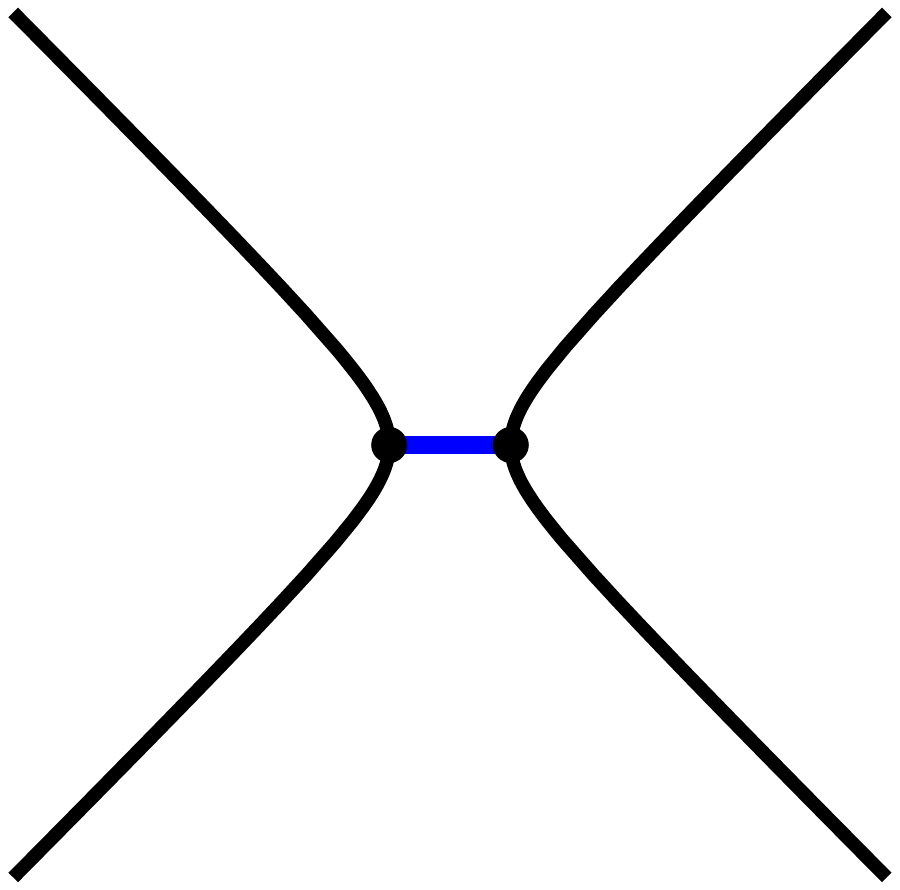}}
   \hspace{.15in}     
\subfloat[$m_{0}=0$]{\includegraphics[width=0.15\textwidth, height=0.15\textwidth]{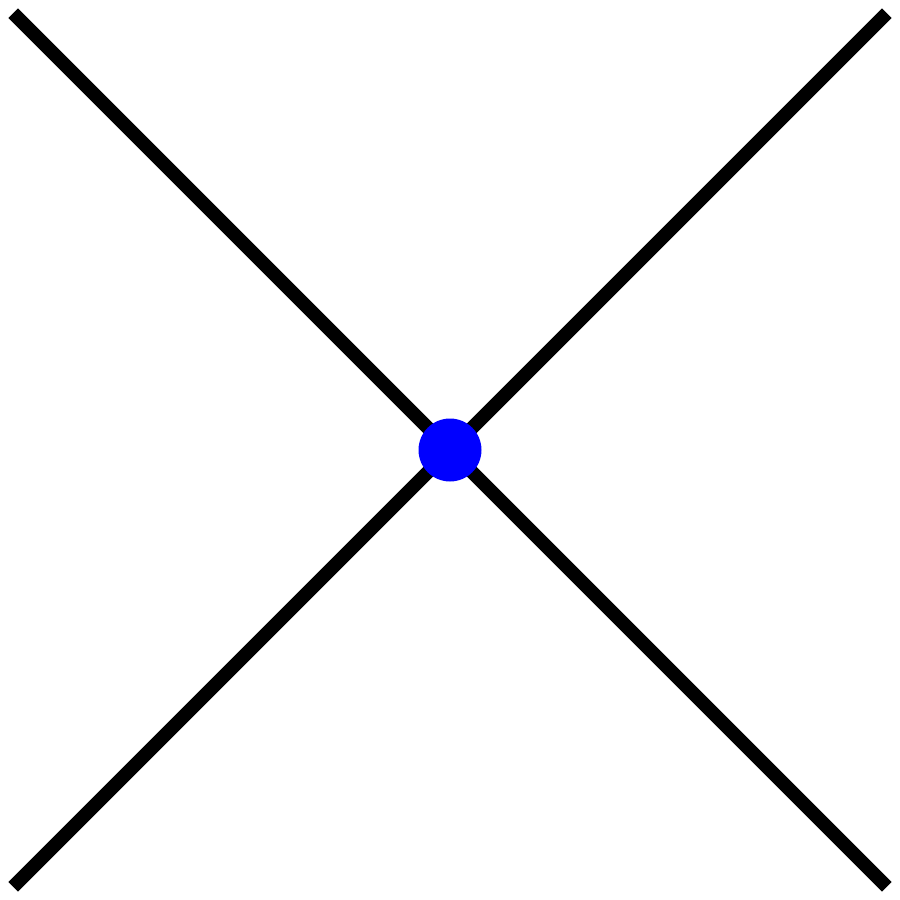}}
   \hspace{.15in}             
\subfloat[$m_{0}<0$]{\includegraphics[width=0.15\textwidth, height=0.15\textwidth]{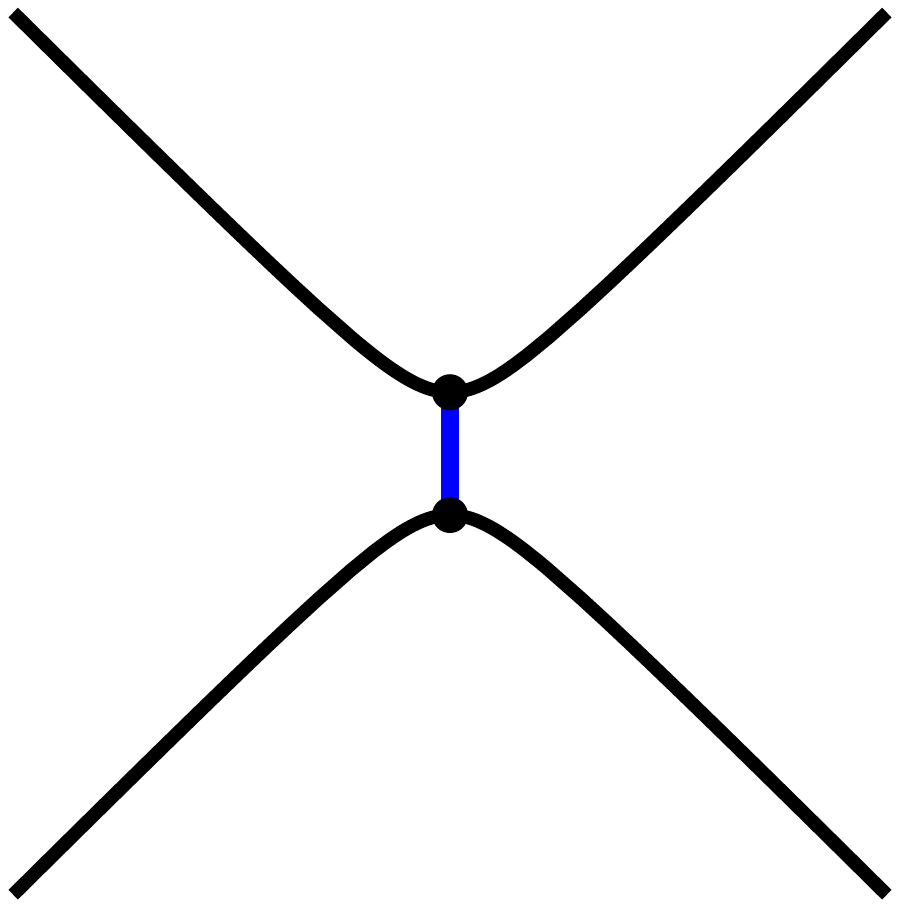}}
   \hspace{.15in}     
   \subfloat[$m_{0}<<0$]{\label{fig:recon2}\includegraphics[width=0.15\textwidth, height=0.15\textwidth]{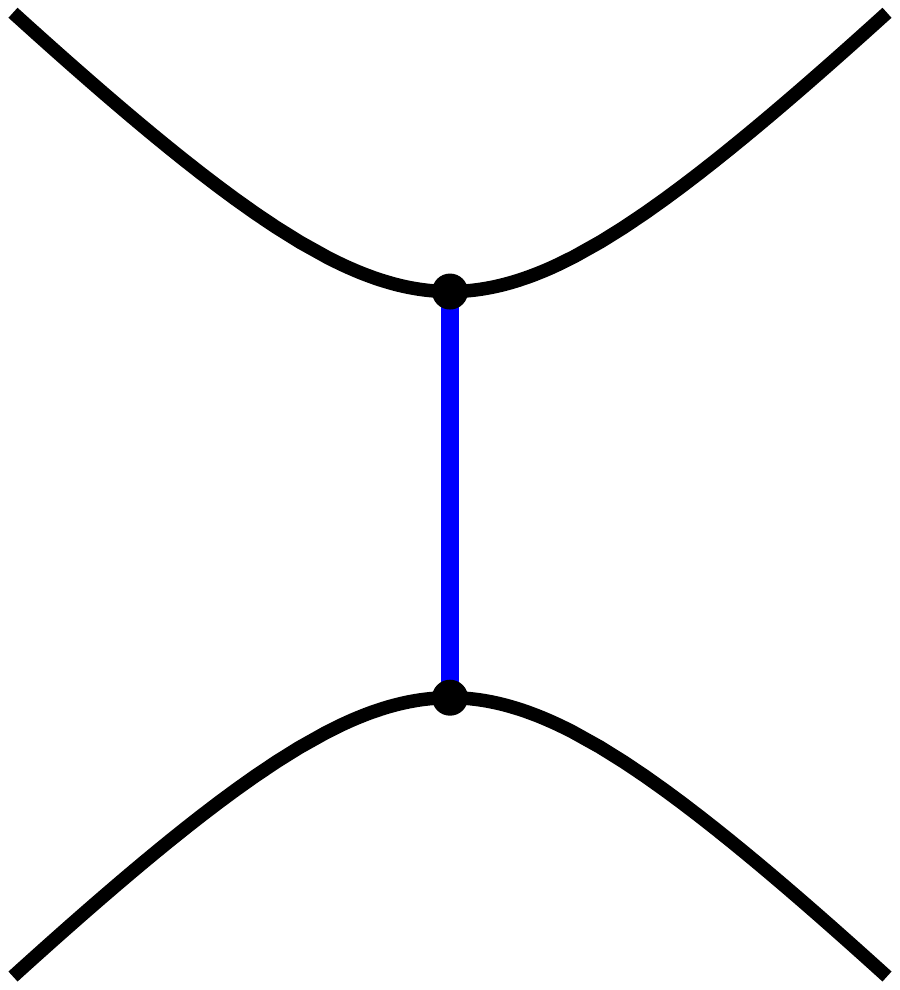}}
  \caption{The reconnection process.  The strands are illustrated in black and the blue line indicates the projection of the boundary of the BPS M2 brane.   In (c), when the mass $m_{0}$ of the particle vanishes, the two strands touch and their individual identity is ambiguous.  As $m_{0}$ becomes negative the strands reconnect.}
  \label{fig:reconb}
\end{figure}

Finally, we can also describe this duality in terms of a geometric transition.  We consider the basic tetrahedron theory encoded in Figure \ref{fig:a1black} and ask what happens as the mass $m_{0}$ of the chiral particle is smoothly taken through 0 to $-m_{0}$.  As studied in section 3, under this process the strands of the braid reconnect as illustrated in Figure  \ref{fig:reconb}.

In terms of the braid diagrams used throughout this section, we can describe this reconnection as follows.  First, as $m_{0} \rightarrow 0$ the braid develops a self-crossing.  Then, as $m_{0}$ becomes negative the original overcross is exchanged with an undercross.  This means that the theory has been acted on by the operator $T^{2}$.  Second, the strands reconnect.  This changes the identification of endpoints which occurs at infinity.  To see this, we compare the topology of the knot obtained by identifying the upper endpoints and the lower endpoints of Figure \ref{fig:recon1}, with the same identification performed in Figure \ref{fig:recon2}.  This changes the topology of the knot which is the signature of the operator $S$.  We conclude that the entire reconnection process is described by acting on the theory with $ST^{2},$ and thus reproduces the black-white duality.

In fact, the above line of reasoning, that is the study of BPS particles with vanishing masses, in some sense explains why it is possible to encode particles in a braid diagram to begin with.  The basic point is simply that when the mass is zero the strands must cross, and a braid diagram is simply a resolution of this situation to account for non-zero masses.
\subsection{$A_{2}$ Domain Walls:  The Bipiramid}\label{sec:domainwalss}
Having investigated the simplest possible example of domain walls in the $A_{1}$ model, we now turn to the $A_{2}$ theory.  This is the simplest 4d theory that exhibits the wall-crossing phenomenon.  In one chamber, there are two particles, while in the second chamber there are three particles.  This fact has dramatic implications for domain walls.  The spectrum in 3d is given by trapped particles from the ambient 4d theory.  Thus the different chambers in 4d yield 3d theories with distinct spectra.  For each such domain wall we must again specify boundary conditions.  A key observation is that the boundary conditions for the two domain walls are related in a non-trivial manner.  Thus, fixing a choice of boundary data in one wall, uniquely fixes the boundary data on the other wall, and hence fully specifies two 3d field theories.  As we will illustrate by example in this section, such pairs of 3d field theories, which are connected by 4d wall-crossing, are mirror pairs.  Thus, equivalence of the parent 4d theory under wall-crossing explains 3d mirror symmetry.
\subsubsection{The Two Chambers and the Pachner Move}
\label{2T}
We begin our analysis with a discussion of the geometry of the manifold $M$.  The $A_{2}$ theory is described by triangulations of a pentagon, and hence this is the front and back face of $M$ as shown in Figure \ref{fig:a2fb}.  In terms of the triangulation on its boundary, the manifold $M$ is therefore a \emph{bipiramid}, that is topologically a solid ball whose boundary is triangulated into six triangles.  

\begin{figure}[here!]
  \centering
 \includegraphics[ height=0.38\textwidth]{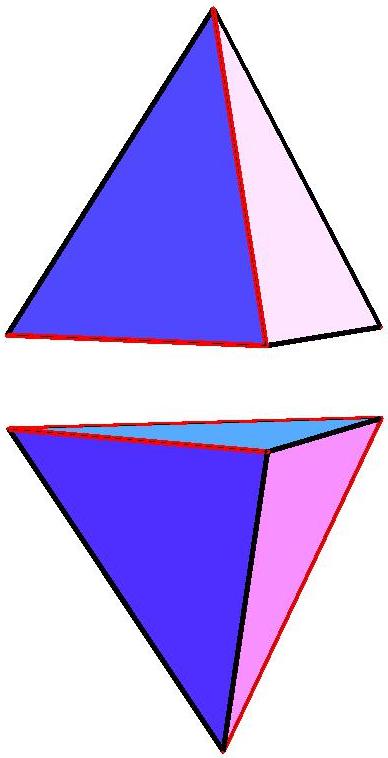}   
  \hspace{.3in}      
\includegraphics[ height=0.38\textwidth]{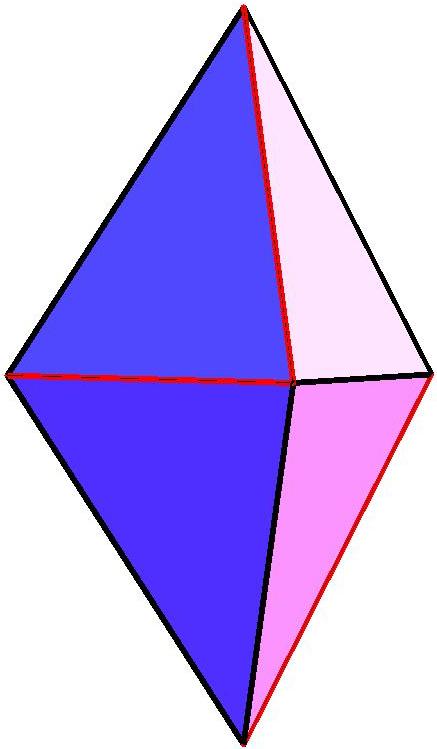}   
  \hspace{.3in}      
  \includegraphics[ height=0.38\textwidth]{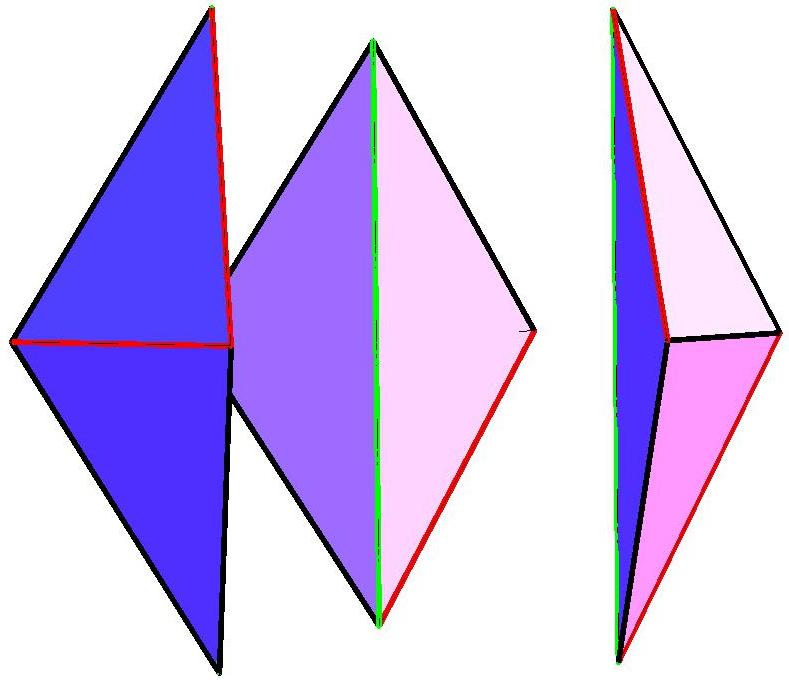}
  \caption{The two chambers of the $A_{2}$ theory give rise to two ways of decomposing the bipiramid.  In the center we see the bipiramid. The pentagon of the $A_{2}$ theory is given by the black edges.  The red diagonals flip as one flows from the front to the back face.  On the left, it is decomposed into two tetrahedra glued along a face.  On the right, it is decomposed into three tetrahedra glued along the green edge.}
  \label{fig:pachner}
\end{figure}

As in the discussion of the tetrahedron, as we flow through time, the triangulation on the front face evolves by a sequence of flips to the triangulation of the back face.  However because the $A_{2}$ theory exhibits wall-crossing, there are now two distinct ways in which the time evolution can occur.  One possibility is that in the course of time evolution, the triangulation will undergo two flips, and hence the 4d theory will support two BPS particles.  The other possibility is that the flow through time produces three flips, and hence three particles.  In each of these cases, a flip encodes a solid tetrahedron and a trapped BPS particle on the wall.  The two possible sequences of flips thus describe two distinct ways of decomposing the bipiramid into tetrahedra as illustrated in Figure \ref{fig:pachner}.

\begin{figure}[here!]
  \centering 
\includegraphics[width=0.6\textwidth]{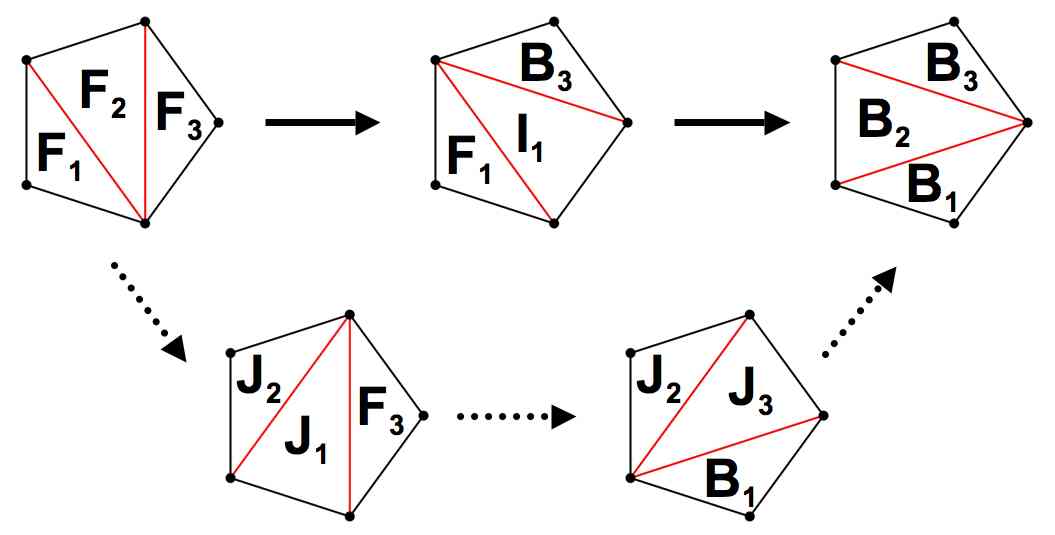}
  \caption{The flips of the $A_{2}$ theory reinterpreted as gluing data for a bipiramid.  In the above, the upper left pentagon with triangles $F_{i}$ is the front face.  The upper-right pentagon with triangles $B_{i}$ is the back face.  As flips happen in time, tetrahedra are created.  Following the solid arrows we see two tetrahedra, while following the dashed arrows we see three tetrahedra.}
  \label{fig:a2movie}
\end{figure}

Above and beyond simply indicating the number of tetrahedra, the sequence of flips on the triangulation completely specifies how the tetrahedra are to be glued together to form the manifold $M$.  Let us illustrate this feature for the case of the bipiramid.  We label the triangles in the front and back faces by $F_{i}$ and $B_{i}$, and let $I_{l}, J_{l}$ denote the possible triangles appearing in the interior of $M$ for the two chambers respectively.  Then, each flip of an edge $E\rightarrow E'$ is associated to four triangles: the two triangles adjacent to $E$  which appear before the flip and the two triangles adjacent to $E'$ which appear after the flip.  These give the four sides of each tetrahedron.  The complete sequences of flips then describes all the faces of all of the tetrahedra as illustrated in Figure \ref{fig:a2movie}.

\begin{figure}[here!]
  \centering
  \subfloat[2 particle chamber]{\label{fig:holoa21}\includegraphics[width=.3\textwidth]{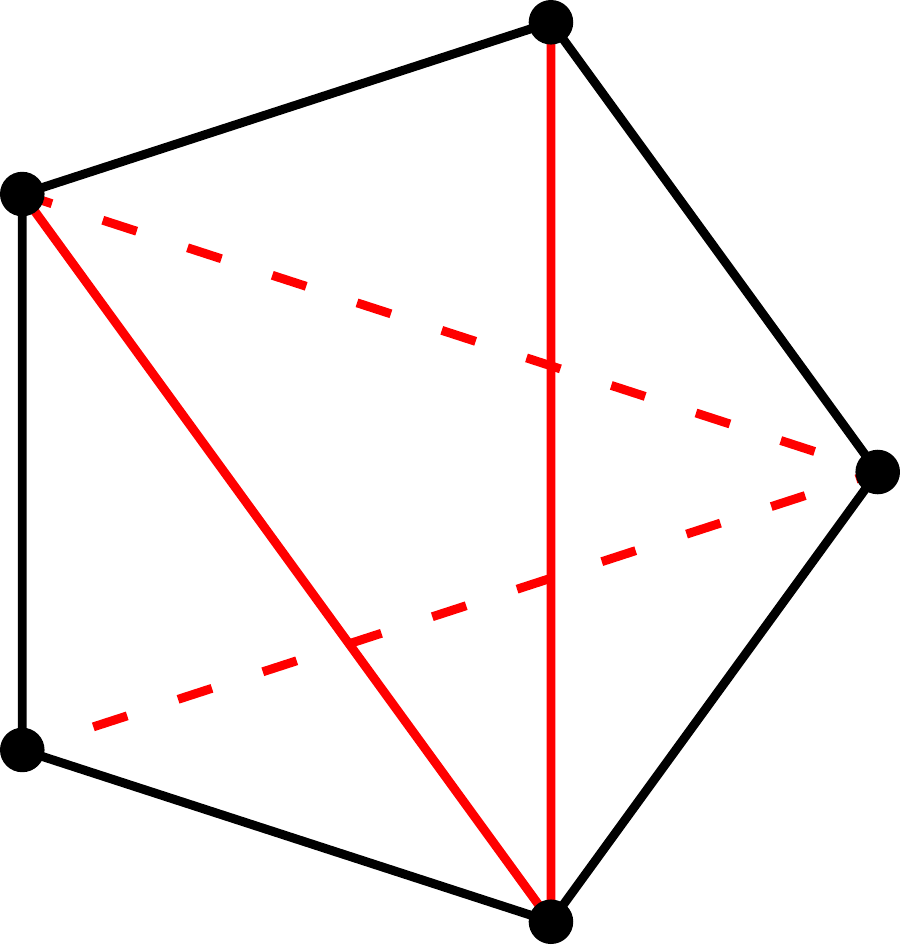}}     
  \hspace{.2in}         
  \subfloat[3 particle chamber]{\label{fig:holoa22}\includegraphics[width=0.3\textwidth]{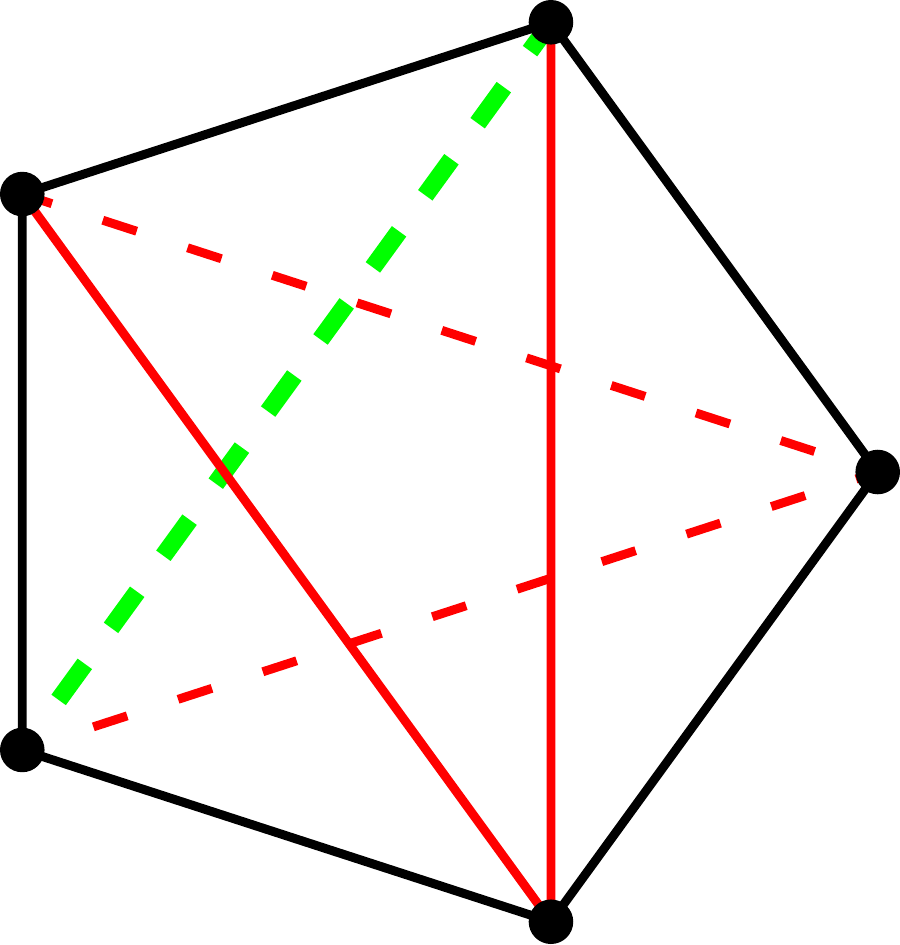}}
  \caption{The two chambers of the $A_{2}$ pentagon model viewed holographically from the front face.  The solid red lines are diagonals on the front face, while the dashed red lines are diagonals on the back face.  The dashed green line in (b) is the internal edge where the three tetrahedra are glued together. }
  \label{fig:holoa2}
\end{figure}

From this sequence of flips we can then easily extract the tetrahedra and their labelled faces.  Thus in the two particle chamber, the tetrahedra are
\begin{equation}
F_{2}F_{3}B_{3}I_{1} \hspace{.3in}\mathrm{and}\hspace{.3in} F_{1}I_{1}B_{1}B_{2}.
\end{equation}
While in the three particle chamber they are 
\begin{equation}
F_{1}F_{2}J_{1}J_{2} \hspace{.3in}\mathrm{and}\hspace{.3in} J_{1}F_{3}J_{3}B_{1} \hspace{.3in}\mathrm{and}\hspace{.3in} J_{2}J_{3}B_{2}B_{3}. 
\end{equation}
The gluing is then specified by simply identifying the shared faces.  As one can easily check, this reproduces the decompositions of the bipiramid shown in Figure \ref{fig:pachner}.  One can view the entire sequence of flips as giving rise to a 'holographic' view of the 3d geometry by drawing all edges that are flipped as in Figure \ref{fig:holoa2}.

Quite generally in the study of triangulated three-manifolds, the operation that we have just described where two tetrahedra glued along a face are replaced with three tetrahedra glued along an edge is known as a 2-3 \emph{Pachner mover}.  In our physical context, the 2-3 Pachner move is a geometric manifestation of the basic 2-3 wall-crossing of the $A_{2}$ theory.  As we will discuss later in this section, in generalizing to the $A_{n}$ model all wall crossing that we encounter is exactly of this 2-3 sort and is thus completely captured in 3d by the Pachner move.

\subsubsection{Boundary Conditions and Braids}
Now that we have addressed the UV geometry, we turn to the solution of the model as encoded by the IR geometry or equivalently its associated braid.  As usual $\widetilde{M}$ is a double cover of the solid ball.  The fastest way to understand its topology is to focus on the double cover of its boundary $S^{2}$.  This cover is branched over exactly the six points which are the zeros of the SW differential on the front and back face of $M$.  Since a double branched cover of the sphere branched at six points is a Riemann surface of genus two, we conclude that $\widetilde{M}$ must be a smooth three-manifold whose boundary is a surface of genus two.

\begin{figure}[here!]
  \centering  
\includegraphics[width=0.6\textwidth]{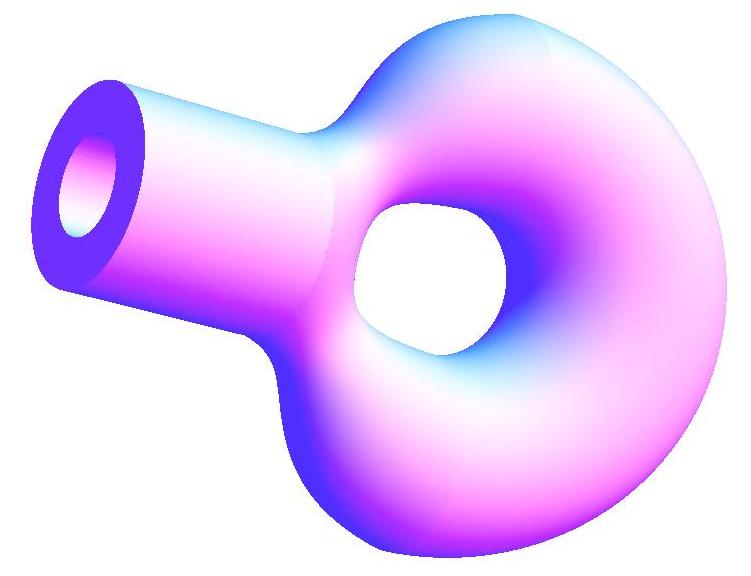}
  \caption{The IR geometry $\widetilde{M}$ for the bipiramid theory.  The SW curve is a punctured torus. $\widetilde{M}$ is a thickening of this to a torus bottle.  The boundary of $\widetilde{M}$ is composed of the exterior torus less a disc which appears on the outside of the bottle, along with the interior of the bottle which is again a torus less a disc.  These are glued together to form a Riemann surface of genus two.}
  \label{fig:torusb}
\end{figure}

We can be more specific about $\widetilde{M}$ by making use of the SW curve of the $A_{2}$ theory.  For the $A_{2}$ model, the SW geometry is a double branched cover of the complex $x$ plane described by the equation 
\begin{equation}
y^{2}=x^{3}+ax +b.
\end{equation}
Where in the above $a, b \in \mathbb{C}$ are parameters of the theory.  This SW curve is a punctured torus, i.e. topologically a torus minus a disc.  Then, the IR geometry $\widetilde{M}$ is a thickening of this Riemann surface and is therefore a  ``torus bottle,'' with boundary a genus two Riemann surface as shown in Figure \ref{fig:torusb}.
This picture allows us the determine the geometry more precisely.  Given the boundary $\partial\widetilde{M}$, we choose generating homology classes $A_{1}, A_{2}$ $B_{1}, B_{2}$ with canonical symplectic relations
\begin{equation}
A_{i}\cdot A_{j}= B_{i}\cdot B_{j}=0, \hspace{.5in}A_{i}\cdot B_{j}=\delta_{ij}.
\end{equation}
Then, the filling $\widetilde{M}$ is specified at the level of homology by choosing a pair of cycles in $\partial \widetilde{M}$ and declaring them to be contractible in the interior.  In our case, as is manifest from Figure \ref{fig:torusb}, the cycles which become contractible in the interior of $\widetilde{M}$ are $A_{1}-A_{2}$ and $B_{1}-B_{2}$.

The fact that $\widetilde{M}$ has boundary given by a surface of genus two makes clear the fact that the 3d field theories that we obtain from such domain walls will naturally be acted on by $Sp(4,\mathbb{Z})$.  Indeed, to completely specify our theory we must now impose boundary conditions on the manifold $\widetilde{M}$.  This means that we must complete this IR geometry to a manifold without boundary.  As $\widetilde{M}$ has boundary a surface of genus two, to remove the boundary we must glue $\widetilde{M}$ to another manifold with boundary a genus two surface.  Now to specify the gluing, we choose an element of the mapping class group of $\partial \widetilde{M}$ and glue the boundaries together.  On considering the action of this mapping class group element in homology of $\partial \widetilde{M},$ we obtain the desired action of $Sp(4,\mathbb{Z})$.\footnote{\ Note that the choice of boundary data intrinsically involves the mapping class group as opposed to merely $Sp(4,\mathbb{Z})$.  It would be interesting to discover a physical phenomenon which is sensitive to the more refined data of the mapping class group element }

From the point of view of the domain wall construction, the action of $Sp(4,\mathbb{Z})$ that we are describing is physically natural.  After the decoupling limit, the 3d wall theory comes equipped with a coupling to two $U(1)$'s, which are the electric and magnetic gauge fields that propagate in the bulk.  As in the case of the tetrahedron model, our choice of boundary conditions involves a specification of whether or not these $U(1)$ are gauged and what their background CS level is.  Then, $Sp(4,\mathbb{Z})$ acts naturally on this data with the various $S$ transformations inducing gaugings and the $T$ transformations changing CS levels.  

\begin{figure}[here!]
  \centering
  \subfloat[2 Particle Chamber]{\label{fig:2part}\includegraphics[width=.4\textwidth]{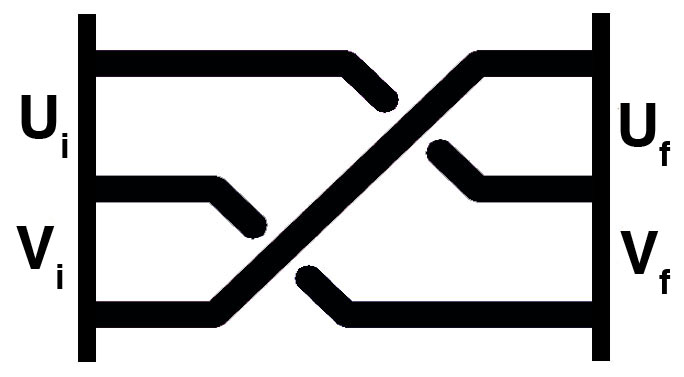}}     
  \hspace{.6in}      
  \subfloat[3 Particle Chamber]{\label{fig:3part}\includegraphics[width=0.4\textwidth]{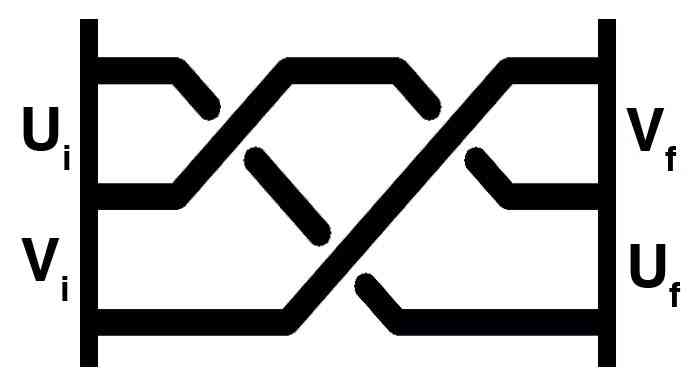}}
  \caption{The two braids associated to the bipiramid theory.  In (a) we see the two particle braid with two moves.  In (b) the three particle chamber with three braid moves.  The labels $u_{i}$ $v_{i}$ denote cycles in the initial and final Riemann surface.  Notice that in the case of the three particle chamber the final cycles differ from the initial ones by a permutation. }
  \label{fig:23braids}
\end{figure}

To really pin down the IR physics, we now turn to a more detailed description of the geometry as defined by its associated braid.  As with our tetrahedral example, the geometric significance of this braid is that it is the branching locus for the double cover.  Since each face of the three-manifold $M$ has three triangles, the braid will be composed of three strands.  However, since the 4d $A_{2}$ theory exhibits wall-crossing there are two distinct braids that we can associate with these domain walls corresponding to the two chambers of the 4d theory.  The first has two particles and hence two braid moves, while the second has three particles and hence three braid moves.  These are shown in Figure \ref{fig:23braids}.
 
 The braids shown in Figure \ref{fig:23braids} have a number of significant properties that demand explanation.  To do that we recall from section 4 that we may describe the BPS spectrum by a sequence of mutations on the $A_{2}$ quiver.  The nodes of this quiver are cycles in the SW curve and hence are associated to a pair of branch points, or equivalently a pair of strands in the braid.  Thus for example in the quiver 
 \begin{equation}
 \xy 
 (-10,0)*+{}*\cir<8pt>{}="a" ; 
 (-10,-8)*+{u_{i}}="c" ; 
(10,0)*+{}*\cir<8pt>{}="b" ; 
 (10,-8)*+{v_{i}}="d" ; 
\ar @{->} "a"; "b" 
 \endxy
\end{equation}
the node labeled $u_{i}$ is associated to the cycle defined by the first and second strand as $t\rightarrow -\infty$.  Similarly, the node labeled $v_{i}$ is associated to the cycle defined by the second and third strand as $t\rightarrow -\infty$.

Now, for each mutation in the sequence defining the BPS spectrum in 4d, we obtain a chiral particle in 3d and hence a braid move.  In the case of a three strand braid relevant for our current example, we first label the strands as 1,2 ,3 going down the page.  Then, the braid group is generated by two elements $b_{12}$ and $b_{23}$ where $b_{ij}$ moves strand $i$ under strand $j$ in time order.  To determine which braid move we do, we look at which node of the quiver is being mutated.  Specifically:
\begin{itemize}
\item If node 1 is mutated do the braid move $b_{12}$.
\item If node 2 is mutated do the braid move $b_{23}$.
\end{itemize}
This determines completely how the sequence of mutations describing the BPS spectrum is mapped to the sequence of braid moves encoding the 3d geometry.  As described in section 4, in the two particle chamber the we recall that the mutation sequence is given by node 2 followed by node 1, and yields the braid shown in Figure \ref{fig:2part}, while in the case of the three particle chamber the mutation sequence is $1, 2, 1$ and determines the braid shown in Figure \ref{fig:3part}.

The fact that the braids are determined by mutation sequences with nodes corresponding to cycles also explains another crucial feature of Figure \ref{fig:23braids}.  We recall from section \ref{sec:Quiver}, that at the conclusion of a sequence of mutations describing a 4d BPS spectrum the cycles have in general undergone a non-trivial permutation $\chi$.  As we saw there, this permutation element depends on the BPS chamber.  In the two particle chamber of $A_{2}$, we found that $\chi$ is the identity, while in the three particle chamber, $\chi$ was the non-trivial element of $S_{2}$.  This explains the labeling of cycles that we have made in Figure \ref{fig:23braids}.  In the two particle chamber the initial basis of cycles denoted $u_{i}, v_{i},$ agrees with the final basis of cycles denoted $u_{f}, v_{f}$.  Meanwhile in the three particle chamber the initial and final basis of cycles disagree, having been acted on by the permutation $\chi$.

\subsubsection{Duality}\label{sec:bkachwithe}

Now we are equipped to study how boundary conditions are imposed on the braids, and thus how we can use the result to extract an explicit Lagrangian description of the resulting field theories.  As in our analysis of the theory of a single tetrahedron, boundary conditions at the level of the braid are a specification of choices for how the braid is closed into a knot.  Then, this knot is the branching locus for the compactified IR geometry $\widetilde{M}_{c}$ presented as a double cover of $S^{3}$.  The most general set of boundary conditions thus involves choosing three pairs of the six endpoints of the braid to glue together.  Then, given any fixed gluing prescription, $Sp(4,\mathbb{Z})$ acts to produce another one by performing various $S$ transformations which change the gauging prescription, and $T$ transformations which act as additional braid moves creating ultra-massive BPS particles and changing the CS levels.  This generates an interesting family of knots in $S^{3}$ which classify in full generality the IR field theories that we obtain from domain walls in the $A_{2}$ theory.

Rather than investigate the general case of such knots, we instead note that our construction of these field theories as domain walls, naturally singles out a simple subclass of boundary conditions which respect the order of time flow.  Indeed, at each boundary $t=\pm \infty$, there exists a pair of cycles.  At $t=-\infty$ these are $u_{i}, v_{i},$ while at $t=+\infty$ these are $u_{f}$ and $v_{f}$.  To impose boundary conditions in general, means to choose cycles to be contractible.  Doing this in a way which preserves the time ordering implies that we choose one cycle from the initial set and one cycle from the final set and declare them to be contractible.  Thus, for example, we may choose $u_{i}$ and $u_{f}$.  Then, given such a choice there is an action on such boundary conditions not by the full $Sp(4,\mathbb{Z})$ group, but rather by the subgroup $Sp(2,\mathbb{Z})\times Sp(2,\mathbb{Z})$ acting on the initial and final trivialized cycles.  Explicitly, given $(g_{i},g_{f})\in Sp(2,\mathbb{Z})\times Sp(2,\mathbb{Z})$ the action is
\begin{equation}
u_{i}\rightarrow g_{i}u_{i}, \hspace{.5in}u_{f}\rightarrow g_{f}u_{f}.
\end{equation}
As with the general action, this induces changes in the gauging prescription and adds various CS levels, but it does so in a way which respects the time order defined by the flow on the geometry.

\begin{figure}[here!]
  \centering
  \subfloat[2 Particle Chamber]{\label{fig:2partg1}\includegraphics[height=.25\textwidth]{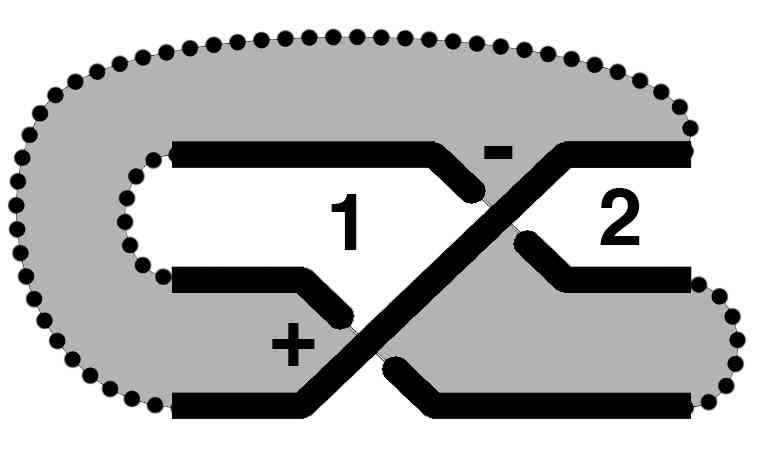}}     
  \hspace{.25in}      
  \subfloat[3 Particle Chamber]{\label{fig:3partg1}\includegraphics[height=0.25\textwidth]{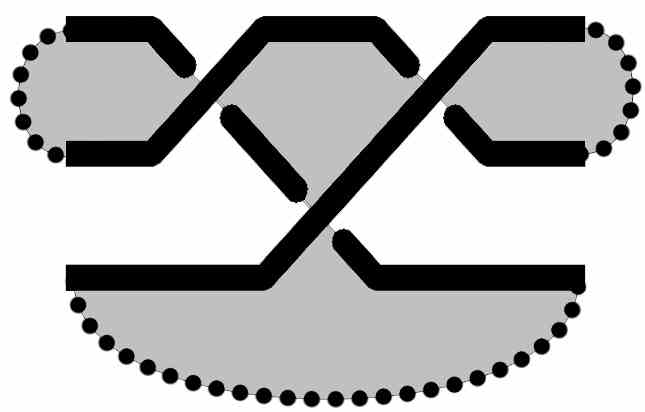}}
  \caption{Closed links and checkerboards for the gluing determined by trivializing $u_{i}$ and $v_{f}$.  In (a), we see the result from the two particle chamber.  In (b), the result from the three particle chamber.  The theory on the left is $N_f=1$ SQED.  The theory on the right is the $XYZ$ model.}
  \label{fig:23braidsg1}
\end{figure}

Let us now see some examples of such boundary conditions applied to the braids of Figure \ref{fig:23braids}.  We declare that $u_{i}$ and $v_{f}$ are contractible and connect the corresponding strands of the braid without introducing additional twists and CS levels.  Further, we give a checkerboard coloring to the resulting knots shown in Figure \ref{fig:23braidsg1}.

Since we have fully specified boundary conditions, we have now fully specified the compactified IR geometry $\widetilde{M}_{c}$ and thus we may now read off the field content and determine the resulting 3d field theories obtained for each chamber.  Let us consider first the theory determined by Figure \ref{fig:2partg1}.  We read off gauge structure by making use of the Seifert surface defined by the checkerboard coloring.  There are two white regions, labeled 1 and 2 in the Figure.  Correspondingly, there is 1 $U(1)$ gauge field in 3d theory on $\mathbb{R}^{1,2}$.  The resulting level $\hat{k}$ of this $U(1)$ is vanishing, since the net number of crossings between regions 1 and 2 vanishes.  Meanwhile, as we will derive later in this section, the two particles in the theory carry opposite $U(1)$ charges.  Thus, the theory encoded by the diagram of Figure \ref{fig:2partg1} is exactly $N_f=1$ SQED.

Similarly, we can read off the IR geometry and field content for the theory encoded by Figure \ref{fig:3partg1}.  There is now only one white region in the checkerboard and hence there are no gauge fields.  The theory supports three BPS chiral multiplets $X$, $Y$, and $Z$ encoded by the crossings in the braid diagram.  However, this theory has one additional crucial feature.  The triangular region of the knot diagram, bounded by the three crossings, is exactly the kind of geometry described in section 2 in which BPS M2-branes yield contributions to the superpotential.  This triangular region should be contrasted with other discs with boundary along the knot that are apparent in the diagram.  As we have previously explained, dashed regions of the knot encode boundary conditions \emph{at infinity}.  Thus, every disc which has its boundary along a dashed component of the knot has infinite volume and hence supports no M2-brane contributions to the superpotential.  However, the triangular region in question has all solid boundaries and hence supports a finite disc.  Thus, this theory also has a superpotential coupling its chiral fields as
\begin{equation}
W=XYZ.
\end{equation}
So defined, the theory described by Figure \ref{fig:3partg1} exactly the so-called $XYZ$ model.

\begin{figure}[here!]
  \centering
  \subfloat[2 Particle Chamber]{\label{fig:2partg2}\includegraphics[height=.25\textwidth]{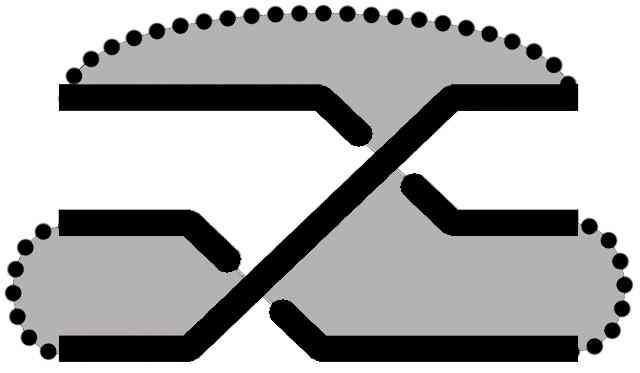}}     
  \hspace{.2in}      
  \subfloat[3 Particle Chamber]{\label{fig:3partg2}\includegraphics[height=0.25\textwidth]{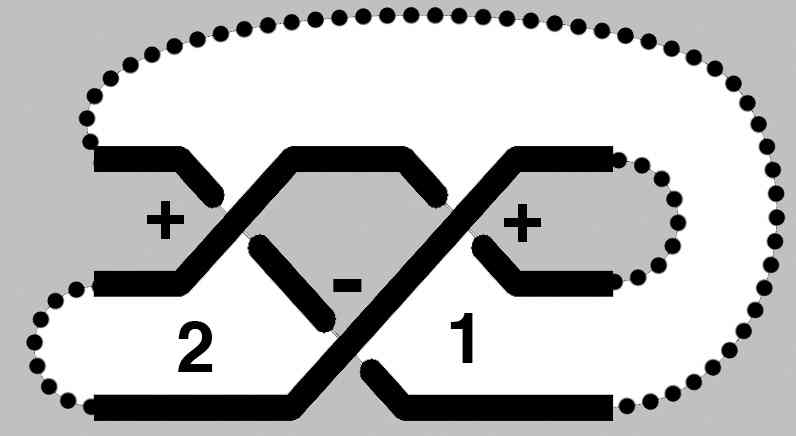}}
  \caption{Closed links for the gluing determined by trivializing $u_{i}$ and $u_{f}$.  In (a), we see the result from the two particle chamber.  In (b), the result from the three particle chamber.  The theory on the left is that of two neutral chirals.  The theory on the right is $\mathcal{N}=4$ $U(1)$ Yang-Mills.}
  \label{fig:23braidsg2}
\end{figure}

Now we observe a striking fact.  The two theories that we have produced via this construction are a mirror pair!  Both  $N_f=1$ SQED and the $XYZ$ model have the same IR dynamics near their conformal fixed points \cite{XYZ}.  This example illustrates a general phenomenon.  The two open braid diagrams of Figure \ref{fig:23braids} admit many different choices of boundary conditions.  However, if we fix boundary conditions in the two particle chamber of  Figure \ref{fig:2part}, then those of Figure \ref{fig:3part} are also fixed automatically by simply demanding that the same cycles are contractible.  Thus, fixing \emph{one} choice of boundary conditions determines \emph{two} theories, and the resulting models are always mirror pairs.  For another familiar example, we may consider trivializing $u_{i}$ and $u_{f}$ which has as a result Figure \ref{fig:23braidsg2}.

What theories are these?  The answer is again obtained by a trivial application of the now familiar rules.  In Figure \ref{fig:2partg2}, we see a theory with no gauge group and two uncharged particles say $X$ and $Y$.  Meanwhile in Figure \ref{fig:3partg2}, we see a theory with a $U(1)$ gauge group, vanishing level, two particles of opposite charge $Q$ and $\widetilde{Q}$ and a neutral particle $\Phi$ with superpotential coupling
\begin{equation}
W=\Phi Q \widetilde{Q}.
\end{equation}
The latter theory is thus exactly the $\mathcal{N}=4$ $U(1)$ gauge theory coupled to a fundamental hypermultiplet.  That this theory is mirror to a theory with just two neutral scalars is in fact the paradigmantic example of three-dimensional mirror symmetry \cite{IS}.

\begin{figure}[here!]
  \centering
  \subfloat[2 Particle Chamber]{\label{fig:qedwd}\includegraphics[height=.25\textwidth]{Figures/QEDblack.jpg}}     
  \hspace{.2in}      
  \subfloat[3 Particle Chamber]{\label{fig:qedbd}\includegraphics[height=0.25\textwidth]{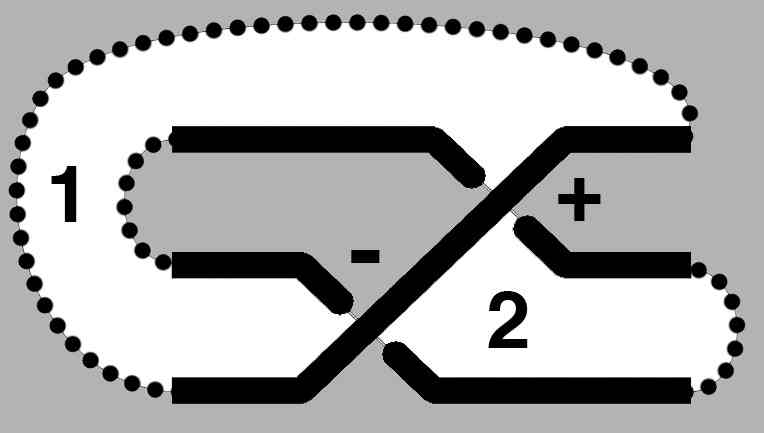}}
  \caption{Black-White duality for $U(1)$ QED.  The theory is self-dual.}
  \label{fig:QEDbwd}
\end{figure}

We can also investigate the role of black-white duality in these theories.  For example, we study first the case of SQED shown in Figure \ref{fig:QEDbwd}.
In terms of the physical content of the theories defined by the knot diagram, the two theories are identical.  Thus under black-white duality, $U(1)$ QED is self-dual

\begin{figure}[here!]
  \centering
  \subfloat[2 Particle Chamber]{\label{fig:XYZwd}\includegraphics[height=.25\textwidth]{Figures/XYZwhite.jpg}}     
  \hspace{.2in}      
  \subfloat[3 Particle Chamber]{\label{fig:XYZbd}\includegraphics[height=0.25\textwidth]{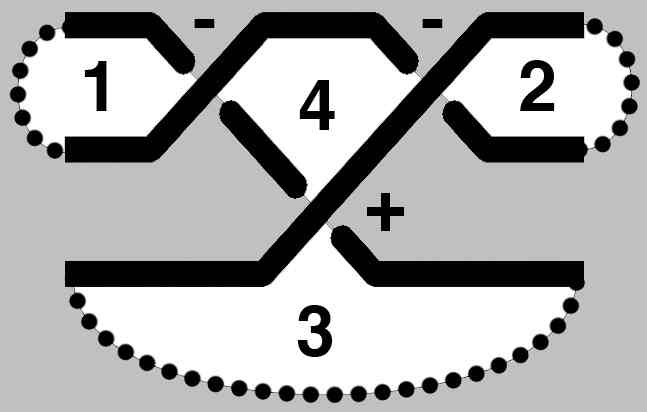}}
  \caption{Black-White duality for the XYZ model.  The theory on the right involves a non-local superpotential which contains the monopole operators.}
  \label{fig:XYZbwd}
\end{figure}

A more interesting case is given by the $XYZ$  model illustrated in Figure \ref{fig:XYZbwd}
In Figure \ref{fig:XYZbd}, there are now four white regions and hence three $U(1)$ gauge fields in the theory on $\mathbb{R}^{1,2}$.  If we take the generators to correspond to the regions labeled $1, 2, 3$, then the matrix of level is given by
\begin{equation}
\hat{k}_{ij}=\left(\begin{array}{ccc}
1 & 0 & 0 \\
0 & 1 & 0 \\
0 & 0 & -1
\end{array}
\right).
\end{equation}
There are three charged particles, $X_{i}$ charged only under $U(1)_{i}$, with charges (-1,-1,+1).  Again the region 4 bounded by undashed components of the knot encodes a superpotential term.  However, now we find a novelty.  Since region $4$ is white, the superpotential also couples to the monopole operators ${\cal M}_i=\exp(\sigma_i +i\gamma_i)$ of the corresponding $U(1)$.  Thus, in this case
\begin{equation}
W=({\cal M}_1X_{1})({\cal M}_2X_{2})({\cal M}_3X_{3}).
\end{equation}
Note that this is gauge invariant because ${\cal M}_i$ carries a $U(1)_i$ charge $\hat k_i$, due to the Chern-Simons terms.   We can see that this is a valid duality by simply invoking the black-white duality for the $A_{1}$ theory three times, once for each of the fields $X_{i}$, replacing ${\cal M}_iX_{i}$ with the dual field ${\tilde X}_i$.

\subsection{General $A_{n}$ Walls}
\label{GT}
Having investigated the two most basic examples, we now state our proposal for the general structure of domain walls in the $A_{n}$ theories. 
\subsubsection{Five-Brane Geometry}
First, there is the UV five-brane geometry $M$.  This is described by a solid ball whose front and back face are triangulations of the $(n+3)$-sidded polygon.  As time evolves, the triangulation of the front face will evolve into the triangulation of the back face by a sequence of flips.  Each flip describes a BPS state of the 4d theory and thus each state gives rise to a trapped 3d particle on the wall.  On the other hand,  each flip naturally describes a solid tetrahedron in $M$.  As in previous sections, the full sequence of flips then encodes a complete tetrahedral decompostion of $M$.  Thus, we have the natural identifications
\begin{equation}  
\mathrm{Flip}\leftrightarrow\mathrm{4d \ BPS \ particle} \leftrightarrow \mathrm{Trapped  \ 3d \ BPS \ particle}\leftrightarrow \mathrm{Tetrahedron}. \nonumber
\end{equation}
In particular, the induced tetrahedral decomposition of $M$ completely captures the 3d BPS spectrum of chiral multiplets.  In the IR, these are all of the matter particles that are physically relevant.

Just as in the $A_{2}$ example, the general $A_{n}$ 4d field theory can exhibit wall-crossing in its BPS spectrum.  This means that there are different chambers of 4d BPS states which in turn describe different possible spectra of trapped BPS states living on the domain wall.  According to our discussion above, this implies that the tetrahedral decomposition of the manifold $M$ is not fixed.  Rather, distinct chambers are related by the primitive 2-3 wall-crossing where in crossing the wall, a single hypermultiplet disappears from the spectrum.  The geometric manifestation of this in $M$ is precisely the 2-3 Pachner move.  In one BPS chamber there are three particles encoded by three tetrahedra glued along an edge.  In the second BPS chammber one of the particles disappear and the three tetrahedra glued along an edge are replaced by two tetrahedra glued along a face.

Next, we may describe the IR geometry $\widetilde{M}$ and the way in which it encodes the solution of the model.  The manifold $\widetilde{M}$ is a branched double cover of the (infinite) solid ball, with branching locus given by the braid determined by the zeros of $\lambda$ or equivalently the evolving SW differential.  At the asymptotic boundary of the ball, there is thus a sphere with $2n+2$ zeros of $\lambda$ describing the initial and final terminal points of the $n+1$ strands in the braid.  Thus, the boundary of $\widetilde{M}$ can be described as a double cover of the sphere branched over $2n+2$ points and is therefore a hyperelliptic Riemann surface of genus $n$.  It follows that  $\widetilde{M}$ is a filling in of this Riemann surface to a three-manifold.  

In fact, our knowledge of the SW curve of the ambient $A_{n}$ theory allows us to be more precise and to specify exactly which filling in is prescribed by the time flow.  Indeed, the SW curve is given by a a certain polynomial of degree $n+1$ in $x$ as
\begin{equation}
y^{2}=P_{n+1}(x).
\end{equation}
There are then two cases determined by the parity of $n$.
\begin{itemize}
\item $n$ even:  

$\widetilde{\Sigma}$ is a surface of genus $n/2$ which has been made non-compact by removing a single disc.  Then, $\widetilde{M}$ is a thickening of this to a genus $n/2$ bottle.  If we label the cycles on the initial surface $\widetilde{\Sigma}_i$ on the outside of the bottle as $(A_{i}, B_{i}),$ and those on the final surface $\widetilde{\Sigma}_f$ on the inside of the bottle as $(\tilde{A}_{i},\tilde{ B}_{i})$,  for $i= 1, \cdots, n/2$, the relations defining the filling in of the boundary surface of genus $n$ to make the manifold $\widetilde{M}$ are
\begin{equation}
A_{i}=\tilde{A}_{i}, \hspace{.5in} B_{i}=\tilde{B}_{i}.
\end{equation}
\item $n$ odd:
\begin{figure}[htb!]
  \centering
  \includegraphics[width=0.8\textwidth]{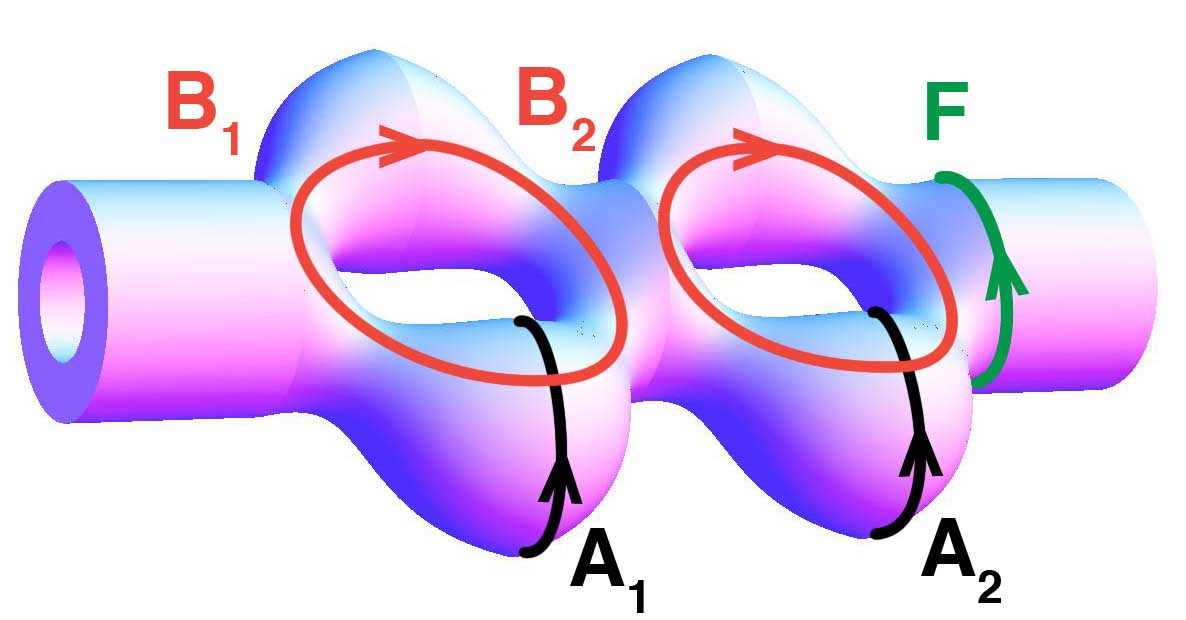}     
    \caption{A thickened genus two pipe describing the IR geometry $\widetilde{M}$ in the case of $A_{5}$.  The SW geometry is a genus two curve with two discs removed. The cycles $A_{i}$, $B_{i}$, $F$, span the homology of the SW curve $\widetilde{\Sigma}$ at $t\longrightarrow-\infty$.  The boundary of $\widetilde{M}$ is a surface of genus five.}
  \label{fig:pipe}
\end{figure}

$\widetilde{\Sigma}$ is a surface of genus $(n-1)/2$ which has been made non-compact by removing two discs.  Then, $\widetilde{M}$ is a thickening of this to a genus $(n-1)/2$ pipe.  Label the symplectically paired cycles on the initial $\widetilde{\Sigma}_i$ on the outside of the pipe as $(A_{i}, B_{i}),$ and those on the final $\widetilde{\Sigma}_f$ on the inside of the pipe as $(\tilde{A}_{i},\tilde{ B}_{i}),$ as  for $i= 1, \cdots, (n-1)/2$.  The remaining two cycles on the boundary are unpaired cycles $F,{\tilde F}$ on the initial and final surfaces.  The relations defining how $\widetilde{M}$ is filled in are then
\begin{equation}
A_{i}=\tilde{A}_{i}, \hspace{.5in} B_{i}=\tilde{B}_{i}, \hspace{.5in} F={\tilde F}.
\end{equation}
This geometry is illustrated in Figure \ref{fig:pipe}.
\end{itemize}

Just as in the explicit examples we have studied thus far, to fully specify a 3d theory we must impose boundary conditions.  These are defined by taking the manifold $\widetilde{M}$ and gluing it to another copy of itself to determine a compact manifold $\widetilde{M}_{c}$ with no boundary.  As the boundary of $\widetilde{M}$ is a Riemann surface of genus $n$, there is a natural action of $Sp(2n,\mathbb{Z})$ acting on our choices of boundary conditions and hence the class of theories constructed in this manner.  Such actions can again be interpreted as changing the gauging prescription and the CS levels of the model.

\subsubsection{$A_n$ Braids and Lagrangians }

To determine the detailed structure of the theory, including charges of fields, superpotentials, and CS levels, we proceed as in our examples to construct a braid canonically associated to each chamber.  The structure of this braid is completely fixed by R-flow.  Indeed, R-flow is specified by the evolution of the 4d central charges which are the periods of the Seiberg-Witten differential, $\phi.$  Then, given the evolution of these central charges, we can in principle invert the period map to determine the evolution of the loci where $\phi=0.$  The strands in $M$ swept out by these zeros during the flow, are then exactly the strands of the braid.  However, even for the simple case of the $A_{n}$ model, inverting the period map explicitly is a non-trivial task.  Nevertheless, for these $A_{n}$ R-flows, we will see that the structure of the braid, and its detailed 3d physical interpretation, can essentially be determined by simple consistency conditions.  Of course, it would still be desirable to invert the period map and verify our results directly. 

First, we address how particles in the theory are visible from the braid diagram.  As we have seen in our analysis of the $A_{1}$ and $A_{2}$ examples, before closing the braid (which may involve $T$ transformations), there is a one-to-one correspondence between braid moves and 3d particles.  In fact, this correspondence holds generally for those 4d chambers, where all the mutating nodes of the 4d BPS quiver have either, all incoming arrows, i.e. \emph{sinks}, or all outgoing arrows, i.e. \emph{sources}.  Note that the $A_{1}$ and $A_{2}$ examples are both of this type.  For more general sequences of mutations which involve nodes which are neither sources nor sinks, what we find is a kind of `non-planar' structure, where each particle corresponds to a specific crossing, but not all crossings correspond to particles.

To begin the investigation, note that BPS states can be viewed as segments connecting a pair of strands in the braid.  This observation provides the basic link between particles and braid moves: when the particle becomes \emph{massless}, the associated pair of strands must meet.  Thus, if we give the particle a small finite mass, we simply resolve the intersection of the strands into a braid move.  It follows that, up to an overcross/undercross prescription to be determined,  each 3d particle will be associated with a braid move.  

\begin{figure}[here!]
  \centering
  \includegraphics[height=0.25\textwidth]{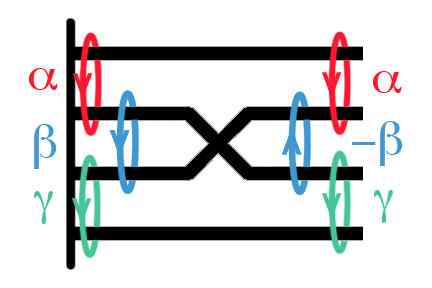}     
    \caption{The effect of mutation at a sink node $\beta$ on the strands.  The undercross/overcross is ambiguous.}
  \label{fig:sinkmute}
\end{figure}

From this basic fact, we can already deduce why source/sink mutation sequences result in a one-to-one correspondence between 3d particles and braid moves.  Indeed, each node of the quiver labels a cycle encircling a pair of strands, and if the corresponding nodes have an arrow between them, then the corresponding pairs share a strand.  Suppose we focus on three adjacent nodes of our quiver, which we label $\alpha, \beta, \gamma$.  Let us consider the mutation of the node $\beta$.  If the node $\beta$ is a sink, i.e., the arrow structure of the quiver is $\alpha\rightarrow \beta \leftarrow \gamma$, then after mutation the quiver has changed to $\alpha \leftarrow -\beta \rightarrow \gamma$.  Therefore, up to the ambiguous overcross/undercross,  the braid and the associated cycles would appear as in Figure \ref{fig:sinkmute}. 
Thus, in this case there is one particle and one braid move.

\begin{figure}[here!]
  \centering
  \subfloat[Source type mutation]{\label{fig:sourcemute1}\includegraphics[height=.25\textwidth]{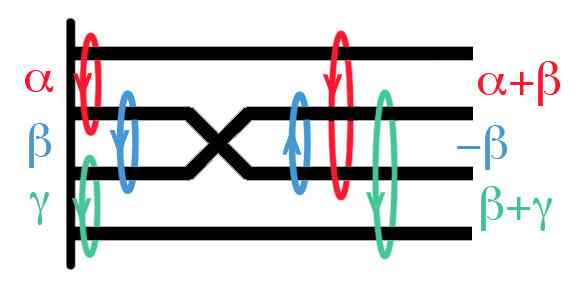}}     
  \hspace{.2in}      
  \subfloat[Change of planar projection]{\label{fig:sourcemute2}\includegraphics[height=0.25\textwidth]{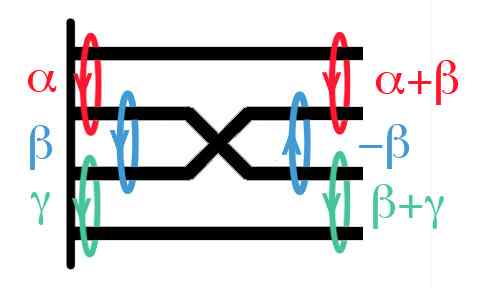}}
  \caption{A source type mutation on the node $\beta$.  In (a) we see the resulting change of basis on the cycles.  In (b), the planar projection is changed by rotating strand two and three.}
  \label{fig:sourcemute}
\end{figure}

Meanwhile, if the node $\beta$ is a source, i.e. the quiver $\alpha\leftarrow \beta \rightarrow \gamma$, then after mutation the classes of the nodes have changed to $(\alpha+\beta)\rightarrow -\beta\leftarrow (\beta+\gamma)$, and thus the corresponding braid and cycles would appear, up to overcross/undercross, as in Figure \ref{fig:sourcemute1}.   
This may look like a complicated structure.  Indeed, if say the next node to mutate is $(\beta+\gamma)$, this means that the second and fourth strand should cross, which cannot be done without an additional crossing involving the third strand, which does not correspond to a physical particle.  To avoid this, we rotate the second and third strand after the mutation, resulting in Figure \ref{fig:sourcemute2}. Then, after this change in our planar projection of the braid, the source type mutation looks the same as the sink type.  The only difference is that the classes we associate to cycles between nearby strands have changed in correspondence with the labeling of charges on the nodes of the quiver.  So again, in this case we see that there is one particle and one braid move.

\begin{figure}[here!]
  \centering
  \includegraphics[height=0.25\textwidth]{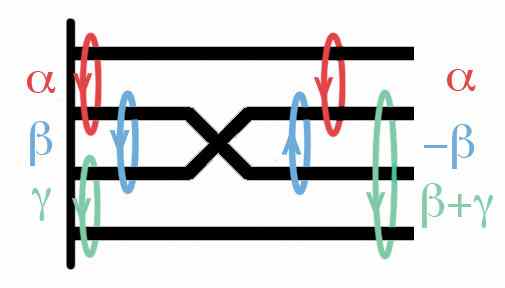}     
    \caption{The effect of mutation at a mixed node $\beta$ on the strands.  }
  \label{fig:mixedmute}
\end{figure}

Finally, consider a mutation on a node which is neither a source nor sink, say mutation of the node $\beta$ for the quiver $\alpha \rightarrow \beta \rightarrow \gamma$.  Then, the mutated quiver would become $\overrightarrow{\alpha \rightarrow -\beta \rightarrow \beta+\gamma}$.  The corresponding braid looks, up to overcross/undercross as in Figure \ref{fig:mixedmute}.  
Suppose next we wish to mutate on $\beta+\gamma$.  This cannot be done without an extra crossing.  However, unlike the source/sink case where we could change our planar projection to avoid the unnecessary crossing, this is not possible to do by any rotation of the strands after mutation.  Continuing with further mutations we will get a `non-planar' braid, for which some crossings will be unphysical in the sense that they do not correspond to chiral particles. 

Thus, for precisely those mutation sequences which involve only sources and sinks, we can achieve a planar projection of the braid where each crossing corresponds to a 3d chiral particle.  For this reason, we restrict our analysis in the remainder of this section to these sink/souce chambers.\footnote{A more general example of a non-planar braid is considered in section 8.6.}  For these braids, the structure is completely determined by the sequence of mutations of the $A_{n}$ quiver describing the parent 4d BPS spectrum.  We label the strands of the braid as $1, 2, \cdots n+1$ going down the page.  Then, the $i$-th node of the quiver labels a pair of adjacent strands $(i,i+1)$, and mutation on the node $i$ corresponds to a crossing involving the pair of strands $(i,i+1)$.    To fully specify the braid we must now fix an overcross/undercross rule.  As we will argue later in this section, this rule is determined by consistency to be that the strand $i+1$ always overcrosses $i$.  We take this as our definition of the braid group generator $b_{i,i+1}$, and thus for source/sink sequences, the braid is completely fixed by:
\begin{equation}
 \mathrm{mutation \ at \ node}\ i\leftrightarrow b_{i,i+1}.
\end{equation}

Next, we specify boundary conditions for the theory by closing the braid with dashed regions encoding the fact that the closure occurs at infinity.  The simplest example of such boundary conditions are those which do not introduce any additional crossings in the diagram.  This preserves the feature that all crossings can be associated to finite mass dynamical 3d particles, and we confine ourselves to such simple examples.  Then, to extract the IR Lagrangian we draw a checkerboard coloring of the resulting knot.  This fixes the gauge multiplet sector of the theory as:
\begin{itemize}
\item The number of $U(1)$'s is one less than the number of white regions in the coloring.
\item The matrix of CS terms is given by computing the Goeritz form for these white regions.
\end{itemize}

Finally, we must fix the superpotentials and charges of particles.   As a consequence of our source/sink assumption, and the simple choice of boundary conditions, each crossing in the diagram corresponds to a particle.  Let $i$ index the white regions in the checkerboard.  Between regions $i$ and $j$ there are some number $\alpha$ of crossings $c_{ij}^{\alpha}$ and associated chiral fields $\Phi_{ij}^{\alpha}$.  The field $\Phi_{ij}^{\alpha}$ carries charge $\pm1 $ under the $U(1)_{i}$ and the $U(1)_{j}$ and vanishing charges for the remaining gauge groups.  We will now determine the sign of these charges by demanding that all apparent superpotential terms are gauge invariant.

To study the superpotential, note that for a given checkerboard coloring of the projected link we have both black and white regions.   For each finite region, black or white, whose boundary does not include any dashed portions of the knot that arise from boundary conditions, we expect a superpotential contribution to our theory.  This superpotential derives from M2-brane instantons ending on the M5-brane.     Thus, each one of the regions corresponds to a superpotential term.  However, depending on the color of the region, white or black, the interpretation is different for the gauge theory on $\mathbb{R}^{1,2}$.   We discuss each of these in turn.
\begin{itemize}
\item \emph{White Regions}

Since white regions $R_{i}$ are associated to gauge groups of the theory, the $i$-th finite white region in the checkerboard describes a superpotential which is proportional to the monopole operator for $U(1)_{i}$
\begin{equation}
{\cal M}_i^{\pm1} ={\rm exp}[\pm( \sigma_i +i\gamma_i)],
\end{equation}
where $\gamma_i$ denotes the corresponding scalar dual to photon, and the sign $\epsilon_{i}=\pm1$ in the exponent of ${\cal M}_{i}$ depends on the sign conventions for the gauge field yet to be determined.   In addition, the M2-brane instanton for white region $i$ will contribute a monomial given by the chiral fields associated to each of the crossings of that region with other white regions. Thus, each finite white region contributes a term
\begin{equation}
W_i={\cal M}_i^{\epsilon_{i}}\prod_{j,\alpha} \Phi_{ij}^\alpha \in \mathcal{W}. \label{whitecontr}
\end{equation} 

\item \emph{Black Regions}

For each finite black region $B,$  we also get a superpotential term.  But this time there is no associated gauge cycle.  Indeed, the one-cycles surrounding the white regions have trivial intersection with any of the black regions (including the neighboring ones), and this implies that the two-cycles defined by the black regions carry no monopole charge.  Thus, for each finite black region we simply get the contribution of the fields at the crossings on the boundary of the region
\begin{equation}
W_B=\prod_{(ij,\alpha)\in \partial B}\Phi_{ij}^\alpha \in \mathcal{W}. \label{blackcontr}
\end{equation}
\end{itemize}

Now, we fix the charges of the fields $\Phi_{ij}^{\alpha}$ by demanding that the superpotential terms \eqref{whitecontr} and \eqref{blackcontr} are gauge invariant.  Let $q_{k,ij}^{\alpha}$ denote the $U(1)_k$ charge of the field $\Phi_{ij}^\alpha$ corresponding to the crossing $c_{ij}^\alpha$.    Each of these fields corresponds to a basic crossing, and hence is $q_{k,ij}^{\alpha}=\pm 1$ for $k=i,j$ and $q_{k,ij}^{\alpha}=0$ for $k\neq i,j$. Note also that the monopole
field ${\cal M}_i^{\epsilon_i}$ carries $U(1)_j$ charge given by $\epsilon_i \hat{k}_{ij}$ induced from the CS term.  

Then, the $U(1)_j$ gauge invariance of the white region contribution $W_i$ implies that
\begin{equation}
\epsilon_i \hat{k}_{ij} + \sum_{\alpha}  q_{j,ij}^\alpha = 0, \label{whiteinv}
\end{equation}
and similarly, the $U(1)_i$ invariance of $W_i$ implies that
\begin{equation}
\epsilon_i \hat{k}_{ii}+\sum_{\alpha,j}q_{i, ij}^\alpha =0.
\end{equation}
Using the $U(1)_i$ invariance of $W_j$ the latter equation we learn 
\begin{equation}
\epsilon_i\hat{k}_{ii}-\sum_j \epsilon_j \hat{k}_{ji}=0.
\end{equation}
And finally, multiplying this by $\epsilon_i$ we see that
\begin{equation}
\hat{k}_{ii}-\sum_{j} \epsilon_i\epsilon_j \hat{k}_{ij}=0
\end{equation}
This is compatible with the definition of the CS matrix \eqref{offdiagchecker} only if, for each pair of $i,j$ with non-vanishing $\hat{k}_{ij}$, we have
\begin{equation}
\epsilon_i\epsilon_j=-1.
\end{equation}
This implies that it must be possible for this class of gauge theories to assign a parity to each $U(1)$ factor of the gauge group, defined by the sign $\epsilon_{i},$ such that gauge fields which have non-vanishing $\hat{k}_{ij}$ have opposite parties.   As a result, we learn that, after deleting a single ungauged node, the Tait graph, defined by the checkerboard coloring as in section 2,  is a bipartite graph for which we can assign opposite $\pm 1$ to vertices which are connected.  This turns out to be true for all the graphs which arise in our constructions for the sink-source sequence of mutations.  

Furthermore, note that equations \eqref{whiteinv} and \eqref{offdiagchecker} can be combined to express the charges of the fields in terms of the sign $\zeta(c_{ij}^{\alpha})$ associated to the crossing
\begin{equation}
\sum_{\alpha}(\epsilon_i\zeta (c_{ij}^\alpha)+q_{i,ij}^\alpha)=0.
\end{equation}
This suggests the canonical solution
\begin{equation}
q_{i,ij}^\alpha=-\epsilon_i\zeta(c_{ij}^\alpha) \label{chargedef}
\end{equation}
Equation \eqref{chargedef} is the key final result which specifies the charges of the theory and completes our description of these models.  Together with the fact that $\epsilon_i\epsilon_j=-1$ for connected vertices, it implies that the each of fields $\Phi_{ij}^{\alpha}$, charged under white regions $i$ and $j$, are bifundamentals which carry opposite charges under $U(1)_{i}$ and $U(1)_{j}$
\begin{equation}
q_{i,ij}^\alpha=-q_{j,ij}^\alpha.
\end{equation}
This means that each of the links in the Tait graph, which corresponds to a crossing $c_{ij}^{\alpha}$ and hence a field $\Phi_{ij}^{\alpha}$, can be oriented by making use of the bipartite structure.  If we make the convention that the link $c_{ij}^{\alpha}$ points out of the node associated to $U(1)_{i}$ if the field $\Phi_{ij}^{\alpha}$ carries charge $+1$ under $U(1)_{i}$, then this makes the Tait graph into the quiver for the resulting gauge theory.

Finally, we observe that these equations uniquely fix the charges in terms of the Chern-Simons levels, up to the choice of $\epsilon_i$.  However, there are only two global choices of $\epsilon_i$ depending on which nodes we assign as even and which one as odd.  A change of an overall sign of $\epsilon_i$ simply flips the oveerall sign of the charges, which gives an equivalent theory, by replacing all gauge fields by their opposites, $A_i\rightarrow -A_i$ 
(which does not affect the Chern-Simons level matrix).  The reverse is also true:  if we assign arrows to the links of the Tait diagram, thus fixing $q_{k,ij}^\alpha,$ we can read off the $\zeta (c_{ij}^\alpha)$ from equation \eqref{chargedef} and hence determine the associated overcross/undercross.   This provides a strong consistency check on our proposal for the charges of the fields, and our identification of mutations with the basic braid move $b_{i, i+1}$.

\subsubsection{Cookbook}
Let us summarize the rules derived in the preceding section into a recipe for extracting the 3d theory.  We confine our description to the simplest examples where the boundary conditions respect the order of time flow, and no additional $T$ transformations are performed.
\begin{itemize}
\item Given an $A_{n}$ BPS quiver, identify a source-sink sequence of mutations describing a chamber of the 4d theory.
\item Construct a braid on $n+1$ strands by reinterpreting the mutation sequence as a sequence of braid moves.  When the node $i$ is mutated do the braid move $b_{i,i+1}$.
\item Impose boundary conditions by choosing cycles to be contractible.  If $n$ is odd, this means contracting $\frac{n+1}{2}$ initial and final cycles.  If $n$ is even, this means contracting $\frac{n}{2}$ initial and final cycles.
\item Equip the resulting knot with a checkerboard coloring and draw the associated Tait graph.  This graph is bipartite except for the presence of one auxiliary framing node.  By framing node, we mean the node that is not gauged.
  For each other node $i$ in the graph, assign a parity $\epsilon_{i}=\pm1$ in such a way that nodes connected by a link have opposite parity.
\item Orient the links in the Tait graph by making use of the parity of the nodes and the parity of the links.  Specifically:
\begin{itemize}
\item If both nodes are not the framing node and the link has parity $+1$ orient the link by pointing it from the node with parity $-1$ to the node with parity $+1$.
\item If both nodes are not the framing node and the link has parity $-1$ orient the link by pointing it from the node with parity $+1$ to the node with parity $-1$.
\item If one node is the framing node, orient the link by having it point out of the framing node if the product of $\epsilon_{i}$ and the link orientation is $+1$ and point into the framing node if the product of $\epsilon_{i}$ and the link orientation is $-1$
\end{itemize}
\item The oriented Tait graph can now be interpreted as the quiver describing the field content and gauge group of the resulting theory on $\mathbb{R}^{1,2}$.  Thus, each node other than the framing node yields a $U(1)$.  Each oriented link defines a bifundamental field.  And the matrix of CS levels $\hat{k}_{ij}$ is determined by computing the Goeritz form the sign $\pm$ assigned to each of the links in the Tait graph.
\item Finally, the superpotential of the theory is given by summing over contributions from finite white and black regions in the checkerboard
\begin{equation}
\mathcal{W}=\mathcal{W}_{\mathrm{Black}}+\mathcal{W}_{\mathrm{White}}.
\end{equation}
At the level of the Tait graph this means the following:
\begin{itemize}
\item For each finite black region $B$, we obtain a contribution to $\mathcal{W}$ in the form of a monomial in elementary fields.  Specifically, each such region $B$ defines a cell in the Tait graph, and we add the cycle in the Tait graph defined by $\partial B$ to the superpotential
\begin{equation}\label{superblack}
\mathcal{W}_{\mathrm{Black}}=\sum_{B}\prod _{\partial B} \Phi_{ij}^{\alpha}.
\end{equation}
\item  For each finite white region $W$,  we add an associated term in the superpotential involving the monopole operator ${\cal M}_{i}$ associated to the i-th gauged node which corresponds to that white region.  Specifically, we take the product over fields charged under the node in question
\begin{equation}\label{superwhite}
\mathcal{W}_{\mathrm{White}}=\sum_{W} {\cal M}_{i}^{\epsilon_{i}}\prod_{i}\Phi_{ij}^{\alpha}.
\end{equation}
\end{itemize}
\end{itemize}

The algorithm defined above can be applied to any source-sink mutated chamber of the 4d $A_{n}$ theories.  To compare the theories defined by two distinct chambers, we keep track permutation $\chi$ which acts on the nodes of the quiver.  Let $\chi_{i}$ for $i=1,2$ denote the two permutations.  We impose boundary conditions at $t=+\infty$ on a given braid by contacting some set of cycles $\gamma_{i}$.  Then, to compare to the second braid we contract the cycles $\chi_{2}\circ \chi_{1}^{-1} (\gamma_{i})$.  Thus, one choice of boundary conditions, fixed for the braid defined by one chamber, determines boundary conditions for the braids defined by all other chambers.  Extracting the physics from the resulting knot as above we obtain a class of mirror 3d theories.  In the next section, we will use the procedure to give new examples of dual pairs.
\subsection{A Final Example: Alternating $A_{2n}$}\label{sec:alternatingA2}
\begin{figure}[here!]
  \centering
  \subfloat[Alternating Triangulation]{\label{fig:zigzag1}\includegraphics[height=.35\textwidth]{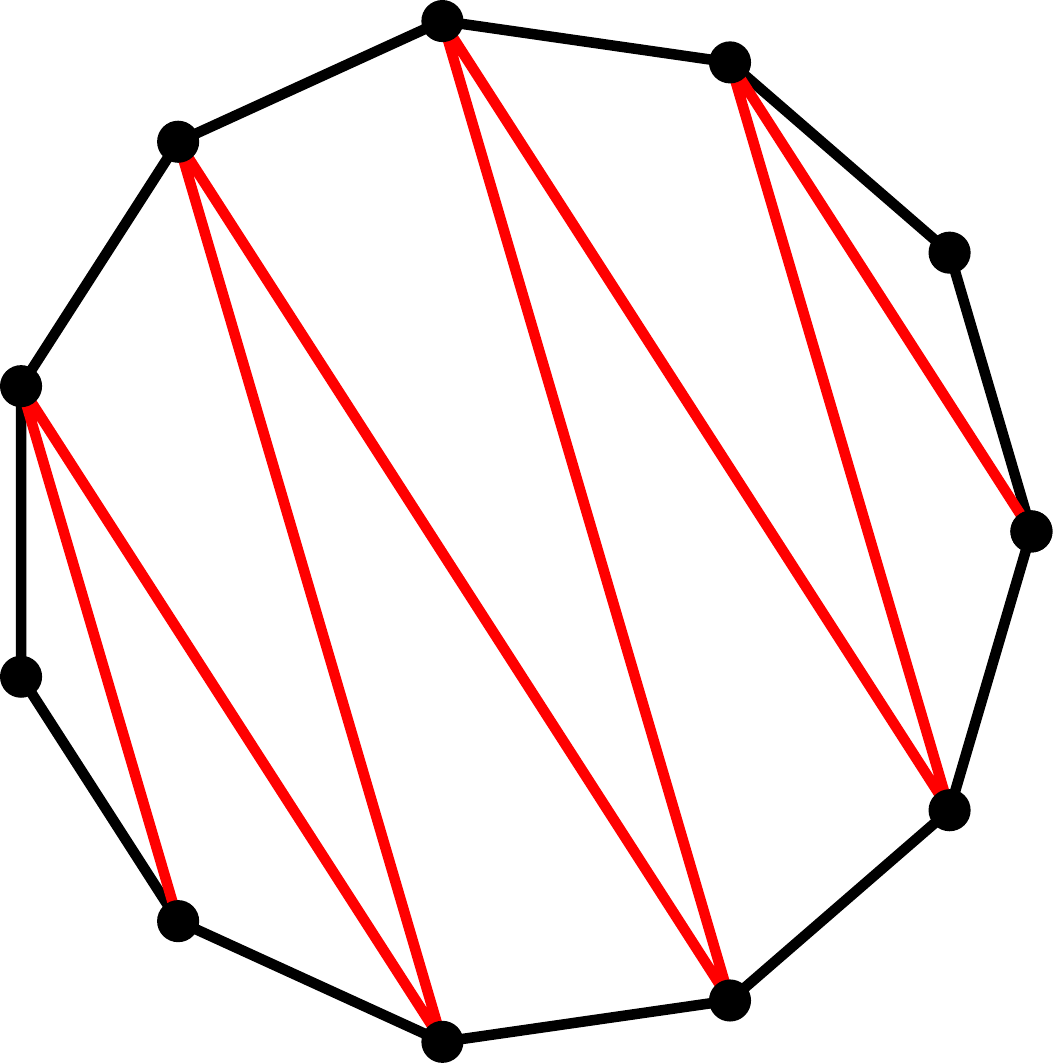}}     
  \hspace{.5in}
 \subfloat[Holographic Tetrahedra]{\label{fig:zigzag2}\includegraphics[height=.35\textwidth]{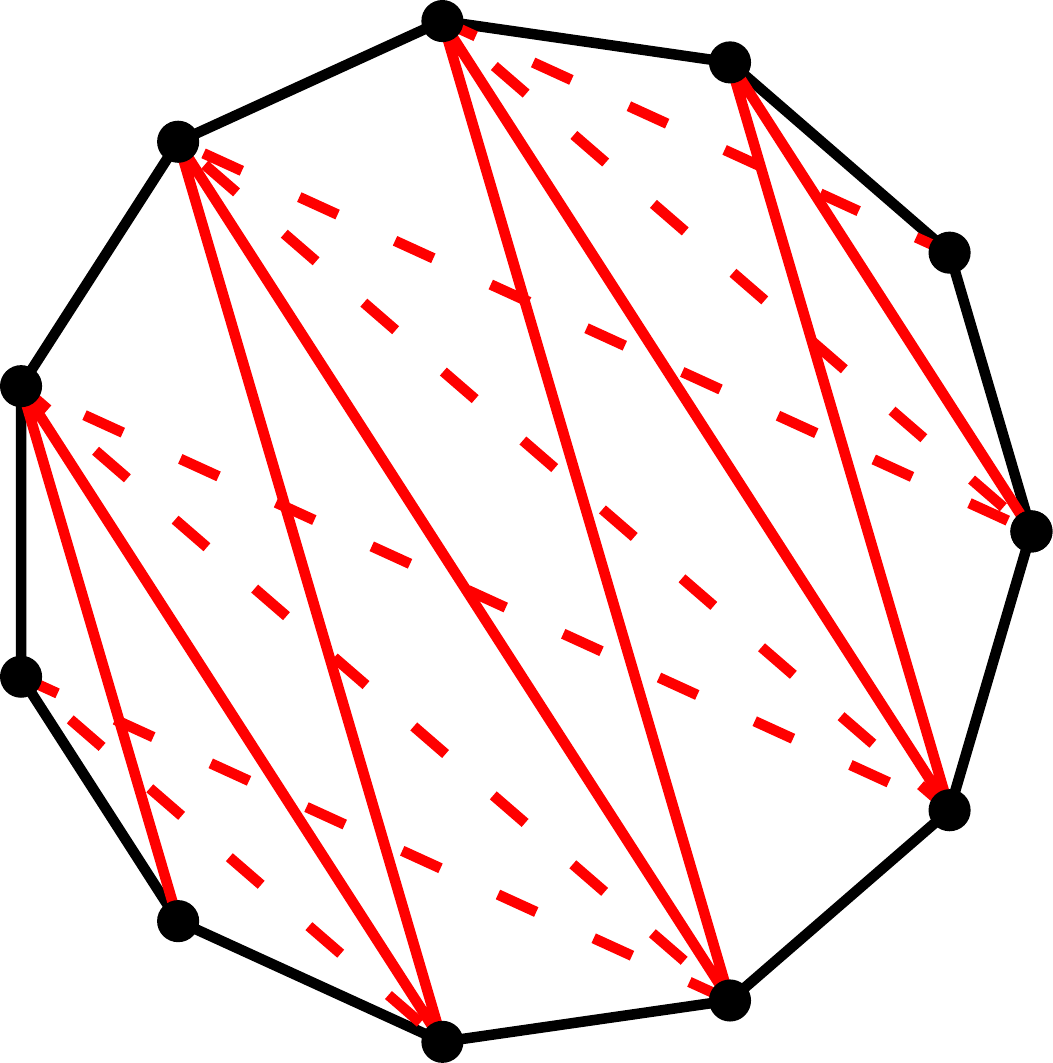}} 
  \caption{Triangulations for the alternating $A_{n}$ quiver in the case of $A_{8}$.  In (a) we see the triangulation of the front face.  In (b) the holographic view of the tetrahedra in the case of the minimal chamber.}
  \label{fig:a2nbraids}
\end{figure}

As an example application of these rules we will consider domain wall theories in the general $A_{2n}$ model.  We consider an alternating orientation of the quiver.
\begin{equation}
\xy 
(-45,0)*+{}*\cir<8pt>{}="a" ; (-30,0)*+{}*\cir<8pt>{}="b" ; (-15,0)*+{}*\cir<8pt>{} ="c" ;(0,0)*+{}*\cir<8pt>{}="d" ;(15,0)*+{\cdots}="i" ; (30,0)*+{}*\cir<8pt>{}="k" ; (45,0)*+{}*\cir<8pt>{}="m"; 
(-45,-8)*+{1}="e" ; (-30,-8)*+{\dot{1}}="f" ; (-15,-8)*+{2}="g" ;(0,-8)*+{\dot{2}}="h" ; (15,-8)*+{\cdots}="j" ;(30,-8)*+{n}="l" ;(45,-8)*+{\dot{n}}="n" 
\ar @{->} "a"; "b" 
\ar @{->} "c"; "b" 
\ar @{->} "c"; "d" 
\ar @{->} "i"; "d" 
\ar @{->} "k"; "m" 
\ar @{->} "k"; "i" 
 \endxy
 \end{equation}
 This quiver corresponds to the zig-zag triangulation of an $(2n+3)$-gon shown in Figure \ref{fig:zigzag1}.  This is the triangulation present on the front and back face of the three-manifold, and as usual, its evolution determines a decomposition of the solid ball into tetrahedra viewed holographically in Figure \ref{fig:zigzag2}.
 
There are two simple chambers of these theories described by their mutation sequences as
 \begin{itemize}
 \item  \emph{Minimal Chamber}
 
 There are $2n$ states.  The mutation sequence proceeds by first mutating on all dotted nodes, and then mutating on all undotted nodes.  The associated permutation element $\chi$ is the identity.
  \item \emph{Maximal Chamber}
  
 The theory has $n(2n+1)$ states.  The mutation sequence proceeds by mutating on all all undotted nodes, then all dotted nodes, then all undotted nodes, etc. for a total of $2n^{2}+n$ mutations.  The associated permutation element is
 \begin{equation}\label{chia2n}
 \chi=(1, \dot{n}) \ (\dot{1}, n)  \ (2, \dot{n-1})  \ (\dot{2}, n-1)\ \cdots.
 \end{equation}
 \end{itemize} 
 We construct the braid by identifying mutations with braid moves as described in the previous section:  when an undotted node $m$ is mutated we do the braid move $b_{2m-1,2m}$, when a dotted node $\dot{m}$ is mutated we do the braid move $b_{2\dot{m}, 2\dot{m}+1}$.  This leads to braids of the form shown in Figure \ref{fig:a2nbraids}.
 
\begin{figure}[here!]
  \centering
  \subfloat[Minimal]{\label{fig:smallbraid}\includegraphics[height=.473\textwidth]{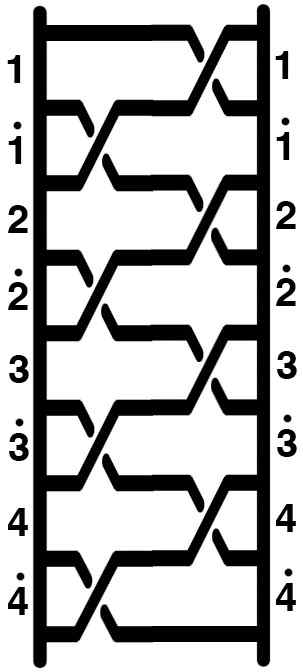}}     
  \hspace{.2in}
 \subfloat[Maximal]{\label{fig:bigbraid}\includegraphics[height=.473\textwidth]{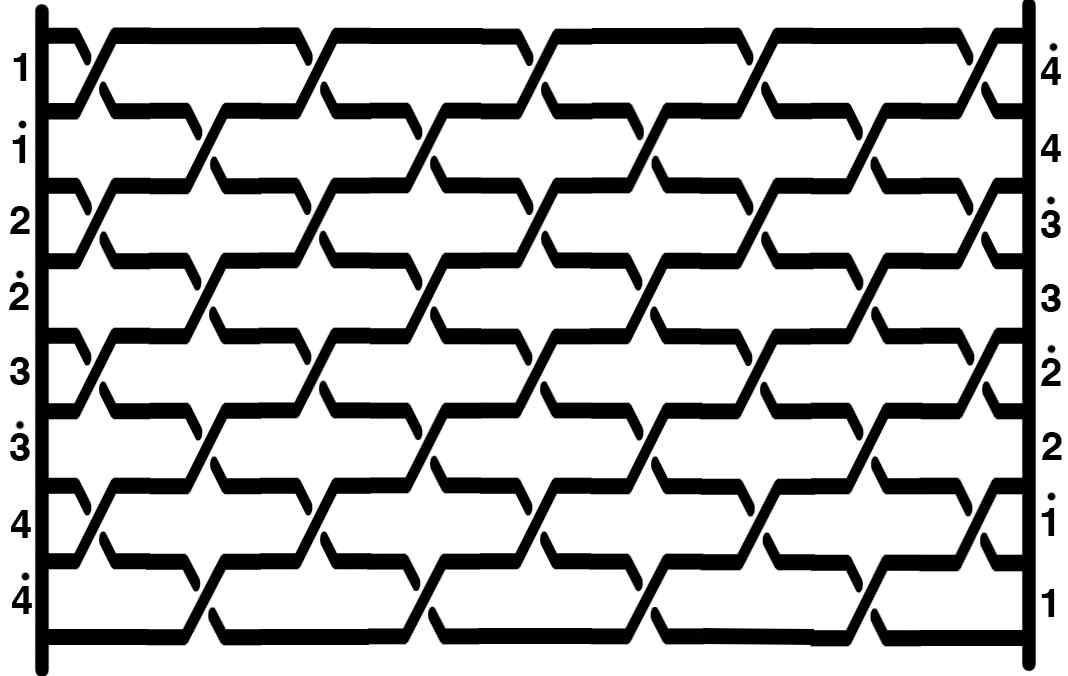}} 
  \caption{Minimal and maximal braids for the $A_{8}$ alternating quiver.}
  \label{fig:a2nbraids}
\end{figure} 

Next, to determine a 3d field theory we impose boundary conditions.  As an illustrative example, we choose to trivialize the undotted cycles at the initial time, and the dotted cycles at the final time.  Of course in doing so, we must also take into account the non-trivial permutation $\chi$ in the maximal chamber.  

\emph{Minimal Chamber}

\begin{figure}[here!]
  \centering
  \subfloat[Checkerboard for the minimal chamber of $A_{8}$]{\label{fig:smallclosedbraid}\includegraphics[width=.45\textwidth]{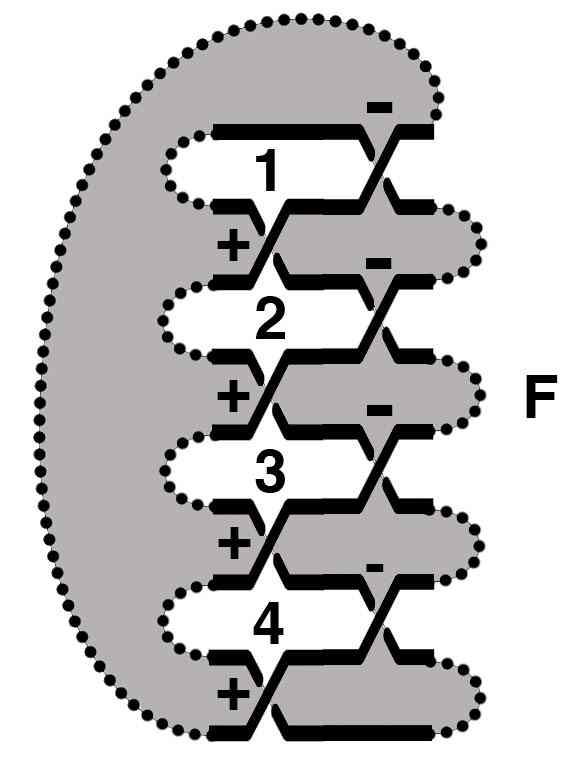}} 
  \hspace{.5in}
   \subfloat[Tait graph for the minimal chamber of $A_{8}$]{\label{fig:a2nsmalltait}\includegraphics[width=.45\textwidth]{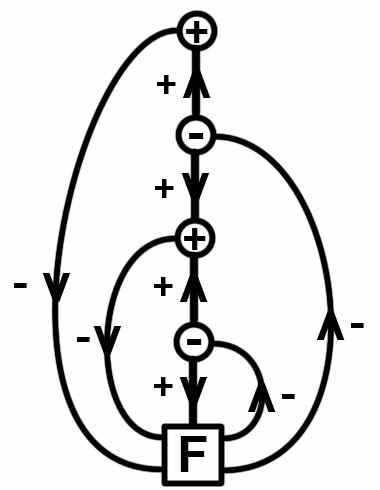}}     
  \caption{Checkerboard coloring and associated Tait graph for the minimal chamber of the alternating $A_{8}$ theory.  The white region denoted by F corresponds to the square framed node in the graph.  The generalization to $A_{2n}$ is a linear Tait graph of length $n$.}
  \label{fig:smallclosed}
\end{figure}

 First, we investigate the physics of the minimal chamber.  We follow the general instructions of the previous section.  We draw a checkerboard coloring of the resulting knot, and its associated Tait graph.  Then, we identify a framing node in the graph which will be ungauged.  All other nodes describe gauge groups in the theory and we assign a sign $\epsilon =\pm1$ to these nodes in such a way that connected node have opposite parity.  Orienting the links using our general rules we obtain a Tait graph of the form shown in Figure \ref{fig:smallclosed}.  
 
 From this Tait graph, reinterpreted as the quiver of the 3d theory, we determine that in the minimal chamber there are $n$ gauge groups, $2n$ particles, and no superpotential terms.  The charges of the fields, and the the associated CS matrix can all be read from the orientation of arrows in the Tait graph and the Goeritz form of the links.

\emph{Maximal Chamber}

\begin{figure}[here!]
  \centering
  \subfloat[Checkerboard for the maximal chamber of alternating $A_{8}$]{\label{fig:bigclosedbraid}\includegraphics[height=.473\textwidth]{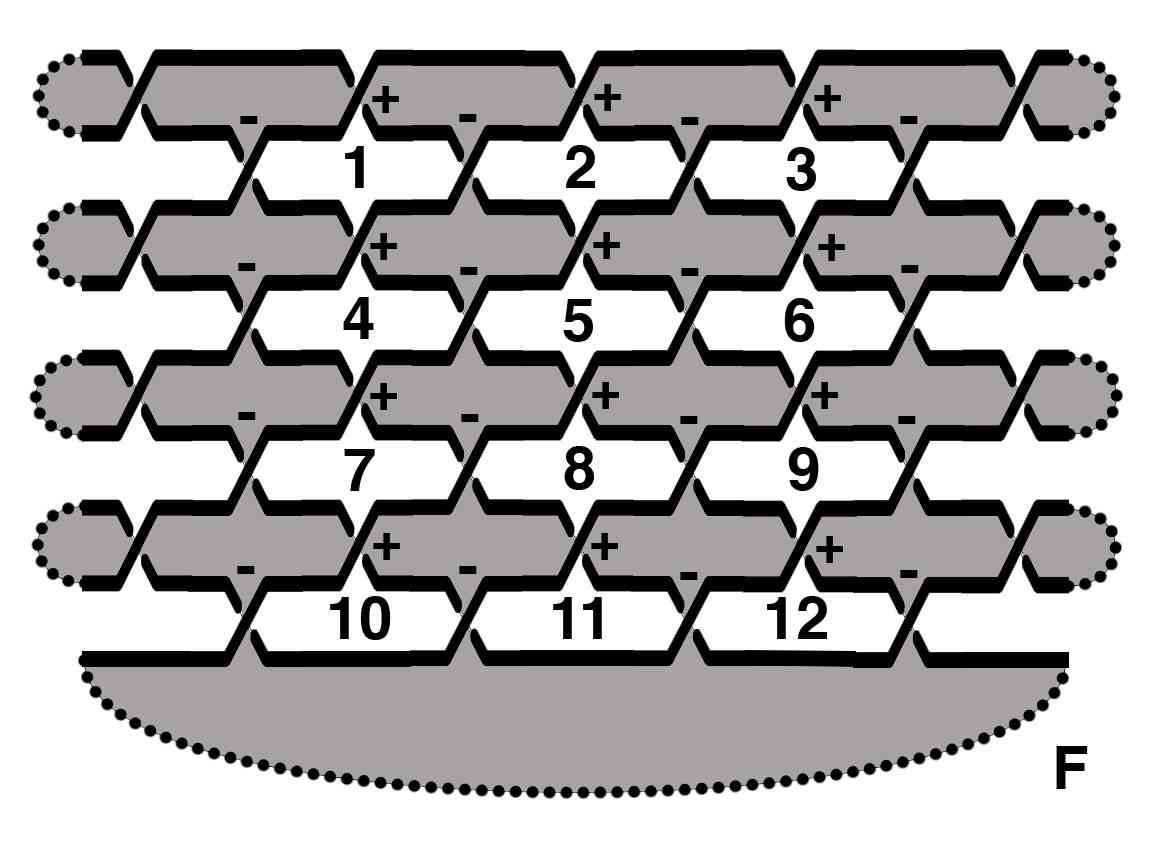}} 
  \hspace{3in}    
  \subfloat[Tait graph for the maximal chamber of alternating $A_{8}$]{\label{fig:biga2ntait}\includegraphics[ height=.473\textwidth]{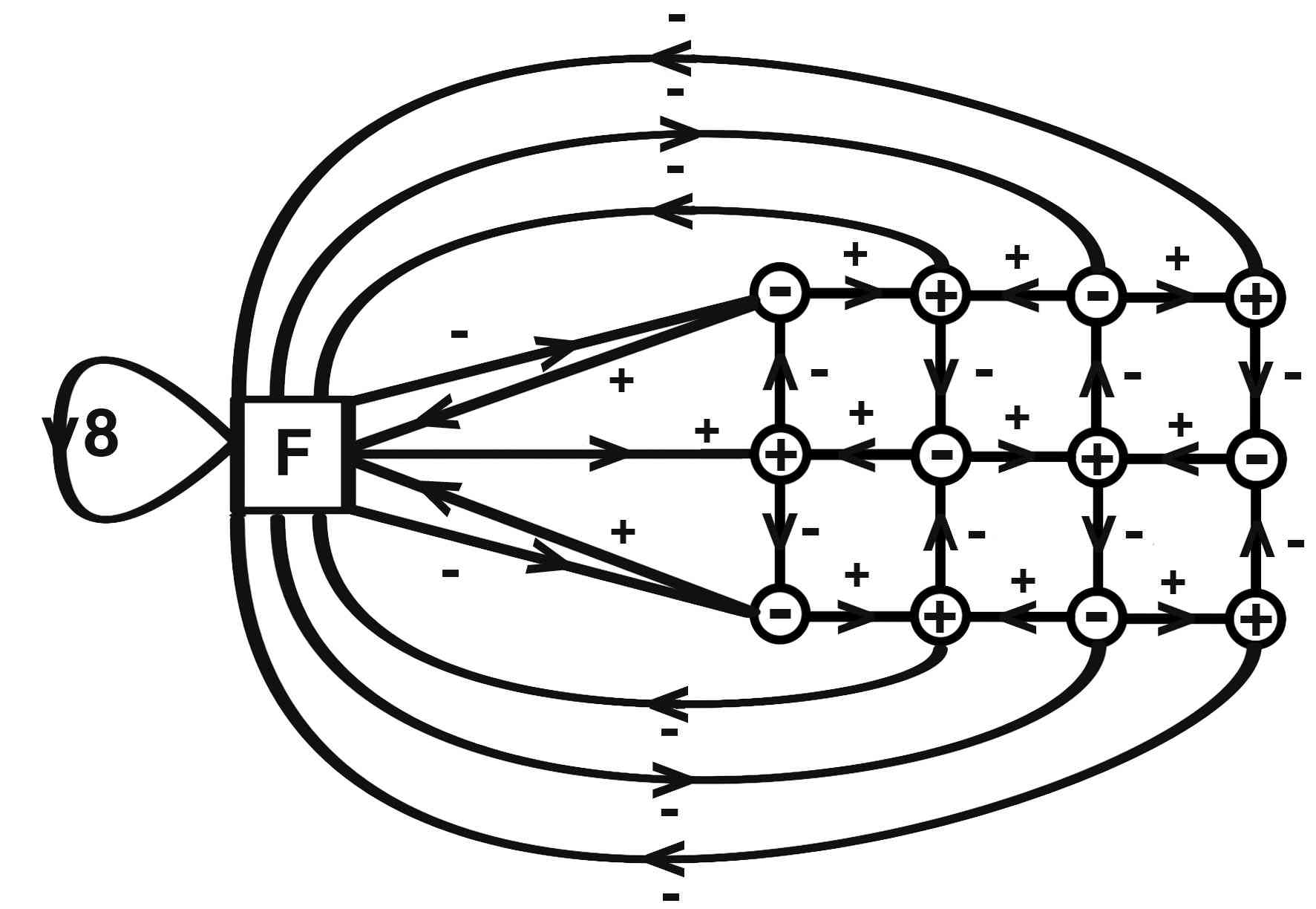}}     
  \caption{Checkerboard coloring and associated Tait graph for the maximal chamber of the alternating $A_{8}$ theory.  The white region denoted by F corresponds to the square framed node in the graph.  The generalization to $A_{2n}$ is an $n(n-1)$ grid Tait graph.}
  \label{fig:bigclosed}
\end{figure}

Next, we consider the maximal chamber.  The checkerboard coloring of the knot, and its associated Tait graph are shown in Figure \ref{fig:bigclosed}.  
From the graph, we read off that there are $n(n-1)$ gauge groups, and $2n^{2}+n$ particle, $2n$ of which are gauge neutral and encoded in the links connecting the framed node to itself.\footnote{\ The neutral links correspond to crossing connecting the framed region with itself.  These do not contribute to the CS levels and hence the links do not have an associated sign.}  Moreover, there is now a superpotential consisting of:
\begin{itemize}
\item Black Terms.  

There are $n^{2}$ finite black regions.  Of these, $n^{2}-n$ yield quartic monomials in $\mathcal{W}$, and $n$ yield cubic contributions.  These can be off from from cells in the Tait graph.
\item White Terms.

All white regions are finite and hence yield monopole contributions to the superpotential.  There are $n(n-1)$ such contributions.
\end{itemize}
The remaining data in the model, such as the charges of the fields and the CS levels, are all encoded by the Tait graph.

In section 8 we will check this proposed duality by comparing partition functions of these two theories.

\section{Flows of General 4d $\mathcal{N}=2$ Theories}\label{sec:flowsgen}
The examples described in the previous section, illustrate domain walls in the simplest possible context of the $A_{n}$ Argyres-Douglas models.  However, the general procedure of extracting a 3d theory from an R-flow of a parent 4d theory can be carried out for an arbitrary $\mathcal{N}=2$ model. For example, the $E_n$ case which does not correspond to 3d geometry will be discussed in section 8.   One could perhaps also consider the R-flow of other ${\cal N}=2$ theories which are not complete, by relaxing the constraint of UV finiteness, though we will not provide examples of that in this paper. Here, we will focus on the case where 4d gauge theory is  defined by wrapping a pair of M5 branes on a punctured Riemann surface $C$ of arbitrary genus $g(C)$.  Such 4d models have a number of interesting geometric features which translate into properties of the resulting three-manifold $M$ which is given as a thickening of $C$.  For example, if we consider the punctures of the surface $C$ there are two basic types \cite{Gaiottodual, GMN1}: 
\begin{itemize}
\item \emph{Irregular Punctures}

These are equivalent to boundary components of the Riemann surface.  For example, the $A_{n}$ Argyres-Douglas model is equipped with such a puncture.  In three dimensions, the boundary data for these punctures is fixed for all time along the R-flow and hence these boundary components are identified on the front and back face of $M$.  Thus, in three dimensions, irregular punctures do not give rise to boundary components, but instead map to pure topology of $M$.

\item \emph{Regular Punctures}

These encode mass parameters of the 4d theory and hence describe first order poles in the SW differential. If we consider a one-parameter family of such punctures then we obtain a line of cusp singularities in $\widetilde{M}$ 

\end{itemize}

Topologically, the manifold $M$ is given by a thickening of $C$ modulo the relation that the boundary components of $C,$ defined by the irregular punctures,  are identified for all time.  It has annular cusp singularities for each regular puncture.  Further, if $C$ has at least one boundary component, then $\partial M$ is a connected Riemann surface obtained by gluing two copies of $C$ along their common boundary.  Specifically, if $C$ has $b\geq1$ boundary components, then the genus of $\partial M$ is determined by a simple computation to be
\begin{equation}
g(\partial M)=2g(C)+b-1.  \label{genusrel}
\end{equation}
The manifold $M$ is then a certain filling in of the boundary $\partial M$.

Given any such surface $C$, its BPS data may be encoded in an ideal triangulation as described in section 4.  As we flow through time, the triangulation of the front face will evolve to the triangulation of the back face and this determines a decomposition of $M$ into tetrahedra.    Each tetrahedron encodes a 3d chiral particle in the theory, and finite 2-3 wall-crossings describe 2-3 Pachner moves on the 3d triangulation.  

Next, we state some general facts about the resulting IR geometries $\widetilde{M}$.  These are branched double covers of $M$ and their structure is encoded by the evolving zeros of the Seiberg-Witten differential $\phi$.  For each triangle in the front face of the triangulation we obtain a zero of $\phi$.  As the zeros evolve, they determine an open knot composed of strands, whose endpoints are fixed at the front and back face of $M$.  In principle, we can find the geometry of these strands using the R-flow.  Indeed, given the evolution of the 4d central charges, which are the periods of $\phi$, we can invert the period map and find the geometry of the branch point flow.  This is quite similar to the case of the $A_{n}$ models studied in detail in section 5.  However, unlike the the examples there, where these strands moved and were tangled in a space with trivial topology of a ball $B^{3}$, now the strands evolve in a space $M$ with non-trivial topology which has as boundary the surface of higher genus \eqref{genusrel}.  Further, the strands may also become braided around the annular cusp singularities descending from the regular punctures in the surface $C$.

Thus, the result of a general R-flow on a punctured Riemann surface is a potentially complicated topological configuration, and some of the technology that we developed for the $A_{n}$ case will be need to be enhanced to study this situation.  Nevertheless, we can still see that some of our general observations hold.  For example, to impose boundary conditions on the resulting theory, we  close the open knot in $M$ into an honest knot and this fixes the compactified geometry $\widetilde{M}_{c}$.  From this description, it is also clear that the resulting 3d theories will be acted on by $Sp(2g(\partial M),\mathbb{Z})$ and, as in our general discussion, this action is physically realized by changing CS levels, and gauging or ungauging some $U(1)'s$.

\subsection{Effective 3d Gauge Theories with Infinite Dimensional Representations}
\label{sec:SL2R}
A major novel feature of the general flows outlined above is the presence of BPS chambers of the 4d theory with infinitely many BPS states.  Indeed, the main examples we have considered involve 3d theories whose 4d parents have finite number of BPS states.  However this is not the typical situation.  For example, the pure $SU(2)$ gauge theory in 4d has infinitely many BPS states in the weak coupling chamber. For $SU(n)$ theories, not only can we have infinitely many BPS states, but in addition, we may have chambers which support BPS states with arbitrarily high spin.  It is thus natural to ask: what would the interpretation of the reductions of such chambers to 3d, and their equivalence to chambers with finitely many states imply?

To gain some insight to what implications these chambers and dualities may have in 3d, let us consider the example of pure $SU(2)$ gauge theory in 4d.  This theory has two chambers.   In the strong coupling regime, we have two states given by $(\mathrm{electric}, \mathrm{magnetic})$ charges $(2,-1)$ and $(0,1)$, and in three dimensions this R-flows to a 3d theory with two chiral multiplets much as in our analysis of the $A_{n}$ models.   Meanwhile, in the weak coupling chamber of 4d, there are infinitely many BPS particles:  the monople and its dyonic descendants, with charges $(2n,1)$, and the vector W-boson which carries charge $(2,0)$.  Consider the R-flow of the weak coupling chamber, where we take the projection to be along the electric charge direction.  In this way, all the 4d dyons will have the same real projection defining the 3d supersymmetry, and hence all the trapped 3d dyons will have equal finite masses.  In addition, the 4d BPS W-boson will result in a \emph{massless} trapped 3d particle.

\begin{figure}[here!]
  \centering
\includegraphics[width=.35\textwidth]{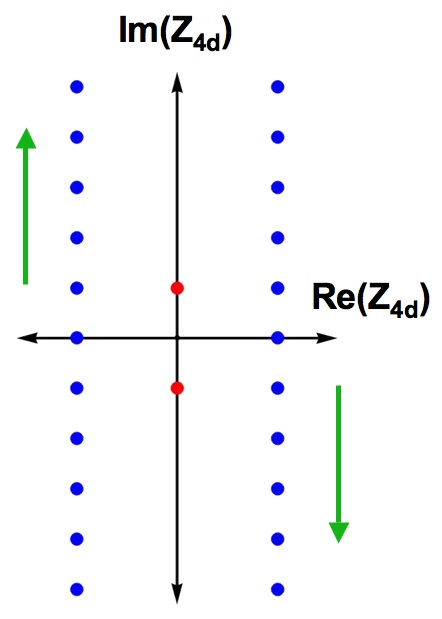}      
  \caption{BPS  spectrum and R-flow for weak coupling $SU(2)$.  The blue dots denote the 4d dyons, the red dots the 4d W-bosons.  The green arrows specify the direction of R-flow.   The 3d trapped W-bosons are massless and give rise to an $Sl(2,\mathbb{R})$ gauge symmetry.  The 3d trapped dyons all have equal mass and form an infinite dimensional representation of $Sl(2,\mathbb{R})$.}
  \label{fig:su2spec}
\end{figure}

From the fact that these 3d particles arise from trapped 4d bulk fields, we can make a number of observations.  First of all, the W-boson must carry vector quantum numbers.  Therefore, we in fact have a massless 3d vector particle.  There are only two possibilities for a non-abelian 3-vector theories:  either we have an $SU(2)$ gauge theory in 3d, or an $SL(2,\mathbb{R})$ gauge theory in 3d.  The first option may look more natural from the 4d perspective, where
in the infinitely weak coupling regime we recover an $SU(2)$ gauge theory.  However, we believe the $SL(2,\mathbb{R})$ is what is realized in 3d, for the following reason:
the infinite number of trapped 3d dyons have the same mass, and this strongly suggests that they form one irreducible object.  Furthermore, note that the W-boson from the bulk can bind to any of these trapped dyons, transforming one to the other. Since the vector particles should form either $SU(2)$ or $SL(2,\mathbb{R})$, and since $SU(2)$ has no infinite dimensional unitary representations with finite Casimir, we conclude that we must have an $SL(2,\mathbb{R})$ theory and that the 3d dyons form a single irreducible representation of $SL(2,\mathbb{R})$ as illustrated in Figure \ref{fig:su2spec}.  Note also that if we tilt the angle of projection to 3d, so that the W-boson has a tiny mass $\epsilon$,
the infinite tower of 3d dyonic states will have BPS masses given by $|m+n\epsilon|$ where $m$ is the real mass and $n \in {\mathbb{Z}}$. This can be interpreted as a deformation to the Coulomb branch of the $SL(2,\mathbb{R})$ theory by giving an expectation value to the adjoint scalar field in the $SL(2,\mathbb{R})$ gauge multiplet.

We can further argue why we may have obtained an $SL(2,\mathbb{R})$ instead of $SU(2)$ by observing that the main difference between these two cases is the sign of the kinetic term for the W-bosons.  How could the sign of the kinetic term for the W have flipped? This actually has a simple explanation:  the W-boson can never become massless in 4d.  No matter how weak we make the 4d coupling, as we come close to making the W massless (by taking the 4d scalar vev to zero) we cross the curve of marginal stability, rendering the W unstable.  However, if we did analytically continue to the region where the W-boson is unstable, it is known that the kinetic term for the W will flip sign \cite{Rocek1, Rocek2}, which is the signature of an $SL(2,\mathbb{R})$ gauge theory.  Of course, the CS level must be non-zero, otherwise we would end up with a non-unitary theory,  and the existence of CS terms would render the gauge particles massive in the IR and make their wrong sign kinetic term irrelevant.  

As in our general discussion at the beginning of this section, the precise theory we get from these R-flows of the weak and strong coupling chambers of $SU(2),$ depends on the boundary conditions.  However, what is clear is that this construction produces  a 3d theory with infinitely many chiral fields, corresponding to the weak coupling chamber, which is mirror to a theory with only two chiral fields, corresponding to the strong coupling chamber.   As for the higher $SU(N)$ gauge theories in 4d,  the chambers which support higher spin BPS states, can lead under R-flow to 3d theories with massless (or nearly massless) particles of higher spin.  One might speculate that this suggests a 3d structure of a higher spin gauge theory.  These are clearly exciting possibilities, and are the subject of active investigation.

\subsection{Accumulating Tetrahedra}

One feature of the resulting 3d geometry which we can see directly from the correspondence between tetrahedra and 3d BPS particles, is that in infinite chambers of the 4d theory, the UV three-manifold $M$ will be partitioned into \emph{infinitely} many tetrahedra.  Further, at the accumulation rays in the 4d BPS spectrum which describe vector multiplets, the resulting tetrahedra will also accumulate.  

\begin{figure}[here!]
  \centering
\includegraphics[width=.4\textwidth]{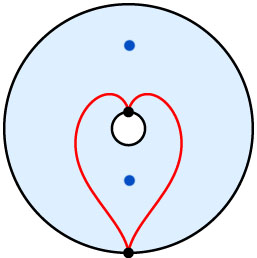}     
  \caption{The $SU(2)$ triangulation of an annulus.  The red lines denote the two internal edges whose flips describe BPS states.  The dark blue dots are the zeros of the SW differential.}
  \label{fig:annulus}
\end{figure}

As an explicit example, we can consider the case of pure $SU(2)$ described above.  Then, the 4d $\mathcal{N}=2$ Riemann surface is given by an annulus with one marked point on each boundary.  The three-manifold $M$ obtained from a flow of this data is therefore a filling of two annuli glued along their common boundary.  A trivial application of the genus formula \eqref{genusrel} shows that $\partial M$ is a torus, and hence $M$ is a solid torus.  The front and back face of this torus are equipped with triangulations, each with two triangles as shown in Figure \ref{fig:annulus}.

\begin{figure}[here!]
  \centering
\includegraphics[height=.2\textwidth]{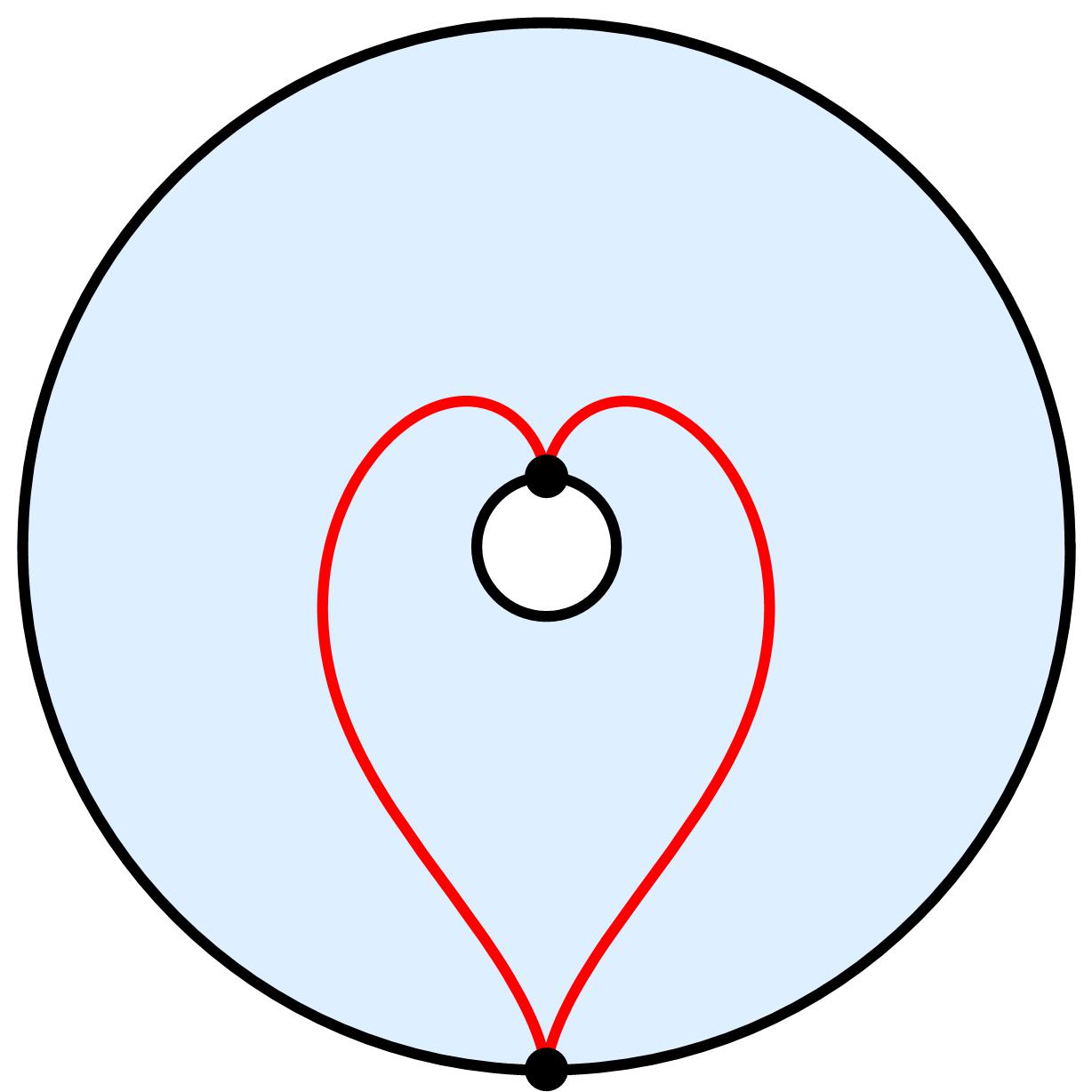}   
  \hspace{.2in}          
\includegraphics[height=.2\textwidth]{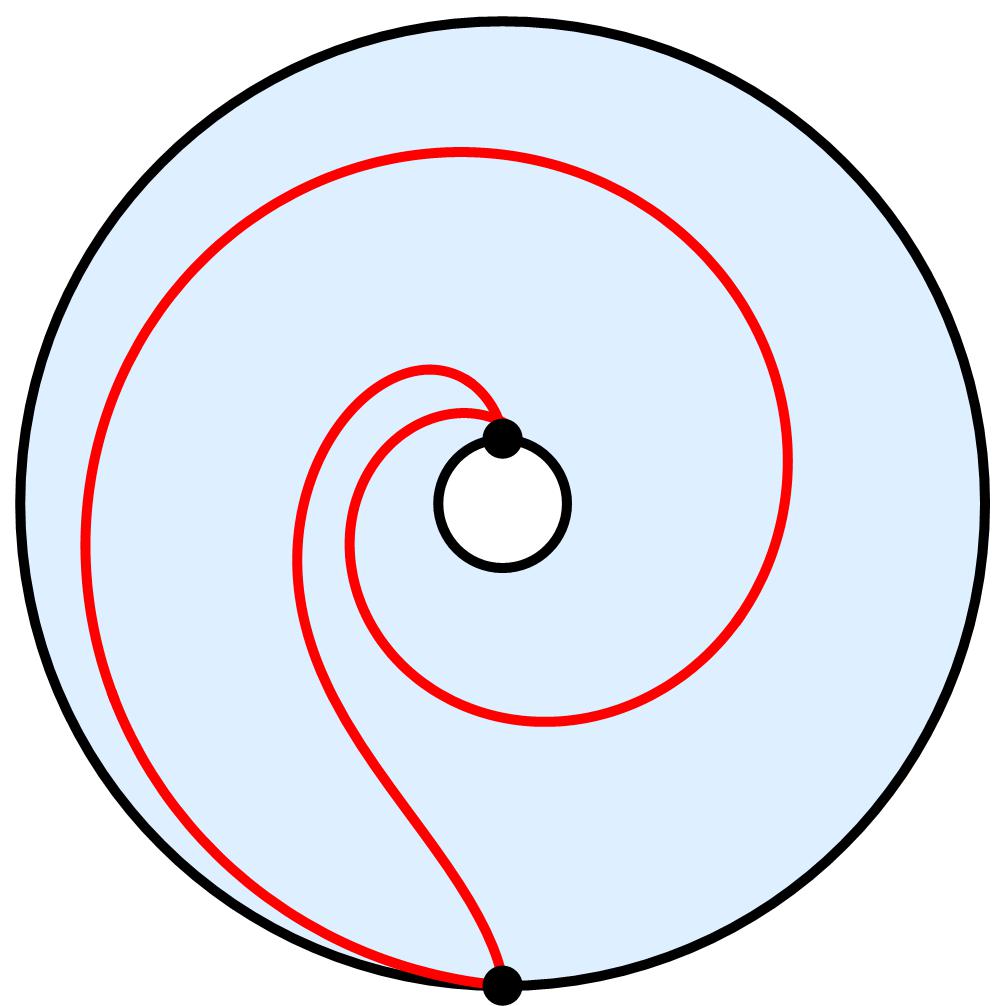}   
  \hspace{.2in}  
  \includegraphics[height=.2\textwidth]{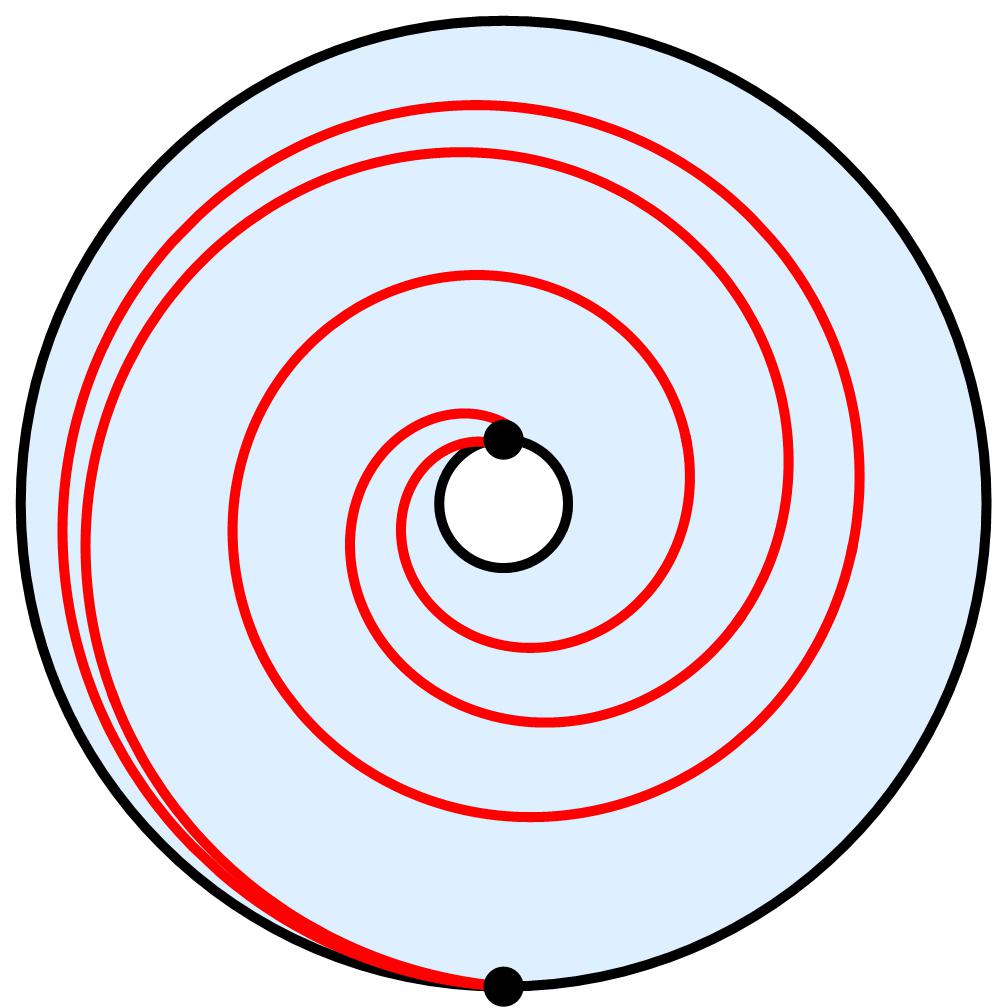}   
  \hspace{.2in}  
  \includegraphics[height=.2\textwidth]{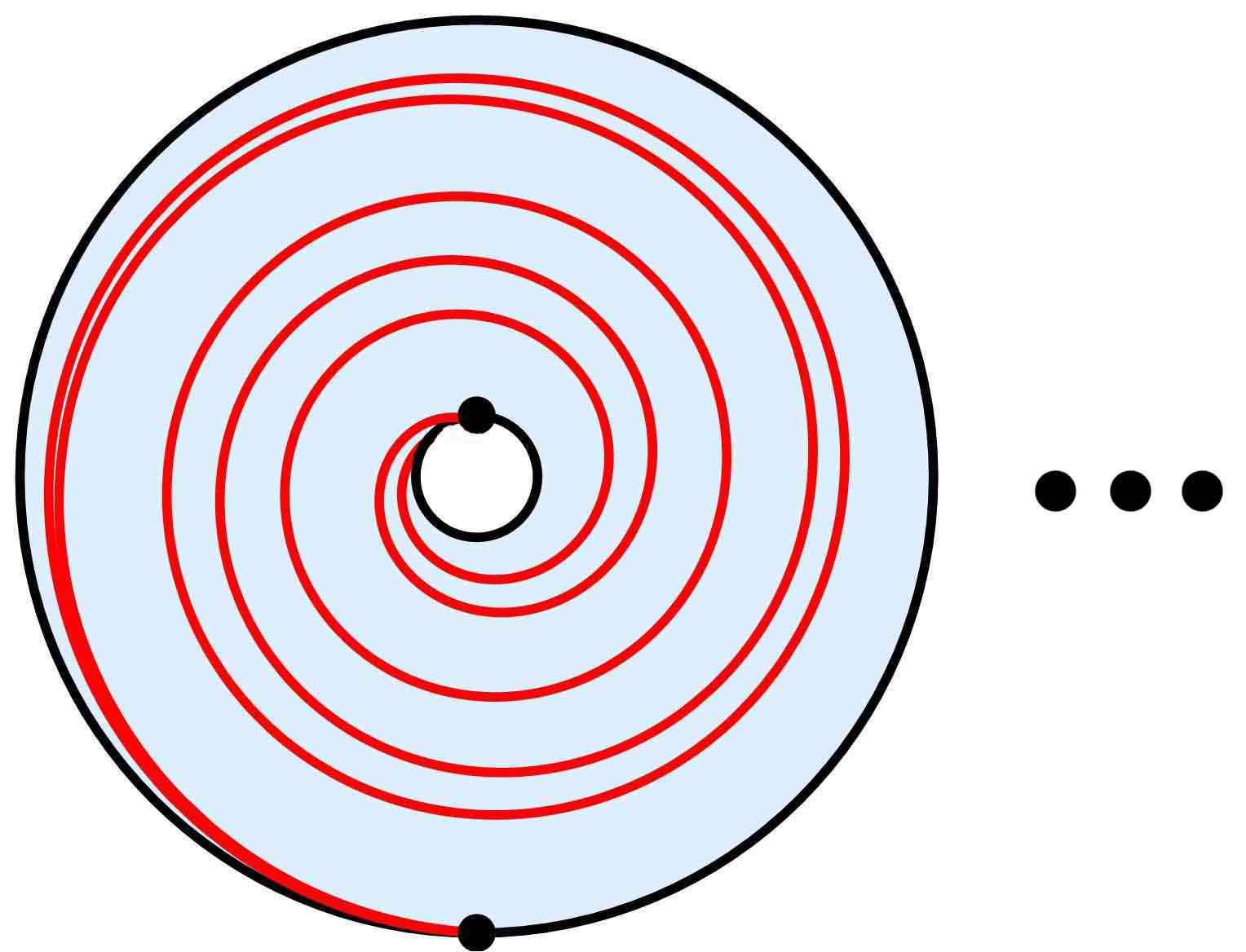}    
  \caption{The sequence of flips for the weak coupling spectrum of $SU(2)$.  There are infinitely many BPS states which give rise to flips that accumulate.  In the corresponding 3d theory these are reinterpreted as gluing data for accumulating tetrahedra.  }
  \label{fig:su2flips}
\end{figure}

As we flow through time, the triangulation of the front face will evolve to that of the back face.  This results in a tetrahedral decomposition of $M$.  In the strong coupling chamber there are two states, and this determines a decomposition of the solid torus into two tetrahedra, much as in the previous section's  description of the $A_{n}$ theories.  However, in the weak coupling chamber, the geometry is much more novel.  Due to the presence of infinitely many BPS particles in 4d, the triangulation undergoes an infinite sequence of flips.  Along these flips the internal edges begin to accumulate as shown in Figure \ref{fig:su2flips}, and in the limit of infinitely many flips the W-boson appears.  Now we can reinterpret this sequence of flips a describing a decomposition of the solid torus into infinitely many tetrahedra which degenerate.  Similar structures have been studied by mathematicians \cite{hodgson}.  

Just as the 2-3 Pachner move is the geometric manifestation of 2-3 wall crossing in 4d, the relationship described above, between two tetrahedra, and infinitely many, can be seen as the geometric version of the wall-crossing governing the decay of a BPS vector multiplet.  The physical meaning, and mathematical consequences of this phenomenon demand further illumination.

\section{Connections with 3d $SL(N)$ Chern-Simons Theories}
\label{sec:CS}
In the previous sections, we have explained how to construct a certain class of ${\cal N}=2$ theories
in 3d, starting from the corresponding parent theory in 4d and its BPS spectrum.  A special instance of these theories
arises from wrapping $N$ M5-branes on a 3-manifold $M$, leading to a 3d ${\cal N}=2$
theory.
We have also seen that two different chambers of a given 4d theory, related by wall-crossing, give rise to two dual theories in 3d.
It is then natural to study the realization of this duality at the level of the partition functions
of the corresponding 3d theories.  In this section and the next, we explore this question and show that compactifying
these theories on a 3d geometry known as Melvin cigar $MC_q$ (which is a twisted
product of a circle with a semi-infinite cigar) leads to computable partition functions.  The equivalence
of the 3d theories defined by different chambers of the same 4d theory then follows from the invariance of the Kontsevich-Soibelman  (KS) monodromy under wall--crossing \cite{ks1,Gaiotto:2008cd,Dimofte:2009bv,Dimofte:2009tm,Cecotti:2009uf,CNV}.
The partition functions of the $\mathcal{N}=2$ models are related to the partition function of $SL(N)$ Chern-Simons theories
expanded near specific backgrounds on the internal 3-manifold $M$. Later, to avoid problems with the boundary condition at infinity and get simpler formulae, we will find it convenient to replace the Melvin cigar by its compact counterpart, the squashed three-sphere $S^3_b$.

We begin by recalling the relation between $N$ copies of M5 branes on $M\times MC_q$ with
$SL(N)$ Chern-Simons theory on $M$, and specialize the discussion to the case that $M$ is a flow
of a punctured Riemann surface $\Sigma$, \textit{i.e.}\! topologically  $M=\Sigma \times \mathbb{R}$, as in section \ref{sec:4d3dL}.
We then explain how this relation can be used to compute the partition function of these theories using
the KS \emph{half}--monodromy operator $\mathbb{K}(q)$, which is the time--ordered product of quantum-dilog operators,
sandwiched between $\langle\mathrm{out}|$, $|\mathrm{in}\rangle$ states,
 where the choice of `in' and `out' states
directly maps to the choice of boundary conditions on the braids of section \ref{sec:T+B}.   We conclude this section with an elementary discussion of some concrete examples
in the context of the flow of $A_2$ theories. In particular, we show how the equivalence of mirror theories
is reflected in their partition functions.  
The more general $ADE$ Argyres-Douglas flow is studied in the minimal/maximal chambers in section \ref{partADE} with the more powerful formalism of \cite{CNV}.   In that section, we also give an example involving R-flows with non-sink/source
mutations and describe the resulting braid.

\subsection{M5-branes and Chern--Simons Theories}\label{M5branesandCS}

Consider M-theory on a Calabi-Yau threefold $Q$.  Furthermore, replace the five-dimensional uncompactified space with a twisted product of Taub-NUT (TN) and $S^1$, where as we go around $S^1$ the TN undergoes a rotation by $\theta$ as in \cite{DVV}. Let $q=\exp(i\theta)$. We consider $N$ M5-branes wrapping the submanifold
\begin{equation}
 M \times MC_q \subset Q \times (TN\times_q S^1),
\end{equation}
where $M\subset Q$ is a (special) Lagrangian submanifold of the Calabi--Yau threefold $Q$, and $MC_q$ denotes the space--time geometry of the 5-brane, which is a Melvin cigar:
\begin{equation}
MC_q=C \times_q S^1 \equiv (C\times \mathbb{R})\Big/\big\{(z,\varphi)\sim (q z, \varphi+2\pi)\big\}
\end{equation}
where $C$ is a semi-infinite cigar subspace of TN,  and $z$ its canonical complex coordinate.
In this paper we are primarily concerned with the case of $N=2$ M5-branes, but for the moment let us be general and consider an arbitrary number $N$ of them. 

Then it is known \cite{CNV,mina} that the partition function of the full M-theory on this geometry, receives an additional contribution due to the presence of the M5-branes given by
\begin{equation}\label{basiceq}
Z_N(M\times MC_q)=Z_{SL(N)}(M)_q
\end{equation}
where $q$ is related to the coupling $1/k$ of the Chern-Simons theory in the usual way,
\begin{equation}
q=\exp(2\pi i/k), 
\end{equation}
except that we do not require the level $k$ to be an integer. We write the CS coupling $1/k$ as $\tau$, and extend both sides of eqn.\eqref{basiceq} to arbitrary complex $\tau$'s.  At the perturbative level there is no ambiguity in the definition of the CS partition function for arbitrary couplings $\tau$.
Non--perturbatively, however, one needs to specify a prescription, as in \cite{WittenCSA}.
 Note that  both sides of the identity \eqref{basiceq} depend on extra data. On the left hand side we have to specify the boundary conditions on $\partial MC_q$, while on the right hand side, to define the CS partition function for non-integral levels $\tau^{-1}$, we need to choose a contour for the path integral \cite{WittenCSA}.  Each choice of contour corresponds to a choice of boundary condition in the $MC_q$ geometry. One way to understand how such a relation emerges is to note that if we take $q\rightarrow 1$ and compactify the M5-brane on the circle, we end up with $D4$ branes on $M$, leading to a 2d theory with $(2,2)$ supersymmetry on $C$.  The  chiral fields of this 2d theory are labeled by holomorphic $SL(N)$ connections on $M$, and the supersymmetric vacua correspond to flat connections, which implies that this $(2,2)$ theory has a superpotential equal to the holomorphic $SL(N)$ Chern-Simons action. To preserve supersymmetry on the cigar $C$, we must do two things: first
  we must add to the action a Warner boundary term \cite{warner}, given by the integral of the superpotential on the boundary, $\int_{S^1}W\,d\sigma$
(note that this is not a superspace integral!). Then, we have to choose a suitable boundary conditions on $\partial C$; this is a special case of the supersymmetric--brane boundary conditions for $(2,2)$ LG theories \cite{HIV}, which requires the chiral fields at the boundary to take value in a half--dimensional real subspace $\mathcal{S}$ of the field space, $\phi|_{\partial C}\in \mathcal{S}$.
Hence, in the limit $\tau\equiv k^{-1}\rightarrow 1$, i.e, $q=e^{2\pi i\tau}\rightarrow 1$,
\begin{multline}\label{expZq->1}
 Z_N(M\times MC_q)\Big|_{q\rightarrow 1}=\\
=\int_\mathcal{S} d\mathcal{A}\: \exp\left\{ \frac{i}{4\pi^2}\int\limits_0^{2\pi} d\varphi \int_M \mathrm{Tr}\Big( \mathcal{A} d\mathcal{A}+ \frac{2}{3} \mathcal{A}^3\Big)\right\}\:\Psi_N(\mathcal{A}; M\times C)\Bigg|_{\tau\rightarrow 1},
\end{multline}
where the functional $\Psi_N(\mathcal{A}; M\times C)$ is given by the path integral over $C$ with Dirichlet boundary condition
$\phi|_{\partial C}=\mathcal{A}$. From eqn.\eqref{expZq->1} it is clear that the choice of boundary condition $\mathcal{S}$ is precisely a choice of an integration contour for the holomorphic $SL(N,\mathbb{C})$ Chern-Simons theory. A slight modification of this construction gives
$Z_N(M\times CM_q)$ for any `physical' value of the CS coupling, $q=\exp(2\pi i/k)$ with $k\in \mathbb{Z}$. Indeed, it suffices to make the identification in $\mathbb{R}\times C$ 
\begin{equation}
(\varphi+2\pi,z)\sim (\varphi, e^{2\pi i/k} z),
\end{equation}
which can be seen as a $\mathbb{Z}_k$ modding out of $S^1\times C$, where now $S^1$ has length $2\pi k$.
Then, \eqref{expZq->1} is replaced by
\begin{multline}\label{tau->1/k}
 Z_N(M\times MC_q)\Big|_{q\rightarrow e^{2\pi i/k}}=\\
=\int_\mathcal{S} d\mathcal{A}\: \exp\left\{ \frac{i}{4\pi^2} \int\limits_0^{2\pi k} d\varphi \int_M \mathrm{Tr}\Big( \mathcal{A} d\mathcal{A}+ \frac{2}{3} \mathcal{A}^3\Big)\right\}\:\Psi_N^k(\mathcal{A}; M\times C),
\end{multline}
for some new operator insertion $\Psi_N^k(\mathcal{A};M\times C)$. As expected, on the boundary we naturally get the holomorphic Chern--Simons at level $k$ integrated along the countour $\mathcal{S}$ in complex field space.
 The analytic continuation of this expression to general $\tau\equiv k^{-1},$
leads to the above relation between functional countours in Chern--Simons theory and boundary conditions for partition functions on the Melvin cigar.

Let us recall how the partition function in the \textsc{lhs} of \eqref{tau->1/k} depends on the contour $\mathcal{S}$. First of all, by holomorphy of the integrand, only the relative homotopy class of $\mathcal{S}$ matters \cite{WittenCSA}. For all $\mathcal{S}$ the path integral is a solution to the same set of differential equations (Ward identitities, Schwinger--Dyson equations); since perturbation theory produces a unique formal solution to these equations, the difference between inequivalent choices of $\mathcal{S}$ is not visible in perturbation theory. In fact distinct choices of $\mathcal{S}$ correspond to taking different linear combinations of the contributions from the non--trivial saddle points which are weighted by the typical non--perturbative factor $\exp(i\, S_\mathrm{class}/\tau)$. Thus, being cavalier with the boundary conditions on $MC_q$ produces an error which is exponentially small at weak coupling.  
\medskip

Having discussed the theory from the point of view of the cigar,  we can also consider this system from the viewpoint of the 3d supersymmetric
gauge theory obtained by compactifying $N$ M5-branes on $M$.  Let us denote this
3d ${\cal N}=2$ theory by $Q_{M,N}$.  The above discussion gives the following identification for the partition function of $Q_{M,N}$ on $MC_q$
\begin{equation}\label{basicidentity}
Z_{Q_{M,N}}(MC_q)=Z_{SL(N)}(M)_q.
\end{equation}

An additional layer of complication arises if the internal three manifold $M$  has
boundaries.  Suppose, as in the theories of main interest for this paper,
that $M$ has the topology\footnote{\ As before, we shall refer to the coordinate in the $\mathbb{R}$ factor of $M$ as `time' denoted by $t$.}
\begin{equation}
M=\Sigma \times \mathbb{R}.
\end{equation}
The Chern-Simons partition function on $M$ (defined with respect to a fixed contour $\mathcal{S}$) becomes an operator,
${\cal O}_{SL(N)}(\Sigma \times \mathbb{R})_q$ mapping the Hilbert space $H_\mathcal{S}(\Sigma)$,
obtained by quantizing the initial conditions on the boundary at ${t=-\infty}$,
to the  Hilbert space $H_\mathcal{S}(\Sigma)$ of final states at ${t=+\infty}$. The identity \eqref{basicidentity} then states that
the partition `function' of the $Q_{M,N}$ theory quantized on $MC_q$ takes values in the algebra of operators
acting on the $SL(N)$ holomorphic Chern--Simons Hilbert space $H_\mathcal{S}(\Sigma)$ 
\begin{equation}
\mathcal{Z}_{Q_{M,N}}(MC_q) \equiv {\cal O}_{SL(N)}(\Sigma \times \mathbb{R})_q\colon H_\mathcal{S}(\Sigma)\rightarrow H_\mathcal{S}(\Sigma).
\end{equation}
It remains to specify the operator ${\cal O}_{SL(N)}(\Sigma \times \mathbb{R})_q$. It is determined by the special Lagrangian geometry $\Sigma\times\mathbb{R}\xrightarrow{\ell} Q$. One case is easy, when the Lagrangian structure is also a product. In this case the image of the fixed time slice $\Sigma_t$ of $M$ in $Q$ is a special Lagrangian submanifold of a hyperK\"ahler subspace $X _t\subset Q$. This special case preserves $8$ supercharges, and the corresponding operator ${\cal O}_{SL(N)}(\Sigma \times \mathbb{R})_q$ is just the identity. Indeed the 3d theory $Q_{M,N}$ is, in this case, the trivial dimensional reduction of the 4d theory defined by the surface $\Sigma$. 

In this paper we have studied a more general situation: the Calabi--Yau threefold $Q$ has two ends each asymptotic to $X\times \mathbb{R}$ for some hyperK\"ahler manifold $X$ and the manifold $M\simeq \Sigma\times \mathbb{R}$ also has two ends, for $t\rightarrow\pm \infty$, which are asymptotic to $\Sigma_{\pm\infty}\times \mathbb{R}\subset X\times\mathbb{R}$ with $\Sigma_{\pm\infty}\subset X$ special Lagrangian. This geometry has the physical interpretation of the two 4d theories corresponding to $\Sigma_{\pm\infty}$ defined, respectively, in the two half--spaces $t\gtrless 0$ and separated by a domain wall on which the degrees of freedom of the 3d theory $Q_{M,N}$ live, see Figure \ref{fig:wall} on page \pageref{fig:wall}. This geometry preserves only half--supersymmetry, and hence from the 4d viewpoint, correspond to a half--BPS configuration. The Chern--Simons operator ${\cal O}_{SL(N)}(\Sigma \times \mathbb{R})_q$ is no longer the identity on $H_\mathcal{S}(\Sigma)$, but still a \textsc{susy} protected object.

The computation of ${\cal O}_{SL(N)}(\Sigma \times \mathbb{R})_q$, being protected, 
 is independent of the scale, and hence we can determine this operator by going to the extreme IR limit.
As discussed in sect.\ref{sec:M5}, in the IR the $N$ M5 branes wrapped on the special Lagrangian three-manifold $M$ may be replaced by a single recombined M5 brane having support on  a manifold $\widetilde{M}$ whose infinite ends have the form $\mathbb{R}\times \widetilde{\Sigma}$, with $\widetilde{\Sigma}$ a $N$-fold cover of $\Sigma$.
This reduces the path integral to that of a $GL(1,\mathbb{C})$ Chern--Simons theory on $\widetilde{\Sigma}$ plus instanton corrections which describe the quantum physics of the brane recombination process. In the two asymptotic ends we have $M \sim \Sigma\times \mathbb{R}$, and $8$ supersymmetries are asymptotically preserved. This means that the instanton contributions vanish as $t\rightarrow \pm\infty$, and hence we have exact isomorphisms at the level of in/out Hilbert spaces 
\begin{equation}
 H_\mathcal{S}(\Sigma; SL(N))\rightarrow H(\widetilde{\Sigma}; GL(1))
\end{equation}
whose explicit form depends on $q$ and $\mathcal{S}$. These isomorphisms may be used to identify the operator
${\cal O}_{SL(N)}(\Sigma \times \mathbb{R})_q$ with some operator acting on the Hilbert space  $H(\widetilde{\Sigma}; GL(1))$ which is easy to describe since the abelian CS theory is free. $H(\widetilde{\Sigma}; GL(1))$ arises from the quantization of the $GL(1,\mathbb{C})$ flat connections on $\widetilde{\Sigma}$. The corresponding classical phase space is given by the complefixied Jacobian of $\widetilde{\Sigma}$, 
$J(\widetilde{\Sigma})\simeq (\mathbb{C}^\ast)^{b_1(\widetilde\Sigma)}$, having coordinates 
\begin{equation}U_a=\exp\left(\int_{\gamma_a} \mathcal{A}\right)\in \mathbb{C}^*\qquad \{\gamma_a\}\ \text{a basis of }H_1(\widetilde\Sigma),\end{equation}
and holomorphic symplectic form
\begin{equation}\label{simplform}
 \Omega = \sum_{a,b}(\gamma_a\circ\gamma_b)\: \frac{dU_a}{U_a}\wedge \frac{dU_b}{U_b},
\end{equation}
where $\gamma_a\circ\gamma_b$ is the intersection form in $H_1(\widetilde\Sigma)$ equal to the electric-magnetic pairing of the 4d theory (in other words, $\gamma_a\circ\gamma_b$ is simply the number of arrows between nodes $a$, $b$ in the 4d quiver, see section \ref{sec:Quiver}). After quantization, the $U_a$ become the generators of the quantum torus algebra $\mathbb{T}_{\tilde\Sigma}$ with commutation relations
\begin{equation}
U_a U_b=q^{\gamma_a\circ\gamma_b}\, U_bU_a.
\end{equation}
The operator ${\cal O}_{SL(N)}(\Sigma \times \mathbb{R})_q$ may then be identified with an element of $\mathbb{T}_{\widetilde\Sigma}$. 

The path integral of the $GL(1,\mathbb{C})$ CS on $\widetilde\Sigma\times\mathbb{R}$ is just the identity operator. Then, the operator $\mathcal{O}$ receives non-trivial contributions only from instantons. Rescaling $t$, we may make the adiabatic approximation as accurate as we wish. In this limit the 3d geometry $\widetilde{M}$, restricted to any finite time interval, will look as a trivial product $\widetilde{\Sigma}\times \mathbb{R}$, preserving $4$ supersymmetries, and hence (seeing the factor $\mathbb{R}$ as part of the 4d space-time) as a half-BPS configuration of the $\mathcal{N}=2$ 4d theory with some BPS angle $\theta$. Moreover, in every finite time interval only the BPS states preserved by a \emph{given} set of $4$ supercharges may contribute; that is, in the adiabatic limit the contributions of instantons associated to BPS states of different angle $\theta=\arg Z$ get widely separated in time \cite{Cecotti:2009uf}. Hence we must have
\begin{equation}
{\cal O}_{SL(N)}(\Sigma \times \mathbb{R})=T\left(\prod_{\alpha} \mathcal{O}_{\alpha}\right),
\end{equation}
where $\mathcal{O}_{\alpha}$ contains contributions from instantons associated to 4d BPS states of given BPS angle $\alpha$. 
Continuity of the $4$ preserved supercharges then requires that the time ordering $T$ is correlated to the (cyclic) ordering in the BPS phase $\alpha$.  This is the basic simplification due to the structure of R-flow: time ordering is phase ordering.

Now, as described in detail in section 2, a BPS state 
is an M2-brane ending on the M5-brane. In fact, the asymptotically stationary configurations relevant in the adiabatic limit are described by the M2-branes with boundary of the form $\gamma\times\mathbb{R}$, with $\gamma$ a closed curve in $\widetilde\Sigma$. The cycle $\gamma$ defines an element of $ H_1(\widetilde\Sigma)$ which specifies the charge of the particle. To such a BPS state, there is associated an element $U_\gamma$  of
$\mathbb{T}_{\widetilde\Sigma}$ defined by
the holonomy along the cycle $\sum_a n_a \gamma_a$: writing $\gamma=\sum_a n_a \gamma_a$, 
\begin{equation}
 U_\gamma= \exp\left(\int_\gamma \mathcal{A}\right)= N\Big[\prod_a U_a^{n_a}\Big]
\end{equation}
where $N[...]$ is the `normal ordering' operation.
Then, the given--angle operator $\mathcal{O}_\alpha\in \mathbb{T}_{\widetilde\Sigma}$ must be a function $\Phi(U_{\gamma(\alpha)};q)$ where $\gamma(\alpha)=\sum_a n_a\gamma_a$ are the charges of the BPS states of angle $\arg(\sum_a n_a Z(\gamma_a))=\alpha$.

The transformation $t\leftrightarrow -t$ inverts the orientation of $M$ and interchanges the in/out Hilbert spaces, mapping the operator ${\cal O}_{SL(N)}(\Sigma \times \mathbb{R})$ to its inverse. On the other hand, in CS theory inverting the orientation is equivalent to flipping the sign of the coupling, $q\leftrightarrow q^{-1}$. Hence the given--angle operator satisfies the functional equation
\begin{equation}
 \Phi(U_\gamma;\,q^{-1})= \Phi(U_\gamma;\,q)^{-1}.
\end{equation}
Strictly speaking, this equation is true up to non--perturbative corrections, since we did not keep tract of the transformation of the contour $\mathcal{S}$.

To determine $\Phi(U;q)$ it is enough to consider the case of a single BPS hypermultiplet, corresponding to $\widetilde\Sigma=\mathbb{C}^*$. From the 4d viewpoint, this is a free theory, and the partition function is really a function, not an operator. Indeed,  $\mathbb{T}_{\mathbb{C}^*}$ is commutative, and we can replace $U$ by a fixed complex number $U=e^{-u}$. 
The partition function contribution from a chiral multiplet on $MC_q$ can be computed using the topological
string amplitude for the contribution of an M2-brane ending on an M5-brane.  This is a local
computation and can be done in the context of the M2-branes ending on Lagrangian branes discussed for $\mathbb{C}^3$ as in section 5.1.2.
 From topological strings \cite{integrable}, we know that the function $\Psi(u;q)=\Phi(e^{-u};q)$ must satisfy a difference equation
(where for convenience we have shifted $u$ by $i\pi \tau$)
\begin{gather}\label{diffequation}
\Big(e^{2\pi i \tau \partial_u}+q^{-1/2}e^{-u}-1\Big)\Psi(u;q)=0,\qquad q=e^{2\pi i\tau},\\
\intertext{that is}
\Psi(u+2\pi i\tau;q)= (1-q^{-1/2}e^{-u})\Psi(u;q). 
\end{gather}
Note that the ratio of two solutions to this equation is a periodic function of period $2\pi i\tau$. Hence, it has a Fourier series of the form $\sum_{k\in\mathbb{Z}}a_n \,e^{nu/\tau}$, which is a constant up to exponentially small non--perturbative corrections. This is an example of the general phenomenon alluded to in the beginning of this section: the solutions to Ward--identities are unique perturbatively, but not necessarily non--perturbatively. However, if we insist that the CS gauge group is
the multiplicative group $GL(1,\mathbb{C})=\mathbb{C}^*$, and not the additive group $\mathbb{C}$, the function $\Psi(u;q)\equiv \Phi(e^{-u};q)$ should be periodic in $u$ of period $2\pi i$. Up to a ($q$ dependent) normalization constant, there is a unique analytic periodic solution, namely the \emph{compact} quantum dilogarithm \cite{fad-compact}. For $|q|<1$ it may be written as
\begin{equation}\label{compquantumdilog}
 \psi(u;q)= \prod_{n=0}^\infty\Big(1-q^{n+1/2} e^{-u}\Big)^{-1},
\end{equation}
which, indeed, satisfies the symmetry condition
\begin{equation}
 \psi(u; q^{-1}) = \psi(u; q)^{-1}.
\end{equation}
Below we shall see that, non--perturbatively, the proper gauge group is $\mathbb{C}$, rather than $\mathbb{C}^*$. To all orders in perturbation theory we may ignore this subtlety, and use just the compact dilogarithm function $\psi(u;q)$.

The case of a single quantum dilogarithm,
\begin{equation}
\mathcal{O}_{SL(N)}(\Sigma\times\mathbb{R})= \Phi(U;q)
\end{equation}
corresponds to $N=2$ M5 branes, with ${\widetilde \Sigma} \times \mathbb{R}$ corresponding
to the $A_1\times \mathbb{R}$ geometry, the IR description of two M5 branes wrapping a tetrahedron. Since we only
get one BPS state from the recombined IR brane, the above relation implies that the partition
operator of $SL(2)$ on the tetrahedron is a single quantum dilog operator.  
This fact was known \cite{tudor,dimogukov,DGG}, but here we have given a derivation of it
based on two equivalent descriptions of the $A_1$ theory, one from the
UV leading to the $SL(2)$ Chern-Simons, and the other from the IR.  For $N=2$ we can identify the $SL(2)$ Chern-Simons
theory as the quantum gravity theory of the metric for three-manifolds \cite{WIT1,WIT2},
which thus sheds some light on why the geometry of tetrahedrons emerge automatically
for such cases, as we have noted before in this paper.
\smallskip
 
More generally, the given--angle operator for a BPS supermultiplet having spin $s$ is, up to non--perturbative corrections, \cite{Cecotti:2009uf}
\begin{equation}
 \Phi^{(s)}(U;q) = \prod_{j=-s}^s \Phi(q^j U;q)^{(-1)^{2s}}.
\end{equation}
  
According to the work of Kontsevich-Soibelman
and its refinements \cite{ks1,Gaiotto:2008cd,Dimofte:2009bv,Dimofte:2009tm,Cecotti:2009uf,CNV}, the operator
$$\mathcal{O}_{SL(N)}(\Sigma\times\mathbb{R})_q= T \prod_{\mathrm{BPS}}\Phi^{(s_\alpha)}(U_{\gamma_\alpha};q)$$
does not depend on the BPS chamber we use to compute the \textsc{rhs}. More precisely, the wall--crossing invariant is the adjoint action
\begin{equation}
 \mathrm{Ad}[\mathcal{O}_{SL(N)}(\Sigma\times\mathbb{R})_q]\colon \mathbb{T}_{\widetilde\Sigma}\rightarrow \mathbb{T}_{\widetilde\Sigma},
\end{equation}
while two realizations of $\mathcal{O}_{SL(N)}(\Sigma\times\mathbb{R})_q$ as operators acting on an arbitrary $\mathbb{T}_{\widetilde\Sigma}$--module $\mathbb{M}_{\widetilde\Sigma}$ will differ, in general, by an element of
$\mathrm{End}(\mathbb{M}_{\widetilde\Sigma})$. This subtlety will be important below.  
Since the equality of the operators in different chambers holds identically in $q$, the perturbative and non--perturbative contributions should be separately wall--crossing invariant. In particular, the perturbative part already determines the BPS spectra on  the two sides of the wall, and the non--perturbative terms then are almost uniquely determined by the condition of being invariant under the very same spectrum discontinuity.

The wall-crossing formula of Kontsevich-Soibelman (and its refinement \cite{ks1,Gaiotto:2008cd,Dimofte:2009bv,Dimofte:2009tm,Cecotti:2009uf,CNV}) may also be understood as stating that 
the different 3d theories $Q_{M,N}$ obtained from the above construction starting from different
chambers of a given 4d ${\cal N}=2$ theory, are in fact equivalent at the quantum level since their (operator valued) partition functions are equal.

For clarity, let us also make a remark about the nature of the two M5-branes wrapping
the surface $\Sigma$. 
 In the case of irregular singularities, such as is the case for $A_n$ theories,
the two M5-branes are never identically on top of each other.  For example, in the $A_1$
theory if we try to make the branes coincide, the closest we get to doing this
is when $y^2=x^2$, which means we have two M5-branes occuping $y=x$ and $y=-x$.
In other words, in the case of irregular singularities we start with a particular
deformed geometry of two M5-branes over the $x$-plane.  Changing the moduli
of the M5 branes can change this to $y^2=x^2-m$ but it will not make it to
$y^2=0$ as this is infinitely far away in deformation parameter.  Thus when we
talk about `$SL(2)$' for tetrahedron, we also mean expanding the Chern-Simons
path-integral around such backgrounds,
which in this case will correspond to the hyperbolic $SL(2,\mathbb{C})$ structure.  In
the case with regular singularities, unlike the tetrahedron case, we expand around trivial $SL(2,\mathbb{C})$ connection (cfr.\! section \,\ref{sec:flowsgen}).

To complete the definition of the 3d theory $Q_{M,N}$ we have to fix the Chern-Simons in/out states making the partition function into an actual function rather than an operator 
\begin{align}
Z(Q_{M,N})&=\langle \mathrm{out}|\,{\cal O}_{SL(N)}(\Sigma \times \mathbb{R})\,|\mathrm{in}\rangle=\langle \mathrm{out}|\,{\cal O}^{inst.}_{GL(1)}({\widetilde \Sigma} \times R)\,|\mathrm{in}\rangle\\
&=\langle \mathrm{out} |\,T \prod_{\mathrm{BPS}}\Phi^{(s_\alpha)}(U_{\gamma_\alpha};q)\, |\mathrm{in} \rangle.
\end{align}
Fixing $\langle\mathrm{out}|$, $|\mathrm{in}\rangle$ corresponds from the perspective of the 3d $\mathcal{N}=2$ theory to completely determining which subgroup $H$ of the symmetries of the theory is gauged, as well as the corresponding CS couplings
$\tau_{ij}=k^{-1}_{ij}$ and FI terms.
As discussed in section \ref{sec:T+B}, the easiest way to specify the gauging is
to tie up the branching strands at each end.  To each pair of strands we can associate
a cycle on the asymptotic surfaces $\widetilde{\Sigma}$ at $t=\pm\infty$.  These pairings can be viewed
as a choice of `A-cycles' for the $\widetilde{\Sigma}$. If the intersection form $(\cdot\circ \cdot)$ in $H_1(\widetilde{\Sigma})$ is non--degenerate, this is also a choice of polarization for the symplectic structure 
\eqref{simplform} and hence, after quantization, a choice of a state in $H_\mathcal{S}(\widetilde{\Sigma}; GL(1,\mathbb{C}))$.  

Now, we have two pairs of
basis for the A-cycles, which we call the `in' and `out' A-cycles:
$(A^\mathrm{out},A^\mathrm{in})$.  To each A--cycle $A_i$ we can associate a corresponding
Wilson loop operator $U_i$.  Saying that we make the $A_i$ contractible amounts
to enforcing the constraint $U_i=1$. Indeed, from the $M$--theory viewpoint
\begin{equation}
U_i={\rm exp} \Big(i\int_{S^1\times A_i} B\Big)={\rm exp} \Big(i\int_{S^1\times D_i} dB\Big)=1
\end{equation}
where $D_i$ is the disc whose boundary is $A_i$ (as we assumed this is contractible) and
$dB=0$.
 Thus, the $|\mathrm{in} \rangle$ and $\langle \mathrm{out} |$ states corresponding to such a reconnection of strands 
are
\begin{equation}
|\mathrm{in} \rangle =\Big|\big\{ U_i=1\: |\: A_i\in A_\mathrm{in}\big\}\Big\rangle,
\end{equation}
$$\langle \mathrm{out} | =\Big\langle \big\{ U_i=1\: |\: A_i\in A_\mathrm{out}\big\}\Big|.$$
In the gauge theory language these special states correspond to turning off all the FI terms and
real masses.

Turning on real masses or FI terms corresponds to shifting $U_i=\exp(u_i)$.  To see this,
note that, as we go down to 2d on a circle parameterized by $
\theta$, the complexified mass  corresponds
to having a non-vanishing $C$-field in the bulk of M-theory Calabi-Yau of the form 
\begin{equation}
C=d\theta \wedge \omega,
\end{equation}
where $d\theta $ is the one-form on $S^1$
and $\omega$ is a two-form which pairs up with the corresponding BPS charge as discussed in section 2.  Furthermore,
in the presence of the $C$ field, the equation for $B$ gets modified to
\begin{equation}
dB+C=0.
\end{equation}
We find
\begin{equation}
U_i={\rm exp}\Big(i\int_{S^1\times A_i} B\Big)={\rm exp}\Big(i\int_{S^1\times D_i} dB\Big)={\rm exp}\Big(-i\int_{S^1\times D_i} C\Big)=\exp(u_i),
\end{equation}
where $\int_{S^1\times D_i} C=-u_i$ is the imaginary part of complexified mass or FI term.
By holomorphy this also leads to the relation between the real part of the FI terms and the
norm of $U_i$.
Then we end up, more generally, with the following $|\mathrm{in}
\rangle$ and $\langle \mathrm{out}|$ states:
\begin{equation}
|\mathrm{in} \rangle =\Big|\big\{ U_i=\exp(u_i^\mathrm{in})\: |\: A_i\in A_\mathrm{in}\big\}\Big\rangle,
\end{equation}
$$\langle \mathrm{out} | =\Big\langle \big\{ U_j=\exp(u_j^\mathrm{out})\: |\: A_j\in A_\mathrm{out}\big\}\Big|.$$
For fixed $\langle\mathrm{out}|$, $|\mathrm{in}\rangle$ states, the wall-crossing formula implies the equality of the partition functions of the 3d theories obtained from the reduction of different chambers of the same parent ${\cal N}=2$ theory in 4d.

Now we present some elementary examples of this in the context of $A_2$ Argyres-Douglas theory and its R--flow.
The partition functions of a more general class of Argyres-Douglas  models are discussed in section \ref{partADE}.

\subsection{$A_2$ Movie and 3d dual theories}
Consider the case of 3d theory obtained from the R--flow of the $A_2$ Argyres-Douglas theory and its compactification on the
Melvin cigar. Let $U,V$ denote the loop operators around the $S^1$ of the Melvin cigar associated to the nodes of the $A_2$ quiver.   As noted
before, they generate a quantum torus algebra with relation
\begin{equation}
UV=q\, V U.
\end{equation}
In other words, writing\footnote{\ Here and below, \textit{sans serif} letters $\mathsf{u}$, $\mathsf{v}$, $\mathsf{y}$, $\dots$ stand for quantum operators, while ordinary ones $u$, $v$, $y$, $\dots$ denote their respective eigenvalues.} $U=\exp(-\mathsf{u}),V=\exp(-\mathsf{v})$, we have
\begin{equation}
[\mathsf{u},\mathsf{v}]=2\pi i\tau\qquad \text{with }q=\exp(2\pi i\tau).
\end{equation}

The 4d $A_2$ theory has two
chambers, one with two hypermultiplets and one with three.  The KS wall-crossing formula relating these two chambers gives the operator
identity 
\begin{equation}\label{opide}
{\cal O}=\Phi(U;q)\,\Phi(V;q)=\Phi(V;q)\,\Phi(N[VU];q)\,\Phi(U;q)\end{equation}
which holds identically in $q$. Here $N[VU]\equiv -q^{-1/2}UV$ is  the normal ordered product\footnote{\ The minus sign just says that in defining the normal order $N[\cdots]$ we take the opposite determination of the square root of $q$ with respect to the one used in the definition of the compact quantum dilogarithm \eqref{compquantumdilog}. This difference is related to the quadratic refinement of the KS algebra, see ref.\!\cite{Gaiotto:2008cd} for a discussion.}. Equivalently, 
\begin{equation}\label{opide2}
\Psi(\mathsf{u}+i\pi;q)\,\Psi(\mathsf{v}+i\pi;q)=\Psi(\mathsf{v}+i\pi;q)\,\Psi(\mathsf{v}+\mathsf{u}+i\pi;q)\,\Psi(\mathsf{u}+i\pi;q).\end{equation}
This equation should be valid in perturbation theory as well as non--perturbatively.

Fixing the $\langle\mathrm{out}|$ and $|\mathrm{in}\rangle$ states, we determine which $3d$ theory we obtain from
the 4d $A_2$ theory, and then each BPS chamber corresponds to a different dual description of the given 3d theory. Depending on how we choose the $\langle\mathrm{out}|$ and $|\mathrm{in}\rangle$ states, we get inequivalent 3d theories.  As described in section 5.2, there are
two well known 3d $\mathcal{N}=2$ theories which arise in this way from the 4d $A_2$ model.  One is the $XYZ$ model, consisting of three 
chiral superfields with superpotential $W=XYZ$, and the other is the ${\cal N}=4$ $U(1)$ theory with 1 fundamental hypermultiplet.
Following section \ref{sec:domainwalss}, to get these two theories we choose the states as follows
\begin{equation}
XYZ:\qquad \langle \mathrm{out}|=\langle \mathsf{v}=v|, \qquad |\mathrm{in}\rangle =|\mathsf{u}=u\rangle,
\end{equation}
$${\cal N}=4:\qquad \langle \mathrm{out}|=\langle \mathsf{v}=v|, \qquad |\mathrm{in}\rangle =|\mathsf{v}=v\rangle.$$

For each of these two theories, the choice of the chamber gives two dual
descriptions of the \textit{same} 3d theory, as we have
observed.
For the $XYZ$ model, the two particle chamber corresponds to the SQED with
1 flavor, while the three particle chamber corresponds to three chiral fields $X,Y,Z$ with
the superpotential $XYZ$.  In the ${\cal N}=4$ theory, the 3 particle chamber is the
gauge theory description whereas the 2 particle chamber is its mirror with two chiral fields. 
 We thus obtain two expressions for the partition function of each of these theories
on the Melvin cigar:
\begin{equation}\label{ZXYZ}
Z_{XYZ}(v,u)=\langle v|\Psi(\mathsf{u})\Psi(\mathsf{v})|u\rangle=\langle v|\Psi(\mathsf{v})\Psi(\mathsf{u}+\mathsf{v})\Psi(\mathsf{u})|u\rangle
\end{equation}
\begin{equation}
Z_{{\cal N}=4}(v,v)=\langle v|\Psi(\mathsf{u})\Psi(\mathsf{v})|v\rangle=\langle v|\Psi(\mathsf{v})\Psi(\mathsf{u}+\mathsf{v})\Psi(\mathsf{u})|v\rangle
\end{equation}
where we used the short--hand $\Psi(\mathsf{u})\equiv \Psi(\mathsf{u}+i\pi;q)$.
\smallskip

The basic property of the function $\Psi(x)$ is the Ward identity \eqref{diffequation}, which fixes it up to non-perturbative contributions
\begin{gather}\label{diffequation2}
\Psi(x+2\pi i\tau)= (1+q^{-1/2}e^{-x})\Psi(x).
\end{gather}
For any solution to this equation, the function
\begin{equation}
 \widehat{\Psi}(x;\mathcal{C})
 =\exp\!\left\{i\,\frac{(x+\pi i(1+\tau))^2}{4\pi \tau}\right\}\int_\mathcal{C} dy\, \exp\!\left\{i\,y\,\frac{x+\pi i(1+\tau)}{2\pi \tau}\right\} \Psi(y),
\end{equation}
where $\mathcal{C}$ is any contour such that the integrand vanishes rapidly at its infinite ends, is also a solution to equation \eqref{diffequation2}. Then, $\widehat{\Psi}(x)= f(e^{x/\tau};\mathcal{C})\cdot \Psi(x)$, for some non--perturbative form--factor
$f(e^{x/\tau};\mathcal{C})$. Neglecting exponentially small non--perturbative corrections, we may replace $f(e^{x/\tau};\mathcal{C})$ by a constant, and then
the function $\Psi(x)$ is essentially its own Fourier transform, up to a Gaussian prefactor and a shift of the argument:
\begin{multline}\phi(x)\equiv\int_\mathcal{C} dy\: \exp[iyx/(2\pi\tau)]\: \Psi(y) =\\
=f(-e^{(x-\pi i)/\tau};\mathcal{C})\: e^{-i x^2/(4\pi\tau)}\:\Psi(x-\pi i (\tau+1)).\end{multline}

To get a feeling for the non--perturbative form--factors $f(e^{x/\tau};\mathcal{C})$, let us consider the particular case of the periodic solution $\psi(u;q)$ of equation \eqref{compquantumdilog}. Taking the contour $\mathcal{C}=\{a+i\,\mathbb{R}, a>0\}$ one gets (for $\mathrm{Im}\,\tau>0$)
\begin{multline}
\int_\mathcal{C} \frac{dy}{2\pi i} \exp\big[i(y^2+2xy+2i\pi y)/(4\pi\tau)\big] \:\psi(y+i\pi\tau;q)=\\
= \frac{q^{1/24}\:\theta((x+i\pi)/(2\pi i\tau),-1/\tau)}{\eta(q)}\: \psi(x;q)
\end{multline}
where $\theta(w,\tau)=\sum_{n\in\mathbb{Z}} e^{i\pi \tau n^2+2\pi i n w}$.
\medskip

For general states $\langle \mathrm{out}|$, $|\mathrm{in}\rangle$ and CS operators $\mathsf{z}$ one has
\begin{equation}
\langle \mathrm{out}|\, \Psi(\mathsf{z})\, | \mathrm{in}\rangle =\int \frac{dw}{4\pi^2\tau}\, \phi(w)\: \langle \mathrm{out}|\, e^{-i w\, \mathsf{z}/(2\pi\tau)}\,| \mathrm{in}\rangle.
\end{equation}
\smallskip

Evaluating the \textsc{rhs} of equation \eqref{ZXYZ} with the help of the above identities leads to
\begin{equation}
\begin{split}
&Z_{XYZ}^R(v,u)=\Psi(v)\,\Psi(u)\,\langle v|\,\Psi(\mathsf{u}+\mathsf{v})\,|u\rangle=\\
&= \Psi(v)\,\Psi(u)\, \int \frac{dw}{4\pi^2\tau}\, \phi(w)\: \langle v|\, e^{-i w(\mathsf{u}+\mathsf{v})/(2\pi\tau)}\, |u\rangle= \\ 
&= \Psi(v)\,\Psi(u)\, \int \frac{dw}{4\pi^2\tau}\, \phi(w)\, e^{- i w^2/(4\pi\tau)}\: \langle v|\, e^{-i w\,\mathsf{v}/(2\pi\tau)} e^{-iw\,\mathsf{u}/(2\pi\tau)}\, |u\rangle \\ \nonumber
&=\Psi(v)\,\Psi(u)\, \int  \frac{dy}{2\pi \sqrt{\tau}}\: e^{i(y-u-v)^2/(4\pi\tau)}\, \Psi(y)=\\ \nonumber
&= \Psi(v)\,\Psi(u)\,\Psi(-v-u+i\pi\tau) \times e^{i(v+u-i\pi \tau)^2/(4\pi\tau)}f(e^{-(u+v+i\pi )/\tau}) \nonumber
\end{split}
\end{equation}
for some function $f(\cdot)$. So, up to non--perturbative corrections, the partition function corresponding to this choice of in/out states is just the product of three decoupled free functions.
\smallskip
 
On the other hand, the \textsc{lhs} of equation \eqref{ZXYZ} leads to
\begin{multline}Z^L_{XYZ}=\int \frac{dw\, dz}{(4\pi^2\tau)^2}\, \phi(w)\,\phi(z)\, \langle v|\,e^{-i w\,\mathsf{u}/(2\pi\tau)} e^{-iz\,\mathsf{v}/(2\pi\tau)}\,|u\rangle=\\
=\int \frac{dw\, dz}{(4\pi^2\tau)^2}\, \phi(w)\,\phi(z)\, 
e^{-i(wz+wu+zv)/(2\pi\tau)}=\\
= \int \frac{dz}{4\pi^2\tau}\,\Psi(z+u)\, \phi(z)\,e^{-izv/(2\pi\tau)}=\\
=\int \frac{dz}{4\pi^2\tau}\,\Psi(z+u)\, \Psi(z-\pi i(\tau+1))\,e^{-i(z^2+2zv)/(4\pi\tau)}\, \tilde f(e^{(z-\pi i)/\tau})
\end{multline}
for some non--perturbative form--factor $\tilde f(\cdot)$. 
The equivalence of the two expressions leads to a perturbative duality equality for the partition function of the $XYZ$ model and $N_f=1$ $\mathcal{N}=2$ SQED. We do not write the perturbative result here, since we are going to state the full non--perturbative equality in the next section.

\subsection{$MC_q$ versus  $S^3_b$, and the non--compact quantum dilog}

The discussion in the previous subsection is not completely satisfactory due to the presence of the non--perturbative form factors $f(e^{z/\tau})$. Without a knowledge of these factors, the 3d dualities can be established only as perturbative equalities. It is desirable to have a more complete setup where the dualities are exact non--perturbative statements. Since the form--factors arise from the boundary conditions at infinity on the Melvin cigar, this amounts to finding good duality--covariant boundary conditions. The best boundary condition is \emph{no} boundary, that is, the natural strategy is to close the space by gluing  some manifold $Y$ to $MC_q$ along the boundary. 

Before looking for a suitable $Y$, let us discuss the algebraic aspects of the non--perturbative form--factors \cite{fad-compact}. At the perturbative level the holomorphic CS theories with gauge groups $\mathbb{C}^*$ and $\mathbb{C}$ are indistinguishable. The operator algebra of the first theory, quantized on $\widetilde\Sigma$, corresponds to the quantum torus algebra $\mathbb{T}_{\widetilde\Sigma}$, which, in the non--degenerate case, is
\begin{equation}
U_i V_j =q^{\delta_{ij}}\,V_j U_i.
\end{equation}
Such an algebra may be realized in terms of canonical variables
\begin{equation}
U_i=\exp \mathsf{u}_i,\ V_i=\exp \mathsf{v}_i,\quad [\mathsf{u}_i,\mathsf{v}_j]=2\pi i \tau \delta_{ij},
\end{equation}
and then the quantum torus algebra $U_i, V_j$ acts on the Hilbert space of the $L^2$--functions of, say, the $u_i$'s in the usual fashion.
However, the canonical Hilbert space is not an irreducible representation of the torus algebra. 
  Indeed, one 
can construct a dual quantum torus algebra $\widehat{\mathbb{T}}_{\widetilde\Sigma}$ which commutes with the $U_i,V_j$:
$${\hat U}_i=\exp(u_i/\tau),\quad {\hat V}_j=\exp(v_j/\tau)$$
\begin{equation}{\hat U}_i{\hat V}_j=({\hat q}^{-1})^{\delta_{ij}}\, {\hat V}_j{\hat U}_i
\end{equation}
$$[U_i,{\hat U}_j]=[V_i,{\hat V}_j]=[U_i,{\hat V_j}]=[V_i,{\hat U}_j]=0$$
where ${\hat q}$ is the `modular transform' of $q$
\begin{equation}
{\hat q}={\rm exp}(-2\pi i/\tau).
\end{equation}
The canonical Hilbert space is an irreducible representation of the total algebra generated by $U_i,V_j,{\hat U}_k,{\hat V}_\ell$, and in this sense these operators form a complete algebra.  

Passing  from the single quantum torus algebra
$\mathbb{T}_{\widetilde\Sigma}$ to the doubled one 
\begin{equation}
\mathbb{T}_{\tilde\Sigma}\times \widehat{\mathbb{T}}_{\tilde\Sigma},
\end{equation}
 precisely corresponds to replacing, as gauge group of the holomorphic Chern--Simons theory, $\mathbb{C}^*$ with $\mathbb{C}$, since the only difference between the two theories is that in the first case the holonomies are the exponentials $U_i=e^{u_i}, V_j=e^{v_j}$, and  we have the identifications
\begin{equation}
u_i\sim u_i+2\pi i, \quad v_j\sim v_j+2\pi i, 
\end{equation}
while in the second case the holonomies are the $u_i$, $v_j$ themselves.

The partition function we found in the previous section for the $A_1$ model on the Melvin cigar $MC_q$ had the form
\begin{equation}\label{partbc}
f(e^{u/\tau}; \mathcal{S})\: \psi(u;q)\equiv f(\hat U;\mathcal{S})\,\psi(u;q)
\end{equation}
where $\psi(u;q)$ is the periodic compact quantum dilogarithm of equation \eqref{compquantumdilog} corresponding to $G=\mathbb{C}^*$. From equation \eqref{partbc} we see that all non--perturbative effects and dependences on the boundary condition $\mathcal{S}$ are encoded in the action of an operator of the dual torus algebra $\widehat{\mathbb{T}}_{\widetilde\Sigma}$ whose presence cannot be detected in perturbation theory. It also shows that the most natural gauge group for the analytically continued $U(1)$ CS theory is $\mathbb{C},$ rather than $GL(1,\mathbb{C})\equiv \mathbb{C}^*$.

Now, gluing another manifold along the boundary of the Melvin cigar, amounts to choosing a specific boundary condition, and hence the partition function on the corresponding closed manifold is expected to be an operator of the general form
\begin{equation}
T\prod_{\alpha\in\text{BPS}} \Big(f(\hat U_i;\alpha)\,\psi(U_\alpha; q)\Big)^{\pm 1},
\end{equation}
where the operators $f(\hat U_i;\alpha)$ depend on the manifold we glue in at infinity. As we have already mentioned, the operators $f(\hat U_i;\alpha)$ 
are severely restricted by the requirement that the KS formula holds non--perturbatively. 
The simplest solution to this constraint is
\begin{equation}\label{canchoice}
f(\hat U_i,\alpha)= \psi\big(\hat U_\alpha; \hat q\big)^{-1}.
\end{equation}
This solution corresponds to gluing  to the Melvin cigar $MC_q$ some particular three--manifold $Y,$ determined implicitly by the above equation. As already noted, the $-1$ in the exponent of \eqref{canchoice} is just the inversion of orientation of $Y$ required to match the orientation of the boundaries in the gluing. Then $\overline{Y}$ will have the same (perturbative) partition function as the Melvin cigar with $q$ replaced by its dual $\hat q$. It is natural to interpret this as the result of gluing two Melvin cigars having dual $q$'s. 
Geometrically, this makes sense: In the coordinates $(z,\theta)$, the infinite end of the Melvin cigar $MC_q$ may be seen as
$\mathbb{C}\times \mathbb{R}$ subject to the two identifications
\begin{equation}
(\zeta,\theta)\sim (\zeta+1,\theta)\sim (\zeta+\tau, \theta+2\pi),
\end{equation} 
where $z=e^{2\pi i\zeta}$. Thus the end of $MC_q$ is a $S^1$ bundle over a torus of period $\tau$. Then we may glue the manifold $\overline{MC}_{\hat q}$ by identifying the base tori by a $S$ modular transformation, and the respective fibers over each point (with opposite orientations). 

This prediction should be compared with the partition functions of the 3d theory on a squashed $3$--sphere $S^3_b$ \cite{squashed}
\begin{equation}
b^2|z_1|^2+b^{-2}|z_2|^2=1,
\end{equation}
equipped with the metric induced by the embedding in the flat $\mathbb{C}^2$ space of coordinates $(z_1,z_2)$, as well as a twisting background connection for the $R$--symmetry
\begin{equation}
V= \frac{1}{2}\big( d\arg z_1-d\arg z_2\big)+\frac{1}{2\, f(|z_1|)}\big(b\,  d\arg z_1-b^{-1}\,d\arg z_2\big),
\end{equation}
where $f^2=b^{-2}|z_1|^2+b^2|z_2|^2$. The partition function of a chiral multiplet of twisted (real) mass $m$ and $R$--charge $q$ is given by \cite{squashed}
\begin{equation}\label{parS3b}
Z(m,q;{S_b^3})=s_b\big(iQ(1-q)/2-m\big),
\end{equation}
 where $s_b(x)$ is the double--sine function
 \begin{equation}\label{quadionncom}
 s_b(x)=s_b(-x)^{-1}= e^{-i\pi x^2/2}\;
 \frac{\prod\limits_{n=0}^\infty (1+e^{(2n+1)\pi i b^2+2\pi b x})}{\prod\limits_{n=0}^\infty (1+e^{(2n+1)\pi i b^{-2}+2\pi b^{-1} x})}	\equiv e^{-i\pi x^2/2}\, e_b(x),
 \end{equation}
 and $Q=b+b^{-1}$. In eqn.\eqref{quadionncom} the function $e_b(x)$ is the non--compact quantum dilogarithm \cite{fadnoncomp}. Moreover, a level $\hat k$ CS term for the $U(1)$ vector multiplet associated to the twisted mass $m$ will produce a factor \cite{DGG}
 \begin{equation}
 \exp(-i\pi \hat k\, m^2/2).
 \end{equation}
Then, under the identifications
\begin{equation}
u=2\pi b\, m+i\pi, \qquad \tau=b^2,
\end{equation}
our prediction for the non--perturbative partition function for the $3d$ theory corresponding to the tetrahedron generated by the flow of the $A_1$ theory precisely matches to the partition function of a $3d$ chiral multiplet with $R$--charge $1$ together with a Chern Simons of level $\hat k=1$ CS\footnote{\ The different sign of the CS term with respect to ref.\cite{DGG} and our conventions in section 5 is due to a different orientation of tetrahedron. This orientation is more standard in the cluster framework of section \ref{partADE}. The two choices are of course equivalent,  in terms of the squashed sphere $S_b^3$ this is the symmetry $b\leftrightarrow b^{-1}$. } \cite{DGG}. In particular, the partition function 
\begin{equation}\label{noncompactdi}
e_b(x)^{-1}=e^{-\pi i x^2/2}\, s_b(-x)=\frac{\psi(2\pi bx+i\pi; e^{2\pi i b^2})}{\psi(2\pi x b^{-1}+i\pi; e^{-2\pi i b^{-2}})}
\end{equation} 
differs from the compact dilogarithm $\psi(2\pi b x+i\pi)$ by the non--perturbative denominator which is the expected  contribution from the dual quantum torus algebra in the gluing of two dual Melvin cigars.

From the definition, equation \eqref{noncompactdi}, it is clear that $e_b(x)^{-1}$ satisfies the Ward identities of both Melvin cigars. In particular, from equation \eqref{opide2} one has the identity
\begin{equation}e_b(\mathsf{u})^{-1}\,e_b(\mathsf{v})^{-1}=e_b(\mathsf{v})^{-1}\,e_b(\mathsf{u}+\mathsf{v})^{-1}\,e_b(\mathsf{u})^{-1},
\end{equation}
where we have rescaled both canonical operators $\mathsf{u}$, $\mathsf{v}$ by a factor $(2\pi b)^{-1}$ so that
\begin{equation}
[\mathsf{u},\mathsf{v}]= \frac{i}{2\pi}.
\end{equation}
The functions $e_b(x)^{\pm 1}$ are their own Fourier transforms up to a Gaussian prefactor and a shift of the argument \cite{fadnoncomp}
\begin{equation}\label{PhiForurie}
\int_\mathbb{R} e_b(x)^{\pm 1}\, e^{2\pi i w x}\, dx= C^{\pm 1}\, e^{\mp \pi i w^2}\, e_b\big(\pm w\pm iQ/2\big)^{\pm 1},
\end{equation}
for some constant $C$.

We thus conjecture that the doubled quantum dilog is the relevant operator for
the partition function of the squashed sphere.
An explanation of how this arises is currently being
investigated \cite{progress}.
Thus, quite generally, the partition operator has the form \cite{CNV}
\begin{equation}
\mathcal{Z}_{2,M}= T \prod_{\alpha\in\text{BPS}} e_b(\gamma_\alpha\cdot \mathsf{y})^{-1},
\end{equation}
where the quantum operators $\mathsf{y}_i$ are related to the holonomies as $U_i=\exp(2\pi b\, \mathsf{y}_i)$.

We can check this conjecture, and ask how the partition functions behave when we consider the non--perturbative completion of the theory,  by replacing the compact quantum dilogarithm $\psi(\cdot)$ by
its non--compact counterpart $e_b(\cdot)^{-1}$.  It turns out that we indeed get the partition function of the corresponding
3d theory on $S^3_b$.
For example, if we consider the $A_2$ theory we obtain the partition function
\begin{equation}Z_{A_2}(v,u) =\langle v|\,e_b(\mathsf{v})^{-1}\,e_b(\mathsf{u}+\mathsf{v})^{-1}\,e_b(\mathsf{u})^{-1}\,
 |u\rangle
=\langle v|\,e_b(\mathsf{u})^{-1}\,e_b(\mathsf{v})^{-1}\,|u\rangle\end{equation}
which, using equation \eqref{PhiForurie}, leads to  two dual expressions for the $A_2$ partition function
\begin{equation}
\begin{split}
&\exp\!\left(-\pi i u v+\frac{\pi Q}{2}(u+v)\right)\: Z_{A_2}(v,u)=\\
&\qquad=C\,s_b(iQ/2+u+v)\, s_b(-u)\, s_b(-v)=\\
&\qquad=C^\prime \int dz\, e^{-2\pi i z \{u+(v+iQ/2)/2\}}\,
s_b(z-y/2+iQ/2)\, s_b(-z-y/2+iQ/2)
\end{split}
\end{equation}
for some constants $C, C^\prime$. Comparing with equation \eqref{parS3b}, the intermediate expression corresponds to the partition function of three chiral field $X$, $Y$, $Z$, with quantum numbers such that the composite operator $XYZ$ is invariant under the global $U(1)^2$ symmetry and has $R$--charge $2$, which means that it should be identified with the superpotential, $\mathcal{W}=XYZ$. Then, the last expression is the dual realization of this theory as $\mathcal{N}=2$ SQED with $N_f=1$ flavor. A more systematic analysis of the resulting partition functions is presented in section \ref{partADE}.

\subsection{$SL(N,\mathbb{C})$ on tetrahedron and generalizations}
So far, we have mainly concentrated on the $SL(2,\mathbb{C})$ theory
on tetrahedron.  However we can consider wrapping $N$
branes around tetrahedron.  In this context we will
have again to expand near a particular $SL(N,\mathbb{C})$ flat connection
$\mathcal{A}= A+i \phi_N$.
In the IR we get a $3$--fold $M_N$ which is an $N$--fold cover of the tetrahedron.
The simplest and most canonical such manifold $M_N$ will be the 5-brane geometry
corresponding to the movie of the $A_{N-1}$ curve
\begin{equation}
\det\!\big[y-\phi_N\big]\equiv y^N+\text{lower order terms}=x^2.
\end{equation}
Interchanging the role of $x$ and $y$, we may see this as the $SL(2,\mathbb{C})$ partition function on a set of $N-1$ glued tetrahedrons. In fact, $N-1$ is the minimal possible number of  tetrahedra; we may increase it by Pachner moves up to the maximal number $N(N-1)/2$. As already discussed, these different ideal triangulations correspond to distinct BPS chambers of the parent 4d theory.

In particular, the $SL(3,\mathbb{C})$
partition function on the tetrahedron should have a branch which
corresponds to the $SL(2,\mathbb{C})$ partition function on doubly glued
tetrahedrons.  Here, we also expect to have potentially other
branches which are more complicated than that
of the $SL(2,\mathbb{C})$ theory.

This construction may be further generalized. We can consider not just the movie of the $A_{N-1}$ curve, but also the movie of any other $ADE$ curve, see table \ref{tableADE}.  The parent 4d theories are the $ADE$ Argyres--Douglas models, whose BPS quivers may be chosen in the form of an orientation of the corresponding $ADE$ Dynkin diagram, all orientations being mutation equivalent.

\begin{table}
\begin{equation*}
\begin{array}{|c|c|c|c|c|}\hline
\text{movie of} & \text{curve }\tilde\Sigma & SL(2,\mathbb{C})\ \text{CS}  & \min\#\text{(particles)} & \max\#\text{(particles)}\\\hline
D_n &x^2y+y^{n-1}+\cdots=0 & \text{yes} & n & n(n-1)\\\hline
E_6 & x^3+y^4+\cdots=0 & \text{no} & $6$ & $36$\\\hline
E_7 & x^3+xy^3+\cdots=0 & \text{no} & $7$ & $63$\\\hline
E_8 & x^3+y^5+\cdots =0 & \text{no} & $8$ & $120$\\\hline
\end{array}
\end{equation*}
\caption{\label{tableADE} Movies of $DE$ curves. Elipsis stands for lower order terms.}
\end{table}

In the table we have specified if the R-flow may be seen as the IR limit of an $SL(2,\mathbb{C})$ Chern--Simons theory (which requires the curve $\widetilde\Sigma$ to be a double cover of some $\Sigma$) as well as the minimal and maximal number of BPS particles equal, respectively, to the rank and the number of positive roots of the Lie algebra with the same name.

The partition functions of the 3d theories arising from the movies of $ADE$ curves, in the two dual representations corresponding to a minimal and a maximal 4d BPS chamber, are discussed in section \ref{partADE} using the quiver mutation technology introduced in section \ref{sec:Quiver}.

\subsection{Interpretation of the R-twisted partition function}
Even though our emphasis in this paper has been for theories which admit mass deformations
there is one special class of 3d theories which arise from R-twisting starting from massless ${\cal N}=2$ superconformal theories in 4d.  This is the idea explored in \cite{CNV}.   One compactifies
the 4d theory on 
\begin{equation}
S^1\times MC_q,
\end{equation}
where as we go around $S^1$ we twist by the symmetry 
\begin{equation}
g=(-1)^F \exp(2\pi iR),
\end{equation}
where $R$ is the R-symmetry action of the superconformal theory.  This leads to a theory
in three dimensions with ${\cal N}=1$ supersymmetry.
Then, the discussion above suggests that the non-perurbative completion of the theory results in the partition function of the superconformal theory on $S^1_R\times S^3_b$ where
$S^1$ is R-twisted, is equal to the trace of the 4d quantum monodromy  computed in \cite{CNV}.  For example,
it was shown that if we start with the $A_{2n}$ Argyres-Douglas conformal theories
given by $y^2=x^{2n+1}$ then
\begin{equation}Z^{A_{2n}}[S^1_R\times S^3_b]=\sum C^{\alpha\beta}\, \chi_\alpha(q)\, \chi_\beta(\hat q),\label{charact}\end{equation}
where $\chi_\alpha(q)$ are characters of $(2,2n+3)$ minimal conformal models in 2d. The factorized form of the \textsc{rhs} of eqn.\eqref{charact} reflects the fact that our Hilbert space carries a representation of the doubled quantum torus algebra $\mathbb{T}_{\tilde\Sigma}\times \widehat{\mathbb{T}}_{\tilde\Sigma}$. 
Such a result is in need of a deeper explanation!  The traces of  powers of the 4d quantum monodromy
are also closely related to the computation of the superconformal index in 4d \cite{superconfindex}.
 
We can also consider modding out the above geometry, in a way
consistent with supersymmetry as was done in \cite{CNV}.  For example we can mod out by a
$\mathbb{Z}_2$ which acts by a rotation by $\pi$  in the $S^3_b$ at the same time
as modding out the $S^1$ by $\exp(i\pi R)$. Now, the partition functions on
$S^1_R\times S^3_b/\mathbb{Z}_2$  are made of characters of
the 2d CFT corrsponding to the coset constructions
\begin{equation}
SU(n+1)_2/U(1)^{n}.
\end{equation}
Similar results are obtained when we consider modding out by suitable $\mathbb{Z}_k$ subgroups.
 It would be interesting to better understand the emergence
of these 2d CFT characters in these computations.

\section{3d Partition Functions of Mirror $ADE$ Argyres-Douglas Flows}
\label{partADE}

In this section, we compute the 3d partition functions of the theories arising from the movies of $ADE$ curves, in the two dual representations associated to a minimal and maximal BPS chambers of the corresponding 4d Argyres--Douglas models.
The $A$ case has been discussed in detail before in this paper.  The $D$ case is again related
to two M5-branes, except that now we have an additional cusp--line in the 3d geometry.  The $E$ case does not
correspond to an M5-brane picture and in particular will not have a description as an $SL(2)$ CS partition function
of some three-manifold.  Nevertheless, the considerations of this paper still apply and provide mirror pairs whose 4d
parents are related by wall-crossing.

We make use of the quiver mutation technology of section \ref{sec:Quiver}.
We denote by $\mu_i$ the effect of the mutation in eqn.\eqref{eqmutation}. More precisely, to each node of the quiver $Q$ we attach a canonical quantum operator, $\mathsf{y}_j$, satisfying the commutation relations
\begin{equation}\label{CCR}
\big[\mathsf{y}_i,\mathsf{y}_j\big]= \frac{i}{2\pi}\, B_{ij},
\end{equation} 
where $B_{ij}\equiv \gamma_i\circ \gamma_j$ is the exchange matrix of the quiver $Q$ counting, with sign, the arrows going from node $i$ to node $j$.  The mutation $\mu_k$ acts on the quantum operators as in equation (4.4) ($j\neq k$)
\begin{equation}
\begin{aligned}
&\mathsf{y}_k\ \longrightarrow\ \mu_k(\mathsf{y}_k)=-\mathsf{y}_k\\
&\mathsf{y}_j\ \longrightarrow\ \mu_k(\mathsf{y}_j)=\begin{cases}
\mathsf{y}_j+B_{kj}\mathsf{y}_k & \text{if }B_{kj}>0\\
\mathsf{y}_j & \text{if }B_{kj}\leq 0.
\end{cases}
\end{aligned}
\end{equation}
Notice that this transformation preserves the canonical commutation relation \eqref{CCR} in the sense that the $\mu_k(\mathsf{y}_j)$ satisfies the same relations but with respect to the exchange matrix of the mutated quiver $\mu_k(Q)$.

We shall denote by $\cq_k$ the elementary quantum mutation, which is the composition of the mutation $\mu_k$ with the adjoint action of the quantum operator $e_b(\mathsf{y}_k)^{-1}$. $\cq_k$ is an involution, $\cq_k^2=1$. For a detailed discussion of the properties of this operation and its relation to cluster algebras we refer to the literature 
\cite{CV11, CNV, MR1887642,MR2004457,fominIV,MR2383126,MR2132323,cluster-intro,qd-cluster,clqd2,qd-pentagon,kel}.

For the $ADE$  movies we may choose the quiver $Q$ to be an orientation of the corresponding Dynkin graph $\mathfrak{g}$. For convenience, we consider the alternating orientation of
$\mathfrak{g}$: each node is either a sink or a source. For $A_n$, this choice corresponds to the braids analyzed in sect.\,\ref{sec:alternatingA2}. We adopt the same convention: sources are labeled by undotted integers $a,b,\dots =1,2,\dots$, while sinks by dotted ones $\dot a, \dot b,\dots =\dot 1, \dot 2, \dots$. See figure \ref{altquivers} for the $A_{2n}$ and $E_6$ examples. With this notation, the exchange matrix $B$ has the properties
\begin{equation}
B_{ab}=B_{\dot a \dot b}=0\qquad B_{a\dot b}=-B_{\dot a b} \geq 0,
\end{equation}
and, in particular, the quantum operators $\mathsf{y}_a$ (resp.\! $\mathsf{y}_{\dot a}$) commute between themselves. Correspondingly, the quantum mutations $\cq_a$ (resp.\! $\cq_{\dot a}$) also commute. The set of operators $\{\mathsf{y}_a,\mathsf{y}_{\dot b}\}$ will be collectively denoted as $\{\mathsf{y}_j\}$, where $j=1,2,\dots, \mathrm{rank}\,\mathfrak{g}$.

\begin{figure}
\begin{align*}
&A_{2n}\colon
&&\begin{gathered}
\xymatrix{1\ar[r] & \dot 1 & 2\ar[l]\ar[r] & \dot 2 & \cdots\cdots\ar[l] \ar[r] & \dot n}
\end{gathered}\\
&E_6\colon &&
 \begin{gathered}
  \xymatrix{& & \dot 3 & &\\
1\ar[r] &\dot 1 & 2\ar[l]\ar[r]\ar[u] & \dot 2 & 3\ar[l]}
 \end{gathered}
\end{align*}
\caption{\label{altquivers} Two examples of alternating Dynkin quivers. Sources (undotted nodes) and sinks (dotted ones) alternate.}
\end{figure}

\subsection{Zamolodchikov Identities}\label{zamodidentities}

The 4d half--monodromy of an $ADE$ Argyres--Douglas system is\footnote{\ Note that we don't need to specify the order of the $\cq_a$'s (resp.\! $\cq_{\dot b}$'s) since they commute between themselves.} \cite{CNV}
\begin{equation}
\mathbb{K}(q)= \prod_a \cq_a\,\prod_{\dot b}\cq_{\dot b}.
\end{equation}
Roughly speaking, the eigenvalues of $\mathbb{K}(q)^2$ correspond to $\{\exp(2\pi i q_\alpha)\}$, where $\{q_\alpha\}$ are the $R$--charges of the chiral primaries at the UV superconformal point. For the  critical $ADE$ Argyres--Douglas models one has \cite{CNV}
\begin{equation}
q_\alpha= \frac{\gcd(h+2,2)}{h+2}\:n_\alpha, \qquad n_\alpha\in\mathbb{N},,
\end{equation}
where $h$ is the Coxeter number of the Lie algebra $\mathfrak{g}$. Hence we must have
\begin{equation}
(\prod_a \cq_a\,\prod_{\dot b}\cq_{\dot b})^{h+2}=\text{identity}. 
\end{equation}
This equation maps to the Zamolodchikov periodicity for $ADE$ $Y$--systems \cite{Zamo,tadeo,keller-periodicity}. Since the $\cq$'s are involutions we may rewrite this in the form
\begin{equation}\label{zamol}
(\prod_a \cq_a\,\prod_{\dot b}\cq_{\dot b})^2= (\prod_{\dot b}\cq_{\dot b}\,\prod_a \cq_a)^h 
\end{equation}
which is interpreted as the Kontsevich--Soilbelman WCF relating a minimal BPS chamber with $\mathrm{rank}\,\mathfrak{g}$ hypermultiplets in the \textsc{lhs} to a maximal chamber in the \textsc{rhs} consisting of $\mathrm{rank}\,\mathfrak{g}\cdot h/2$ hypermultiplets in one--to--one correspondence with the positive roots of $\mathfrak{g}$.

Equation \eqref{zamol} is an equality between two expressions for the full 4d monodromy $\mathbb{K}(q)^2$. For the application to the movies we are interested in the half--monodromy, so we have to take a `square--root' of the Zamolodchikov identity.

Consider the following quiver mutation
\begin{equation}\label{mutomega}
\omega=\begin{cases}\prod_{a} \mu_a\Big(\prod_{\dot b}\mu_{\dot b}\prod_{c}\mu_c\Big)^{\!(h-1)/2} & h\ \text{odd}\\
\Big(\prod_{\dot b}\mu_{\dot b}\prod_{c}\mu_c\Big)^{\!h/2} &h\ \text{even.}
\end{cases}\end{equation} 
It acts on the quantum operators $\mathsf{y}_j$ as
\begin{equation}
\omega\cdot \mathsf{y}_j = w_{jh}\,\mathsf{y}_h,
\end{equation}
where the matrix $w_{jh}$ is the element $\mathrm{Weyl}(\mathfrak{g})$
\begin{align}\label{weee1}
&\text{for } \mathfrak{g}=A_{2n} &&w=\prod_{a} s_a\Big(\prod_{\dot b}s_{\dot b}\prod_{c}s_c\Big)^{\!n}\equiv\Big(\prod_{\dot b}s_{\dot b}\,\prod_{a}s_a\Big)^{\!n}\prod_{\dot e} s_{\dot e},\\
&\text{for } \mathfrak{g}\neq A_{2n}
&& w=\Big(\prod_{\dot b} s_{\dot b} \, \prod_a s_a\Big)^{h/2},
\label{weee2}\end{align} 
which has the property that $w^2=1$. Thus, the mutation $\omega$ is a square root of the identity \eqref{zamol}. In fact, $w=-\chi$, where $\chi$ is the permutation defined in section \ref{sec:Quiver} which specifies how the boundary conditions in distinct chambers are related. For $A_1$, $D_{2n}$, $E_7$ and $E_8$ $\chi$ is the identity. For $A_{2n+1}$, $D_{2n+1}$ and $E_6$ gives back the same quiver up to a $\mathbb{Z}_2$ involution of the quiver. For $A_{2n}$ the $\mathbb{Z}_2$ involution of the Dynkin graph inverts all the arrows, and hence the mutated quiver is the opposite of the initial one, with dotted and undotted nodes interchanged (cfr.\! eqn.\eqref{chia2n}). Thus, for $A_{2n}$ $w$ invertes the commutation relations \eqref{CCR}.
Let $\mathscr{T}$ be the isometry of Hilbert space $\ch_{Q}$ corresponding to $-w=\chi$ ($\mathscr{T}$ is an involution which is unitary for $\mathfrak{g}\neq A_{2n}$ and anti--unitary for $\mathfrak{g}=A_{2n}$); for alternating Dynkin quivers the Zamolodchikov identity for \emph{half}--monodromy $\mathbb{K}(q)$ takes the form\footnote{\ Here $I$ is the inversion unitary operator, acting as $I\,\mathsf{y}_j\,I=-\mathsf{y}_j$. See \cite{CNV} for details.}
\begin{align}\label{WCFA2nK}
&\mathfrak{g}=A_{2n}  &&I\,\mathbb{K}(q)\equiv I\,\prod_{a}\cq_{a} \prod_{\dot b}\cq_{\dot b}= \mathscr{T}\, 
\prod_{a} \cq_a\Big(\prod_{\dot b}\cq_{\dot b}\prod_{c}\cq_c\Big)^{\!n},\\
&\mathfrak{g}\neq A_{2n} &&I\,\mathbb{K}(q)\equiv I\,\prod_{a}\cq_{a} \prod_{\dot b}\cq_{\dot b}= \mathscr{T}\, 
\Big(\prod_{\dot b}\cq_{\dot b}\prod_a\cq_a\Big)^{h/2}.\label{WCFGK}
\end{align}
Both sides are now operators $\ch_{Q}\rightarrow \ch_{Q}$, where $\ch_{Q}$ is the canonical $L^2$ Hilbert space defined by the commutation relations \eqref{CCR}, and in fact a product of quantum dilog operators \cite{CNV}.

For $A_{2n+1}$, $D_{2n+1}$, and $E_7$ the exchange matrix B has one zero eigenvector; for $D_{2n}$ there are two linearly independent zero eigenvectors. These zero eigenvectors describe flavor symmetries of the associated 4d theroy.  It is easy to check that the non--zero entries of these eigenvectors are either all dotted or all undotted. By reversing all the arrows, if necessary, we assume they are all dotted, and write the zero--eigenvectors as $v_{\dot a}^{(\alpha)}$, where $\alpha$ takes the value $1$ for $A_{2n+1}$, $D_{2n+1}$, $E_7$, and the values $1,2$ for $D_{2n}$. It follows from equation \eqref{CCR} that the quantum operators $\zeta^\alpha\equiv v_{\dot a}^{(\alpha)}\mathsf{y}_{\dot a}$ commute with everything.

We are interested in taking the matrix elements of the identities \eqref{WCFA2nK} and \eqref{WCFGK} between generic states in $\ch_{Q}$.
In order to do this, we introduce an explicit Schroedinger representation
\begin{align}\label{sch1}
 & \mathsf{y}_a=x_a,\\
& \mathsf{y}_{\dot a}=-\frac{1}{2\pi i}\,B_{\dot a b}\,\frac{\partial}{\partial x_b}+v_{\dot a}^{(\alpha)}\,c_\alpha\equiv - B_{\dot a b}\, p_b+v_{\dot a}^{(\alpha)}\,c_\alpha,\label{sch2}
\end{align}
where the $c$--numbers $c_\alpha$ represent the central elements $\zeta^\alpha$ of the quantum operator algebra.

\subsection{3d $ADE$ Partition Function from the Minimal Chamber}

From the discussion in section \ref{M5branesandCS}, we know that to get the partition function of the 3d movie of an $ADE$ curve we must compute the matrix elements of the 4d half--monodromy operator $\mathbb{K}(q)$ between suitable
$\langle \text{out}|$, $|\text{in}\rangle$ states. Generalizing the choice adopted in sect.\,\ref{M5branesandCS} for the $A_2$ movie, we take the $|\text{in}\rangle$ state to be an eigenstate of `position', that is of the quantum operators associated to the undotted nodes
\begin{equation}
|\text{in}\rangle\equiv |x\rangle,\qquad \text{with }\mathsf{y}_a\,|x\rangle= x_a\,|x\rangle,
\end{equation}
while the $\langle\text{out}|$ states is taken to be an eigenstate of `momentum', \textit{i.e.}\! of the quantum operators at the dotted nodes
\begin{equation}
\langle\text{out}|\equiv \langle p|, \qquad\text{with } \langle p|\, \mathsf{y}_{\dot a}= (-B_{\dot ab}\,p_b+ v_{\dot a}^{(\alpha)} c_\alpha)\: \langle p|.
\end{equation}
For the $AD$ cases this choice may be directly related to a specific pairing of strands. In fact each node in the quiver is associated with an arc of the ideal triangulation of the in/out asymptotic surface. Each arc is associated to a pair of strands. In the in (resp.\! out) surface we connect the pairs associated to the undotted (resp.\! dotted) arcs.

The expression for $\mathbb{K}(q)$ in the minimal chamber is given by  the \textsc{lhs} of equation \eqref{WCFA2nK}.  Then, $\langle -p|\, \mathbb{K}(q)\,|x\rangle$ is simply,
\begin{equation}\label{LHS}
 \langle -p|\,\prod_{a}\cq_a \prod_{\dot b}\cq_{\dot b}\, |x\rangle=
 \langle p|\,\prod_{a} e_b(\mathsf{y}_a)^{-1} \prod_{\dot b}e_b(\mathsf{y}_{\dot b})^{-1}\: |x\rangle.\end{equation}

 Let $\phi_-(y)$  be the Fourier transform of $e_b(x)^{-1}$ \cite{fadnoncomp}. One has
 \begin{equation}
 \begin{aligned}
&  \langle- p|\,\mathbb{K}(q)\: |x\rangle=\\
  &= \int  \langle p|\, e^{-2\pi i z_a \mathsf{y}_a}\, e^{-2\pi i w_{\dot b} \mathsf{y}_{\dot b}}\, |x\rangle\;\prod_c dz_c\,\phi_-(z_c) \prod_{\dot d} dw_{\dot d}\, \phi_-(w_{\dot d})\\
 &= e^{-2\pi i p_a x_a}\int e^{-2\pi i B_{a\dot b} z_a w_{\dot b}-2\pi i z_ax_a-2\pi i (B_{a\dot b}p_a+v_{\dot b}\cdot c)w_{\dot b}}\;\prod_c dz_c\,\phi_-(z_c) \prod_{\dot d} dw_{\dot d}\, \phi_-(w_{\dot d})\\
& = e^{-2\pi i p_a x_a}\int e^{-2\pi iz_ax_a}\: \prod_{\dot b}\,e_b\big(B_{a\dot b}(p_a+z_a)+ v_{\dot b}\cdot c\big)^{-1} \;\prod_c dz_c\,\phi_-(z_c).
\end{aligned}
 \end{equation}
Using the identities \cite{fadnoncomp} (here $C$, $C^\prime$ are irrelevant overall constants which shall be omitted in all formulae)
\begin{gather}
\phi_-(w)= C\, e^{i\pi w^2}\: e_b(-w-iQ/2)^{-1}\\
s_b(x)= e^{i\pi x^2/2} \:e_b(-x)^{-1},
\end{gather}
one gets the final expression for the 3d partition function as defined by the minimal chamber
\begin{multline}\label{minchamber}
e^{-2\pi i x_a p_a}\int \left(\prod_d dz_d\right)\prod_a s_b\!\left(\frac{iQ}{2}+z_a\right)\: \prod_{\dot b} s_b\big(-B_{a\dot b}(p_a+z_a)-v_{\dot b}\cdot c)\big) \times\\
\times\: \exp\Big\{-2\pi i z_a x_a - \frac{\pi i}{2}\big(B_{a\dot b}(p_a+z_a)+v_{\dot b}\cdot c\big)\big(B_{c\dot b}(p_c+z_c)+v_{\dot b}\cdot c\big)+\frac{\pi i}{2} (z_a-iQ/2)^2\Big\}
\end{multline}

Equation \eqref{minchamber} has the interpretation of a 3d $\mathcal{N}=2$ theory with two kinds of chiral multiplets, called \emph{source--type} $X_a$ and \emph{sink--type} $Y_{\dot b}$, in one--to--one correspondence with the sinks and the source nodes of $Q$, respectively. They are coupled to a set of $U(1)$ vector multiplets, in one--to--one correspondence with the source nodes, whose scalars are identified with the integration variables $z_a$. The chiral multiplets are also charged with respect to a number of $U(1)$ flavor symmetries --- in one--to--one correspondence with the sink nodes  --- with associated real masses $p_a$, $c_\alpha$.
The quantum numbers of the chiral multiplets are as in the table

\vglue 12pt
\begin{center}
\begin{tabular}{|c|c|c|}\hline
& $X_a$ & $Y_{\dot b}$ \\\hline
$d$--th gauge $U(1)$ & $\delta_{ad}$ & $-B_{d\dot b}$\\\hline
$p_c$ flavor & $0$ & $-B_{c\dot b}$\\\hline
$c_\alpha$ flavor & $0$ & $-v_{\dot b}^{(\alpha)}$\\\hline
$R$--charge & $0$ & $1$\\\hline
\end{tabular}\vskip9pt

\textsc{Table:} quantum numbers of the minimal chamber\\ chiral multiplets for the $\mathfrak{g}=ADE$ movie.
\end{center}
\vglue 12pt
 
The diagonal entries of the Chern--Simons level--matrix are
\begin{equation}
\hat{k}_{aa}= \sum_{\dot b} (B_{a\dot b})^2-1 = \Big(\text{valency of source }a -1\Big)
\end{equation} 
which is also one--half the difference of the number of negatively and positively charged chiral fields with respect to the $a$--th $U(1)$ gauge symmetry.

For instance, in the $A_8$ case
\begin{equation}
\begin{gathered}
\xymatrix{1\ar[r] & \dot 1 & 2\ar[r]\ar[l] & \dot 2 & 3\ar[r]\ar[l] & \dot 3 & 4\ar[r]\ar[l] & \dot 4}
\end{gathered}
\end{equation}
the eight chiral multiplets carry four gauge charges $e_1,e_2,e_3,e_4$, three flavor charges $f_1,f_2,f_3,f_4$, and no $c$--flavor charge (which
corresponds to shift symmetry of dual photons). The quantum numbers are

\begin{center}
\begin{tabular}{|c|c|c|c|c|c|c|c|c|}\hline
chiral mult. & $e_1$ & $e_2$ & $e_3$ & $e_4$ & $f_1$ & $f_2$ & $f_3$ & $f_4$ \\\hline
$X_1$ & $1$ & $0$ & $0$ & $0$ & $0$ & $0$ & $0$ & $0$ \\\hline
$X_2$ & $0$ & $1$ & $0$ & $0$ & $0$ & $0$ & $0$ & $0$ \\\hline
$X_3$ & $0$ & $0$ & $1$ & $0$ & $0$ & $0$ & $0$ & $0$ \\\hline
$X_4$ & $0$ & $0$ & $0$ & $1$ & $0$ & $0$ & $0$ & $0$ \\\hline
$Y_{\dot 1}$ & $-1$ & $-1$ & $0$  & $0$& $-1$ & $-1$& $0$ & $0$\\\hline
$Y_{\dot 2}$ & $0$ & $-1$ & $-1$ & $0$ & $0$ & $-1$ & $-1$ & $0$ \\\hline
$Y_{\dot 3}$ & $0$ & $0$ & $-1$ & $-1$ & $0$ & $0$ & $-1$ & $-1$ \\\hline
$Y_{\dot 4}$ & $0$ & $0$ & $0$ & $-1$ & $0$ & $0$ & $0$ & $-1$ \\\hline
\end{tabular}
\end{center}

For the $E_6$ movie the quantum numbers are (labeling of nodes as in figure \ref{altquivers})
\begin{center}
\begin{tabular}{|c|c|c|c|c|c|c|}\hline
chiral mult. & $e_1$ & $e_2$ & $e_3$ & $f_1$ & $f_2$ & $f_3$\\\hline
$X_1$ & $1$ & $0$ & $0$ & $0$ & $0$ & $0$\\\hline
$X_2$ & $0$ & $1$ & $0$ & $0$ & $0$ & $0$\\\hline
$X_3$ & $0$ & $0$ & $1$ & $0$ & $0$ & $0$\\\hline
$Y_{\dot 1}$ & $-1$ & $-1$ & $0$& $-1$ & $-1$  & $0$\\\hline
$Y_{\dot 2}$ & $0$ & $-1$ & $-1$ & $0$ & $-1$ & $-1$\\\hline
$Y_{\dot 3}$ & $0$ & $-1$ & $0$ & $0$ & $-1$ & $0$\\\hline
\end{tabular}
\end{center}
 
There are no gauge/flavor invariant monomials in these chiral fields, consistent with the fact that, for $A_{2n}$, no superpotential is expected to be present in this chamber, see figure \ref{fig:smallclosed} on page \pageref{fig:smallclosed}.  Thus, we have verified the expected field content, superpotentials, and CS terms derived geometrically for this chamber in section 5.4.\footnote{In comparing to the conventions set there, some gauge fields must be redefined as $A\rightarrow-A$.}

\subsection{3d Partition Function from the Maximal Chamber}

In section \ref{zamodidentities}, we saw that the combinatorics of the quiver mutation for the $\mathfrak{g}=ADE$ model is controlled by the  Weyl group $\mathrm{Weyl}(\mathfrak{g})$. We begin by recalling the properties of $\mathrm{Weyl}(\mathfrak{g})$ that we need. 
We denote by $w_-$ (resp.\! $w_+$) the element of $\mathrm{Weyl}(\mathfrak{g})$ corresponding to the product of the reflections with respect to the simple roots associated with the undotted (resp.\! dotted) nodes of the Dynkin quiver. One has
$w_-^2=w_+^2=1$, and the Coxeter element is $c=w_+w_-$.
The following properties hold\footnote{\ These relations may also be understood as a consequence of the 2d/4d correspondence \cite{CNV}. Indeed let $D_{ij}$ the diagonal matrix which is $+1$ on the undotted entries and $-1$ in the dotted one. $S=Dw_+$ is the 2d Stokes matrix and hence $B=S-S^t$, $c=-(S^{-1})^tS$ \cite{CV92}.}
\begin{align}\label{cul1}
 (w_-)_{a j}&= -\delta_{a j} & (w_-)_{j\dot a}&= \delta_{j\dot a}\\
 (w_+)_{\dot a j}&= -\delta_{\dot a j} &(w_+)_{j a}&= \delta_{j a}
\label{cul2}\\
(w_+)_{a\dot b}&= B_{a\dot b} & (w_-)_{\dot a b}&=-B_{\dot a b},
\end{align}
which imply for $c^{-1}=w_-w_+$
\begin{align}\label{ide1}
(c^{-1})_{ab}&=-\delta_{ab} &
 (c^{-1})_{\dot a b}&= -B_{\dot a b}\\
 (c^{-1})_{a\dot b}&= -B_{a\dot b} &
 (c^{-1})_{\dot a\dot b}&= - B_{\dot a c}B_{c\dot b}-\delta_{\dot a\dot b}.\label{iden}
\end{align}
Moreover, we have the property
\begin{align}
&h\ \text{even:}  &&(c^k)_{a,j}\geq 0,\ \text{for } 0\leq k\leq \frac{h}{2}-1;  & &(c^k)_{\dot a,j}\leq 0,\ \text{for } 1\leq k\leq \frac{h}{2};\\
&h\ \text{odd:}  &&(c^k)_{a,j}\geq 0,\ \text{for } 0\leq k\leq \frac{h-1}{2};  & &(c^k)_{\dot a,j}\leq 0,\ \text{for } 1\leq k\leq \frac{h+1}{2};
\end{align}
In particular, for $A_{2n}$, the set $\{(c^n)_{aj}\,\mathsf{y}_j\}$ is equal to the set $\mathsf{y}_{\dot b}$ (up to permutation).

In the maximal chamber the half--monodromy $\mathbb{K}(q)$ is given by the \textsc{rhs} of the Zamolodchikov identities \eqref{WCFA2nK}-\eqref{WCFGK}.
Computing the mutations in eqn.\eqref{mutomega} with the help of eqns.\eqref{weee1}-\eqref{weee2}, we get
\begin{multline}
I\,\mathbb{K}(q)= \Psi \prod_{\dot b}e_b(-c^{[h/2]}_{\dot b j}\mathsf{y}_j)^{-1}\prod_a e_b(c^{[h/2]-1}_{a j}\mathsf{y}_j)^{-1}\cdots\times\\
\times\cdots \prod_{\dot b}e_b(-c^2_{\dot b j}\mathsf{y}_j)^{-1}\prod_a e_b(c_{a j}\mathsf{y}_j)^{-1} \prod_{\dot b} e_b(-c_{\dot b j}\mathsf{y}_j)^{-1}\prod_a e_b(\mathsf{y}_a)^{-1}
\end{multline}
where
\begin{equation}
\Psi=\begin{cases} \prod_{\dot b}e_b(\mathsf{y}_{\dot b})^{-1} & \mathfrak{g}=A_{2n}\\
1 & \text{otherwise.}
\end{cases}
\end{equation}

The Coxeter element $c$ has the two crucial properties. First of all $c$ is an invariance\footnote{\ This is obvious from the 2d/4d correspondence \cite{CNV}. Let $S$ be the $2d$ Stokes matrix \cite{CV92}. Since $c$ is \emph{minus} the $2d$ quantum monodromy,
\begin{align*}
&B= S^t-S, && c=- (S^{-1})^tS.
\end{align*}
Then
\begin{equation*}
c^t Bc= S^t S^{-1}(S^t-S)(S^t)^{-1}S=S^t-S=B.
\end{equation*}}  of the skew--symmetric form $B$
\begin{equation}\label{sumpl}
c^t B c = B,
\end{equation}
secondly, a zero--eigenvector $v_{\dot a}^{(\alpha)}$ of $B$ is automatically an eigenvector of $c$ associated to the eigenvalue $-1$.
For $\mathfrak{g}=A_{2n}$, $E_6$, and $E_8$ the exchange matrix
$B_{ij}$ is a non--degenerate symplectic pairing,  and from eqns.\eqref{CCR}\eqref{sumpl} it follows that there exists a unitary operator $\mathscr{C}$ such that 
\begin{equation}
 c_{j\ell}\,\mathsf{y}_\ell = \mathscr{C}\,\mathsf{y}_j\, \mathscr{C}^{-1},
\end{equation}
whereas for $\mathfrak{g}=A_{2n+1}$, $D_r$, and $E_7$, there is no such unitary operator\footnote{\ $\mathscr{C}$ would exist if we specialize the central elements $c_\alpha$ to zero.} (since the adjoint action is trivial on the central subalgebra  generated by $v_{\dot a}^{(\alpha)}\mathsf{y}_{\dot a}$, while $c$ acts as multiplication by $-1$ on them). However, there still exists a unitary operator $\mathscr{S}$ such that
\begin{equation}
- c_{j\ell}\,\mathsf{y}_\ell = \mathscr{S}\,\mathsf{y}_j\, \mathscr{S}^{-1}.
\end{equation}
Clearly, for $\mathfrak{g}=A_{2n}$, $E_6$, $E_8$ one has 
\begin{equation}
e_b(\pm c^k_{j\ell}\mathsf{y}_\ell)= \mathscr{C}^k\,e_b(\pm \mathsf{y}_j)\, \mathscr{C}^{-k},
\end{equation}
and $\mathscr{C}$ may be used to rewrite $\mathbb{K}(q)$ in the convenient form
\begin{equation}\label{simpleexP}
I\,\mathbb{K}(q)= \Psi\: \mathscr{C}^{[h/2]}\,\Big(\prod_{\dot b} e_b(-\mathsf{y}_{\dot b})^{-1}\, \mathscr{C}^{-1}\, \prod_{a} e_b(\mathsf{y}_a)^{-1}\Big)^{[h/2]}
\end{equation}
For $\mathfrak{g}=A_{2n+1}, D_r$, and $E_7$
the operator $I\,\mathbb{K}(q)$ has a slightly more complicated expression
\begin{equation}
 \Xi\; \mathscr{S}^{h/2}\, \Big(\prod_{\dot b} e_b(-\mathsf{y}_{\dot b})^{-1}\, \mathscr{S}^{-1} \prod_{a} e_b(-\mathsf{y}_a)^{-1}\prod_{\dot b}e_b(\mathsf{y}_{\dot b})^{-1}\, \mathscr{S}^{-1} \prod_{a} e_b(\mathsf{y}_a)^{-1}\Big)^{[h/4]}
\end{equation}
where
\begin{equation}
\Xi= \begin{cases} 1 & \text{for } A_{4k+3}\, D_{2s+1}\\
\prod_{\dot b}e_b(\mathsf{y}_{\dot b})^{-1}\,\mathscr{S}^{-1} \prod_a e_b(-\mathsf{y}_a)^{-1} & \text{for } A_{4k+1},\ D_{2s},\ E_7.
\end{cases}
\end{equation}

For simplicity, for the rest of this section we limit ourselves to the non--degenerate case, $\mathfrak{g}=A_{2n}, E_6, E_8$, so that the simpler expression \eqref{simpleexP} applies. The extension to the general case is straightforward.

%%%%%%%%

\subsubsection{The Basic Kernel}

In the non--degenerate case, equation \eqref{simpleexP} reduces the computation of the 3d partition function $\langle -p|\,\mathbb{K}(q)\,|x\rangle$ to the  determination of  the basic integral kernel
\begin{multline}
\Big\langle p\, \Big|\,\prod_{\dot b} e_b(-\mathsf{y}_{\dot b})^{-1}\,\mathscr{C}^{-1}\, \prod_{a} e_b(\mathsf{y}_a)^{-1}\,\Big|\,x\Big\rangle=\\
=\Big\langle p\, \Big|\,\mathscr{C}^{-1}\, \Big|\,x\Big\rangle\;
\prod_{\dot a} e_b\big(B_{\dot ab}\,p_b\big)^{-1}\prod_a e_b\big(x_a\big)^{-1},\phantom{aaaaaaaaa}
\end{multline}
where we used our Schroedinger representation \eqref{sch1}-\eqref{sch2}.

The extra factor for the $A_{2n}$ case, $\Psi$, is diagonal in the momentum representation 
\begin{equation}
\langle p|\, \Psi\,| q\rangle =  \delta(p-q)\: \prod_{\dot b} e_b(-B_{\dot b c}\,p_c)^{-1}.
\end{equation}

To conclude the computation of the partition function, we need only the kenrels of the unitary operators $\mathscr{C}^{-1}$ and $\mathscr{C}^{[h/2]}$.

By definition, the kernels satisfy the equations
\begin{align}\label{firsteee}
& -\left((c^k)_{a,b}\,\frac{1}{2\pi i}\,\frac{\partial}{\partial p_b}+(c^k)_{a\dot b}\,B_{\dot bc}\,p_c \right)\,\langle p|\, \mathscr{C}^k\,|x\rangle= x_a\: \langle p|\,\mathscr{C}^k\,|x\rangle\\
& -\left((c^k)_{\dot a,b}\,\frac{1}{2\pi i}\,\frac{\partial}{\partial p_b}+(c^k)_{\dot a\dot b}\,B_{\dot bc}\,p_c \right)\,\langle p|\, \mathscr{C}^k\,|x\rangle=\frac{1}{2\pi i}\,\langle p|\,\mathscr{C}^k\,|x\rangle\: B_{\dot a b}\,\overleftarrow{\frac{\partial}{\partial x_b}},\label{secondeee}
\end{align}

We specialize these equations to $k=-1$ usings eqn.\eqref{ide1}--\eqref{iden}. We  get
\begin{equation}
\langle p|\,\mathscr{C}^{-1}\,|x\rangle=\exp\!\Big\{2\pi i\, p_a x_a- \pi i\, B_{a\dot c}B_{\dot c b}\,p_ap_b+\pi i\, x_a x_a\Big\}
\end{equation}
Then the basic kernel is
\begin{multline}\label{basicsy}
\Big\langle p\, \Big|\,\prod_{\dot a}e_b(-\mathsf{y}_{\dot a})^{-1}\,\mathscr{C}^{-1}\, \prod_{a}e_b(\mathsf{x}_a)^{-1}\,\Big|\,x\Big\rangle=\\
=\exp\!\Big\{2\pi i\, p_a x_a- \pi i\, B_{a\dot c}B_{\dot c b}\,p_ap_b+\pi i\, x_a x_a\Big\}\:
\prod_{\dot a} e_b\big(B_{\dot ab}\,p_b\big)^{-1}\prod_a e_b\big(x_a\big)^{-1}.\phantom{aaaaaaaaa}
\end{multline}

Finally, we have to compute $\langle p|\, \mathscr{C}^{[h/2]}\,|p^\prime\rangle$. For $E_8$  one has
\begin{equation}\label{mate8}
\langle p|\,\mathscr{C}^{15}\,|p^\prime\rangle =\delta(p+p^\prime)
\end{equation}
while for $E_6$ we have
\begin{equation}\label{mate6}
\langle p|\,\mathscr{C}^6\,|p^\prime\rangle =\delta(\chi(p)+p^\prime)
\end{equation}
where $\chi$ is the $\Z_2$ automorphism of the $E_6$ alternating quiver.

For $A_{2n}$ one plugs into equations \eqref{firsteee}-\eqref{secondeee} the expressions
\begin{gather}
(c^n)_{2s-1,j}= \delta_{j,2n-2s+2},\\
(c^n)_{2s,j}=-\delta_{j,2n-2s}-\delta_{j,2n-2s+1}-\delta_{j,2n-2s+2},
\end{gather}
yielding
\begin{equation}\label{mata2n}
\langle q|\,\mathscr{C}^n\,|p\rangle=
\exp\!\Big\{i\pi\Big[\sum_{a=1}^n(q_a+q_{a+1})^2+ 2\sum_{a=1}^np_a\big(q_{n-a+1}+q_{n-a+2}\big)\Big]\Big\}. 
\end{equation}

\subsubsection{The Partition Function and its Dualities}\label{partanddualities}

Putting the various pieces together, one has
\begin{multline}
\langle-p|\, \mathbb{K}(q)\,|x\rangle= 
\langle p| \,\Psi\,| p\rangle\,  \int \prod_{k=1}^{[h/2]} dq_a^{(k)}
\, \prod_{\ell=1}^{[h/2]-1} d x_a^{(\ell)}\,\exp\!\Big\{2\pi i\sum_{\ell=1}^{[h/2]-1} x_a^{(\ell)} q^{(\ell+1)}_a\Big\}\times\\
\times\,\Big\langle p\Big|\, \mathscr{C}^{[h/2]}\,\Big| q^{(1)}\Big\rangle \prod_{r=1}^{[h/2]} 
\Big\langle q^{(r)}\, \Big|\,\prod_{\dot b}e_b(-\mathsf{y}_{\dot b})^{-1}\:\mathscr{C}^{-1}\, \prod_{a} e_b(\mathsf{y}_{a})^{-1}\,\Big|\,x^{(r)}\Big\rangle\\
= 
\langle p| \,\Psi\,| p\rangle\,  \int \prod_{k=1}^{[h/2]} dq_a^{(k)}
\, \prod_{\ell=1}^{[h/2]-1} d x_a^{(\ell)}\,\exp\!\Big\{2\pi i\sum_{\ell=1}^{[h/2]-1} x_a^{(\ell)} q^{(\ell+1)}_a\Big\}\,\Big\langle p\Big|\, \mathscr{C}^{[h/2]}\,\Big| q^{(1)}\Big\rangle\times\\
\times \prod_{r=1}^{[h/2]}\left( 
\exp\!\Big\{2\pi i\, q_a^{(r)} x_a^{(r)}+ \pi i\, B_{\dot c a}B_{\dot c b}\,q_a^{(r)}q_b^{(r)}+\pi i\, x^{(r)}_a x^{(r)}_a\Big\}\:
\prod_{\dot a} e_b\big(B_{\dot ab}\,q^{(r)}_b\big)^{-1}\prod_a e_b\big(x_a^{(r)}\big)^{-1}\right)\label{parFunct}
\end{multline}
with the convention $x_a^{([h/2])}=x_a$.

Replacing in the \textsc{rhs} of equation \eqref{parFunct}  the non--compact dilogarithms $e_b(z)^{-1}$ by  double--sine function $s_b(x)$ using the identity
\begin{equation}
e_b(x)^{-1} = e^{-\pi i x^2/2}\: s_b(-x),
\end{equation}
we can, at the face value, interpret the above expression as the partition function on the squashed $3$-sphere $S^3_b$ of a system of $h\cdot \mathrm{rank}(\mathfrak{g})/2$ chiral multiplets of unit $R$--charge coupled to a system of\footnote{\ Here $n_s$ is the number of the sources in the alternating quiver, equal to half the rank $r$ for $A_{2n}, E_6, E_8$.} $n_V\equiv (2[h/2]-1)n_s$ dynamical $U(1)$ vector multiplets whose scalar components are identified with the integration variables $q^{(k)}_a$, $x^{(\ell)}_a$.  

Although this is true at the level of the partition function, physically it is not the most convenient viewpoint since it does not correspond to a description of the theory in terms of a minimal set of independent fields to be identified with the physical degrees of freedom. To achieve such minimality, one has to integrate out (roughly) half of the vector multiplets. This produces a description of the 3d theory in terms of a convenient set of fields, making the physics much more transparent.

However, here we have two choices: we may either integrate away the
$q^{(k)}_a$'s or the $x^{(\ell)}_a$'s. Apart for `boundary' terms for $k,\ell=1,[h/2]$, the two sets of variables enter in equation \eqref{parFunct} symmetrically as we can see by the change of variables
\begin{equation}\label{fieldred}
\tilde q_{\dot a}^{(\ell)}= B^{-1}_{\dot a b}\, x_b^{([h/2]-\ell+1)},\qquad
\tilde x_{\dot a}^{(\ell)}= B_{\dot a b}\, q^{([h/2]-\ell+1)}_b, 
\end{equation}  
whose effect is, essentially, to interchange the role of the dotted and undotted nodes of the Dynkin quiver.
(Note that the linear redefinition of the $U(1)$ vector multiplets in eqn.\eqref{fieldred} will not change the integral lattice of the $U(1)^{n_V}$, since $B_{\dot a b}$ is an integral $r/2\times r/2$ matrix with  $\det B=\pm 1$).

Integrating away one set of vector fields or the complementary one we get two formulations of the theory which differ (up to some  boundary correction for special values of $k,\ell$) just by a \textit{dotted $\longleftrightarrow$ undotted} duality of the alternating Dynkin quiver which may be seen as the manifestation in the present language of the \textit{black/white} duality of the checkerboard coloring discussed in section \ref{sec:T+B}. Of course, the black/white duality applies only to the geometrical 3d models, and hence only to $A_{2n}$ case. The comparison of dotted/undotted versus black/white for $A_{2n}$ is presented in section 8.5 below.  

There is an important difference between the case of $A_{2n}$ and $E_6,E_8$ (or, more generally, between odd and even Coxeter number $h$). In the $E_6,E_8$ cases, $\langle p|\,\mathscr{C}^{[h/2]}\,|q^{(1)}\rangle$ is a delta--function, eqns.\eqref{mate8}\eqref{mate6}, and the integration in $q^{(1)}_a$ has the effect of setting in the integrand 
\begin{equation} E_6,E_8\colon \qquad q^{(1)}_a=-\chi(p)_a,\end{equation}
so that the $q^{(1)}_a$'s become real masses associated to global symmetries. Then, for $h$ even,  the duality map
\eqref{fieldred} extends also to the boundary values of the indices, mapping the twisted masses $x_a^{[h/2]}$ into the twisted masses $q^{(1)}_a$ and otherwise mapping dynamical vectors into dynamical vectors. So the only effect of the above dotted/undotted duality is to interchange the two kinds of quiver nodes.

In the $A_{2n}$ case, it is convenient to rewrite equation \eqref{mata2n} in the form
\begin{equation}
A_{2n}\colon\qquad\langle p|\,\Psi\,|p\rangle\:\langle p|\,\mathscr{C}^{n}\,|q^{(1)}\rangle= e^{\pi i\sum_a (x_a^{(0)})^2}\, e^{2\pi i \sum_a x_a^{(0)}q_a^{(1)}}\,\prod_a e_b(x_a^{0})^{-1}
\end{equation}
where we use the convention
\begin{equation}
A_{2n}\colon\qquad x_a^{(0)}= \chi(p)_a = p_{n-a+1}+p_{n-a+2}= -B_{\dot b c}\,p_c\Big|_{\dot b=n-a+1},
\end{equation}
and the net effect of the two extra factors present for $A_{2n}$ is to extend, in that case, the upper index $\ell$ of
$x_a^{(\ell)}$ down to zero, except that there is no integration over $x_a^{(0)}$, which is the real mass parameter associated to a flavor charge. 

The integral in $dq^{(k)}_a$ or, respectively in $dx^{(\ell)}_a$, is computed
using the identity
\begin{equation}\label{integralidentity}
\int dy \exp\!\Big[\pi i y^2+2\pi i y x\Big]\: e_b(y)^{-1}= C\, e^{-\pi ix^2}\: e_b\!\left(i\frac{Q}{2}-x\right)
\end{equation}
where $C$ is some constant.

Then, up to an irrelevant overall constant, $\langle -p|\, \mathbb{K}(q)\,|x\rangle$ for $E_6,E_8$ (or, more generally, for all $\mathfrak{g}$ with $h$ even, as long as we specialize the central parameters $c_\alpha$ to zero) is
\begin{align}
&\bullet\ \text{for }E_{6}, E_8\colon &&\left[\:
\begin{aligned}
\int  &\prod_{k=2}^{h/2} d q_a^{(k)}\,
\left(\prod_{\ell=1}^{h/2}\prod_{\dot b} 
e^{\pi i (B_{\dot bc}\,q_c^{(\ell)})^2/2}\: s_b(-B_{\dot b c}\,q_c^{(\ell)})\right)\times\\
&\ \times
\left(\prod_{r=1}^{h/2-1}\prod_{a} e^{-\pi i(q_a^{(r)}+q_a^{(r+1)}+iQ/2)^2/2}\: s_b\!\left(i\frac{Q}{2}-q_a^{(r)}-q_a^{(r+1)}\right)\right)\times\\
&\ \times \prod_a e^{2\pi i\, x_a q_a^{(h/2)}}\, e^{\pi i x_a^2/2}\: s_b(-x_a)\\
&\text{with the convention: }\ q_a^{(1)}=-\chi(p)_a\end{aligned} \right.\label{e6e6}
\end{align}
with the understanding that the dual expression is obtained by inverting dotted with undotted.

Instead for $A_{2n}$ we get
\begin{align}\label{a2npart}
&\bullet\ \text{for }A_{2n}\colon &&\left[\:\begin{aligned}
\int  &\prod_{k=1}^{n-1} d y_a^{(k)}\,
\left(\prod_{\ell=0}^n\prod_{a=1}^n 
e^{\pi i ((\mathsf{M}\cdot y^{(\ell)})_a)^2/2}\: s_b\big((\mathsf{M}\cdot y^{(\ell)})_a\big)\right)\times\\
&\ \times
\left(\prod_{r=1}^n\prod_{a=1}^n e^{-\pi i(y_a^{(r-1)}+y_a^{(r)}+iQ/2)^2/2}\: s_b\!\left(i\frac{Q}{2}-y^{(r-1)}_a-y_a^{(r)}\right)\right)\\
&\text{where }\mathsf{M}_{ab}=\delta_{a,b}+\delta_{a,b+1}\\
&\text{with the conventions: } y_a^{(n)}=(\mathsf{M}^{-1}\cdot x)_a, \ y_a^{(0)}=p_{n-a+1}\end{aligned} \right.
\end{align}
where we  redefined the sign of $y_a^{(k)}$.
\smallskip

The shifts by $iQ/2$ in the exponential factors, means that we have to redefine the integration variables $q^{(k)}_a$ (resp. $y^{(k)}_a$ to absorb the linear term in the exponent. This has the effect of changing the definition of the $R$--charge by mixing it with the gauge and flavor charges. However, for simplicity, we shall classify the states according to the `naive' $R$--charge, which is still, of course, a good quantum number.

With respect to this $R$--charge, we adopt the sign convention that the gauge/flavor charges of chiral multiplets with $R=1$ are positive, whereas those of the chiral multiplets with $R=0$ are negative. That this convention is consistent may be seen from equations \eqref{e6e6}-\eqref{a2npart}.

\subsection{Physical Predictions for $E_6$, $E_8$ ($h$ even)}

We start extracting the physical predictions of the above 3d partition function from the simpler case of $h$ even and, in particular, for $E_6$ and $E_8$. These models do not have a geometrical interpretation in terms of two M5--branes wrapping some three-manifold, and hence to them braid analysis does not apply.
Nevertheless, the R-flow construction yields a theory, whose partition function
we have computed.

The
maximal chamber representation of the $E_6$ (resp.\! $E_8$) movie partition function corresponds to $36$ (resp.\! $120$) chiral multiplets, denoted as follows
\begin{equation}
\begin{array}{c| c|c}\hline
 & E_6 & E_8\\\hline
X_{\ell,a} & a=1,2,3;\ \ell=1,\dots,5 & a=1,2,3,4;\ \ell=1,\dots 14\\
Y_{m,\dot a}& \dot a=1,2,3;\ m=0,\dots,5 & \dot a=1,2,3,4;\ m=0,\dots,14 \\
Z_a & a=1,2,3 & a=1,2,3,4\\\hline
\end{array}\label{chiralEE}
\end{equation} 
coupled to
$15$ (resp.\! $56$) $U(1)$ vector multiplets with scalar components
$q^{(k+1)}_a$, $a=1,2,3$, $k=1,\dots,5$ (resp.\! $k=1,\dots, 14$), and also charged under a $U(1)^6$ (resp.\! $U(1)^6$) flavor symmetry associated to the real masses $x_a$, $q^{(1)}_a$, $a=1,2,3$. We write the charges corresponding to $q^{(k+1)}_a$, $x_a$ as, respectively, $e^{(k)}_a$, $f_a$;
$e^{(0)}_a$, $f_a$ being flavor charges.
With these conventions the quantum numbers of the chiral multiplets are as follows
\begin{center}
\begin{tabular}{|c|c|c|c|}\hline
{chiral mult.} &  $e^{(k)}_a$ & $f_a$ & $R$--charge\\\hline
$X_{b,\ell}$ & $-\delta_{ab}(\delta_{k,\ell-1}+\delta_{k,\ell})$ & $0$  & $0$\\\hline
$Y_{\dot b,m}$& $\delta_{k,m} B_{a\dot b}$ &$0$  & $1$\\\hline
$Z_b$ & $0$ & $-\delta_{ab}$ & $1$\\\hline
\end{tabular}
\end{center}

It is important to count the number of chiral multiplets positively and negatively charged with respect to the various gauge $U(1)$ charges. Since, in our convention, the $Y$'s are positively charged and the $X$'s negatively, the number of chiral fields with charge $+1$ with respect to $U(1)$ charge $e^{(k)}_a$, $k\geq 1$, is
\begin{equation}\label{charpus}
\#\{\text{chirals with }e^{(k)}_a\ \text{charge }+1\}= \sum_{\dot b} B_{a\dot b}= \text{(valency node }a) 
\end{equation}
while 
\begin{equation}\label{charneg}
\#\{\text{chirals with }e^{(k)}_a\ \text{charge }-1\}= \begin{cases}
2 & k\neq h/2-1\\
1 & k=h/2-1.
\end{cases}
\end{equation}
Notice that in the dotted-undotted dual picture, the number of positively charged chirals is counted by the valency of the dotted nodes $\dot a$.
\medskip

For $E_6$ (resp.\! $E_8$) there are $15$ (resp.\! $56$) gauge and flavor invariant monomials in these chiral fields which we organize in a $3\times 5$ (resp.\! $4\times 14$) matrix $W_{\dot a, \ell}$
\begin{equation}
W_{\ell,\dot a}= \left(\prod_{d} X_{\ell,a}^{\,B_{d\dot a}}\right)\left(\prod_{m} Y_{m,\dot a}^{\,(\delta_{m+1,\ell}+\delta_{m,\ell})} \right)
\end{equation}
all of which have precisely two $Y$'s, and hence have $R$--charge $2$ and are legitimate superpotential terms. 
The number of chiral fields appearing in each monomial is
\begin{equation}
\#(W_{\dot a,\ell})= 2+\sum_{d=1}^3B_{d\dot a} = 2+\text{(valency of  }\dot a\ \text{node)}.
\end{equation}
For instance, for $E_6$, making reference to the quiver in  Figure \ref{altquivers}, we have
\begin{equation}
\#(W_{\dot a,\ell})=\begin{cases} 4 & \text{for }\dot a=1,2\\
3 & \text{for } \dot a=3
\end{cases}
\end{equation}
and thus the $E_6$ ordinary superpotential contains $5$ cubic and $10$ quartic terms.
However, in the dual picture discussed in section \ref{partanddualities}, the roles of dotted and undotted node interchange, and the $15$ invariant monomials of the $E_6$ theory
contain a number of chiral fields equal to
\begin{equation}\label{quintic}
\#(W_{\dot a,\ell})=2+\text{(valency of  }a\ \text{node)}=\begin{cases} 3 & \text{for } a=1,3\\
5 & \text{for } a=2,
\end{cases}
\end{equation}
and we have $10$ cubic and $5$ quintic terms.

In conclusion, the $E_6$ superpotential without monopole operators contains $5$ cubic and $10$ quartic terms.
From the dotted/undotted duality we expect other contributions to the superpotential containing the monopole operators similar to the ones we find for the $A_{2n}$ case. The two kinds of superpotential terms are expected to be interchanged by the duality, and hence in correspondence.
In particular we expect there to be in addition 15 superpotential terms involving the monopole operators,
in 1-1 correspondence with the 15 gauge factors.

\medskip

In equation \eqref{chiralEE} we wrote the relevant chiral fields as two matrices $X$ and $Y$ of sizes, respectively, $(h/2-1)\times (r/2)$ and $(h/2)\times (r/2)$, where $r$ is the rank of the corresponding Lie algebra. The relevant charges $e$ where also written as a
$(h/2)\times (r/2)$ matrix. In this compact notation, the charge of the chiral fields, as tabled in the previous section, become   
\begin{align}
&e(X)= -\Big((E_1\:|\:\mathsf{M})\otimes \boldsymbol{1}_{r/2}\Big)\cdot e\\
&e(Y)= (\boldsymbol{1}_{h/2}\otimes B^t)\cdot e,
\end{align}
where $\mathsf{M}$ is the $(h/2-1)\times (h/2-1)$ matrix defined in eqn.\eqref{a2npart}, $E_1$ is the column vector $(1,0,\dots, 0)^t$,
and $\boldsymbol{1}_n$ stands for the unit matrix of size $n$.

In this compact notation
the CS level--matrix  $\hat k$ is (we omit the CS terms involving flavors) 
\begin{equation}
\hat k= \boldsymbol{1}_{h/2}\otimes BB^t- \left(\begin{array}{c|c}
1 & E_1^t\,\mathsf{M}\\\hline
\mathsf{M}^t\, E_1 & \mathsf{M}^t\,\mathsf{M}\end{array}\right)\otimes \boldsymbol{1}_{r/2}
\end{equation}

\subsection{Physical Predictions for $A_{2n}$}

For $A_{2n},$ equation \eqref{a2npart} represents the partition function on the squashed three-sphere $S_b^3$ of an $\mathcal{N}=2$ system of $2n^2+n$ chiral multiplets coupled to $n(n-1)$ $U(1)$ gauge vector multiplets, whose scalar components  --- in the dotted--undotted duality frame we are using --- are denoted as
\begin{equation}\label{listU(1)}
y_{\dot a}^{(\ell)},\quad \dot a=1,2,\dots, n\qquad \ell=1,2\dots, n-1.
\end{equation}
The chiral fields also carry a $U(1)^{2n}$ flavor symmetry whose associated twisted masses are denoted as $y_{\dot a}^{(0)}$ and $y_{\dot a}^{(n)}$, $\dot a=1,2,\dots n$. The corresponding gauge and flavor charges will be denoted as $e^{(k)}_{\dot a}$, and we organize them in an $(n+1)\times n$ matrix, where the first and last rows correspond to flavor charges. 

\begin{figure}[here!]
  \centering
  \subfloat{\label{fig:bigclosedbraid2}\includegraphics[height=.7\textwidth]{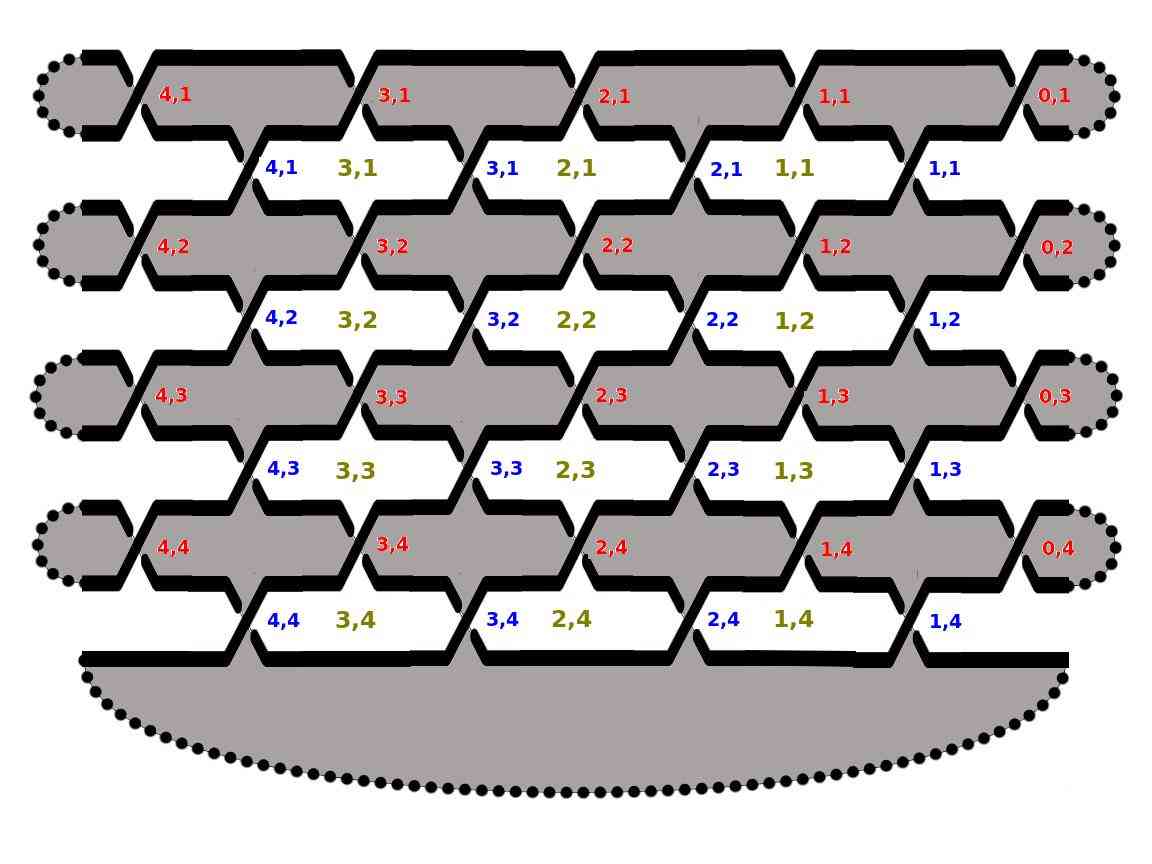}} 
  %\hspace{3in}      
  \caption{Comparison with Figure \ref{fig:bigclosed}. The crossing corresponding to the chiral field $Y_{k,a}$ (resp.\! $X_{\ell,\dot a}$) are indicated in red (resp.\! blue). The white region associated to the $(k,a)$ vector multiplet is indicated in green.}
  \label{fig:bigclosed2}
\end{figure}

Likewise, we organize the $2n^2+n$ chiral multiplets into two matrices $Y_{k,a}$ and $X_{\ell,\dot a}$ of sizes $(n+1) \times n$ and $n\times n$, respectively.  For simplicity, we shall omit the dots over the indices, identifying the two kinds of indices. In this convention, the quantum numbers of the chiral multiplets are as in the following table while their correspondence with the checkerboards of section \ref{sec:T+B} is shown in Figure \ref{fig:bigclosed2}

\begin{center}
\begin{tabular}{|c|c|c|c|}\hline
chiral field $\phantom{\Big|}$& index range & charge vector & $R$--charge\\\hline
$Y_{k,a}\phantom{\Big|}$ & $0\leq k\leq n$,\  $1\leq a\leq n$ & $e^{(k)}_a+e^{(k)}_{a-1}$ & $1$\\\hline
$X_{\ell,\dot a}\phantom{\Big|}$ & $1\leq \ell\leq n$,\  $1\leq \dot a\leq n$ & $-e^{(k)}_{\dot a}-e^{(k-1)}_{a}$ & $0$\\\hline
\end{tabular}\vskip 8pt

\textsc{Table:} quantum numbers of chiral multiplets\\ in the maximal chamber for the alternating $A_{2n}$ \\
\end{center}

As it is evident from the Figure \ref{fig:bigclosed2}, at most \emph{four} chiral multiplets carry a given charge $e^{(k)}_a$, namely
\begin{align}\label{chargesmull}
&Y_{k,a}\ \textbf{(+1)}, &&Y_{k,a+1}\ \textbf{(+1)},
&&X_{k,\dot a}\ \textbf{(-1)}, && X_{k+1,\dot a}\ \textbf{(-1)}
\end{align} 
(the number in parenthesis being the value of the charge for the given chiral field which, in our conventions, are positive for $R=1$ chiral fields and negative for the $R=0$ ones). In fact, the only exceptions to the above `four rule' happens when $k=n$ or $a=n$. The $e^{(n)}_a$ charges are flavor, and hence the only gauge charges carried by only \emph{three} chirals are the ones in the last column of the charge matrix, namely the $e^{(k)}_n$, $1\leq k\leq n-1$. With respect to $e^{(k)}_n$, two chirals have charge $-1$ and one has charge $+1$, and hence the charge asymmetry, $\sum_\text{chirals} e^{(k)}_n$, is $-1$ for these boundary $U(1)$'s.

\bigskip

\underline{\textit{Black/white duality}}\smallskip

From  Figure \ref{fig:bigclosed2} we see that, with the exception of the regions at the boundary of the checkerboard, interchanging black and white regions is equivalent to interchange red and blue crossings (up to a relabeling of the second index $a\leftrightarrow n+1-a$) which is precisely the dotted/undotted duality. However, the black/white and dotted/undotted dualities are not exactly the same, as we see by noticing that in the $A_1$ theory, which has only one undotted node, the dotted/undotted duality is necessarily  trivial.
In fact, by comparing with the $A_2$ case discussed in section \ref{sec:bkachwithe}, we see that the black/white duality is the composition of the dotted/undotted duality with the $A_1$ duality performed on the $2n$ boundary chiral fields. In this way we get
$n^2+2n$ vectors which is the right number in the black/white dual of Figure \ref{fig:bigclosed2}, as one easily verifies.

\bigskip

\underline{\textit{Superpotential terms}}\smallskip

There are two types of contributions to the superpotential $W$, the ones from the white regions which involve the monopole operators,
and those from the black regions which are polynomials in elementary fields.  To determine the terms which do not involve the monopole operator, we need to identify
the subring $\mathcal{R}^G$ of the chiral ring $\mathcal{R}$ generated by all the monomials in $Y_{k,a}$, $X_{\ell,\dot a}$
which are flavor and gauge invariant. The number of algebraically independent such monomials is  
\begin{equation}
 \#\{\text{chiral multiplets}\}-\#\{\text{charges}\}= n^2.
\end{equation}
We explicitly describe $\mathcal{R}^G$ by giving $n^2$ invariant monomials which generate $\mathcal{R}^G$. We organize them as a $n\times n$ matrix of monomials, $W_{ij}$, where $i,j=1,2,\dots,n$. Explicitly,
\begin{equation}\label{Wij}
W_{ji}= \left(\prod_{\dot a=1}^n X_{j,\dot a}^{\,(\delta_{i,\dot a}+\delta_{i,\dot a+1})}\right)\left(\prod_{k=0}^n Y_{k,i}^{\,(\delta_{j,k}+\delta_{j,k+1})}\right).
\end{equation}
Inspection of Figure \ref{fig:bigclosed2} shows that these invariant monomials are in one--to--one correspondence with the the compact black regions, the product in the \textsc{rhs} being on the chiral fields associated to the crossings on the boundary of the given black region. In fact, equation \eqref{Wij} coincides with the prediction \eqref{superblack} of section \ref{sec:T+B}.
 
Aa a further check, we note that all $n^2$ invariant monomials in \eqref{Wij} have  $R$--charge $2$, and hence they appear in the superpotential $W$. In fact, since the $X$'s have $R$--charge zero and the $Y$'s $R$--charge $1$, the $R$--charge of a monomial is just the number of $Y$'s it contains
\begin{equation}
q(W_{ji})= \sum_{k=0}^n (\delta_{j,k}+\delta_{j,k+1})=2.
\end{equation}
Moreover, all the $W_{ij}$ are quartic in the chiral field, except for the $n$ $W_{1j}$'s which are cubic. Indeed, the total number of fields in an invariant monomial is
\begin{equation}
\#(W_{ji})= 2+\sum_{a=1}^n (\delta_{i,a}+\delta_{i,a+1})=\begin{cases} 3 & \text{if }i=1\\
4 & \text{otherwise,}
\end{cases}
\end{equation}
in agreement with Figure \ref{fig:bigclosed2}.
Of course, this is just a manifestation of the general rule established above that the number of chiral fields in an invariant monomial is $2$ plus the valency of the corresponding node: in the duality--frame we are using the nodes labelling the $Y$'s are the undotted ones, and the only undotted node in the first quiver of Figure \ref{altquivers} having valency $1$ is precisely the one labeled $1$.

Thus, we recover the fact the part of the superpotential not involing the monopole operators is $W=\sum_{ij}\lambda_{ij}\,W_{ij}$ is the sum of $n$ cubic and $n(n-1)$ quartic terms as expected from the braid analysis of section \ref{sec:T+B}.

%%%%%%%%%%%%%%%%
From the black/white duality we expect a contribution to the superpotential from each $(k,a)$ white region containing the monopole operator, see equation \eqref{superwhite}. In the present notation, \eqref{superwhite} corresponds to the operators
\begin{equation}\label{monopoleW}\begin{split}
&\widehat{W}_{\ell,\dot a}= \mathcal{M}_{\ell,a}^{\epsilon}\, 
\left(\prod_{k=1}^n Y_{k,a}^{\,(\delta_{k,\ell}+\delta_{k,\ell+1})}\right)\left(\prod_{b=1}^n X_{\ell,b}^{\,(\delta_{b,a}+\delta_{b,a+1})}\right)\\
\dot a= 1,2,\dots,n\quad \ell=1,\dots, n-1,
\end{split}\end{equation}
 where $\epsilon=\pm 1$. A part for the presence of the monopole $\mathcal{M}^\epsilon_{\ell,a}$, these operators are related to the $W_{ij}$ by the dotted/undotted duality, which is related to the black/white one as discussed above.

Consistency requires these operators to be gauge and flavor invariant for a suitable choice of the sign $\epsilon$.\footnote{
In fact, we are going to show that they are invariant if $\epsilon$ is chosen to be $+1$.  In comparing to the general analysis of section 5.4, this means that exactly half of the gauge fields should have their signs flipped $A\rightarrow -A$. }

In order to compute the charges of the monopole operators $\mathcal{M}_{\ell,a}$ we have to determine the Chern--Simons level--matrix $\hat{k}$ which may be read from the Gaussian factor in equation \eqref{a2npart}. Explicitly,
\begin{equation}
 \begin{aligned}
& y^t\cdot \hat{k} \cdot y = \\
 &=2\sum_{s,k=1}^n \Big(y^{(k)}_s\, y^{(k-1)}_s-y^{(k)}_s\,y^{(k)}_{s-1}\Big)+ && \left|
\:\begin{matrix}\text{\begin{small}$\pm 1$ for adjacent vectors in the charge matrix\end{small}}\\
\text{\begin{small} sign $+$ if adjacent vertically, $-$ if horizontally\end{small}}\end{matrix}\right.\\
&+\sum_{k=1}^n (y^{(k)}_n)^2- && \Big|\:\text{\begin{small}
diagonal terms for vectors coupled to 3 chirals\end{small}}\\
&- \sum_{s=1}^n (y_s^{(n)})^2-\frac{1}{2}\sum_{s=1}^n(y^{(0)}_s)^2 +\frac{1}{2} (y_n^{(0)})^2 && \Big|\:\text{\begin{small}
purely flavor terms\end{small}}
 \end{aligned}
\end{equation}
Note that the diagonal terms are equal to the sum of the charges of the chiral multiplets, and are non--zero only for the last column vectors. Note that the diagonal entries corresponding to the `corner' charges, $y^{(0)}_n$ and $y^{(n)}_n$ vanish.

Now it is elementary to check that the all operators \eqref{monopoleW} are gauge and flavor invariant, and that they have physically consistent $R$--charges.

\subsection{Intermediate $A_4$ With Mixed Mutations}

All examples discussed up to now correspond to 4d $ADE$ Argyres--Douglas theories which involve mutations on sink/source nodes only.
 Although these are the situations with simpler rules
for constructing their braids, as noted in section 5, our methods are not limited to them. 
Therefore one may wish to consider examples of R-flow which include mutations on nodes which are not sink/source type.  For such cases we expect
the braids to be `non-planar' in the sense that some crossings
do not correspond to particles.

The simplest such example with an even number of nodes would be the five hypermultiplets chamber of the $A_4$ theory described by the linear quiver 
$\vec A_4$
\begin{equation}\label{linearquiver}
\vec A_4\colon\qquad \begin{gathered}
\xymatrix{1 \ar[r] & 2\ar[r] & 3\ar[r] &4}
\end{gathered}
\end{equation}   
This intermediate chamber is produced, in the sense of section \ref{sec:Quiver}, by the sequence of mutations at nodes
\begin{equation}
2,\ 4,\ 3,\ 1,\ 2.
\end{equation}
For example, the first mutation on $2$ does not involve a source/sink node.
In the cluster--mutation notation of  section \ref{zamodidentities}, the corresponding 3d partition function will take the form
\begin{equation}
Z_{\vec A_4}= \langle \widetilde{\text{out}}|\, \cq_2\cq_1\cq_3\cq_4\cq_2\, |\text{in}\rangle,
\end{equation}
where the tilde means that the out state is defined in a different (mutated) Hilbert space where the canonical commutation relations are specified by mutated quiver
\begin{equation}\label{exmutation}
\mu_2\mu_1\mu_3\mu_4\mu_2(\vec A_4)\colon\qquad \begin{gathered}
\xymatrix{1 \ar[r] & 3\ar[r] & 2\ar[r] &4}
\end{gathered}
\end{equation}  
In other words, we must keep track of the permutation $\chi$ which has acted on the nodes.

In particular, we are interested in the partition function with boundary condition corresponding to the following $ \langle \widetilde{\text{out}}|$, $|\text{in}\rangle$ states,
\begin{equation}\label{linemono}\begin{split}
Z_{\vec A_4}&= \langle \widetilde{y}_1,\widetilde{y}_2|\, \cq_2\cq_1\cq_3\cq_4\cq_2\, |y_1, y_3\rangle\\
&= \langle y_1^\prime, y^\prime_3|\,
 e_b(\mathsf{y}_1)^{-1}\, e_b(\mathsf{y}_3)^{-1}\,
e_b(\mathsf{y}_2+\mathsf{y}_3)^{-1}\, e_b(\mathsf{y}_2)^{-1}\, e_b(\mathsf{y}_4)^{-1}\, |y_1,y_3\rangle,
\end{split}\end{equation}
where we made the identification 
\begin{equation}\label{mapmap}
y_1^\prime=-\widetilde{y}_1,\qquad y_3^\prime=-\widetilde{y}_2,
\end{equation}
and in the last line of equation \eqref{linemono} both in/out states are in the same Hilbert space with commutation relations specified by the initial quiver \eqref{linearquiver}, the two Hilbert spaces being related by the map \eqref{mapmap} which is the permutation $\chi$ induced by the mutation sequence \eqref{exmutation}. From the second line, it is also evident that the present five hypermultiplet chamber is obtained from the minimal four hypers one by a simple Pachner move on the two central tetrahedra.

We adopt the following Schoerendiger representation
\begin{align}
&\mathsf{y}_1=y_1 &&\mathsf{y}_3=y_3\\
&\mathsf{y}_2= \frac{1}{2\pi i} \left(\frac{\partial}{\partial y_1}-\frac{\partial}{\partial y_3}\right) && \mathsf{y}_4= \frac{1}{2\pi i} \,\frac{\partial}{\partial y_3},
\end{align}
so that
\begin{equation}
\langle y_1, y_3| y_2, y_4\rangle= \exp\!\Big\{2\pi i \Big(y_1y_2+(y_1+y_3)y_4\Big)\Big\}.
\end{equation}

We have
\begin{multline}
Z_{\vec A_4} 
=e_b({y}_1^\prime)^{-1}\, e_b({y}^\prime_3)^{-1}\:
\int du\, dy_2\, dy_4\, \phi_-(u)\, e_b(y_2)^{-1}\, e_b(y_4)^{-1}\,\times\\
\times\,
\langle y^\prime_1, y^\prime_3|\, e^{-2\pi i u(\mathsf{y}_2+\mathsf{y}_3)}\, |y_2,y_4\rangle\: \langle y_2, y_4|y_1, y_3\rangle =\\
=e_b({y}_1^\prime)^{-1}\, e_b({y}^\prime_3)^{-1}\:
\int du\, dy_2\, dy_4\, e^{-i\pi u^2} \phi_-(u)\, e_b(y_2)^{-1}\, e_b(y_4)^{-1}\,e^{-2\pi i u y_3^\prime}\, e^{-2\pi i u y_2}\times\\
\times\,
\exp\!\Big\{2\pi i \Big((y_1^\prime-y_1) y_2+(y^\prime_1+y_3^\prime-y_1-y_3)y_4\Big)\Big\}=\\
=C\,e_b({y}_1^\prime)^{-1}\, e_b({y}^\prime_3)^{-1}\:
\int dy_2\, dy_4\, e_b(y_2)^{-1}\, e_b(y_4)^{-1}\, e_b(-iQ/2-y_3^\prime-y_2)^{-1}\, e^{\pi i (y_3^\prime+y_2+iQ/2)^2}\times\\
\times\,
\exp\!\Big\{2\pi i \Big((y_1^\prime-y_1) y_2+(y^\prime_1+y_3^\prime-y_1-y_3)y_4\Big)\Big\}.
\end{multline}
It is convenient to integrate out $y_4$ with the result (up to an overall constant)
\begin{multline}
e_b({y}_1^\prime)^{-1}\, e_b({y}^\prime_3)^{-1}\, \phi_-(y_1^\prime+y_3^\prime-y_1-y_3)\:\times\\
\times\:\int dy_2\, e_b(y_2)^{-1}\,\, e_b(-iQ/2-y_3^\prime-y_2)^{-1}\, e^{\pi i (y_3^\prime+y_2+iQ/2)^2}
\exp\!\big\{2\pi i (y_1^\prime-y_1) y_2\big\}
\end{multline}
Using the identities
\begin{equation}
e_b(x)^{\pm 1}= e^{\pm \pi i x^2/2}\, s_b(\pm x), \qquad \phi_-(x)= C\, e^{i\pi (x-iQ/2)^2/2}\: s_b(x+iQ/2),
\end{equation}
The partition function may be written (always up to an overall constant)
\begin{multline}
e^{\pi i[(y_1^\prime+y_3^\prime-y_1-y_3-iQ/2)^2-(y_1^\prime)^2-(y_3^\prime)^2]/2}\: s_b(-y^\prime_1)\, s_b(-y^\prime_3)\, s_b(y_1^\prime+y_3^\prime-y_1-y_3+iQ/2)\, \times\\
\times\, \int dy_2\, e^{2\pi i(y^\prime_1-y_1)y_2}\,e^{\pi i[(y_3^\prime+y_2+iQ/2)^2- y_2^2]/2}\: s_b(-y_2)\, s_b(iQ/2+y^\prime_3+y_2)
\end{multline}
which is interpreted as a $\mathcal{N}=2$ system of five chiral fields coupled to a dynamical $U(1)$ (of CS level zero) and charged under $U(1)^3$ flavor symmetry $f_a$.
The quantum numbers are as in the table\footnote{\ We redefine the overall signs to adhere to our general sign covention. } \vglue 12pt

\begin{center}
\begin{tabular}{|c|c|c|c|c|c|}\hline
& $X_1$ & $X_2$ & $Y_1$ & $X_3$ & $Y_2$ \\\hline
$e$ & 0 & $0$ & $0$ & $1$ & $-1$  \\\hline
$f_1$ & $1$ & $0$ & $-1$ & $0$ & $0$ \\\hline
$f_2$ & $0$ & $1$ & $-1$ & $0$ & $-1$ \\\hline
$f_3$ & $0$ & $0$ & $1$ & $0$ &$0$ \\\hline
$R$ & $1$ & $1$ & $0$ & $1$ & $0$ \\\hline
\end{tabular}
\end{center}\vglue 12pt

The only invariant monomial is \begin{equation}X_2X_3Y_2\end{equation} which has $R$--charge $2$ and so it is a superpotential term.
Furthermore, this theory does not include a monopole superpotential.

\begin{figure}[htb!]
  \centering
  \subfloat[Non-planar Braid for $A_{4}$]{\label{fig:nonplanar0}\includegraphics[width=.48\textwidth]{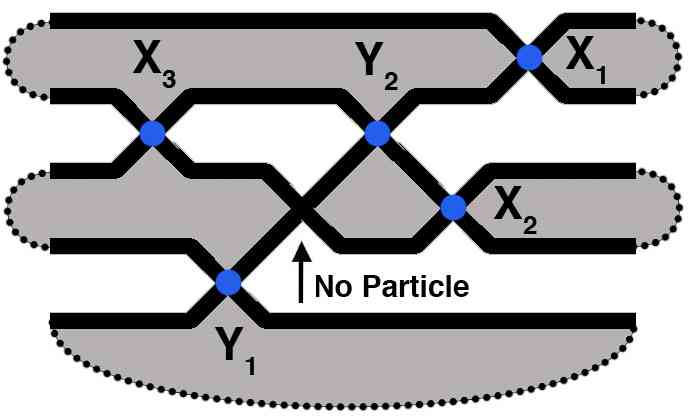}}     
  \hspace{.2in}         
  \subfloat[Cubic Superpotential]{\label{fig:nonplanar1}\includegraphics[width=0.48\textwidth]{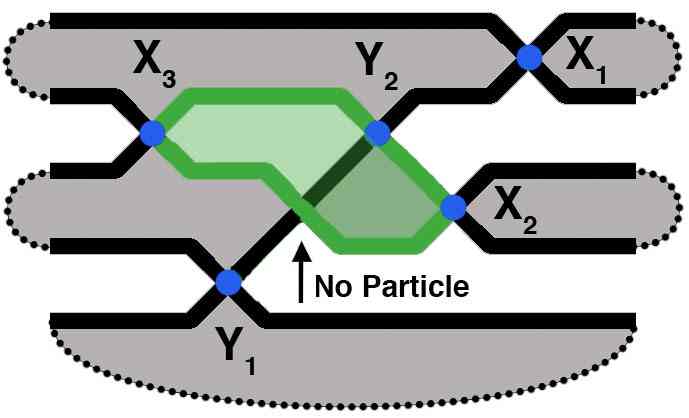}}
  \caption{The braid for the intermediate chamber of $A_{4}$.  The braid is non-planar in that it has crossings which do not correspond to particles. In (b), we see the green shaded region which gives rise to the cubic superpotential.}
  \label{fig:nonplanar}
\end{figure}

Using the braid description discussed in section 5, we can reproduce
much of this structure, even without specifying the overcross/undercross
structure of the braid. In particular, we obtain the braid represented in
Figure \ref{fig:nonplanar0}.  The five particles each correspond to strands coming
together.  However, now we also have a crossing which does not lead to a physical
particle.  We can immediately see that we have only a $U(1)$ gauge theory
(given by the bounded white region) with two charged fields $X_3,Y_2$.
Moreover there is only one finite triangle shown as the shaded green region in Figure \ref{fig:nonplanar1}  whose boundary is on the braids
and thus gives rise to the superpotential term $X_2X_3Y_2$.  The rule for
these more general mutations will be presented elsewhere.

\section*{Acknowledgements}

We thank M. Aganagic, M. Alim, R. Dijkgraaf, T. Dimofte, S. Espahbodi, S. Gukov, C. Hodgson, D. Jafferis, A. Kapustin, J. Morgan, A. Rastogi, D. Sullivan, and M. Yamazaki for helpful discussions. We would also like to thank the 2010 and 2011 Simons workshop in Mathematics and Physics and the Simons
Center for Geometry and Physics for hospitality during the inception of this
work. CV would also like to acknowledge the MIT physics department for hospitality.  The work of CC and CV
is supported in part by NSF grant PHY-0244821.

\end{document}